\documentclass[a4paper,fleqn,usenatbib]{mnras}

\usepackage{txfonts}

\usepackage[T1]{fontenc}
\usepackage{ae,aecompl}

\usepackage{psfig}   
\usepackage{graphicx}
\usepackage{amssymb}
\usepackage{subfigure}

\title[]{The evolution of protoplanetary disc radii and disc masses in star-forming regions}

\author[]{Bridget Marchington and Richard  J. Parker\thanks{E-mail: R.Parker@sheffield.ac.uk}\thanks{Royal Society Dorothy Hodgkin Fellow} \vspace*{0.1cm}\\
Department of Physics and Astronomy, The University of Sheffield, Hicks Building, Hounsfield Road, Sheffield, S3 7RH, UK}

\begin{document}

\date{}
                             
\pagerange{\pageref{firstpage}--\pageref{lastpage}} \pubyear{2020}

\maketitle

\label{firstpage}

\begin{abstract}
  Protoplanetary discs are crucial to understanding how planets form and evolve, but these objects are subject to the vagaries of the birth environments of their host stars. In particular, photoionising radiation from massive stars has been shown to be an effective agent in disrupting protoplanetary discs. External photoevaporation leads to the inward evolution of the radii of discs, whereas the internal viscous evolution of the disc causes the radii to evolve outwards. We couple $N$-body simulations of star-forming regions with a post-processing analysis of disc evolution to determine how the radius and mass distributions of protoplanetary discs evolve in young star-forming regions. To be consistent with observations, we find that the initial disc radii must be of order 100\,au, even though these discs are readily destroyed by photoevaporation from massive stars. Furthermore, the observed disc radii distribution in the Orion Nebula Cluster is more consistent with moderate initial stellar densities (100\,M$_\odot$\,pc$^{-3}$), in tension with dynamical models that posit much higher inital densities for the ONC. Furthermore, we cannot reproduce the observed disc radius distribution in the Lupus star-forming region if its discs are subject to external photoevaporation. A more detailed comparison is not possible due to the well-documented uncertainties in determining the ages of pre-main sequence (disc-hosting) stars. 

\end{abstract}

\begin{keywords}   
methods: numerical – protoplanetary discs – photodissociation region (PDR) – open clusters and associations: general.
\end{keywords}

\section{Introduction}

Most stars form in groups with between $10 - 10^4$ stellar siblings \citep{Lada03,Gutermuth09,Bressert10}. Some of these groups coalesce to form long-lived bound star clusters \citep{Kruijssen12b}, or unbound associations \citep{Wright20,Wright22}. Whilst the evolution and fate of star-forming regions is likely to be more nuanced than this dichotomy, at early ages ($<10$\,Myr) the stellar density in all of these regions significantly exceeds the stellar density in the Milky Way's disc by several orders of magnitude \citep{Korchagin03}.

Planets are observed to be forming \citep{Haisch01,Richert18}, and in some cases are thought to have already formed \citep{Alves20} when their host stars are still embedded in these stellar groups, implying that the birth environment of stars has a direct influence on the planet formation process \citep[e.g.][]{Adams10,Parker20}.  

The most dense star-forming regions ($\tilde{\rho} > 10^4$\,M$_\odot$\,pc$^{-3}$) can cause the truncation of protoplanetary discs due to interactions with passing stars \citep{Vincke18,Winter18a}, and moderately dense star-forming regions ($\tilde{\rho} > 10^2$\,M$_\odot$\,pc$^{-3}$) can lead to planetary orbits being altered or disrupted \citep{Bonnell01b,Parker12a,Cai17a}. Intriguingly, even star-forming regions with modest-to-low stellar densities  ($\tilde{\rho} \geq 10$\,M$_\odot$\,pc$^{-3}$) can affect planet formation if these regions contain stars massive enough to produce photoionising Extreme Ultraviolet (EUV) and Far Ultraviolet (FUV) radiation. EUV and FUV radiation readily evaporates the gas component of protoplanetary discs \citep[e.g.][and many others]{Johnstone98,Storzer99,Hollenbach00,Scally01,Adams04,Fatuzzo08,Nicholson19a,Winter18b,Parker21a}, and to a lesser extent the dust \citep{Haworth18b}.

In addition to external photoevaporation from EUV/FUV radiation produced by massive stars, many other processes affect the evolution of protoplanetary discs, including internal photoevaporation from the host star, and both internal and external X-ray driven evaporation \citep{Picogna19,Coleman22}. Notably, even in the absence of external effects, the disc is expected to evolve due to viscous spreading \citep{LyndenBell74,Hartmann98}, leading to rapid outward evolution of its radius \citep{ConchaRamirez19}.

In previous work \citep{Parker21a}, we have shown that the stellar density is the main factor in determining how many discs are destroyed by external photoevaporation. In some of those models we also included a prescription for viscous spreading, and the net effect of this viscous evolution was to accelerate the mass lost from the disc due to photoevaporation. The reason for this is that the disc mass lost from photoevaporation, combined with spreading of the disc due to viscous evolution, acts to reduce the surface density of the disc and make the material more susceptible to further photoevaporation.

These simulations used the \texttt{FRIED} grid of models \citep{Haworth18b}, which show that the gas component of the discs is readily evaporated in radiation fields more than 10 times that of the background Interstellar Medium, but the dust component largely remains. However, work by \citet{Sellek20} showed that external FUV photoevaporation would still alter the dust radius, with this roughly following the gas radius in their simulations. 

There has also been a huge proliferation in observational data on protoplanetary discs \citep[see][for a recent review]{Miotello22}, in part due to the ALMA telescope which has provided dust continuum measurements of disc masses and radii \citep{Eisner18,Otter21}, and in some cases detections of CO to trace the gas component of the discs.

In this paper, we compare the evolution of protoplanetary disc radii and disc masses in simulations in which we calculate the effects of mass-loss due to photoevaporation, and the subsequent inward evolution of the disc radius. In the same models we also implement viscous evolution, which causes the disc radius to move outwards. We compare our disc radii, disc masses, and the overall fraction of discs, to recent observational data.

The paper is organised as follows. In Section~2 we describe the observational data with which we will compare our models. In Section~3 we describe the set-up of the $N$-body simulations and the implementation of external photoevaporation, and viscous evolution, of the protoplanetary discs. We present our results in Section~4, we provide a discussion in Section~5 and we conclude in Section~6. 

\section{Observational data}

We will compare our simulated disc evolution to observational data from nearby star-forming regions. The observational data are collated from various sources, and the disc masses are usually derived from the continuum measurements of the dust component. Where the original papers quote disc masses, these are usually given in Earth masses (M$_\oplus$). We assume a gas-to-dust ratio of 100, and multiply the dust masses by this factor to make a comparison with our simulations.

In some of the papers, the continuum fluxes at 1.3mm, $F_{\rm 1.3mm}$, are given and for $\rho$~Oph we follow \citet{Cieza19} and convert this into a dust mass \citep[this is a simplified version of the method presented in][with various assumptions about the values used in the Planck function]{Andrews13}:
\begin{equation}
M_{\rm dust} = 0.58\frac{F_{\rm 1.3mm}}{\rm mJy}{\rm M_\oplus}.
\end{equation}
    For $\lambda$~Ori we follow \citet{Ansdell20} and convert the continuum flux at 1.25mm, $F_{\rm 1.25mm}$ to disc mass using
\begin{equation}
M_{\rm dust} = 4\frac{F_{\rm 1.25mm}}{\rm mJy}{\rm M_\oplus}.
\end{equation}
All other disc masses are taken directly from the papers cited in the fifth column of Table~\ref{obs_info}.

Similarly, the disc radii are usually taken directly from the literature sources, although the $\rho$~Oph disc radii from \citet{Cieza19} are estimated from the FWHM assuming a distance to the star-forming region of $d = 140$\,pc.

\begin{figure*}
  \begin{center}
\setlength{\subfigcapskip}{10pt}
\hspace*{0.3cm}\subfigure[Disc radii]{\label{disc_evolution-a}\rotatebox{270}{\includegraphics[scale=0.35]{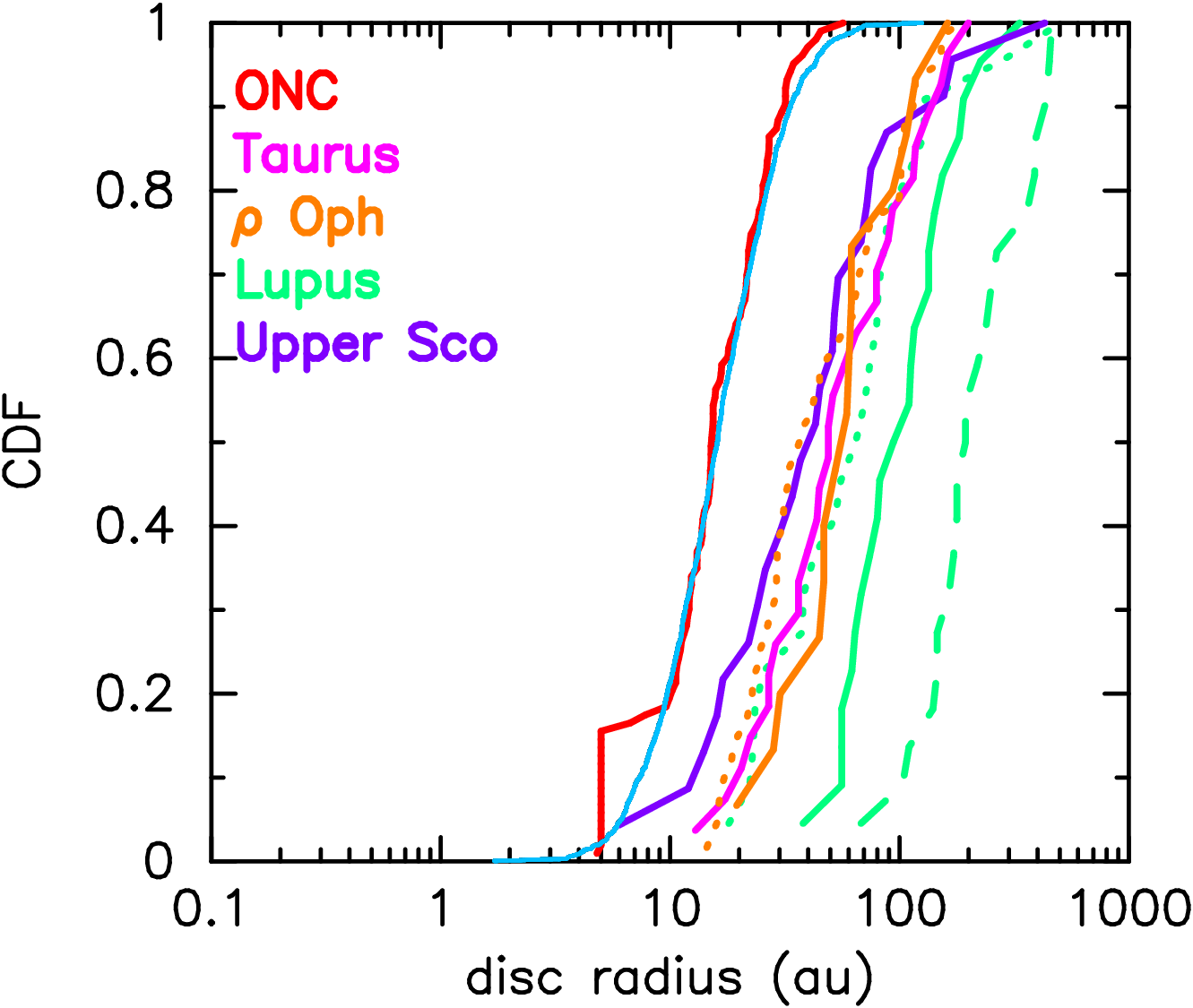}}}
\hspace*{0.3cm}\subfigure[Disc masses]{\label{disc_evolution-b}\rotatebox{270}{\includegraphics[scale=0.35]{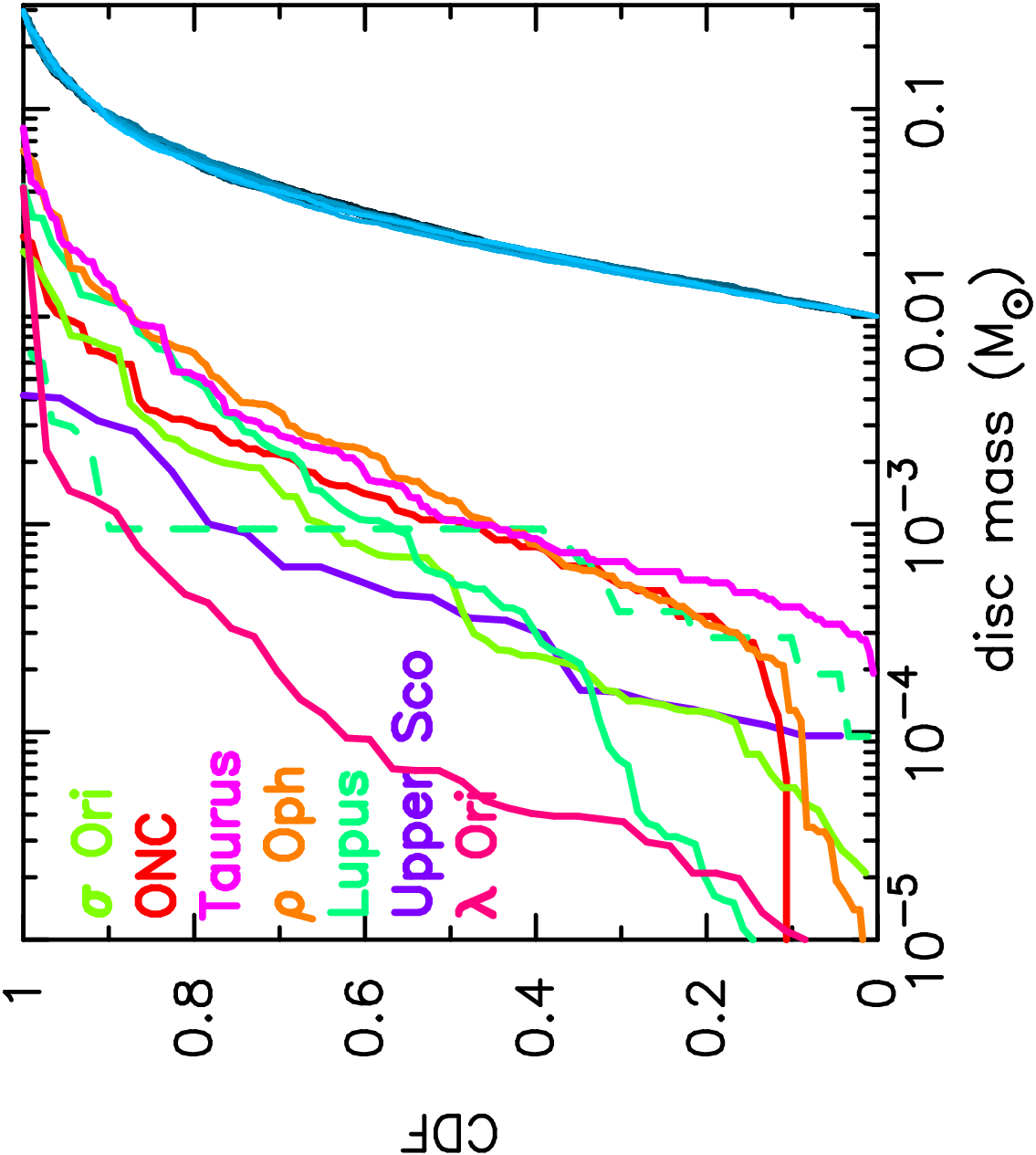}}}

\caption[bf]{The observational data we will use to compare with our simulations. The blue lines in each panel are the initial conditions for our simulations. In panel (a) these blues lines are a Gaussian fit to the ONC data (the red line). In each panel, the mint green dashed line represents the respective gas distribution, whereas solid lines represent the dust distributions. Dotted lines indicate an alternative dust radius in $\rho$~Oph (orange line) and Lupus (mint green line).}
\label{disc_obs}
  \end{center}
\end{figure*}

 The observational data are shown in Fig.~\ref{disc_obs}. In panel (a) we show the disc radii cumulative distributions for the ONC \citep[red line,][]{Eisner18}, Taurus \citep[pink line,][note this sample is likely to be incomplete]{Tripathi17} and Upper Sco \citep[purple line,][]{Barenfeld17}. We show data for $\rho$~Oph from two different studies, \citep[][the orange solid line]{Tripathi17} and \citep[][the orange dotted line]{Cieza19}. For Lupus, data on both the dust radii (mint green dashed line) and gas radii (mint green solid line) come from \citet{Ansdell18} and the dust radii from \citet[][the mint green dotted line, note this sample is also likely to be incomplete]{Tazzari17}. The initial radii in one set of our simulations are shown by the blue lines, designed to mimic the \citet{Eisner18} distribution with mean 16\,au and variance $\sigma_{{\rm log}\,r_{\rm disc}} = 0.25$.

 In panel (b) we show the disc mass cumulative distributions for $\sigma$~Ori \citep[the apple green line,][]{Ansdell17}, the ONC \citep[red line,][]{Eisner18}, Taurus \citep[pink line,][]{Andrews13}, $\rho$~Oph \citep[the orange line,][]{Cieza19}, Upper Sco \citep[purple line,][]{Barenfeld17} and $\lambda$~Ori \citep[the raspberry line,][]{Ansdell20}. Data for both the gas masses and dust masses are shown for Lupus \citep[the mint green lines,][]{Ansdell16}. The initial mass distribution in all of our simulations (see next section) is shown by the blue lines.

 We will compare our simulated disc populations to the observational data at various ages. Stellar ages are notoriously difficult to estimate \citep{Bell13}, and many of the star-forming regions we will compare our data with have more than one age estimate, or a broad range. For this reason, we will compare our simulations to the observations for a broad range of ages.

 As an example, the current best estimate for the age of the ONC is around 2.5\,Myr \citep{Jeffries11}, but there is some debate about the range of ages, or age spread, in this region \citep{Reggiani11b,Beccari17}. We will therefore compare our simulated disc radii, masses, and overal disc fraction to the observations for the ONC at both 1 and 5\,Myr in the simulations. We adopt this practice for all of the observed star-forming regions.

 \begin{table*}
  \caption[bf]{A summary  of the observational data with which we will compare our simulated disc radii and disc mass distributions. The table lists the star-forming region, the parameter (either mass or radius), the observational type (gas or dust), the line colour/style used in our figures, and the reference for these data. We also list the estimated age of each region, and the reference for this age.}
  \begin{center}
    \begin{tabular}{|l|l|l|l|l|l|l|}
      \hline

Region & parameter & obs. type & line & Disc Ref. & Age & Age Ref.  \\
\hline
ONC & radius & dust & solid red & \citet{Eisner18} & 2.5\,Myr & \citet{Jeffries11,DaRio10}  \\
ONC & mass & dust & solid red & \citet{Eisner18} & & \\
\hline
Taurus & radius & dust & solid pink & \citet{Tripathi17} & 1 -- 6.3\,Myr & \citet{Krolikowski21} \\
Taurus & mass & dust & solid pink & \citet{Andrews13} & & \\
\hline
$\rho$~Oph & radius & dust & solid orange & \citet{Tripathi17} & & \\
$\rho$~Oph & radius & dust & dotted orange & \citet{Cieza19} & 0.3 -- 6\,Myr & \citet{Rigliaco16,Grasser21} \\
$\rho$~Oph & mass & dust & solid orange & \citet{Cieza19} & & \\
\hline
Lupus & radius & dust & solid mint green & \citet{Ansdell18} & & \\
Lupus & radius & gas & dashed mint green & \citet{Ansdell18} & & \\
Lupus & radius & dust & dotted mint green & \citet{Tazzari17} & 6\,Myr & \citet{Luhman20}\\
Lupus & mass & dust & solid mint green & \citet{Ansdell16} & & \\
Lupus & mass & gas & dashed mint green & \citet{Ansdell16} & & \\
\hline
Upper Sco & radius & dust & solid purple & \citet{Barenfeld17} & 5 -- 11\,Myr & \citet{Preibisch02,Pecaut12} \\
Upper Sco & mass & dust & solid purple & \citet{Barenfeld17} & & \\
\hline
$\sigma$~Ori & mass & dust & solid apple green & \citet{Ansdell17} & 3 -- 5\,Myr & \citet{Osorio02a};\\
& & & & & & \citet{Oliveira02,Mayne08} \\
\hline
$\lambda$~Ori & mass & dust & raspberry & \citet{Ansdell20} & 3 -- 10\,Myr & \citet{Mayne08,Bell13} \\
      \hline
    \end{tabular}
  \end{center}
  \label{obs_info}
\end{table*}

 \begin{table*}
  \caption[bf]{Stellar density and estimated radiation field (in $G_0$) for the star-forming regions in this work. The columns are the region, mass density, OB star content, $G_0$ (calculated in 2D) and the reference for the census (and in the case of Lupus, the estimate for $G_0$ from disc modelling).}
  \begin{center}
    \begin{tabular}{|l|l|l|l|l|}
      Region & Density & OB star content & $G_0$ & ref.\\
      \hline
      \hline
      ONC & $33^{+132}_{-12}$\,M$_\odot$\,pc$^{-3}$ & 3 O-type stars, 8 B-type stars & $14103^{+85221}_{-4326}$ & \citet{Hillenbrand98} \\
      \hline
      Taurus & $0.8^{+3.9}_{-0.1}$\,M$_\odot$\,pc$^{-3}$ & None & 1 (background ISM)  & \citet{Guedel07} \\
      \hline
      $\rho$~Oph & $13^{+25}_{-6}$\,M$_\odot$\,pc$^{-3}$ & 2 B-type stars & $57^{+202}_{-31}$ & \citet{Parker12c} \\
      \hline
      Lupus & $0.2^{+1.9}_{-0.02}$\,M$_\odot$\,pc$^{-3}$ & None & 4 (from nearby [$\sim10$\,pc] OB assoc.) & \citet{Galli20,Cleeves16} \\
      \hline
      Upper Sco & $0.6^{+1.2}_{-0.3}$\,M$_\odot$\,pc$^{-3}$ & 17 B-type stars & $47^{+108}_{-29}$& \citet{Luhman20} \\
      \hline
      $\sigma$~Ori & $0.9^{+2.0}_{-0.4}$\,M$_\odot$\,pc$^{-3}$ & 6 B-type stars & $557^{+1786}_{-323}$& \citet{Caballero19} \\ 
      \hline
      $\lambda$~Ori & $2^{+5}_{-1}$\,M$_\odot$\,pc$^{-3}$ & 1 O-type star, 6 B-type stars &  $1713^{+4577}_{-696}$ & \citet{Bayo11} \\
      \hline
    \end{tabular}
  \end{center}
  \label{obs_info_regions}
\end{table*}

To facilitate a comparison with our simulations, in Table~\ref{obs_info_regions} we calculate the stellar density and radiation field (in $G_0$) for each region. We calculate the local density for each star, $\tilde{\rho}$, which is the local volume density out to the tenth nearest neighbour. We first obtain the local surface density to the tenth nearest neighbour, then assume a depth of 1\,pc to calculate the number density. We then convert this to a mass density by multiplying by the average mass in the IMF, i.e.\,\,0.2\,M$_\odot$. We quote the median density in each region, but also show the interquartile range in the format $\tilde{\rho} = {\rm [median]}^{+\rm 75~percentile}_{-\rm 25~percentile}$. We do not have accurate mass determinations for all of the stars in each region, and so we multiply the number densities by this average mass. Note that these densities are the \emph{median} in these regions; the central densities, or the densities in the areas of highest clustering will be much higher \citep[e.g.][]{King12a}.

In Table~\ref{obs_info_regions} we also show the numbers of stars that emit photoionising radiation (i.e.\,\,those $\geq$5\,M$_\odot$) and the median and interquartile ranges of the FUV fluxes calculated in each star-forming region. To do this, we take FUV luminosities from \citet{Armitage00} and then calculate the flux at each $<$5\,M$_\odot$ star. We then convert this flux into \citet{Habing68} units, $G_0 = 1.6 \times 10^{-3}$\,erg\,s$^{-1}$\,cm$^{-2}$, where $1G_0$ is the background flux in the interstellar medium.

\section{Method}

In this Section we describe the $N$-body simulations we use to model star-forming regions, the post-processing analysis we use to follow the evolution of protoplanetary discs, and the observational data with which we will compare our simulations.

\subsection{Star-forming regions}

We model the dynamical evolution of young star-forming regions using the \texttt{kira} integrator within the \texttt{Starlab} environment \citep{Zwart99,Zwart01}. This allows us to accurately calculate the FUV and EUV radiation fields used to determine the mass-loss due to photoevaporation from our protoplanetary discs. 

We adopt several different sets of initial conditions for these simulations in order to model a comprehensive range of star-forming regions. Our models contain either $N = 1500$ or $N = 150$ stars, and we vary the initial radius so that the median local stellar density in the simulations varies between 10 -- 1000\,M$_\odot$\,pc$^{-3}$.

We draw stellar masses from the IMF described in \citet{Maschberger13}, which has a probability density function of the form
\begin{equation}
p(m) \propto \left(\frac{m}{\mu}\right)^{-\alpha}\left(1 + \left(\frac{m}{\mu}\right)^{1 - \alpha}\right)^{-\beta},
\label{maschberger_imf}
\end{equation}
Here, $\mu = 0.2$\,M$_\odot$ is the average stellar mass, $\alpha = 2.3$ is the \citet{Salpeter55} power-law exponent for higher mass stars, and $\beta = 1.4$ describes the slope of the IMF for low-mass objects \citep*[which also deviates from the log-normal form;][]{Bastian10}. We randomly sample this distribution in the mass range 0.1 -- 50\,M$_\odot$, such that brown dwarfs are not included in the simulations. For simplicity, we also do not include primordial binaries, although these are likely to be ubiquitous in star-forming regions.

In the simulations where we draw $N = 1500$ stars from the IMF we typically obtain between 10 and 50 stars with mass $\geq$5\,M$_\odot$, which would emit photionising radiation. When we select $N = 150$ stars, in some realisations there are no stars $\geq$5\,M$_\odot$, although usually there are at least 1 -- 2 examples \citep{Parker21a,Parker21b}.

Young star-forming regions are observed to exhibit spatial and kinematic substructure \citep{Larson81,Gomez93,Larson95,Cartwright04,Sanchez09,Jaehnig15,Wright16,Buckner19}. To mimic this in our initial conditions, we use a box fractal generator \citep{Goodwin04a}, in which the degree of spatial and kinematic substructure is governed by just one parameter, the fractal dimension $D$. For details of the set-up of these fractals, we refer the interested reader to \citet{Goodwin04a} and \citet{DaffernPowell20}, but we briefly describe them here. 

The fractals are set up by placing a `parent' particle at the centre of a cube of side $N_{\rm div}$, which spawns $N_{\rm div}^3$ subcubes. Each of these subcubes contains a `child' particle at its centre, and the fractal is constructed by determining how many successive generations of `children' are produced. The likelihood of each successive generation producing their own children is given by $N^{D-3}$, where $D$ is the fractal dimension.

Fewer generations are produced when the fractal dimension is lower, leading to a less uniform appearance and hence more substructure.  Star-forming regions with a low fractal dimension (e.g.\,\,$D = 1.6$) have a high degree of substructure, whereas regions with higher values (e.g.\,\,$D = 2.0, 2.6$) have less structure and regions with $D = 3.0$ are approximately uniform.

The velocities of the parent particles are drawn from a Gaussian distribution of mean zero, and the velocities of the child particles inherit this velocity plus a small random component which scales as $N^{D-3}$ but becomes progressively smaller on each generation. This means that physically close particles have very similar velocities, but more distant particles can have very uncorrelated velocites.

In our simulations we adopt $D = 2.0$ for all our simulations, which gives a moderate amount of spatial and kinematic substructure. We scale the velocities to a virial ratio $\alpha_{\rm vir} = T/|\Omega|$, where $T$ and $|\Omega|$ are the total kinetic and potential energies, respectively. The velocities of young stars are often observed to be subvirial along filaments, so we adopt a subvirial ratio ($\alpha_{\rm vir} = 0.3$) in all of our simulations. 

Finally, we scale the radii of the fractals to achieve the required stellar density. For our regions with $N = 1500$ stars, radii of 1, 2.5 and 5.5\,pc lead to median local densities of 1000, 100 and 10\,M$_\odot$\,pc$^{-3}$ respectively, whereas our regions with $N = 150$ stars have radii of 0.75\,pc to establish initial stellar densities of 100\,M$_\odot$\,pc$^{-3}$.

We evolve our simulations for 10\,Myr using \texttt{kira}, in order to encompass the full protoplanetary disc evolutionary phase. We do not include stellar evolution in the calculations. 

\subsection{Disc photoevaporation and evolution}

We follow the evolution of protoplanetary discs in a post-processing analysis, such that the discs are not included in the $N$-body integrations. We calculate the FUV radiation field for each star at each snapshot in the $N$-body simulations using the $L_{\rm FUV}$ luminosities in \citet{Armitage00}:
 \begin{equation}
 F_{\rm FUV} = \frac{L_{\rm FUV}}{4\pi d^2},
 \end{equation}
where $d$ is the distance from each low-mass star to each star  with mass $\geq$5\,M$_\odot$. Below this mass, the FUV flux becomes negligible. In the simulations that contain more than one massive star, we sum these fluxes to obtain the FUV radiation field for each disc-bearing star. This FUV radiation field is then scaled to the \citet{Habing68}  unit, $G_0 = 1.8 \times 10^{-3}$\,erg\,s$^{-1}$\,cm$^{-2}$, which is the background FUV flux in the interstellar medium. 

For each star between $0.1 < M_\star/{\rm M_\odot} < 3$\,M$_\odot$, we assign it a disc of mass
 \begin{equation}
   M_{\rm disc} = 0.1\,M_\star.
 \end{equation}

 We adopt three different distributions for the initial disc radii. First, we adopt a  delta function where all the discs have radii $r_{\rm disc} = 10$\,au, second a delta function where all of the discs have $r_{\rm disc} = 100$\,au,  and finally, we draw discs from a Gaussian distribution of mean  16\,au and variance $\sigma_{{\rm log}\,r_{\rm disc}} = 0.25$, designed to closely mimic the observed distribution of disc radii in the ONC \citep{Eisner18}.

 To calculate the mass loss due to FUV radiation, we use the \texttt{FRIED} grid of models from \citet{Haworth18b}, which uses the stellar mass, $M_\star$, radiation field $G_0$, disc mass $M_{\rm disc}$ and disc radius $r_{\rm disc}$ as an input, and produces a mass-loss rate, $\dot{M}_{\rm FUV}$ as an output. As the \texttt{FRIED} grid produces discrete values, we perform a linear interpolation over disc mass and mass-loss. 

In addition to calculating the mass-loss due to FUV radiation, we also calculate the (usually much smaller) mass loss due to EUV radiation. To calculate the mass loss due to EUV radiation, we adopt the following prescription from \citet{Johnstone98}:
\begin{equation}
\dot{M}_{\rm EUV} \simeq 8 \times 10^{-12} r^{3/2}_{\rm disc}\sqrt{\frac{\Phi_i}{d^2}}\,\,{\rm M_\odot \,yr}^{-1}.
\label{euv_equation}
\end{equation}
Here, $\Phi_i$ is the  ionizing EUV photon luminosity from each massive star in units of $10^{49}$\,s$^{-1}$ and is dependent on the stellar mass according to the observations of \citet{Vacca96} and \citet{Sternberg03}. For example, a 41\,M$_\odot$ star has $\Phi = 10^{49}$\,s$^{-1}$ and a 23\,M$_\odot$ star has $\Phi = 10^{48}$\,s$^{-1}$. The disc radius $r_{\rm disc}$ is expressed in units of au and the distance to the massive star $d$ is in pc.

  We subtract mass from the discs according to the FUV-induced mass-loss rate in the \texttt{FRIED} grid and the EUV-induced mass-loss rate from Equation~\ref{euv_equation}. Models of mass loss in discs usually assume the mass is removed from the edge of the disc (where the surface density is lowest) and we would expect the radius of the disc to decrease in this scenario. We employ a very simple way of reducing the radius by assuming the surface density of the disc at 1\,au, $\Sigma_{\rm 1\,au}$, from the host star remains constant during mass-loss \citep[see also][]{Haworth18b,Haworth19}. If
  \begin{equation}
\Sigma_{\rm 1\,au} = \frac{M_{\rm disc}}{2\pi r_{\rm disc} [{\rm 1\,au}]},
\label{rescale_disc}
  \end{equation}
  where $M_{\rm disc}$ is the disc mass, and $r_{\rm disc}$ is the radius of the disc, then if the surface density at 1\,au remains constant, a reduction in mass due to photoevaporation will result in the disc radius decreasing by a factor equal to the disc mass decrease. 

The decrease in disc radius due to photoevaporation will be countered to some degree by expansion due to the internal viscous evolution of the disc. We implement a very simple prescription for the outward evolution of the disc radius due to  viscosity following the procedure in \citet{ConchaRamirez19a}. 

First, we define a temperature profile for the disc, according to 
\begin{equation}
T(R) = T_{\rm 1\,au} R^{-q},
\end{equation}
where $R$ is the distance from the host star, $T_{\rm 1\,au}$ is the temperature at 1\,au from the host star and is derived from the stellar luminosity. For a 1\,M$_\odot$ star, we derive $T_{\rm 1\,au} = 393$\,K, although $T_{\rm 1\,au} = 300$\,K is more commonly adopted. We assume a main sequence mass-luminosity relation, and use data from \citet{Cox00}. We adopt $q = 0.5$ \citep{Hartmann98}.  

 In the model of \citet{Hartmann98}, the characteristic initial radius, $R_c(0)$ is defined by 
 \begin{equation}
 R_c(0) = R'\left(\frac{M_\star}{{\rm M_\odot}} \right)^{0.5},
 \end{equation} 
 where $R' = 30$\,au. At some time $t$, the characteristic radius $R_c(t)$ at that time is given by \citep{LyndenBell74}
  \begin{equation}
    R_c(t) = \left( 1 + \frac{t}{t_\nu}\right)^{\frac{1}{2 - \gamma}} R_c(0),
    \label{r_crit}
 \end{equation}
 where the viscosity exponent $\gamma$ is unity \citep{Andrews10}. $t_\nu$ is the viscous timescale, and is given by
  \begin{equation}
    t_\nu =   \frac{\mu_{\rm mol}m_pR_c(0)^{0.5 + q}\sqrt{GM_\star}}{3\alpha(2 - \gamma)^2k_BTR^q},
    \label{time_viscous}
 \end{equation}
 where $\mu_{\rm mol}$ is the mean molecular weight of the material in the disc (we adopt $\mu_{\rm mol} = 2$), $m_p$ is the proton mass, $G$ is the gravitational constant, $M_\star$ is the mass of the star, $k_B$ is the Boltzmann constant and $T$ and $R$ are the temperature and distance from the host star, as described above.  $\alpha$ is the turbulent mixing strength \citep{Shakura73} and based on observations of T~Tauri stars, \citet{Hartmann98} adopt $\alpha =  10^{-2}$  \citep[see also][]{Isella09,Andrews10}. This $\alpha$ leads to significant viscous spreading, and we note that some recent work suggests much lower values, of order $\alpha \sim 10^{-3} - 10^{-4}$ \citep{Pinte16,Flaherty20}. One can see from Eqns.~\ref{time_viscous}~and~\ref{r_crit} that reducing $\alpha$ would reduce the size of the outward change in radius. To test this in our simulations, in Appendix~\ref{appendix_viscosity} we present results for simulations with $\alpha = 10^{-4}$. 
 
 We set $r_{\rm disc} = R$ to be the radius of the disc, and following mass-loss due to photoevaporation and the subsequent inward movement of the disc radius according to Equation~\ref{rescale_disc}, we calculate the change in characteristic radius ($R_c(t_n)/R_c(t_{n-1})$) and scale the disc radius {\bf $r_{\rm disc}$} accordingly:
 \begin{equation}
   r_{\rm disc}(t_n) = r_{\rm disc}(t_{n - 1})\frac{R_c(t_n)}{R_c(t_{n-1})}.
   \label{disc_radius_time}
 \end{equation} 

 Both the inward evolution of the disc radius due to mass-loss, and the outward viscous evolution occur on much shorter timescales than the gravitational interactions between stars in the star-forming regions. We therefore adopt a timestep of $10^{-3}$\,Myr for the disc evolution calculations \citep{Parker21a}.

 In the calculations of \citet{Haworth18b}, the mass lost from discs due to FUV radiation is predominantly gas, whereas the discs mostly retain their dust component. However, \citet{Sellek20} find that during mass loss, the dust radius follows the inward evolution of the gas radius. In our analysis, we assume that the mass lost from the disc is in the form of gas, and that the dust radius follows the gas radius (both inwards due to photoevaporation, and outwards due to viscous spreading). However, we note that although the dust radius may follow the gas radius during photoevaporation, the two are by no means the same. \citet{Sanchis21} and \citet{Long22} find that the gas radii of discs in nearby star-forming regions (Lupus, $\rho$~Oph and Upper Sco) are at least a factor of two larger than the dust radii. 

 A summary of the simulations is given in Table~\ref{simulations}.

 \subsection{Comparison with observational data}

 In our simulations, the initial disc radius is either 10 or 100\,au, or drawn from a Gaussian to mimic the \citet{Eisner18} distribution. At subsequent snapshots the disc radius is set by Eqn.~\ref{disc_radius_time}. However, the observed disc radii are defined in different ways. In Lupus \citep{Tazzari17} and Upper Sco \citep{Barenfeld17} the radius is the point at which 95\,per cent of the flux is enclosed, whereas for the ONC \citet{Eisner18} adopt the half-width half-maximum (HWHM) major axes of Gaussian fits. \citet{Eisner18} report a difference of less than a factor of two in these measurements. However, for Taurus and $\rho$~Oph \citet{Tripathi17} adopt the point at which 68\,per cent of the flux is enclosed. 

   \citet{ConchaRamirez19a} derive an expression using the characteristic radius $R_c$ to convert between the full disc radius and the radius $R_{0.95}$ that encloses 95\,per cent of the mass. They obtain $R_{0.95} \simeq 3R_c(t)$, and for the radius that encloses 68\,per cent of the flux, $R_{0.68} \simeq R_c(t)$. As an example, after 10\,Myr of viscous evolution with the parameters defined above, a 1\,M$_\odot$ star with $M_{\rm disc} = 0.1$\,M$_\odot$ and initial disc radius $r_{\rm disc} = 100$\,au expands to a radius of 173\,au. It has $R_c(t) = 56$\,au, so $R_{0.95} \simeq 170$\,au.

   Given the similarities between $R_{0.95}$ and the radius defined by the Gaussian HWHM, we can make a direct comparison between the simulation data and the observed data for the ONC, Lupus, Upper~Sco, $\sigma$~Ori and $\lambda$~Ori. Because the disc radii for $\rho$~Oph and Taurus are derived from $R_{0.68} \simeq R_c(t)$, when comparing our simulation data we will caveat that a direct comparison would likely require a scaling of the observed disc distributions to larger radii. However, in the case of Taurus this region has a low to non-existent $G_0$ field, so any disc photoevaporation would likely be minimal.

\begin{table}
  \caption[bf]{A summary of the different initial conditions of our simulated star-forming regions. The columns show the number of stars, $N_{\rm stars}$, the initial radius of the star-forming region, $r_F$, the initial median local stellar density, $\tilde{\rho}$, and the initial distribution for the radii of protoplanetary discs, $r_d$. }
  \begin{center}
    \begin{tabular}{|c|c|c|c|}
      \hline

$N_{\rm stars}$ & $r_F$ & $\tilde{\rho}$ & $r_d$  \\
\hline
1500 & 1\,pc & $1000$\,M$_\odot$\,pc$^{-3}$ & 100\,au \\
1500 & 2.5\,pc & $100$\,M$_\odot$\,pc$^{-3}$ & 10\,au \\
1500 & 2.5\,pc & $100$\,M$_\odot$\,pc$^{-3}$ & 100\,au \\
1500 & 2.5\,pc & $100$\,M$_\odot$\,pc$^{-3}$ & \citet{Eisner18} \\
1500 & 5.5\,pc & $10$\,M$_\odot$\,pc$^{-3}$ & 100\,au  \\
\hline
150 & 0.75\,pc & $100$\,M$_\odot$\,pc$^{-3}$ & 100\,au \\ 

      \hline
    \end{tabular}
  \end{center}
  \label{simulations}
\end{table}

\section{Results}

In all of our results, we show ten realisations of the same simulation, identical apart from the random number seed used to generate the simulations. This means that statistically, the distribution of stellar masses, stellar positions and stellar velocities are the same. However, each realisation contains statistical variations, including in the numbers of massive stars in each simulation.

\subsection{Evolution of stellar density and $G_0$}

\begin{figure*}
  \begin{center}
\setlength{\subfigcapskip}{10pt}
\vspace*{-0.7cm}
\hspace*{-1.0cm}\subfigure[$\tilde{\rho} \sim 1000$\,M$_\odot$\,pc$^{-3}$, $N_\star = 1500$]{\label{G0_densities-a}\rotatebox{270}{\includegraphics[scale=0.19]{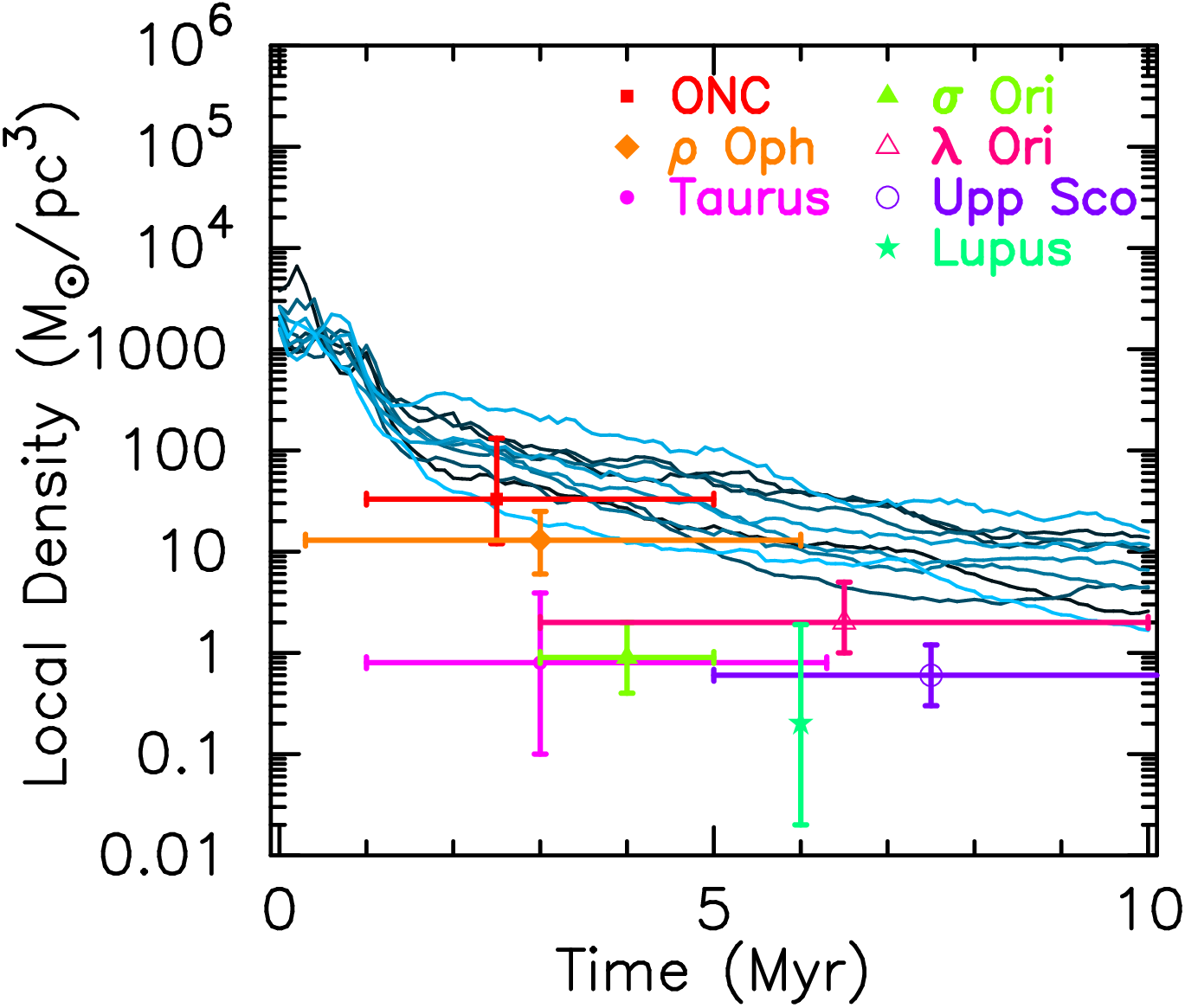}}}
\hspace*{0.2cm}\subfigure[$\tilde{\rho} \sim 100$\,M$_\odot$\,pc$^{-3}$, $N_\star = 1500$]{\label{G0_densities-b}\rotatebox{270}{\includegraphics[scale=0.19]{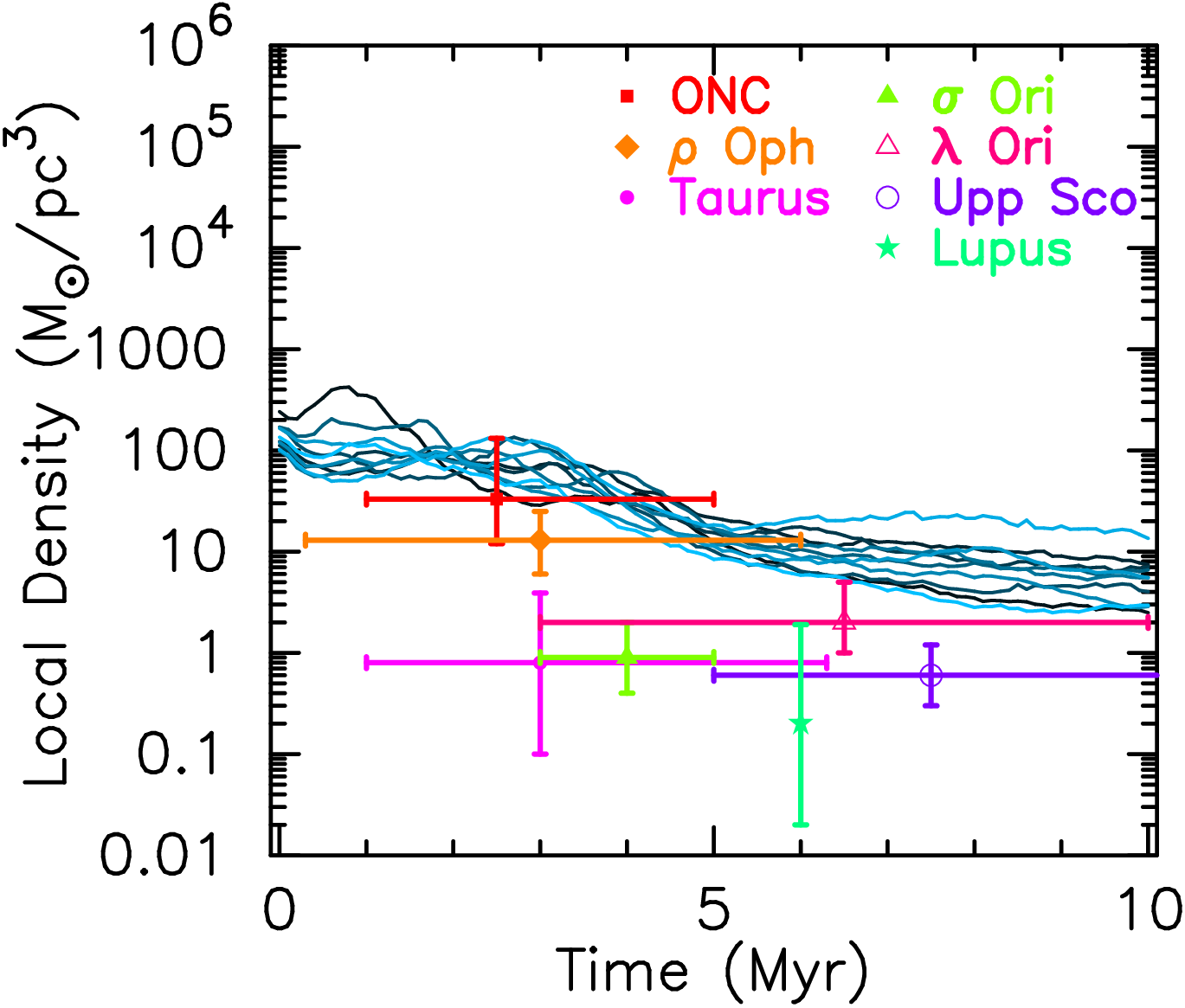}}}
\hspace*{0.2cm}\subfigure[$\tilde{\rho} \sim 10$\,M$_\odot$\,pc$^{-3}$, $N_\star = 1500$]{\label{G0_densities-c}\rotatebox{270}{\includegraphics[scale=0.19]{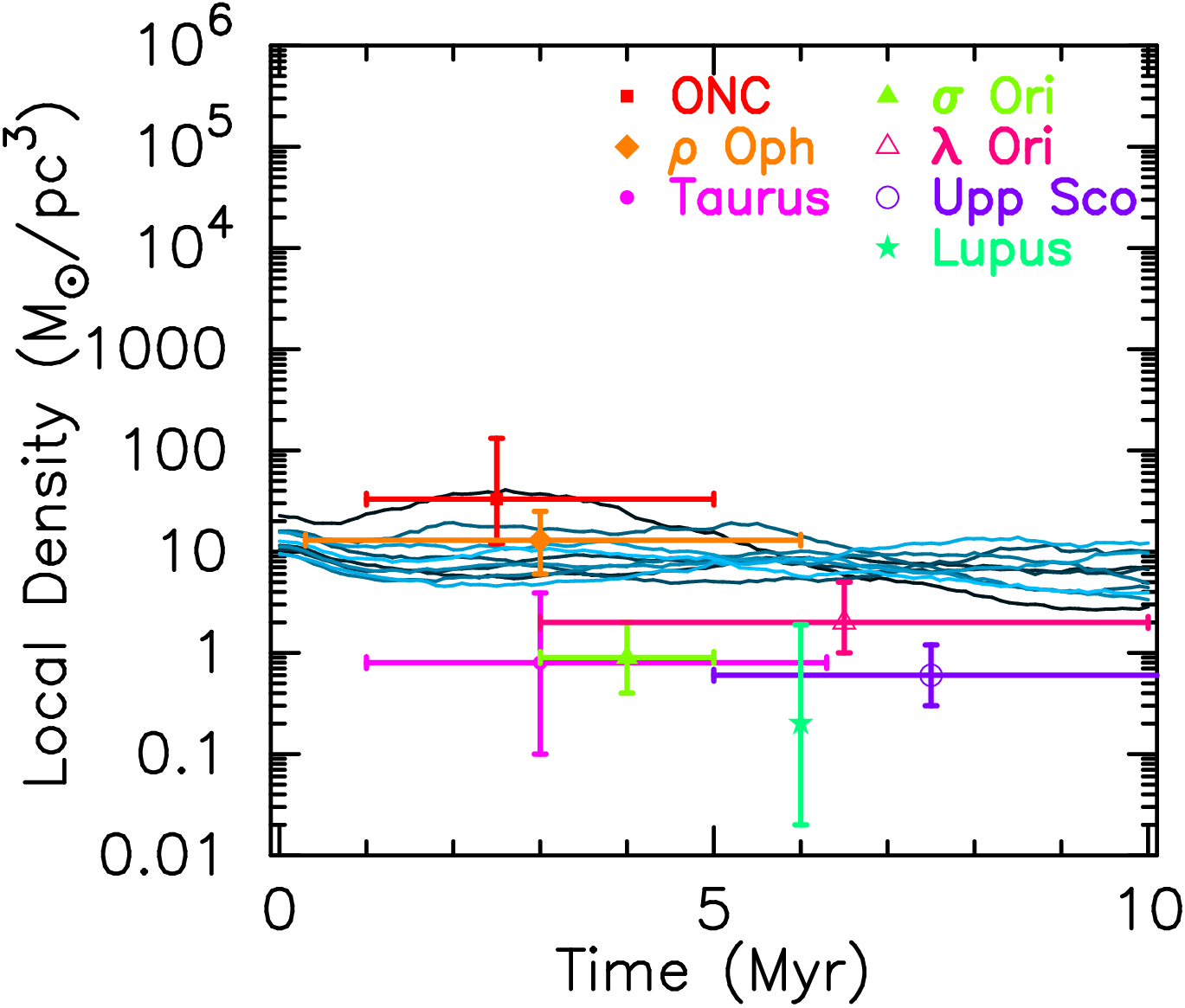}}}
\hspace*{0.2cm}\subfigure[$\tilde{\rho} \sim 100$\,M$_\odot$\,pc$^{-3}$, $N_\star = 150$]{\label{G0_densities-d}\rotatebox{270}{\includegraphics[scale=0.19]{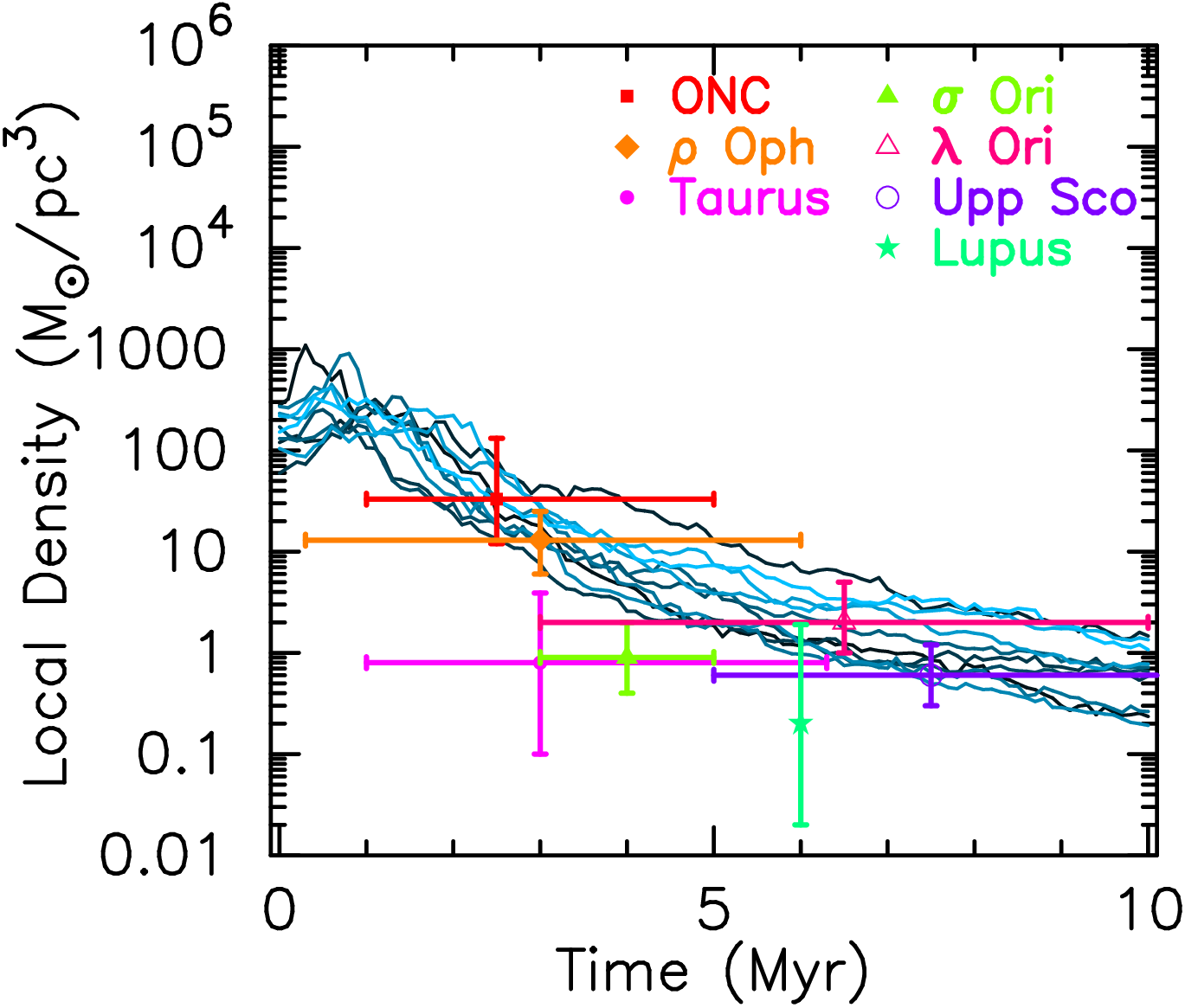}}}

\hspace*{-1.0cm}\subfigure[$\tilde{\rho} \sim 1000$\,M$_\odot$\,pc$^{-3}$, $N_\star = 1500$]{\label{G0_densities-e}\rotatebox{270}{\includegraphics[scale=0.19]{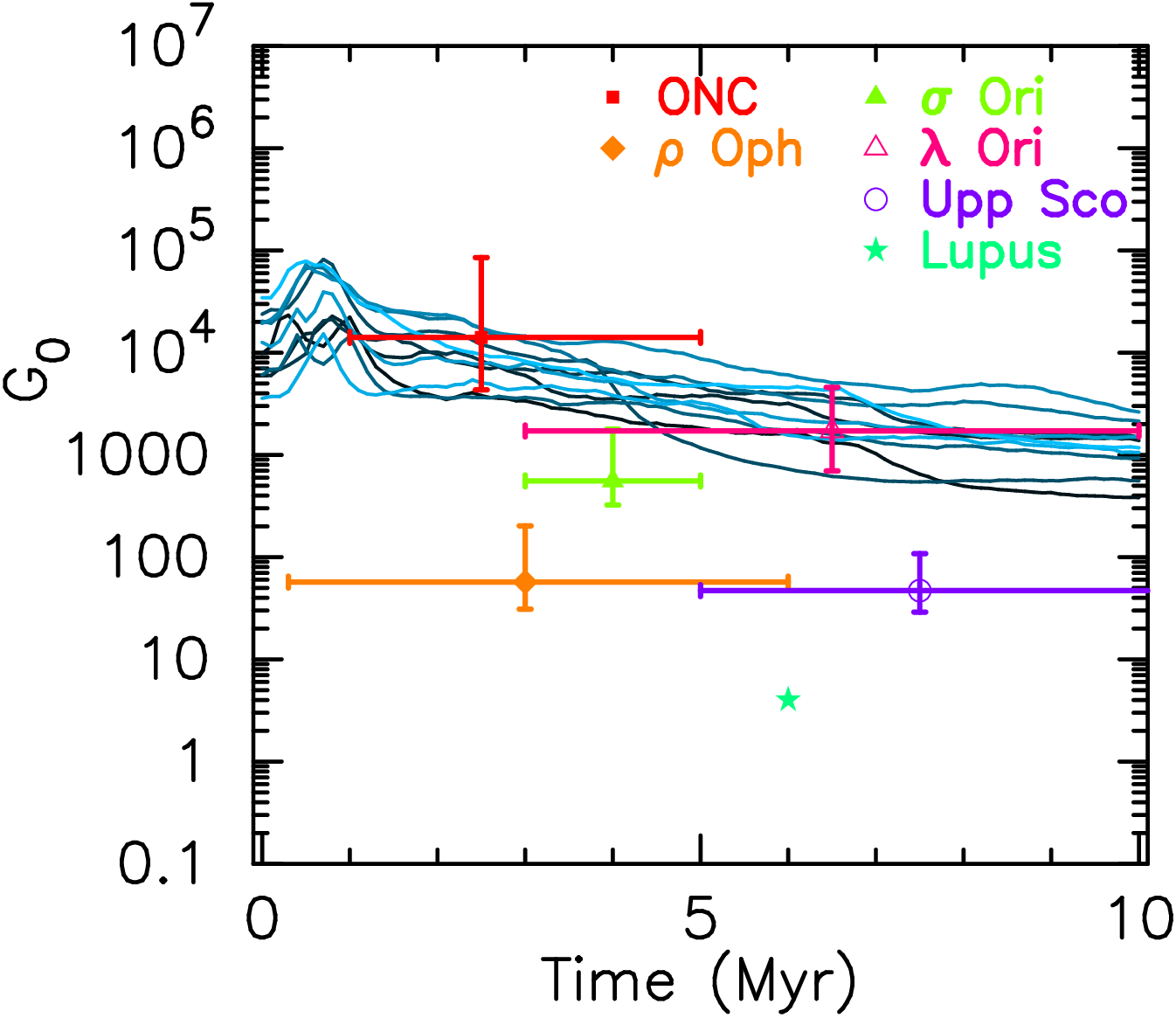}}}
\hspace*{0.3cm}\subfigure[$\tilde{\rho} \sim 100$\,M$_\odot$\,pc$^{-3}$, $N_\star = 1500$]{\label{G0_densities-f}\rotatebox{270}{\includegraphics[scale=0.19]{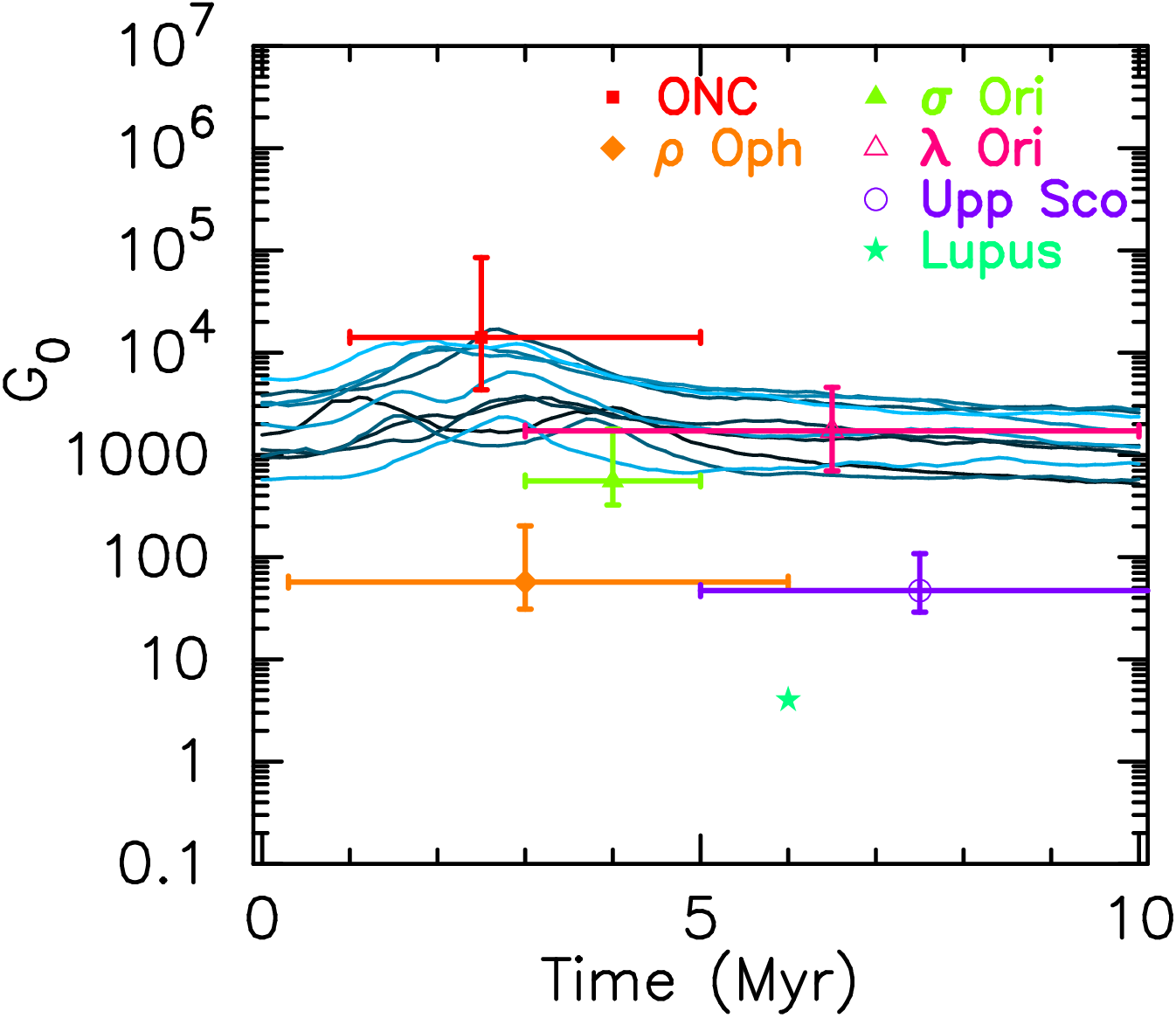}}}
\hspace*{0.3cm}\subfigure[$\tilde{\rho} \sim 10$\,M$_\odot$\,pc$^{-3}$, $N_\star = 1500$]{\label{G0_densities-g}\rotatebox{270}{\includegraphics[scale=0.19]{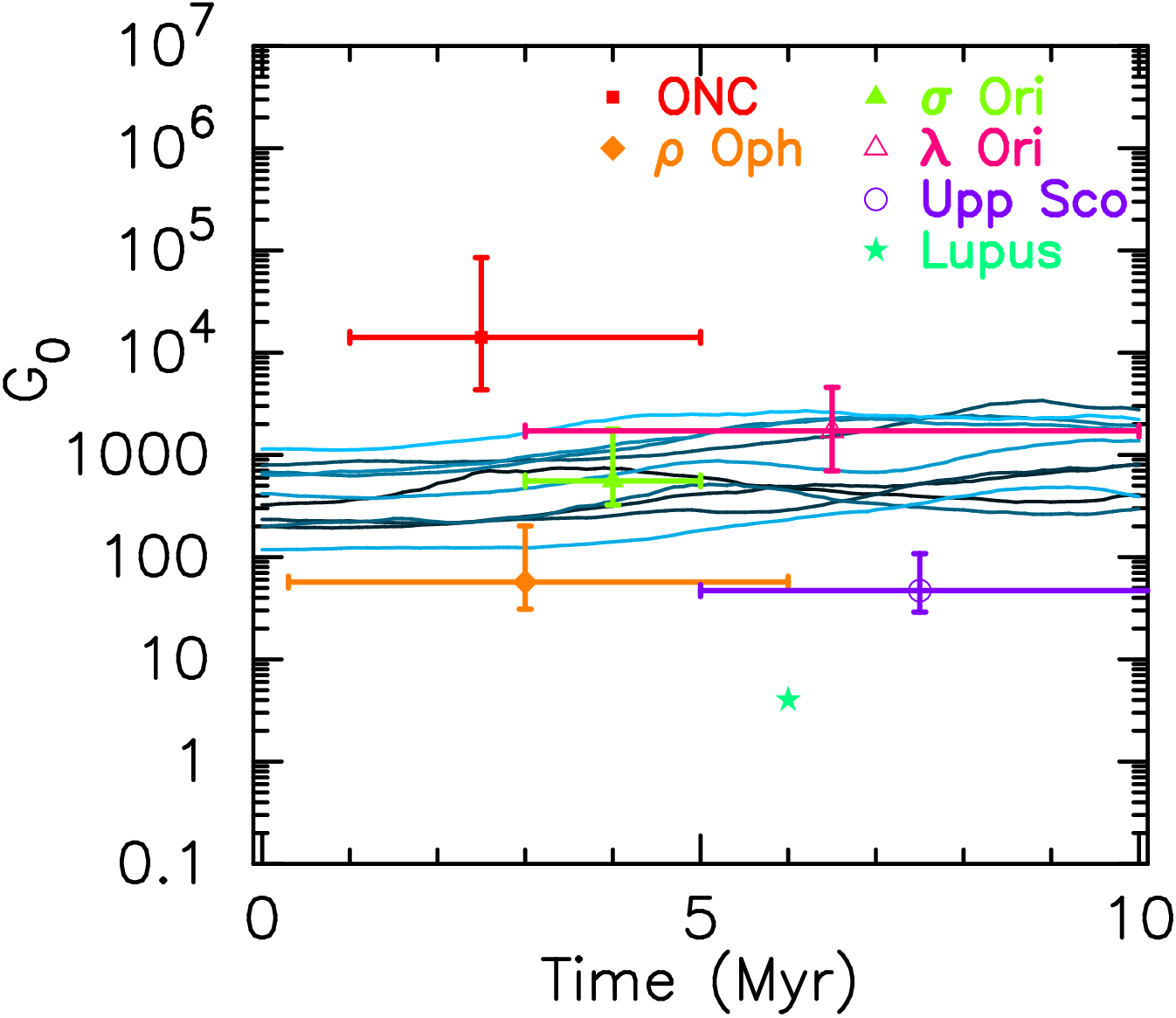}}}
\hspace*{0.3cm}\subfigure[$\tilde{\rho} \sim 100$\,M$_\odot$\,pc$^{-3}$, $N_\star = 150$]{\label{G0_densities-h}\rotatebox{270}{\includegraphics[scale=0.19]{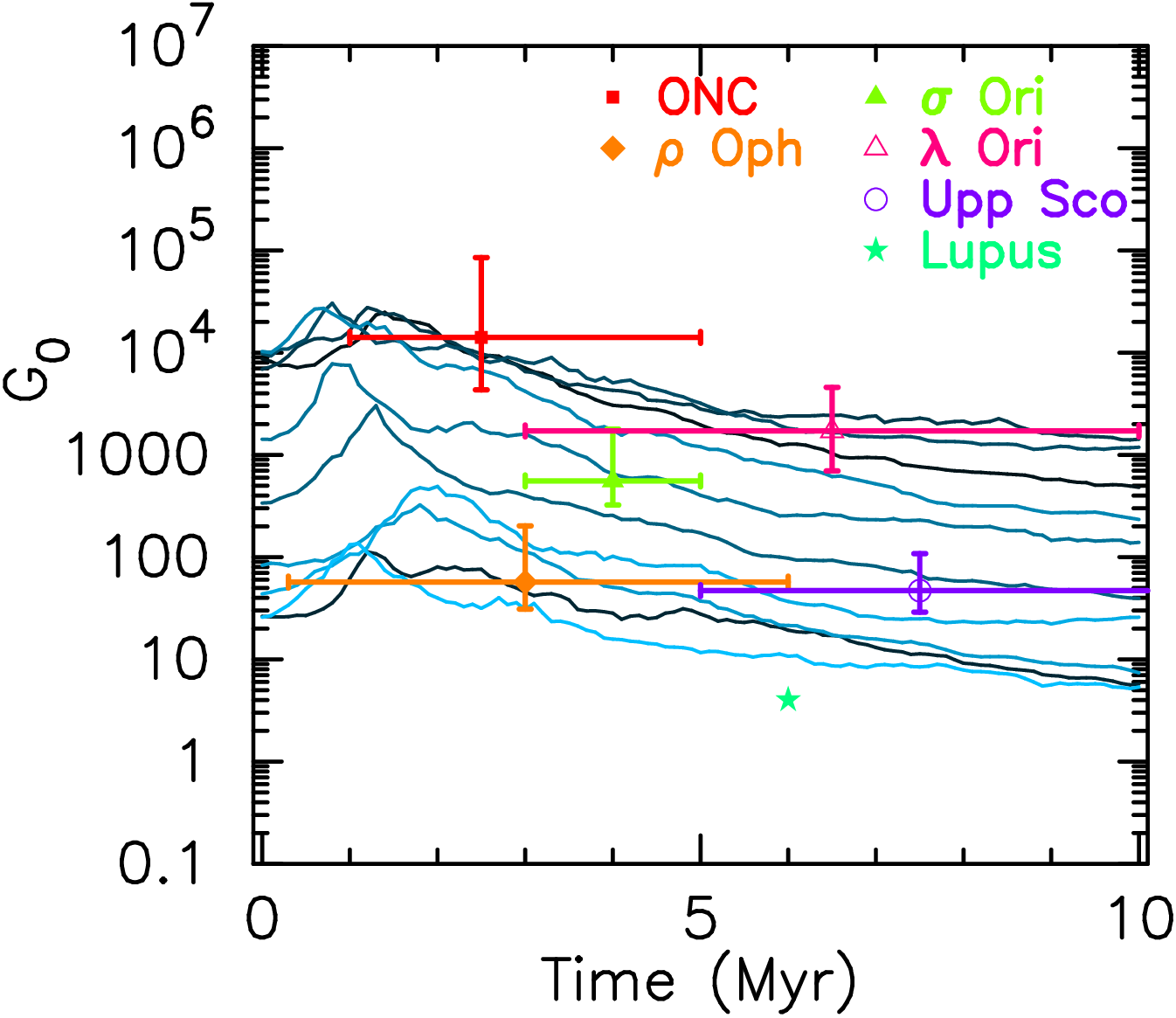}}}
\caption[bf]{The evolution of the local stellar density (the volume density within a sphere that encompasses the ten nearest neighbours to each star) and the median $G_0$ field in our simulations. The top row shows the median local density in ten realisations of the same star-forming region (indicated by the different blue lines). The bottom row shows the median FUV flux, $G_0$, in each simulation, again shown by  blue lines of different hues. The values for the observed star-forming regions are shown. We omit the $G_0$ value for Taurus, as this region does not contain any OB stars, nor is it close enough to an OB association to receive any flux from extra-associated stars, i.e.\,\,stars from a nearby star-forming region.}
\label{G0_densities}
  \end{center}
\end{figure*}

We begin by showing the evolution of the median local stellar density and the median $G_0$ in our simulations in Fig.~\ref{G0_densities}. Each simulation is shown by a blue line of different hue, and the observed values are shown in different colours. The `error' bars on $\tilde{\rho}$ and $G_0$ are the interquartile ranges of the observed values (i.e.\,\,the median, and 25$^{\rm th}$ and 75$^{\rm th}$ percentiles), whereas the error bars on the ages of the observed regions indicate the ranges quoted in the literature.

  Of the observed regions, the ONC is closer to the $N = 1500$ star simulations in terms of the numbers of members, whereas Lupus, Taurus, $\rho$~Oph and $\sigma$~Ori are closer to the $N = 150$ star simulations. $\lambda$~Ori and Upper Sco straddle the two regimes. The present-day densities and $G_0$ fields in the ONC and $\lambda$~Ori are both consistent with moderate to high (100\,M$_\odot$\,pc$^{-3}$ -- 1000\,M$_\odot$\,pc$^{-3}$) initial densities, whereas the other (low-$N$) regions are more consistent with moderately dense (100\,M$_\odot$\,pc$^{-3}$) initial conditions. We will use these constraints when interpreting the evolution of disc masses and radii in the following analysis.

\subsection{Inward radius evolution only}

\begin{figure*}
  \begin{center}
\setlength{\subfigcapskip}{10pt}
\hspace*{0.3cm}\subfigure[1\,Myr]{\label{mod_dens_100au_no_viscous-a}\rotatebox{270}{\includegraphics[scale=0.23]{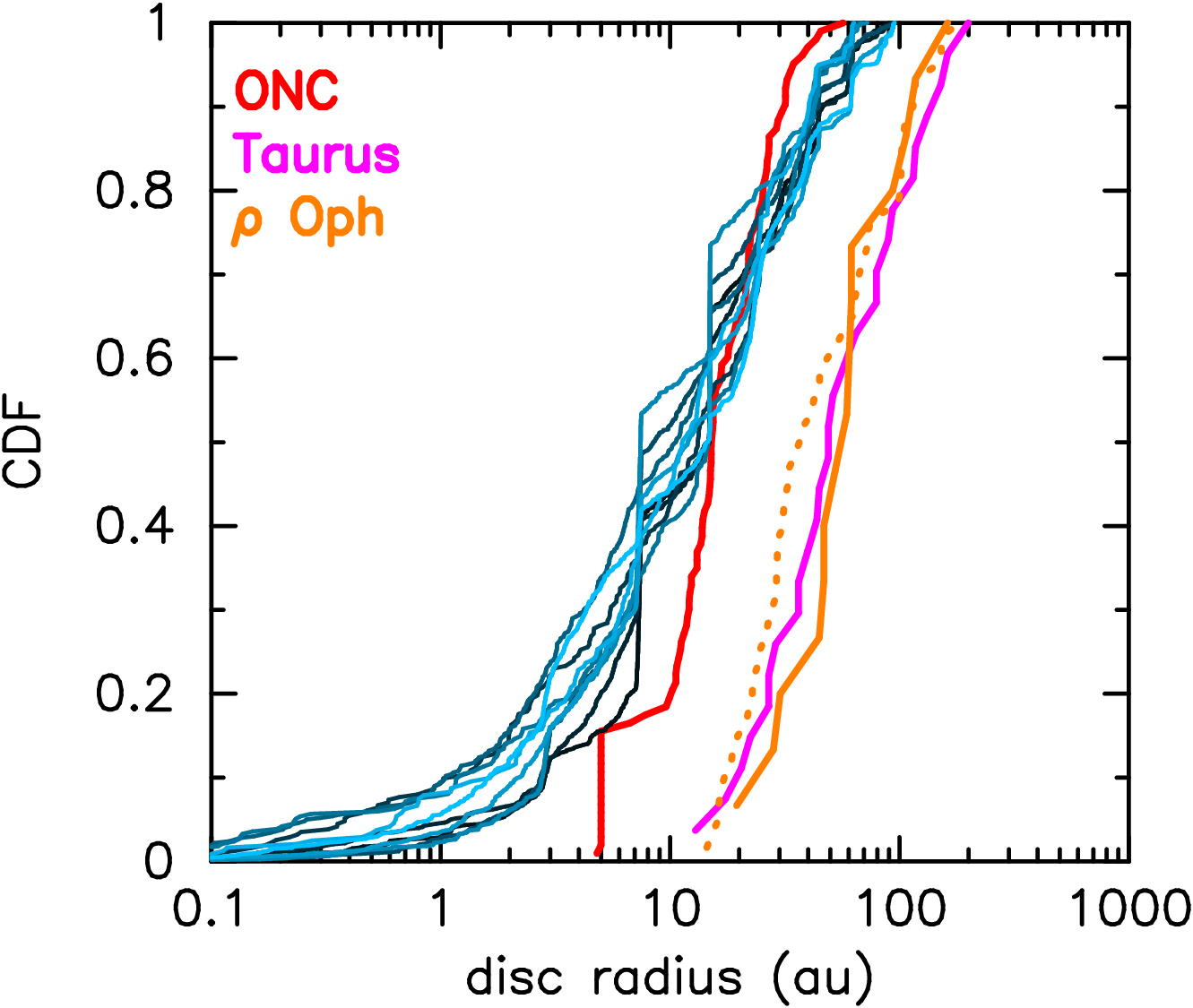}}}
\hspace*{0.3cm}\subfigure[5\,Myr]{\label{mod_dens_100au_no_viscous-b}\rotatebox{270}{\includegraphics[scale=0.23]{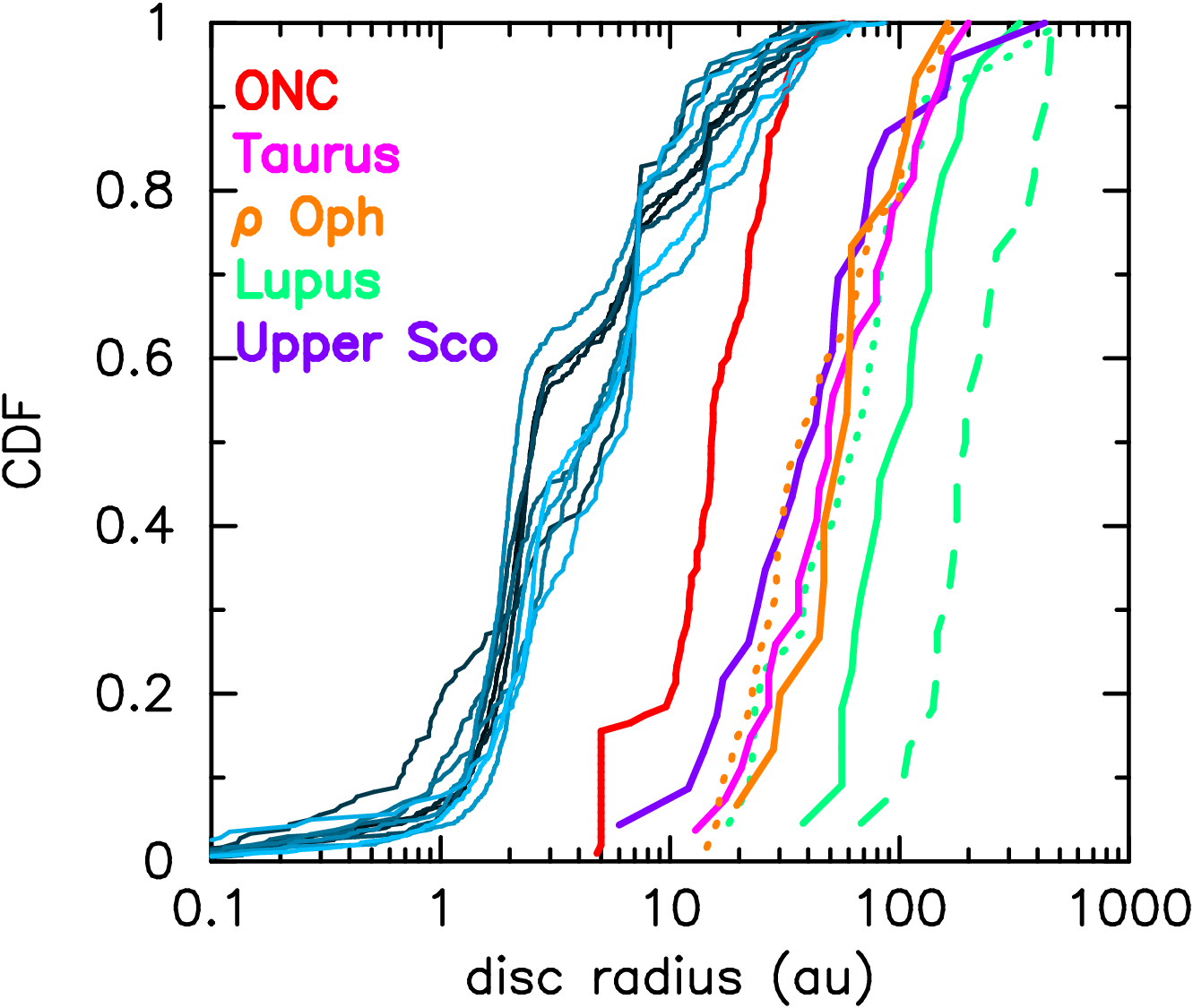}}}
\hspace*{0.3cm}\subfigure[10\,Myr]{\label{mod_dens_100au_no_viscous-c}\rotatebox{270}{\includegraphics[scale=0.23]{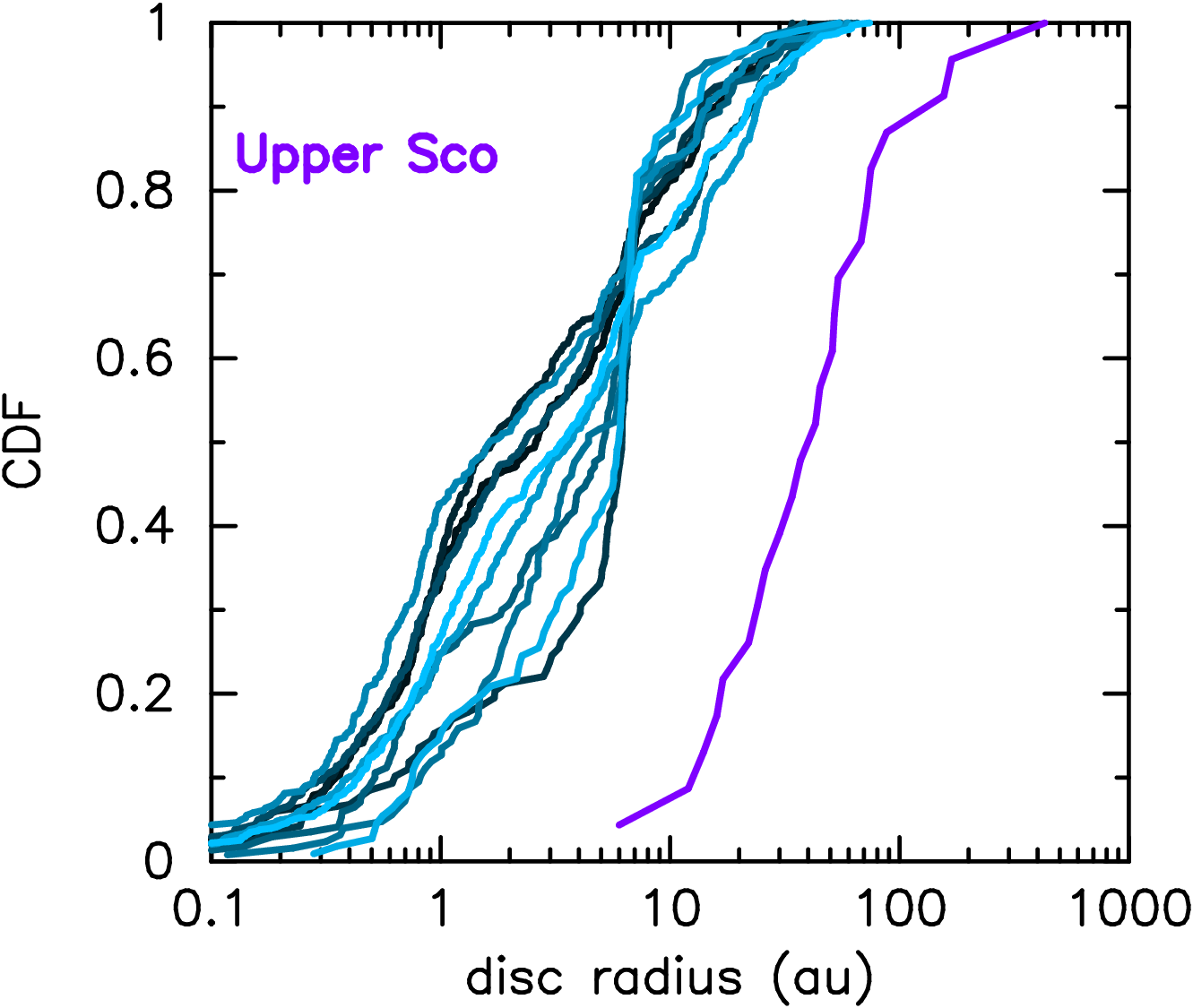}}}
\hspace*{0.3cm}\subfigure[1\,Myr]{\label{mod_dens_100au_no_viscous-d}\rotatebox{270}{\includegraphics[scale=0.23]{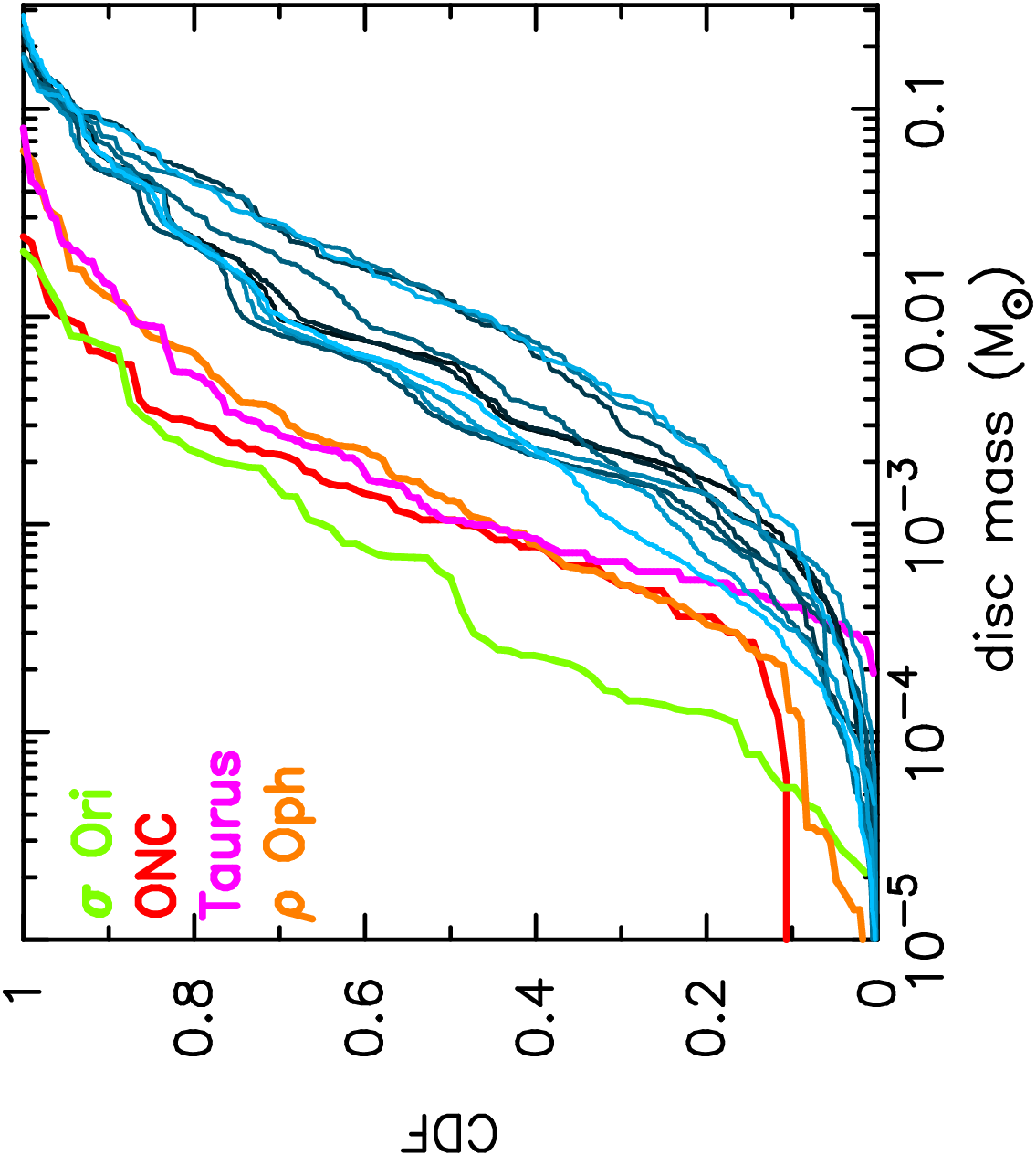}}}
\hspace*{0.3cm}\subfigure[5\,Myr]{\label{mod_dens_100au_no_viscous-e}\rotatebox{270}{\includegraphics[scale=0.23]{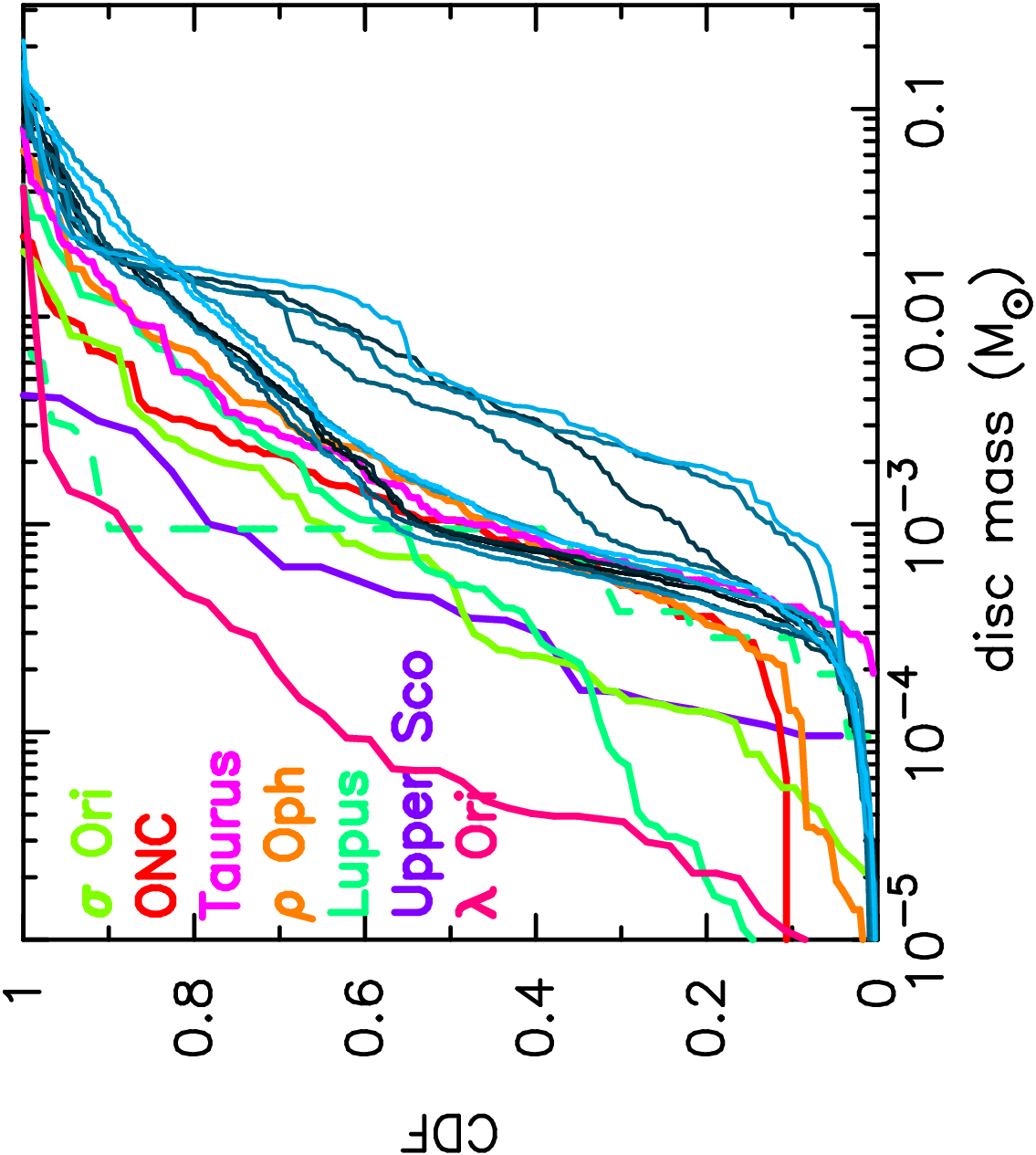}}}
\hspace*{0.3cm}\subfigure[10\,Myr]{\label{mod_dens_100au_no_viscous-f}\rotatebox{270}{\includegraphics[scale=0.23]{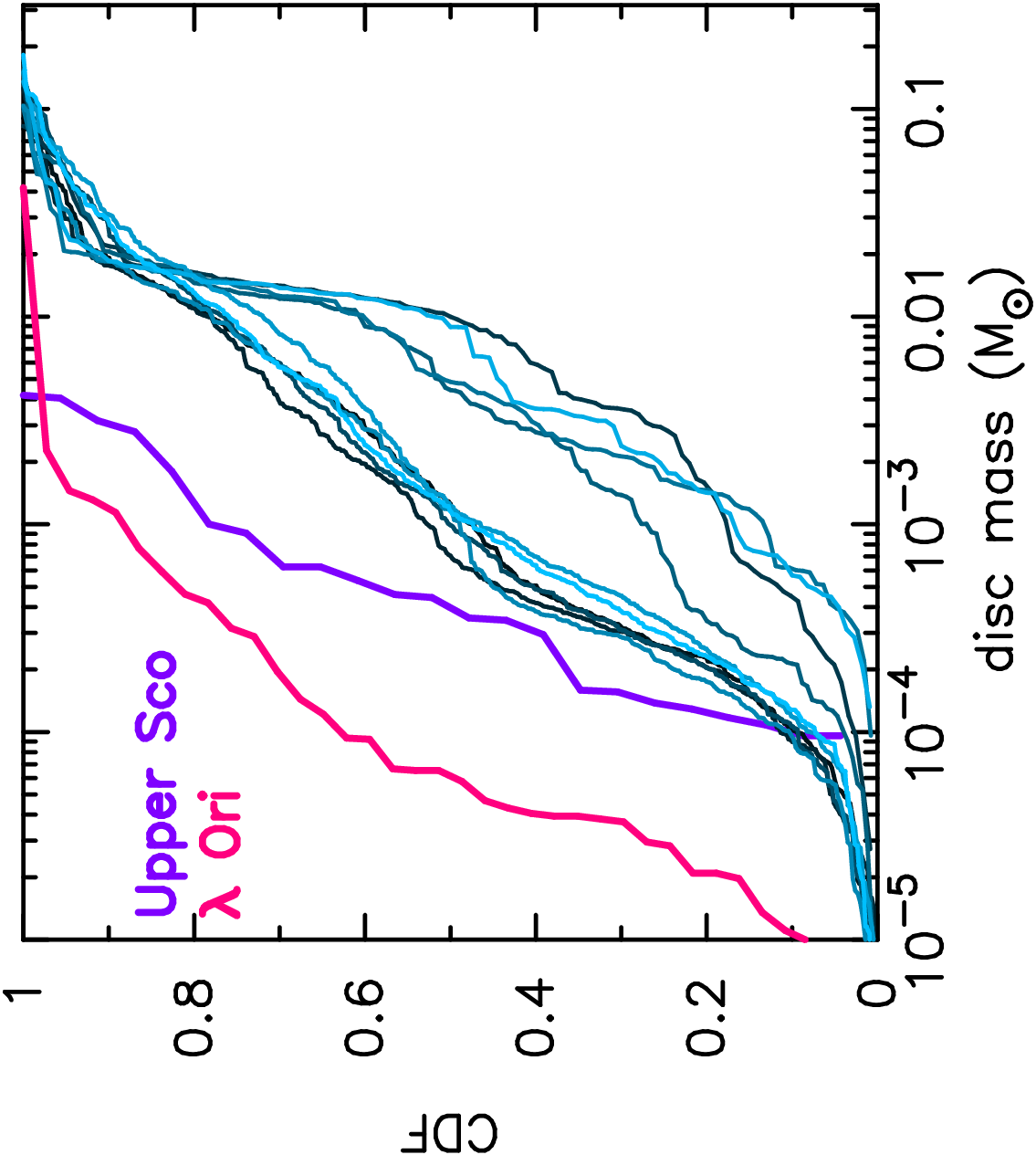}}}
\hspace*{0.3cm}\subfigure[1\,Myr]{\label{mod_dens_100au_no_viscous-g}\rotatebox{270}{\includegraphics[scale=0.23]{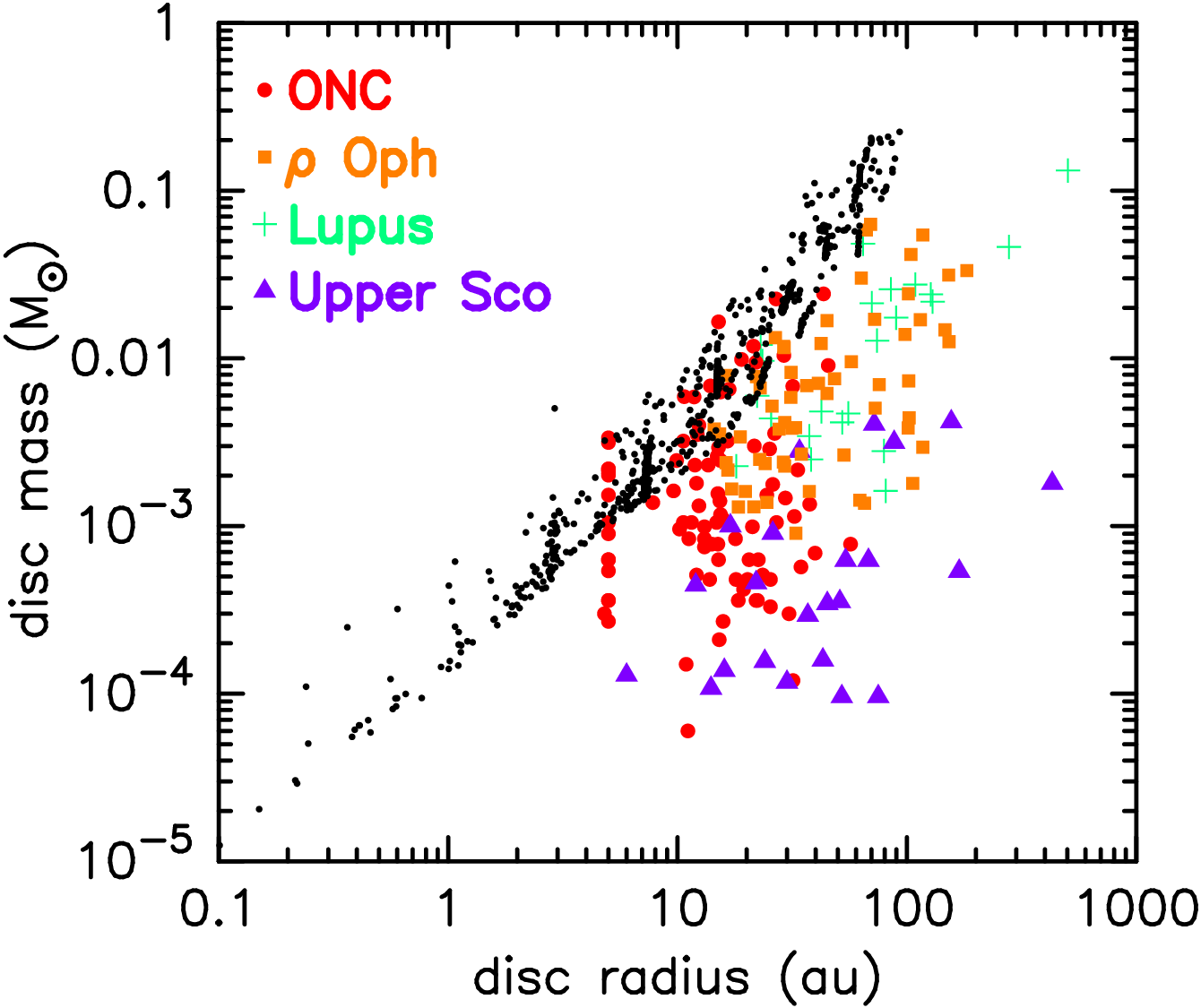}}}
\hspace*{0.3cm}\subfigure[5\,Myr]{\label{mod_dens_100au_no_viscous-h}\rotatebox{270}{\includegraphics[scale=0.23]{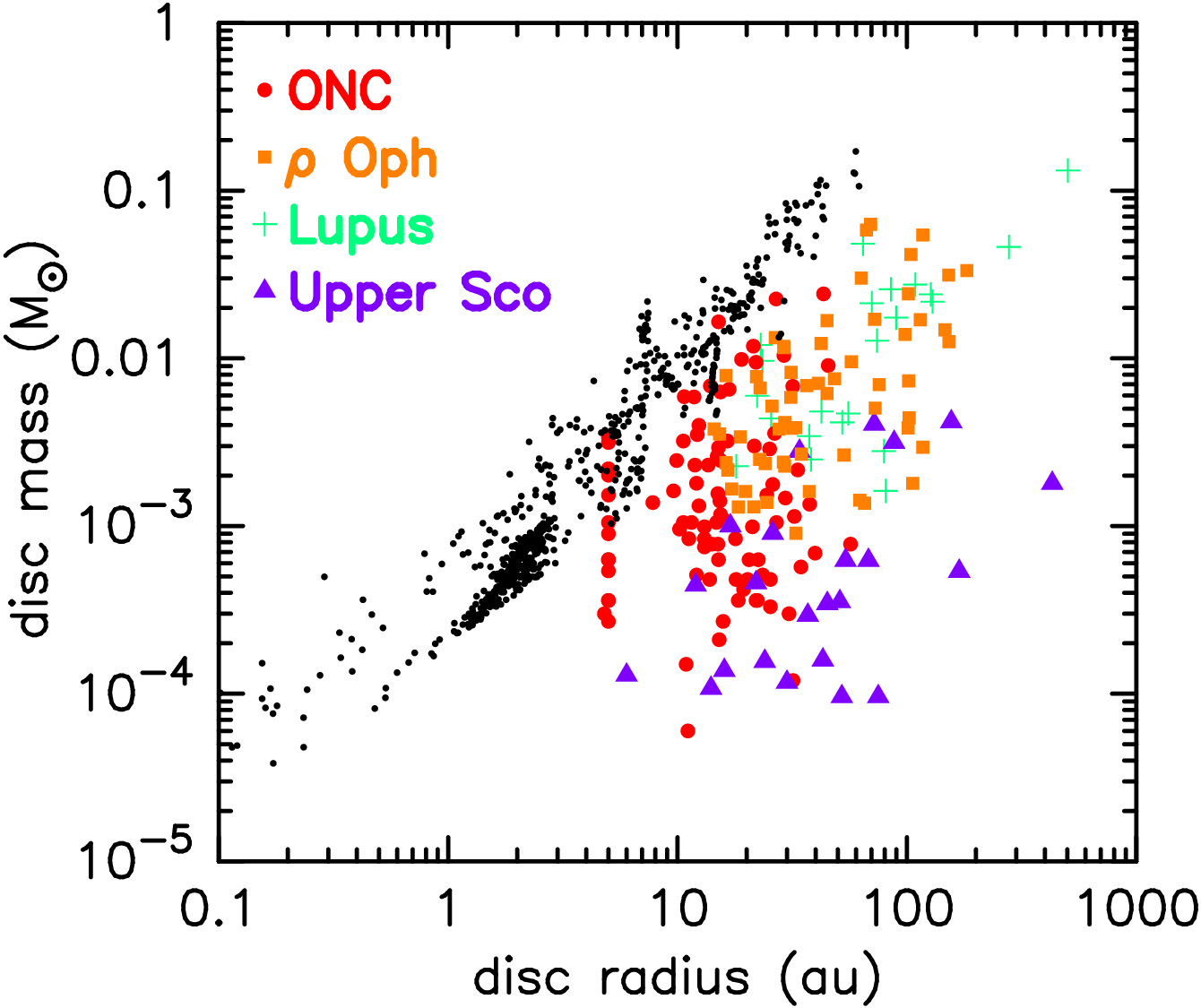}}}
\hspace*{0.3cm}\subfigure[Disc fractions]{\label{mod_dens_100au_no_viscous-i}\rotatebox{270}{\includegraphics[scale=0.23]{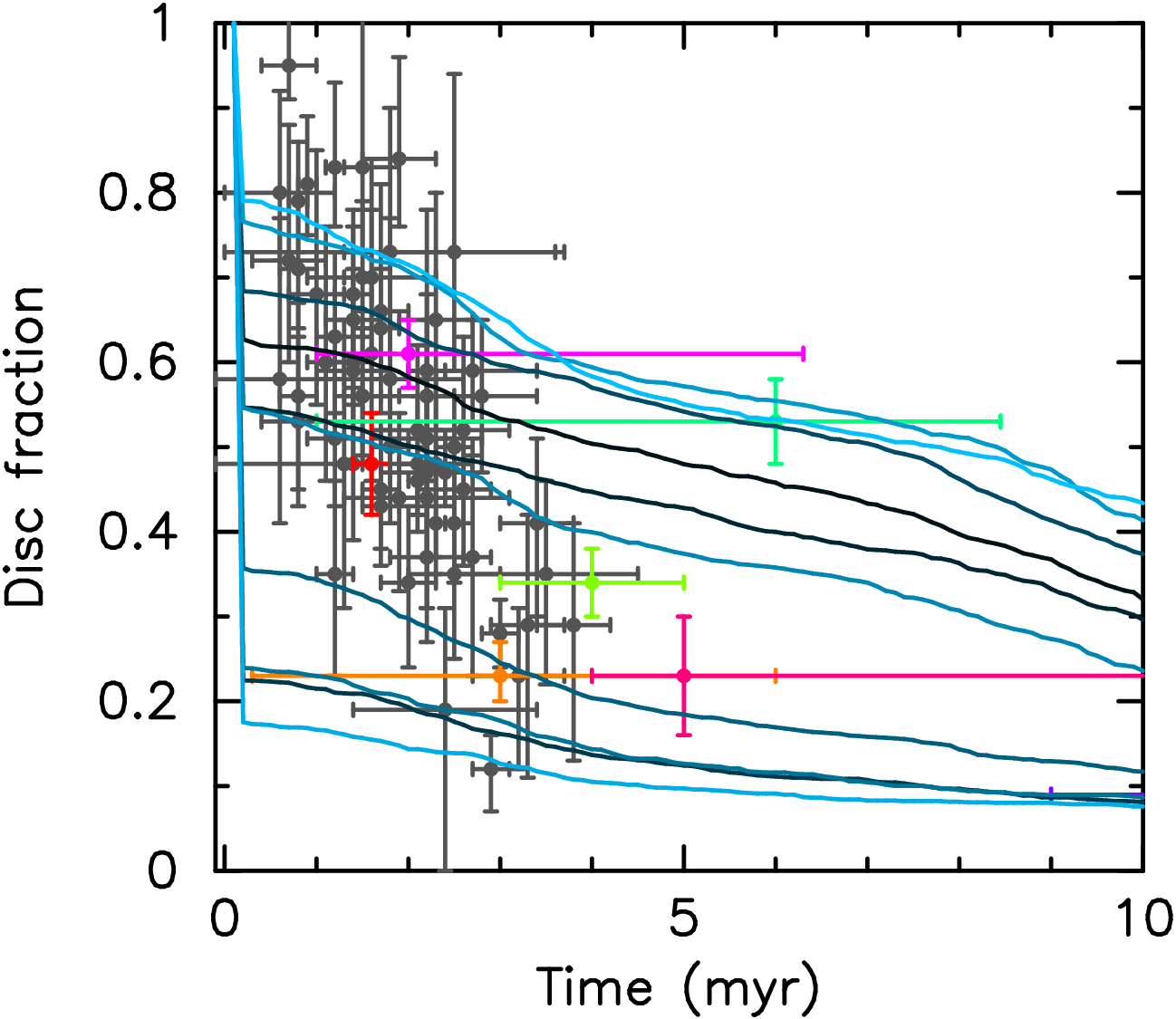}}}
\caption[bf]{Results for simulations in which the initial disc radii are drawn from a delta function with $r_{\rm disc} = 100$\,au, and the discs are only allowed to evolve inwards due to external photoevaporation.  The star-forming regions have moderate initial stellar density ($\tilde{\rho} \sim 100$\,M$_\odot$\,pc$^{-3}$). In panels (a)--(c) we show the cumulative distributions of disc radii at different ages, and in panels (d)--(f) we show the cumulative distributions of disc masses. The blue lines of varying hues show the results from the ten individual $N$-body simulations. The mint green dashed line represents the respective gas distribution in the Lupus star-forming region, whereas solid lines represent the dust distributions. Dotted lines indicate an alternative dust radius for $\rho$~Oph (orange line) and Lupus (mint green line); see Table~\ref{obs_info} for further details. In panels (g) and (h) we show the disc mass versus disc radius for four observed star-forming regions at 1 and 5\,Myr, and data from one post-processed $N$-body simulation are shown by the black points. We plot the disc fractions (defined as when the discs have non-zero mass) as a function of time in our $N$-body simulations in panel (i), with observational data taken from \citet{Richert18} and \citet{Ribas15}.}
\label{mod_dens_100au_no_viscous}
  \end{center}
\end{figure*}

We first show an example where the disc radii evolve inwards according to Eqn~\ref{rescale_disc}, but where there is no viscous spreading. In Fig.~\ref{mod_dens_100au_no_viscous} the initial density of the star-forming region is 100\,M$_\odot$\,pc$^{-3}$, the region contains $N = 1500$ stars and the disc radii are all initially $r_d = 100$\,au. We show the disc radii (top row) and disc mass (middle row) distributions at 1, 5 and 10\,Myr, and show the corresponding observational data for the younger regions (lefthand panels), all regions (middle panels) and the oldest regions (righthand panels).

As mass is lost from the discs due to photoevaporation, the radii move inwards, but at such a rate that the simulated disc radii distributions are not consistent with those in the observed star-forming regions at any age (this discrepancy is worse for the simulations where the discs have even smaller (10\,au) initial radii).

Despite the seemingly dramatic evolution of the discs, the mass loss from the discs due to photoevaporation is relatively modest; no simulations are consistent with the observations of discs at ages of $\sim$1\,Myr. There is some overlap between our simulations and the observed regions at 5\,Myr \citep[determining accurate stellar ages and hence disc ages is notoriously difficult,][]{Bell13}, but at 5\,Myr the disc radii distributions are completely inconsistent with the observed distributions.

We show the individual disc mass versus disc radius for one of our simulations at 1 and 5\,Myr in Figs.~\ref{mod_dens_100au_no_viscous-g}~and~\ref{mod_dens_100au_no_viscous-h}. The structure in the simulation data is caused by the resolution of the \texttt{FRIED} grid. When a disc loses mass due to photoevaporation, its radius moves inwards. However, if the new radius is closer to a different subset of  \texttt{FRIED} models with smaller radii, the rate of photoevaporation decreases and a `pile-up' of points occurs. Both of these plots show that the  masses of the surviving discs are too high to be consistent with the observational data. 

Finally, we plot the evolution of the protoplanetary disc fraction in these simulations (i.e. the number of stars that have some ($M_{\rm disc} > 0)$ disc mass at a given time -- in practice most of the remaining discs are always $\gtrsim 10^{-3}$\,M$_\odot$ -- divided by the total number of stars that had discs initially) in Fig.~\ref{mod_dens_100au_no_viscous-i}. The simulation data are shown by the blues lines of various hues, and in this plot we also show the observed disc fractions in star-forming regions.

The simulations display a wide range of disc fractions; in some models up to 80\,per cent of the discs are destroyed very quickly, whereas in others the destruction rate is much lower (20 - 50\,per cent).

\subsection{Including viscous evolution}

We now present the results of simulations where in addition to inward evolution of the disc radius due to photoevaporation, we allow the discs to spread outwards due to viscous evolution. We keep the other parameters in the simulation fixed (i.e.,\,initial stellar density, which remains at 100\,M$_\odot$\,pc$^{-3}$, and number of stars, $N = 1500$). The amount of viscous spreading depends quite strongly on the chosen viscous parameter, $\alpha$, which is set to $\alpha = 10^{-2}$. In Appendix~\ref{appendix_viscosity} we show results for $\alpha = 10^{-4}$, which are very similar to the simulations where the discs evolve inwards due to photoevaporation only.  

\subsubsection{Initial disc radii $r_{\rm disc} = 100$\,au}

\begin{figure*}
  \begin{center}
\setlength{\subfigcapskip}{10pt}
\hspace*{0.3cm}\subfigure[1\,Myr]{\label{mod_dens_100au-a}\rotatebox{270}{\includegraphics[scale=0.23]{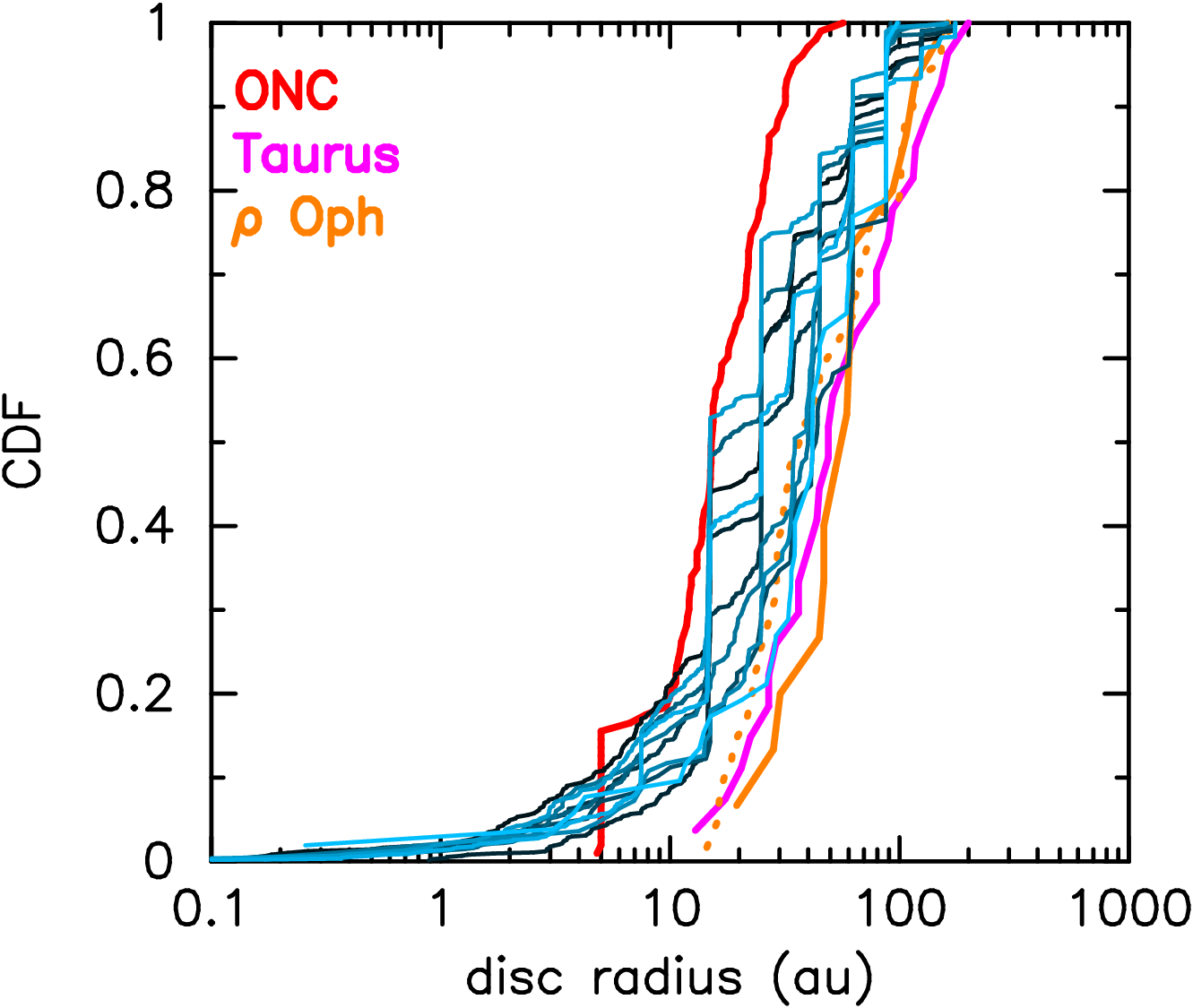}}}
\hspace*{0.3cm}\subfigure[5\,Myr]{\label{mod_dens_100au-b}\rotatebox{270}{\includegraphics[scale=0.23]{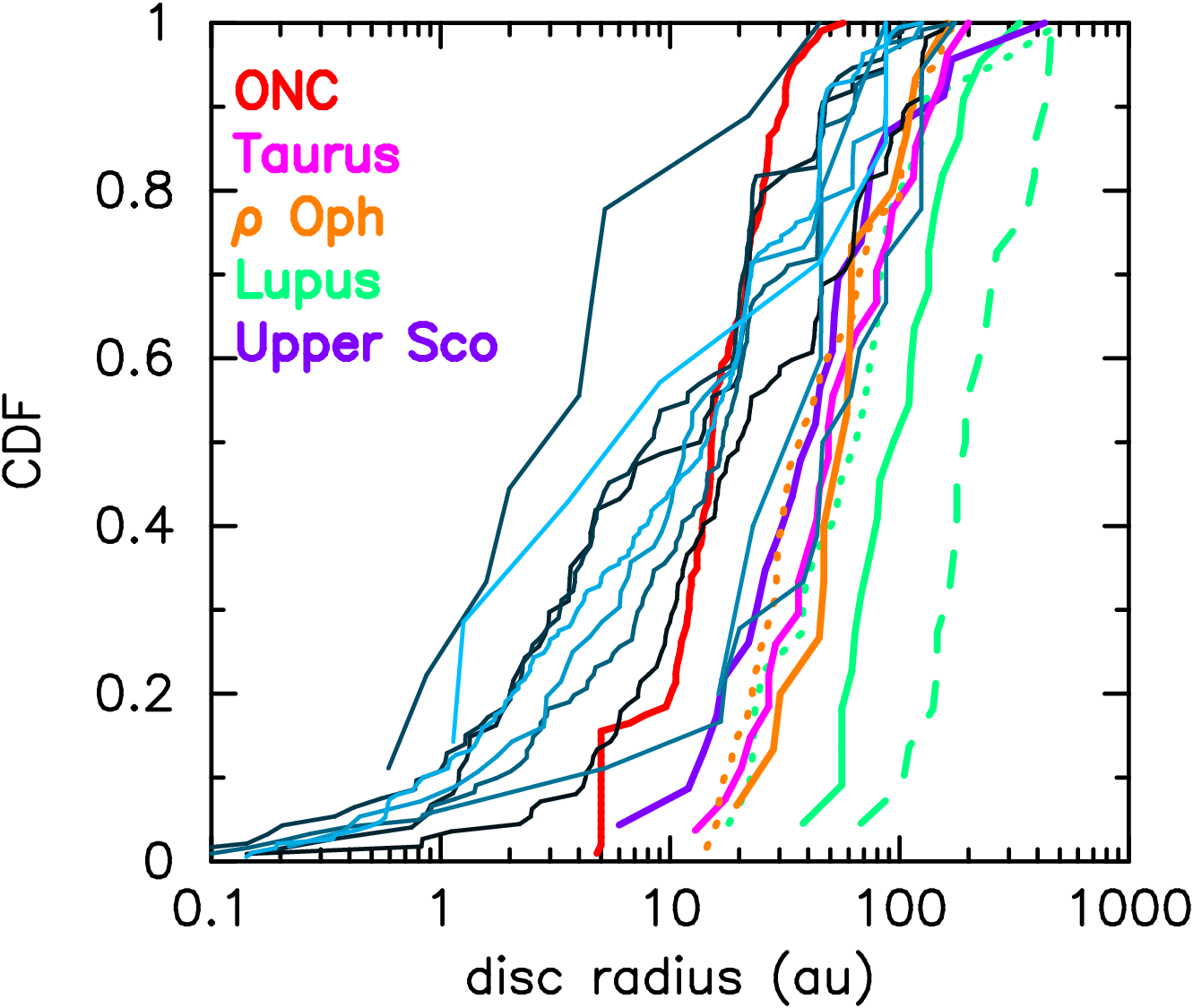}}}
\hspace*{0.3cm}\subfigure[10\,Myr]{\label{mod_dens_100au-c}\rotatebox{270}{\includegraphics[scale=0.23]{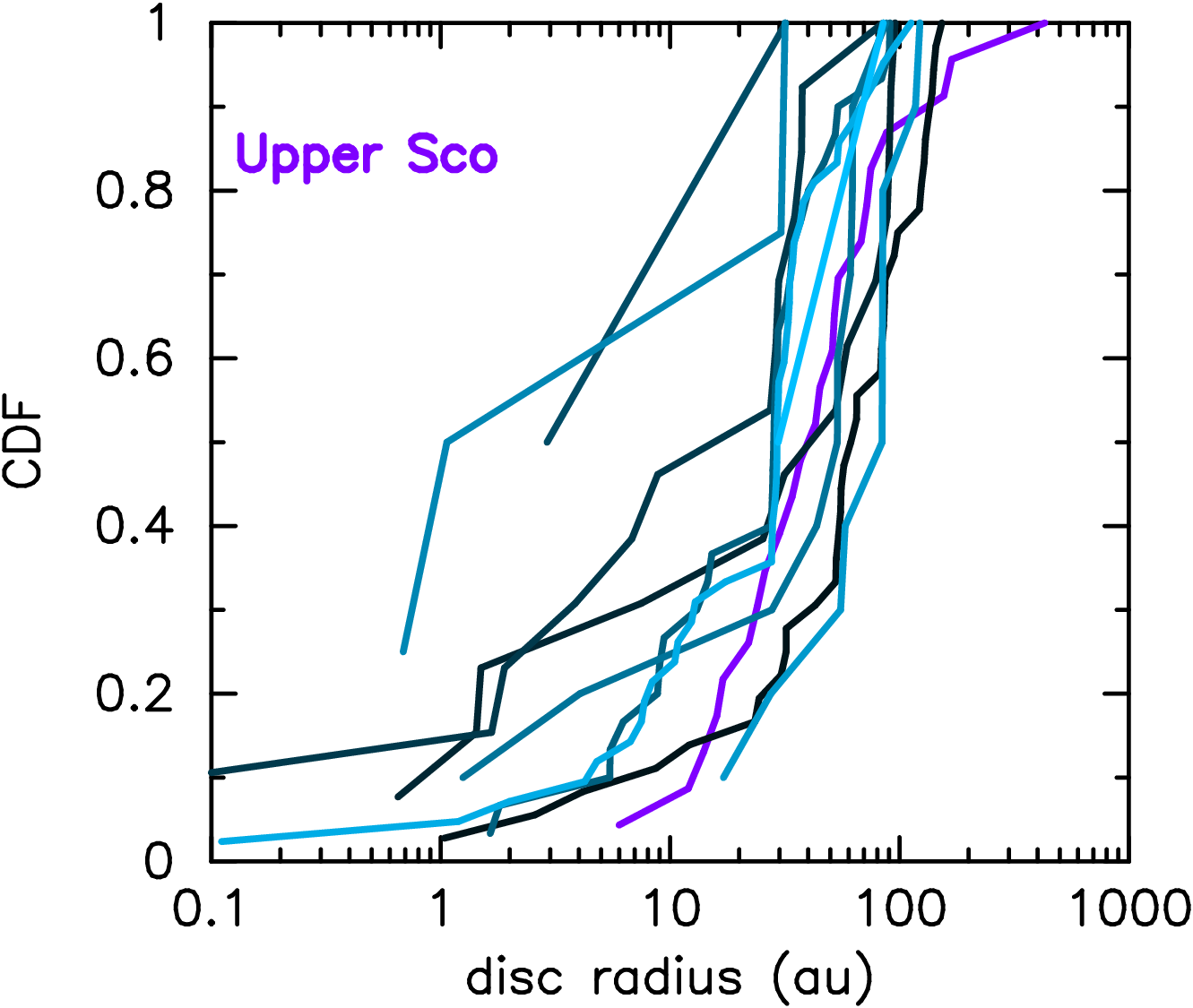}}}
\hspace*{0.3cm}\subfigure[1\,Myr]{\label{mod_dens_100au-d}\rotatebox{270}{\includegraphics[scale=0.23]{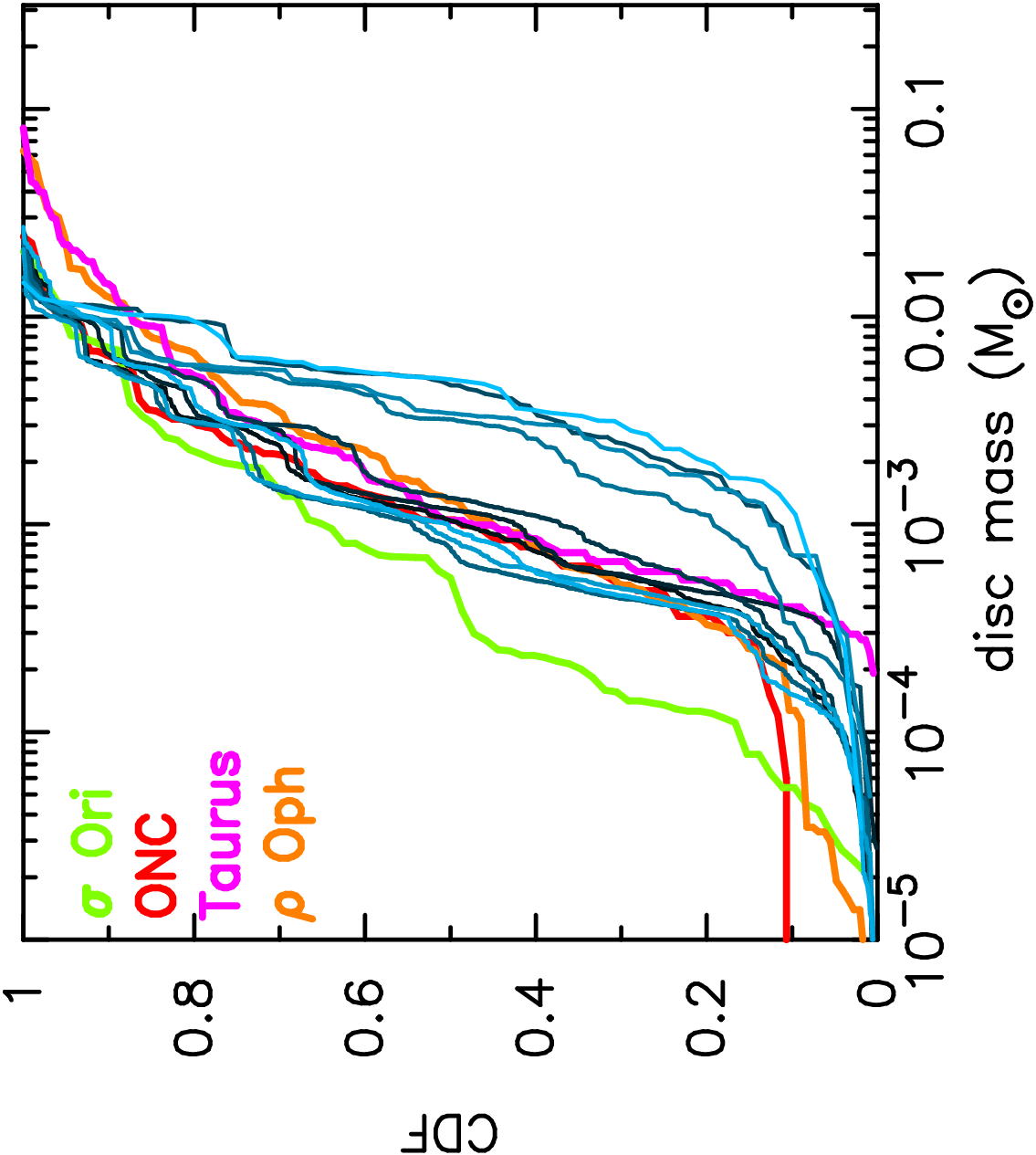}}}
\hspace*{0.3cm}\subfigure[5\,Myr]{\label{mod_dens_100au-e}\rotatebox{270}{\includegraphics[scale=0.23]{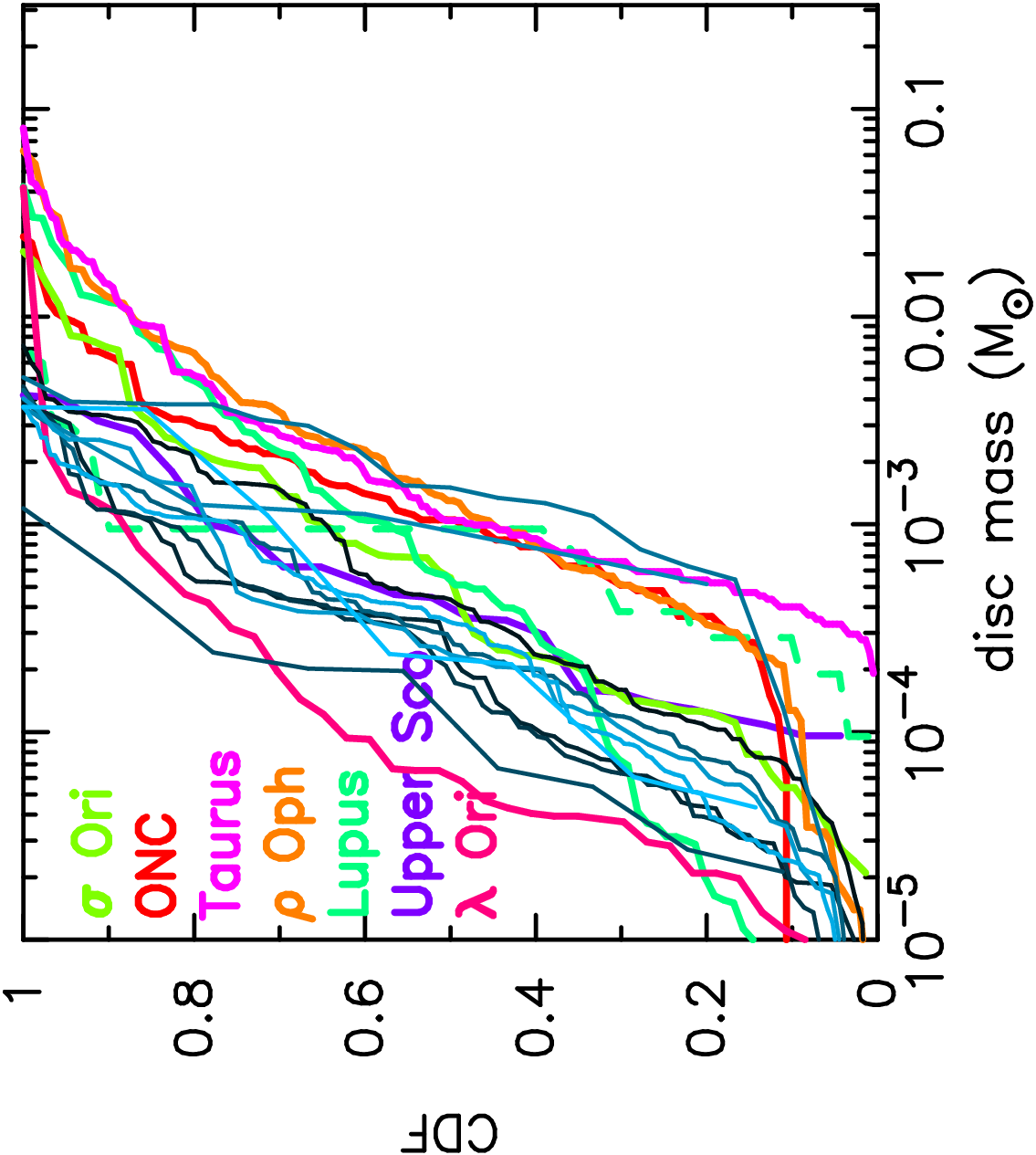}}}
\hspace*{0.3cm}\subfigure[10\,Myr]{\label{mod_dens_100au-f}\rotatebox{270}{\includegraphics[scale=0.23]{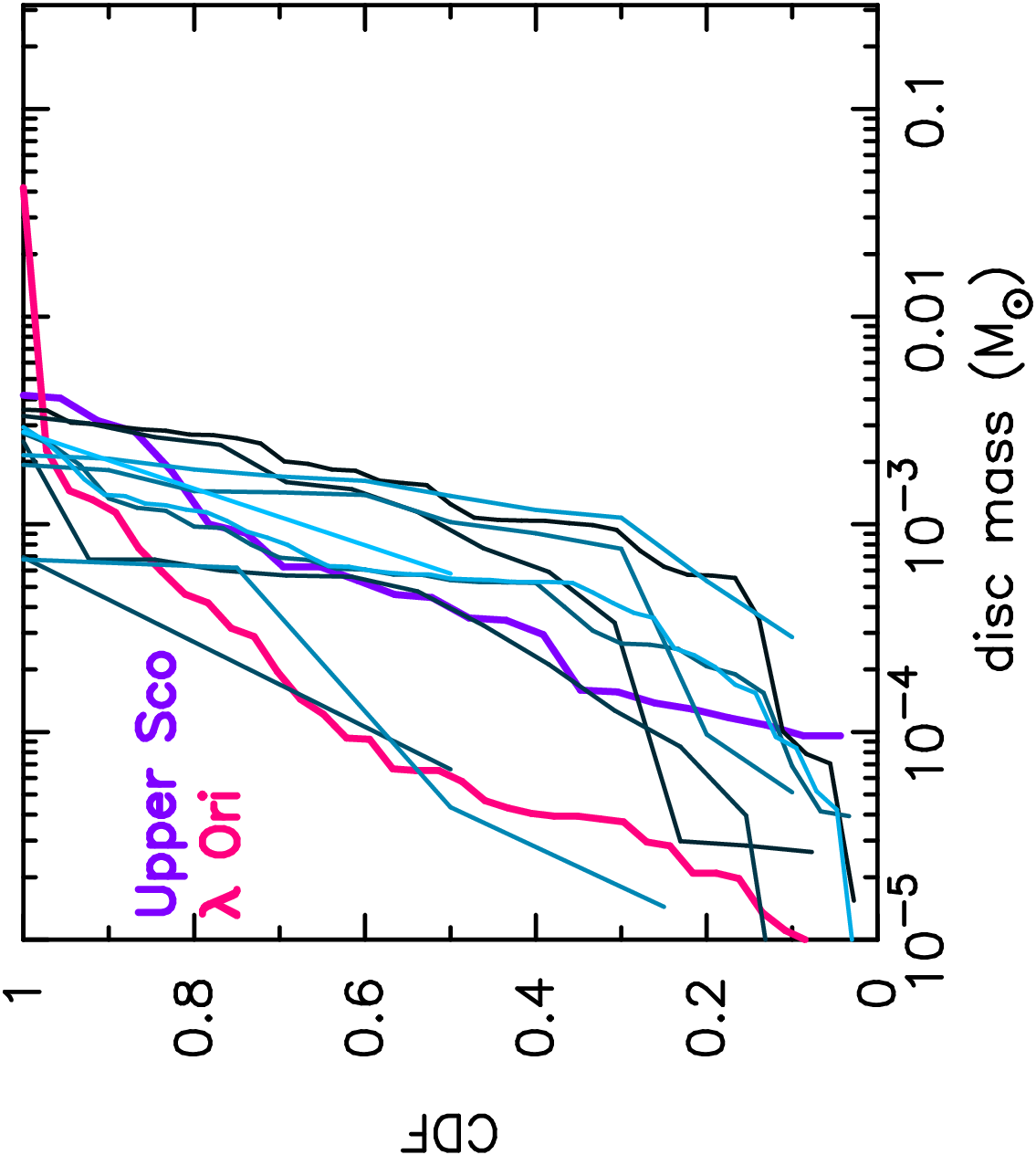}}}
\hspace*{0.3cm}\subfigure[mass vs radius 1\,Myr]{\label{mod_dens_100au-g}\rotatebox{270}{\includegraphics[scale=0.23]{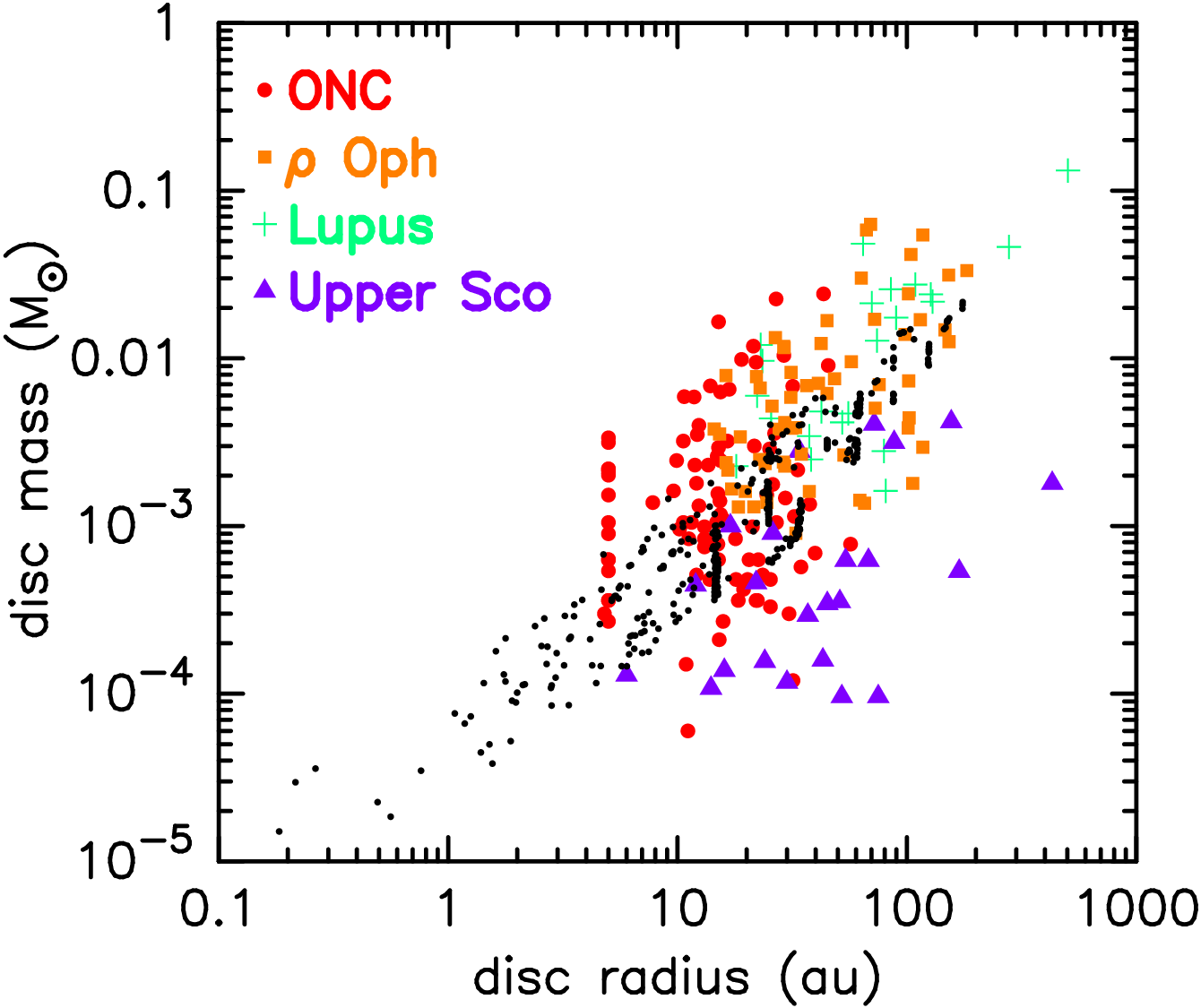}}}
\hspace*{0.3cm}\subfigure[5\,Myr]{\label{mod_dens_100au-h}\rotatebox{270}{\includegraphics[scale=0.23]{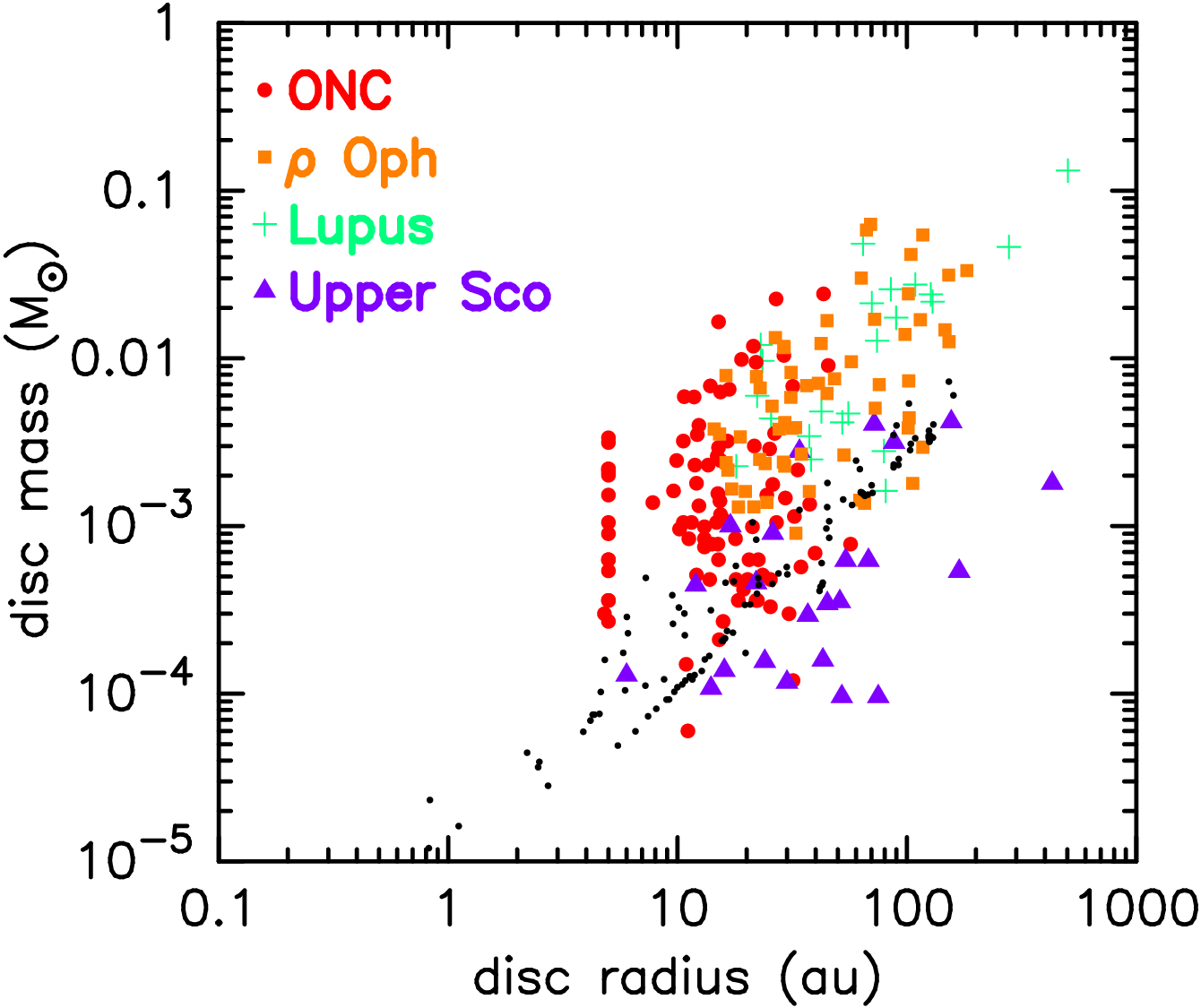}}}
\hspace*{0.3cm}\subfigure[Disc fractions]{\label{mod_dens_100au-i}\rotatebox{270}{\includegraphics[scale=0.23]{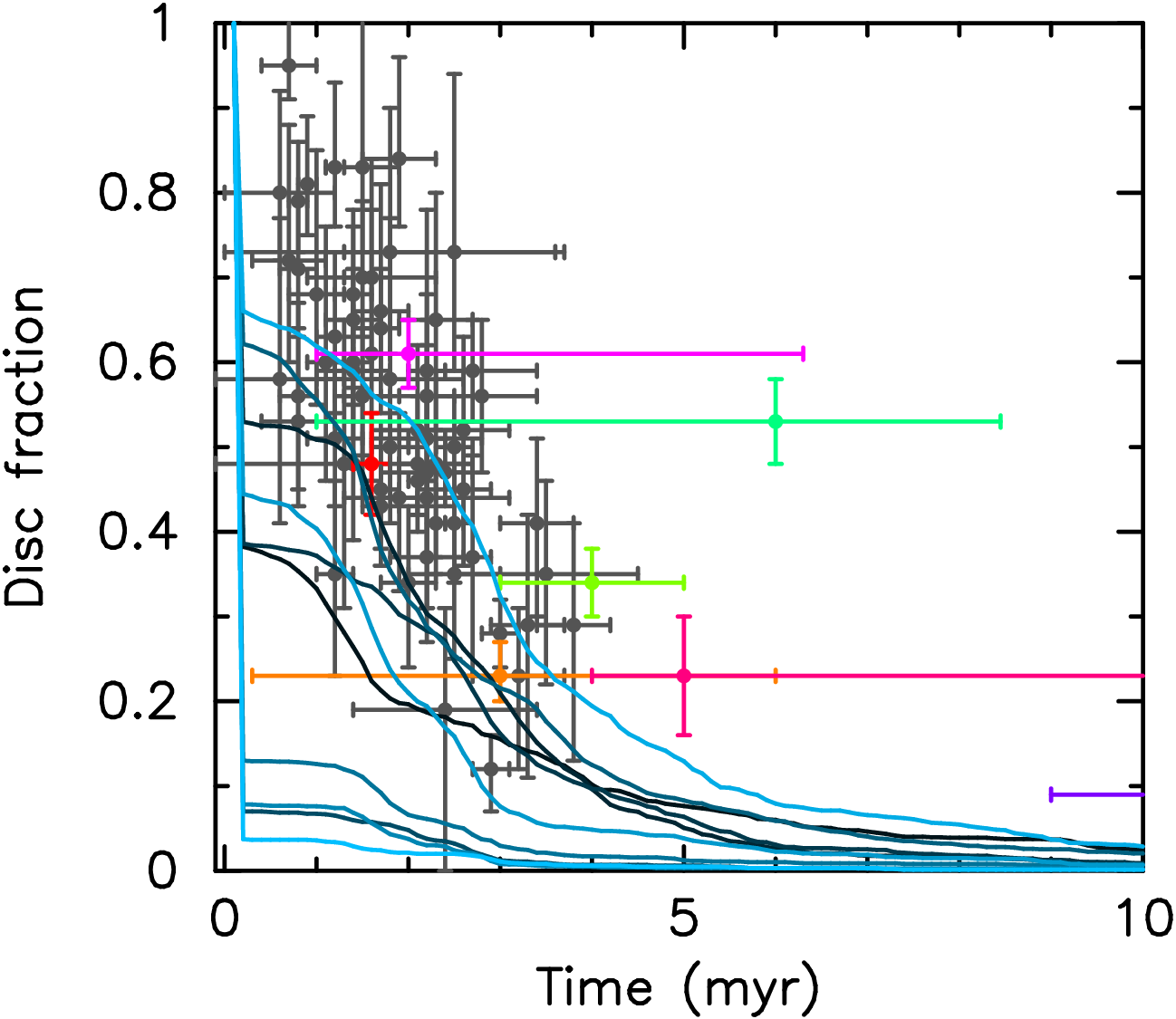}}}
\caption[bf]{Results for simulations in which the initial disc radii are drawn from a delta function with $r_{\rm disc} = 100$\,au, and the discs are allowed to evolve inwards due to external photoevaporation, and outwards due to viscous spreading.  The star-forming regions have moderate initial stellar density ($\tilde{\rho} \sim 100$\,M$_\odot$\,pc$^{-3}$). In panels (a)--(c) we show the cumulative distributions of disc radii at different ages, and in panels (d)--(f) we show the cumulative distributions of disc masses. The blue lines of varying hues show the results from the ten individual $N$-body simulations. The mint green dashed line represents the respective gas distribution in the Lupus star-forming region, whereas solid lines represent the dust distributions. Dotted lines indicate an alternative dust radius for $\rho$~Oph (orange line) and Lupus (mint green line); see Table~\ref{obs_info} for further details. In panels (g) and (h) we show the disc mass versus disc radius for four observed star-forming regions at 1 and 5\,Myr, and data from one post-processed $N$-body simulation are shown by the black points. We plot the disc fractions (defined as when the discs have non-zero mass) as a function of time in our $N$-body simulations in panel (i), with observational data taken from \citet{Richert18} and \citet{Ribas15}.}
\label{mod_dens_100au}
  \end{center}
\end{figure*}

In our first model with viscous evolution, we present the evolution of discs with initial radii all set to $r_{\rm disc} = 100$\,au in Fig.~\ref{mod_dens_100au}. In the first 5\,Myr, the effects of photoevaporation cause the disc radii to decrease, but this is countered by the effects of viscous spreading. When no viscous evolution was included (in Fig.~\ref{mod_dens_100au_no_viscous}), mass loss from photoevaporation caused the inward decrease in radius, which in turn decreased the potency of photoevaporative mass loss in the next timestep (as the surface density of material at the edge of the disc increased). However, when the disc then reacts to viscous spreading the radius increases, lowering the surface density of material and making the disc more susceptible to further mass-loss due to photoevaporation.

In four out of ten of our simulations, the majority ($>$85\,percent) of discs are destroyed within the first Myr (see Fig.~\ref{mod_dens_100au-i}), and in the remaining six simulations the majority of discs are destroyed after 5 -- 6 Myr. This is evident in the cumulative distributions of the disc radii at 5 and 10\,Myr (panels b \& c), where some of the lines only contain a handful of systems.

After 1\,Myr of evolution, the disc radii in the simulations sit between the observed distributions in the ONC (the red line) and in Taurus and $\rho$~Oph (the orange and pink lines), and then after 5\,Myr the majority of disc radii from the simulations are smaller than the observed distribution in the ONC and the other regions (Fig.~\ref{mod_dens_100au-b}). As shown in  Fig.~\ref{mod_dens_100au-i}, most of the discs are destroyed at 10\,Myr but those that survive are generally consistent with the observed distribution for Upper~Sco (the purple line in Fig.~\ref{mod_dens_100au-c}).

The disc mass distributions follow a similar pattern. After 1\,Myr of evolution, most of the simulated disc distributions are consistent with the observed distributions for the ONC, Taurus and $\rho$~Oph. At 5\,Myr, the disc mass cumulative distributions from the simulated data have moved to lower values, but individual simulations are also consistent with the majority of the observed data.

As more of the discs are destroyed between 5 and 10\,Myr, there is a slight systematic shift in the cumulative distributions of the simulated discs to higher masses. The discs are not gaining mass, but rather the low-mass discs present at earlier times have been destroyed, and the cumulative distributions are dominated by the surviving, more massive discs. (At this stage in the simulations, the FUV radiation field is decreasing in strength as the star-forming regions expand, reducing the amount of mass-loss from discs, \citealp{Parker21a}.)

The individual disc mass versus disc radius plots (Figs.~\ref{mod_dens_100au-g}~and~\ref{mod_dens_100au-h}) are consistent with the observations after 1\,Myr, but then display lower masses than the majority of the observed discs, save for the older poulation in Upper~Sco. In addition, the simulated disc masses are also higher than the observed disc fractions (Fig.~\ref{mod_dens_100au-i}).

\subsubsection{Initial disc radii $r_{\rm disc} = 10$\,au}

\begin{figure*}
  \begin{center}
\setlength{\subfigcapskip}{10pt}
\hspace*{0.3cm}\subfigure[1\,Myr]{\label{mod_dens_10au-a}\rotatebox{270}{\includegraphics[scale=0.23]{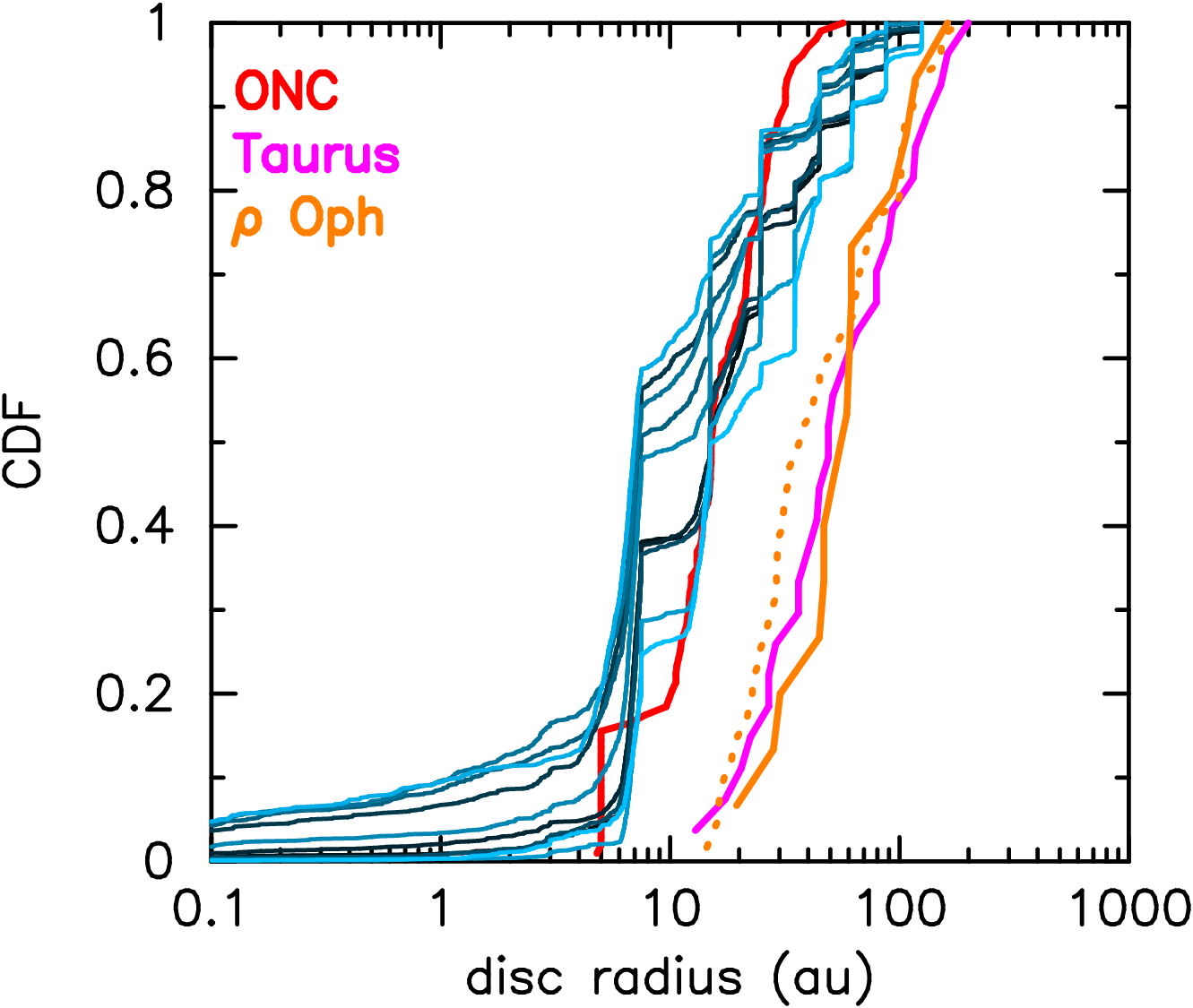}}}
\hspace*{0.3cm}\subfigure[5\,Myr]{\label{mod_dens_10au-b}\rotatebox{270}{\includegraphics[scale=0.23]{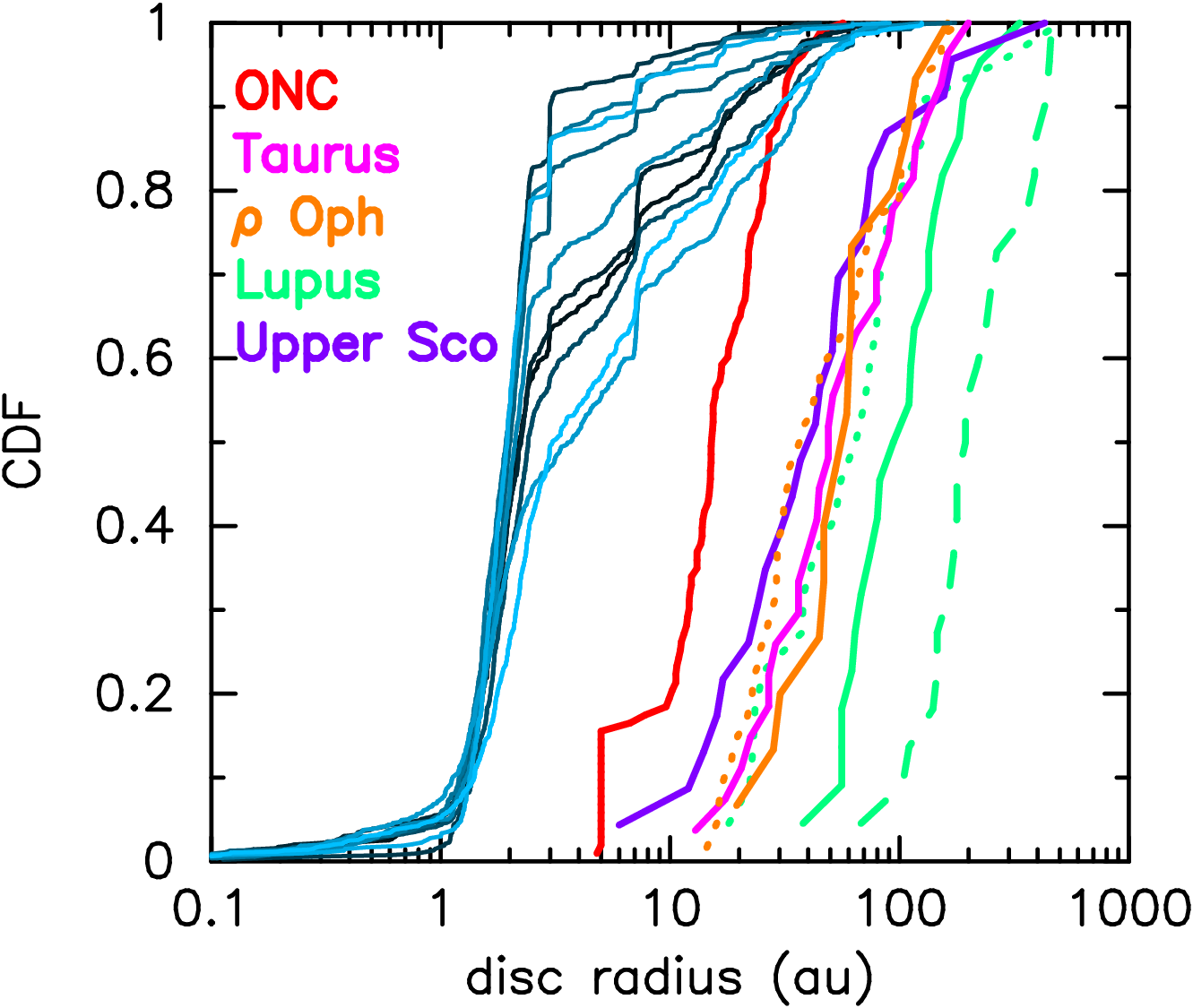}}}
\hspace*{0.3cm}\subfigure[10\,Myr]{\label{mod_dens_10au-c}\rotatebox{270}{\includegraphics[scale=0.23]{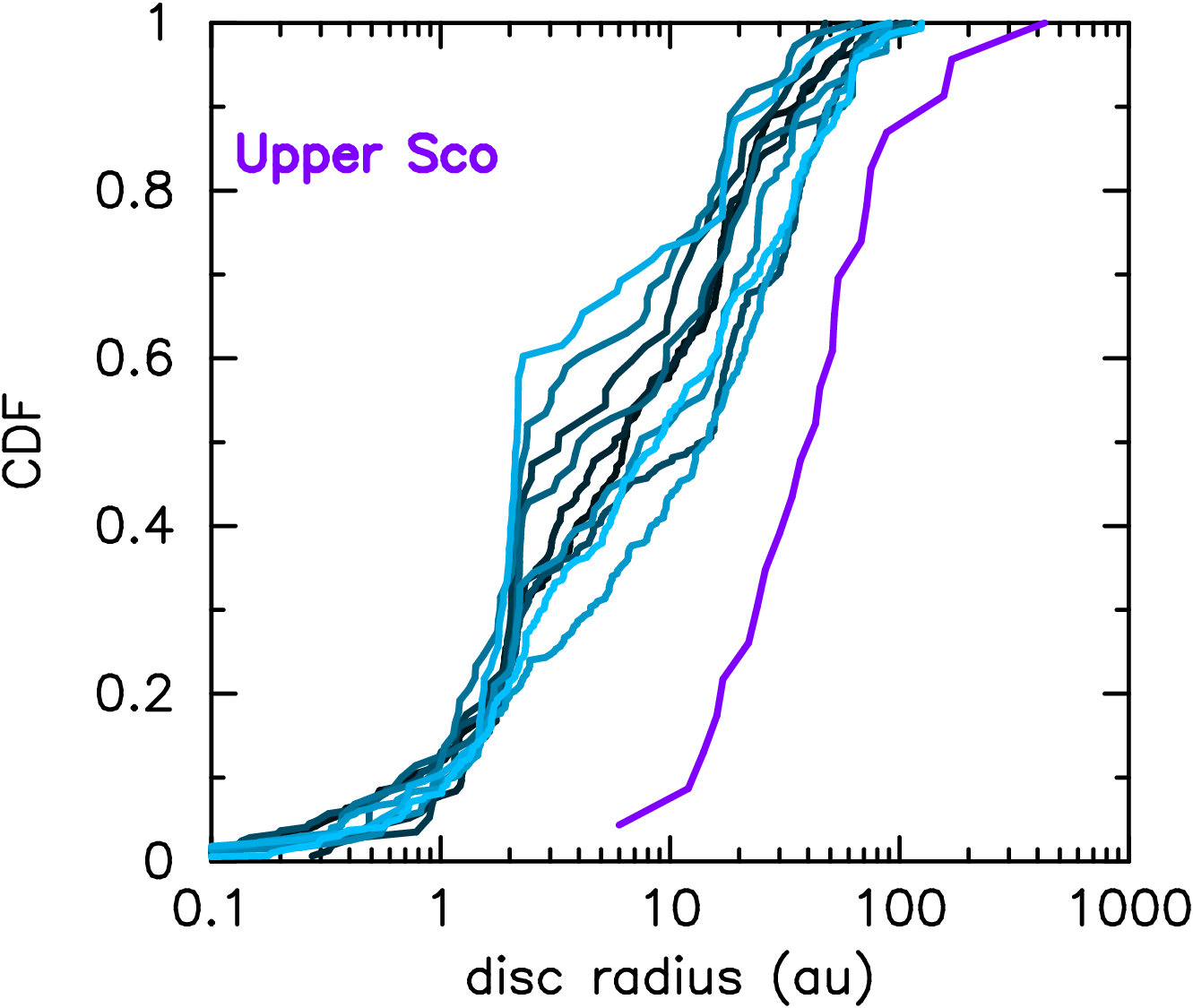}}}
\hspace*{0.3cm}\subfigure[1\,Myr]{\label{mod_dens_10au-d}\rotatebox{270}{\includegraphics[scale=0.23]{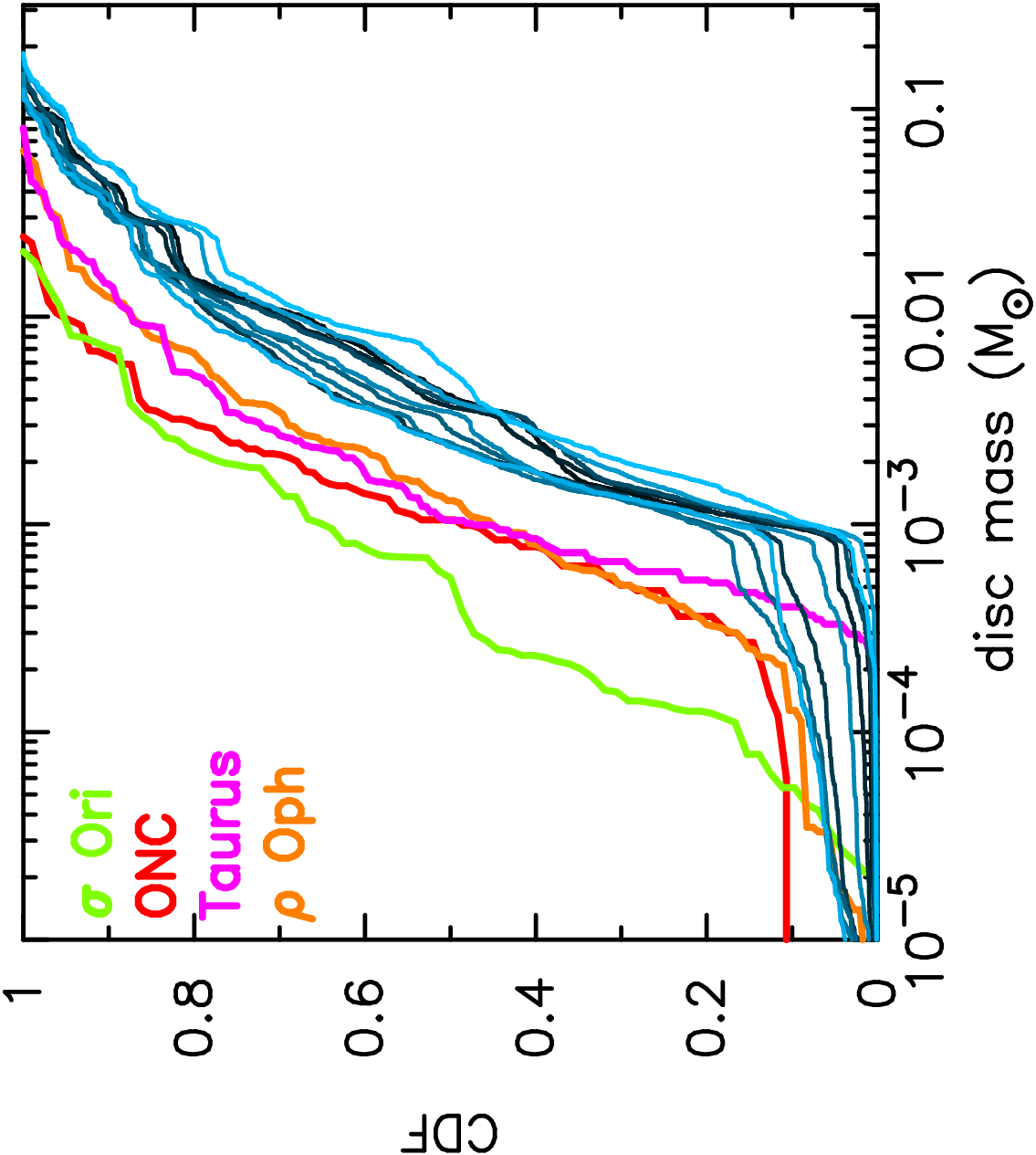}}}
\hspace*{0.3cm}\subfigure[5\,Myr]{\label{mod_dens_10au-e}\rotatebox{270}{\includegraphics[scale=0.23]{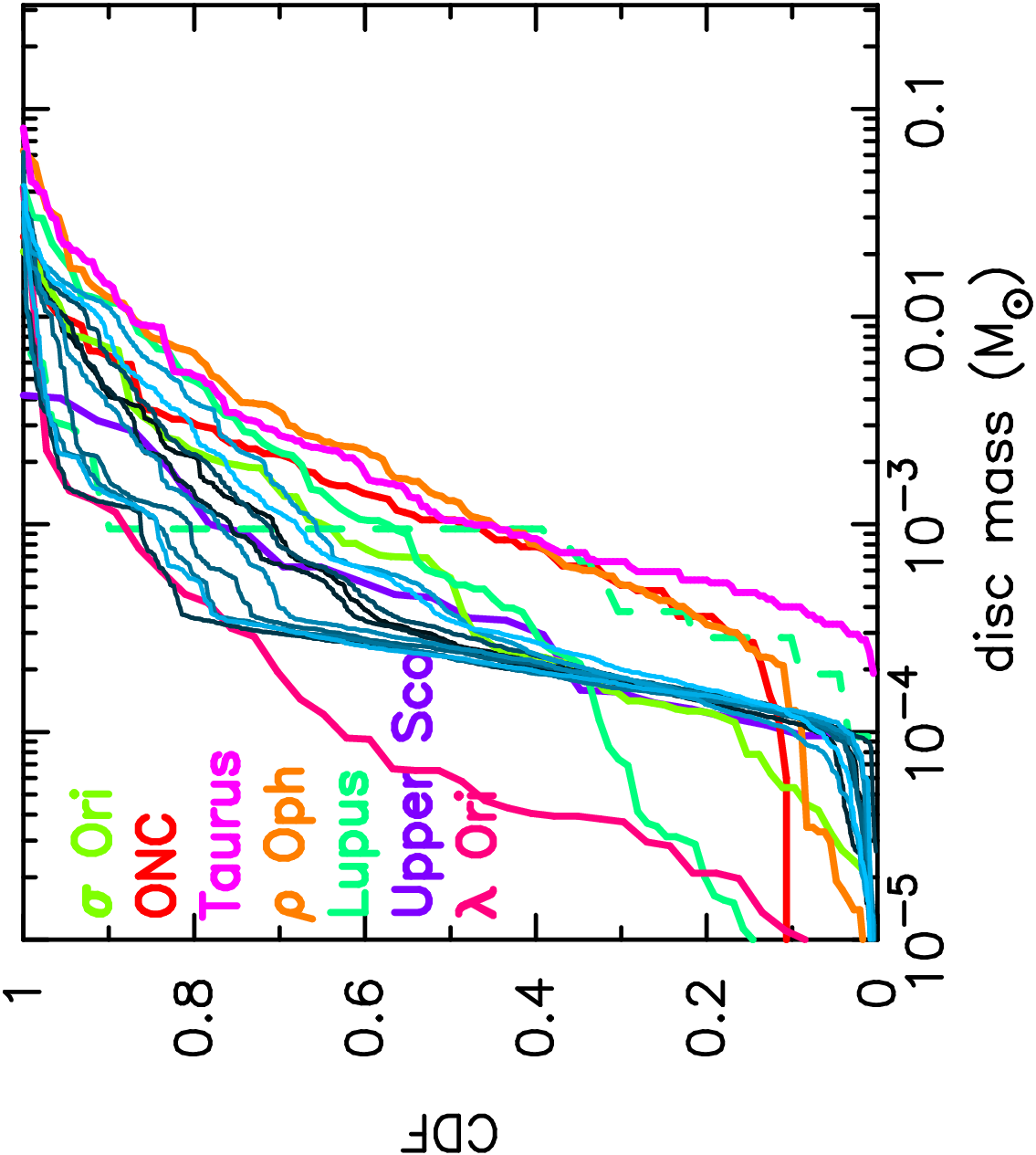}}}
\hspace*{0.3cm}\subfigure[10\,Myr]{\label{mod_dens_10au-f}\rotatebox{270}{\includegraphics[scale=0.23]{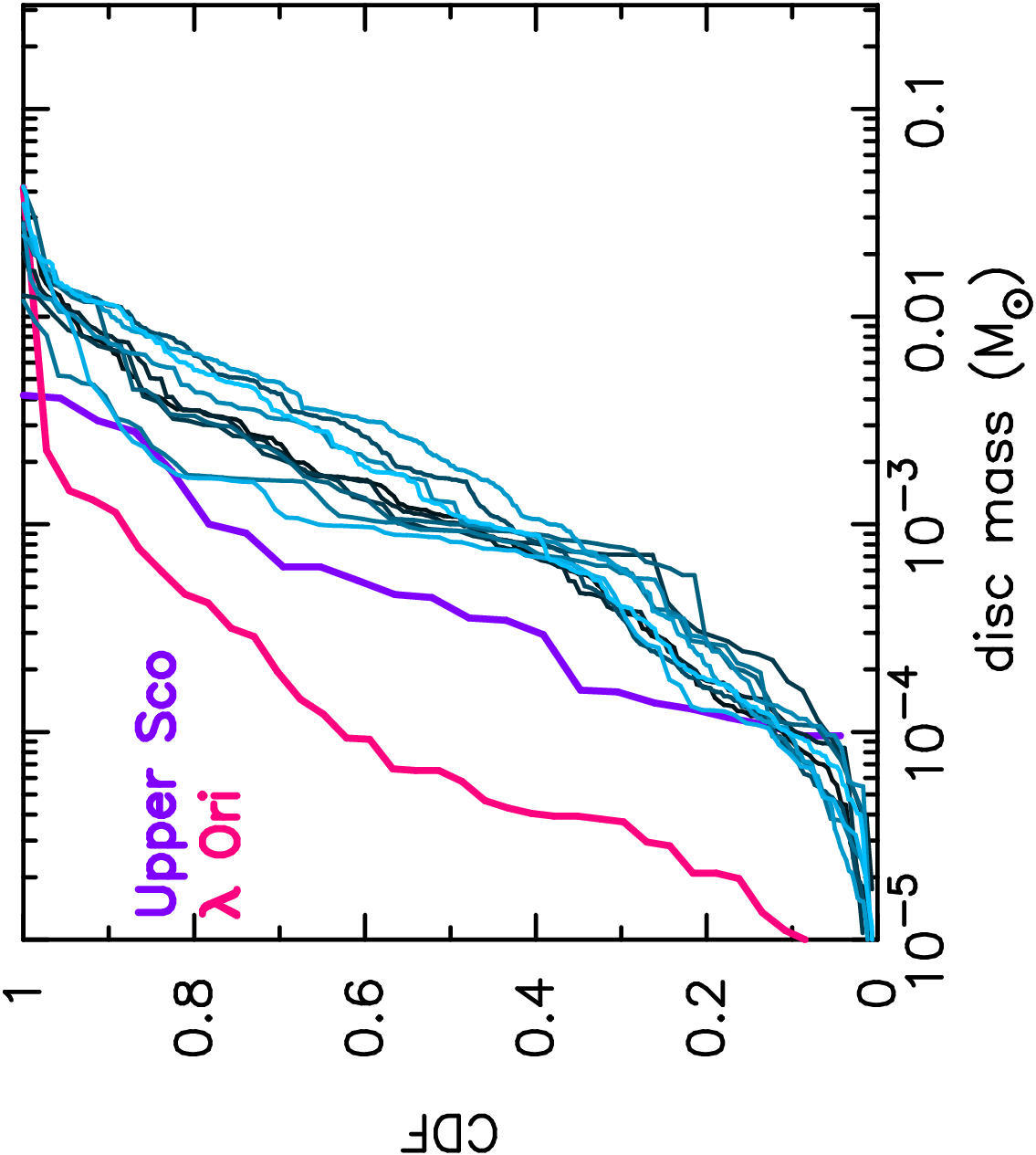}}}
\hspace*{0.3cm}\subfigure[mass vs radius 1\,Myr]{\label{mod_dens_10au-g}\rotatebox{270}{\includegraphics[scale=0.23]{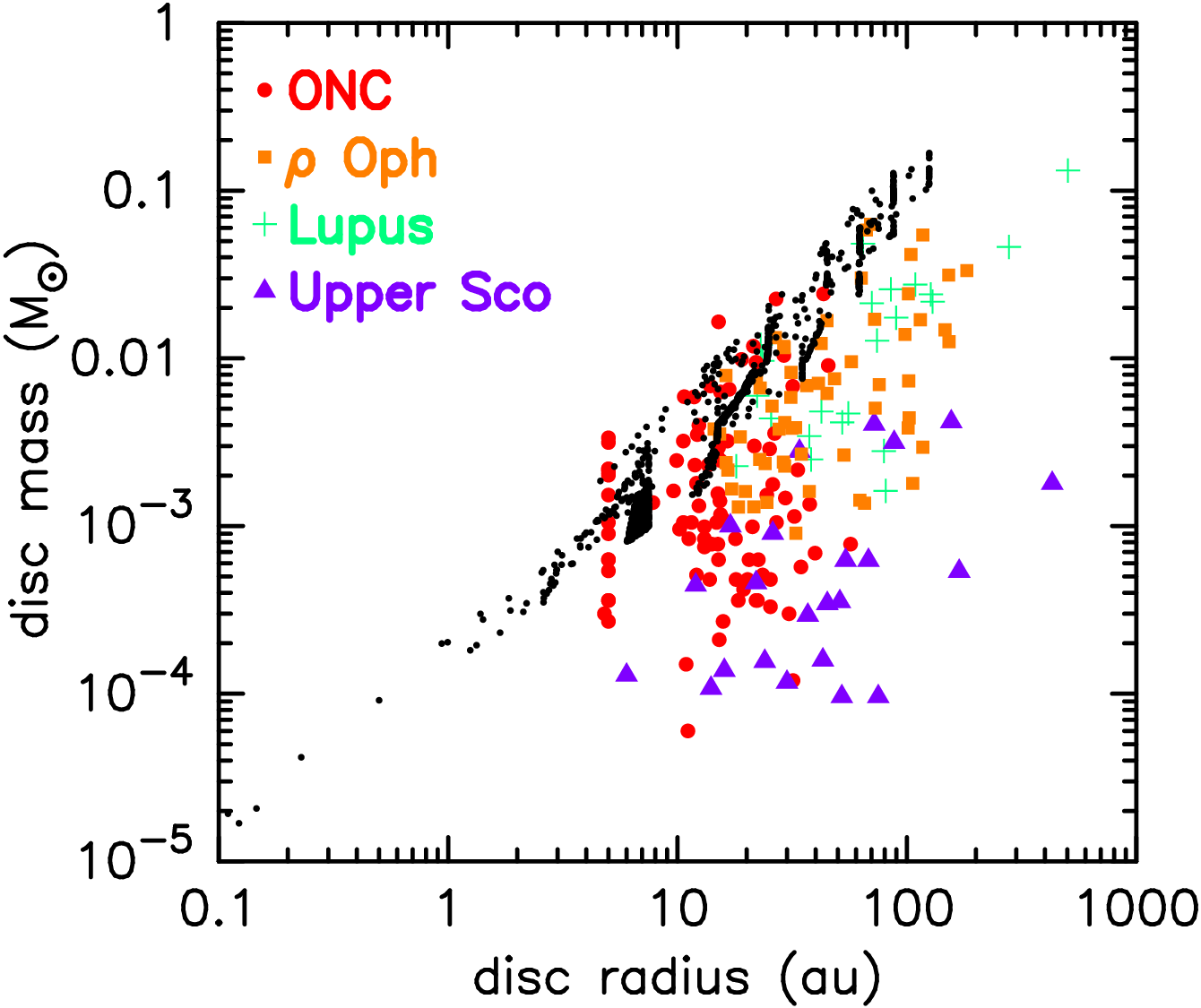}}}
\hspace*{0.3cm}\subfigure[5\,Myr]{\label{mod_dens_10au-h}\rotatebox{270}{\includegraphics[scale=0.23]{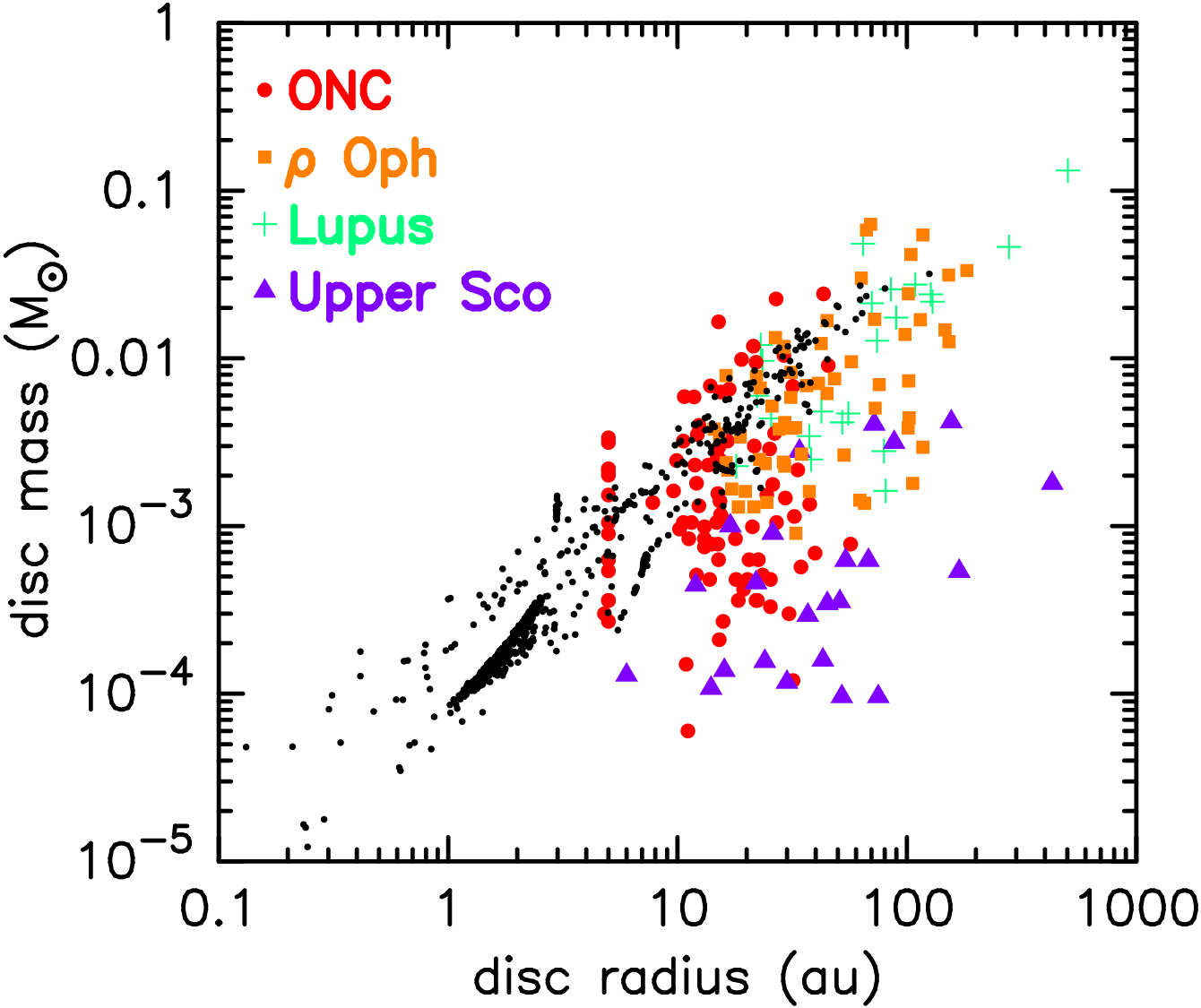}}}
\hspace*{0.3cm}\subfigure[Disc fractions]{\label{mod_dens_10au-i}\rotatebox{270}{\includegraphics[scale=0.23]{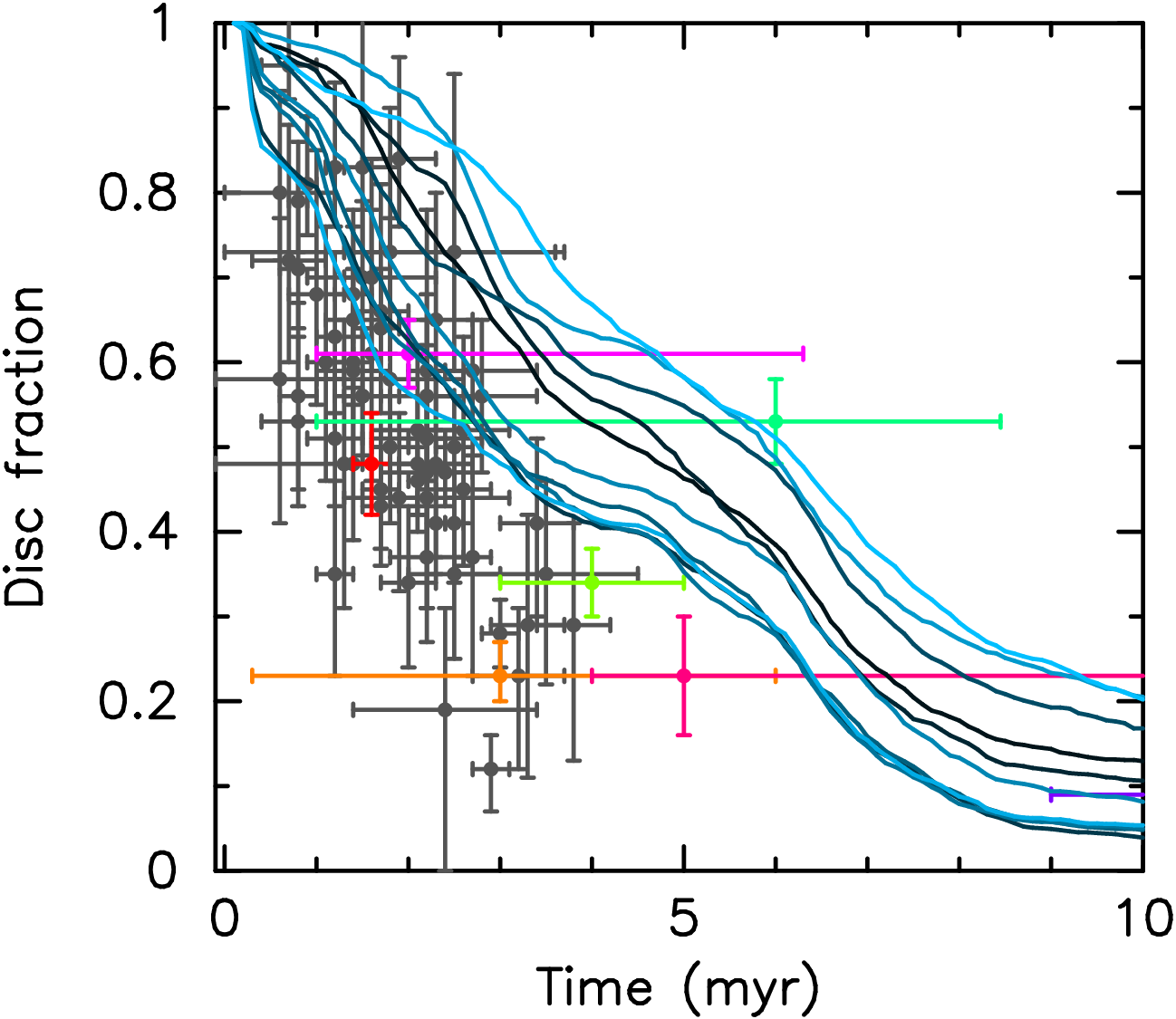}}}
\caption[bf]{Results for simulations in which the initial disc radii are drawn from a delta function with $r_{\rm disc} = 10$\,au, and the discs are allowed to evolve inwards due to external photoevaporation, and outwards due to viscous spreading.  The star-forming regions have moderate initial stellar density ($\tilde{\rho} \sim 100$\,M$_\odot$\,pc$^{-3}$). In panels (a)--(c) we show the cumulative distributions of disc radii at different ages, and in panels (d)--(f) we show the cumulative distributions of disc masses. The blue lines of varying hues show the results from the ten individual $N$-body simulations. The mint green dashed line represents the respective gas distribution in the Lupus star-forming region, whereas solid lines represent the dust distributions. Dotted lines indicate an alternative dust radius for $\rho$~Oph (orange line) and Lupus (mint green line); see Table~\ref{obs_info} for further details. In panels (g) and (h) we show the disc mass versus disc radius for four observed star-forming regions at 1 and 5\,Myr, and data from one post-processed $N$-body simulation are shown by the black points. We plot the disc fractions (defined as when the discs have non-zero mass) as a function of time in our $N$-body simulations in panel (i), with observational data taken from \citet{Richert18} and \citet{Ribas15}.}
\label{mod_dens_10au}
  \end{center}
\end{figure*}

In Fig.~\ref{mod_dens_10au} we show the results for $N = 1500$, $\tilde{\rho} \sim 100$M$_\odot$\,pc$^{-3}$ star-forming regions where the initial disc radii are all drawn from a delta function with $r_{\rm disc} = 10$\,au.

The overall disc fraction decreases steadily throughout the 10\,Myr period of the simulations, and is consistent with some of the higher observed disc fractions, but not the lower disc fractions at earlier ages (i.e.\,\,in panel (i) the [simulation] lines do not overlay the grey [observed] points for ages $<$4\,Myr and disc fractions $<$50\,per cent). As many of the discs survive the duration of the simulation, the cumulative distributions of the simulation data do not display the low-$N$ noise apparent in the data from the $r_{\rm disc} = 100$\,au simulations.

In general, the disc radii evolve to smaller values, although viscous spreading does lead to discs with radii up to 100\,au. After 1\,Myr, some of the simulation data are consistent with the observed disc radius distribution from the ONC, but no other star-forming regions (Fig.~\ref{mod_dens_10au}). At later times the simulation data are not consistent with any of the observed distributions (Fig.~\ref{mod_dens_10au-b} \& \ref{mod_dens_10au-c}).

The disc masses from the simulation data remain too high after 1\,Myr to be consistent with the obervations (Fig.~\ref{mod_dens_10au-d}), and whilst they are consistent with the ONC after 5\,Myr (Fig.~\ref{mod_dens_10au-e}), at this age the disc radii distributions are clearly inconsistent (Fig.~\ref{mod_dens_10au-b}).

This is also seen in the plots that show the individual disc masses and disc radii. After 1\,Myr, the disc masses are higher than those in the observed regions (Fig.~\ref{mod_dens_10au-g}, and after 5\,Myr there are many discs with small radii and masses that cannot be directly compared to observational data (Fig.~\ref{mod_dens_10au-h}).

We therefore suggest that for these initial conditions, the initial disc radii are likely to be much larger than 10\,au, despite these discs being more susceptible to photoevaporation.

\subsubsection{Initial disc radii from \citet{Eisner18}}

We also ran a set of simulations in which the disc radii were drawn from the \citet{Eisner18} data, in order to demonstrate that the present-day disc radius distribution cannot be the initial distribution. Because the median disc radius in the dataset is only a little over 10\,au, we found very similar results to our numerical experiments with  a delta function at $r_{\rm disc} = 10$\,au. We include these plots in Appendix~\ref{appendix:eisner}.

\subsection{High stellar density}

We now focus on star-forming regions with $N = 1500$ stars, but with high initial stellar densities ($\tilde{\rho} \sim 1000$\,M$_\odot$\,pc$^{-3}$). Such high densities have been postulated as the initial conditions for some, and potentially all, star-forming regions \citep{Marks12,Parker14e}. These simulations have high radiation fields, with initial $G_0 \sim 10^4$.

First, we note that in a similar vein to the moderate density initial conditions, an initial disc radius distribution where the discs have small ($r_{\rm disc} = 10$\,au) radii is inconsistent with the observational data. Whilst some of the discs in the simulations survive the 10\,Myr duration, the disc radii distributions rapidly evolve to small ($<$10\,au) radii, even with viscous spreading counteracting (some of) the effects of inward evolution due to photoevaporation. As in the moderate density simulations, the mass distributions of the surviving discs are not consistent with the observed distributions.

\begin{figure*}
  \begin{center}
\setlength{\subfigcapskip}{10pt}
\hspace*{0.3cm}\subfigure[1\,Myr]{\label{high_dens_100au_viscous-a}\rotatebox{270}{\includegraphics[scale=0.23]{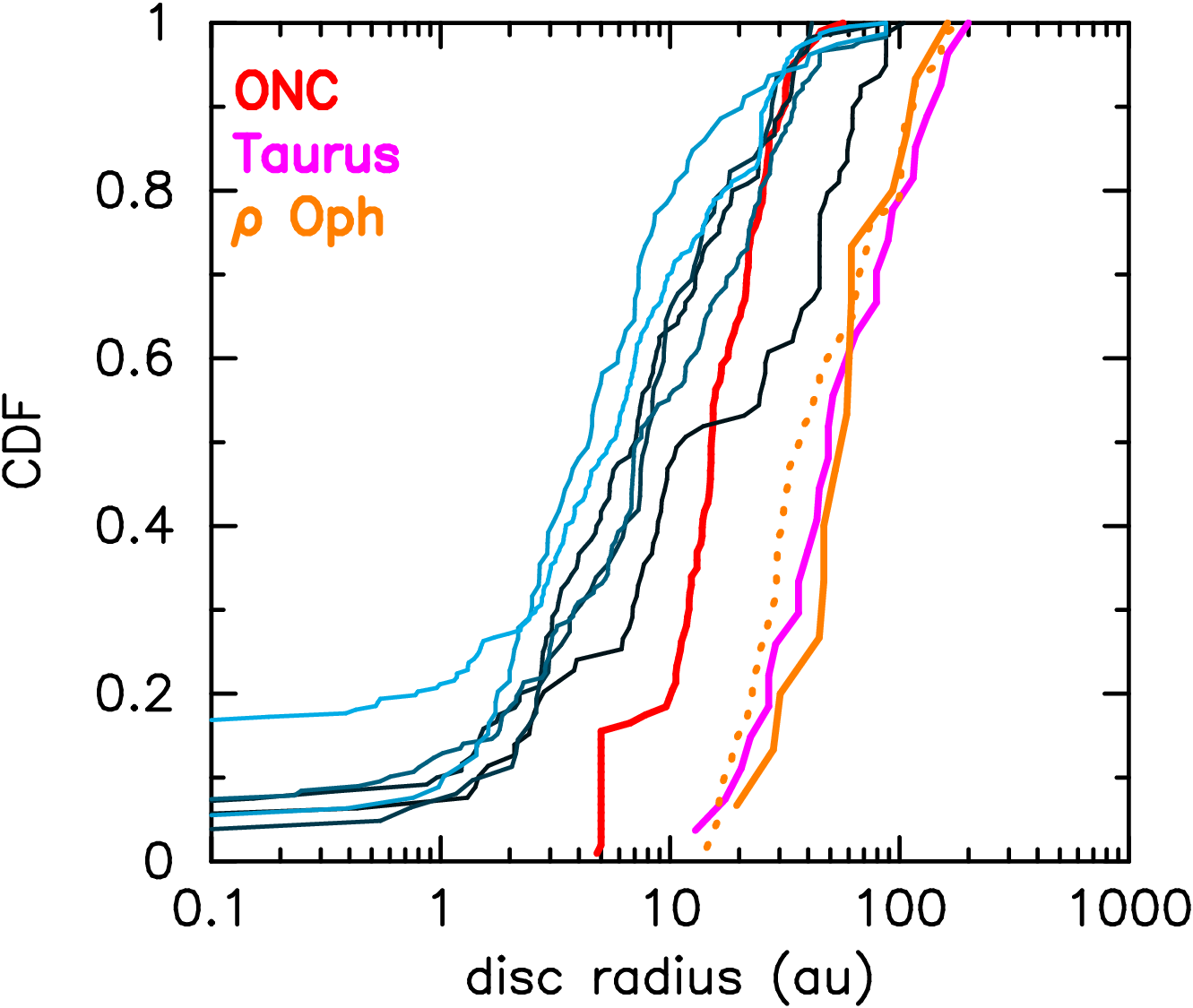}}}
\hspace*{0.3cm}\subfigure[5\,Myr]{\label{high_dens_100au_viscous-b}\rotatebox{270}{\includegraphics[scale=0.23]{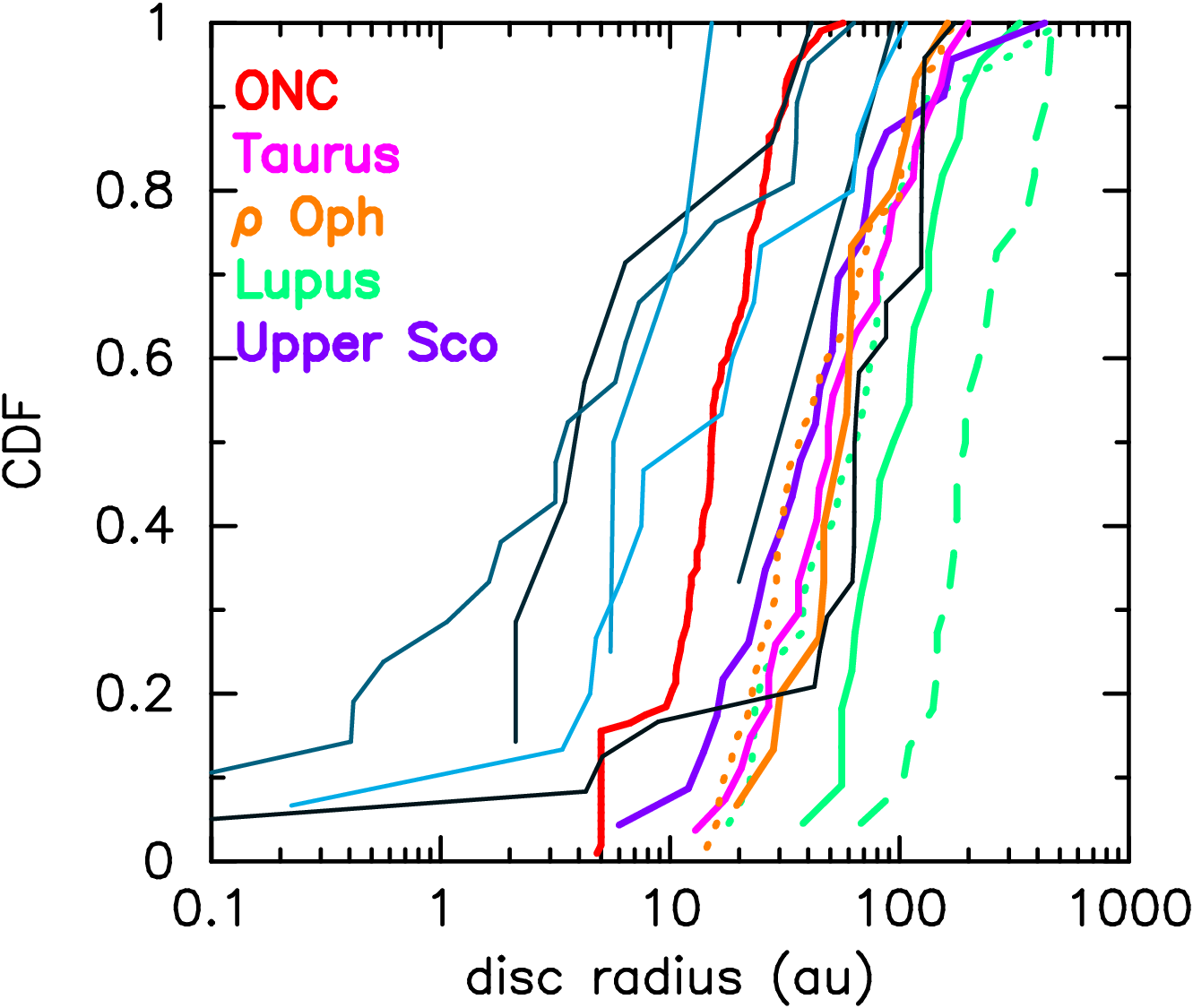}}}
\hspace*{0.3cm}\subfigure[10\,Myr]{\label{high_dens_100au_viscous-c}\rotatebox{270}{\includegraphics[scale=0.23]{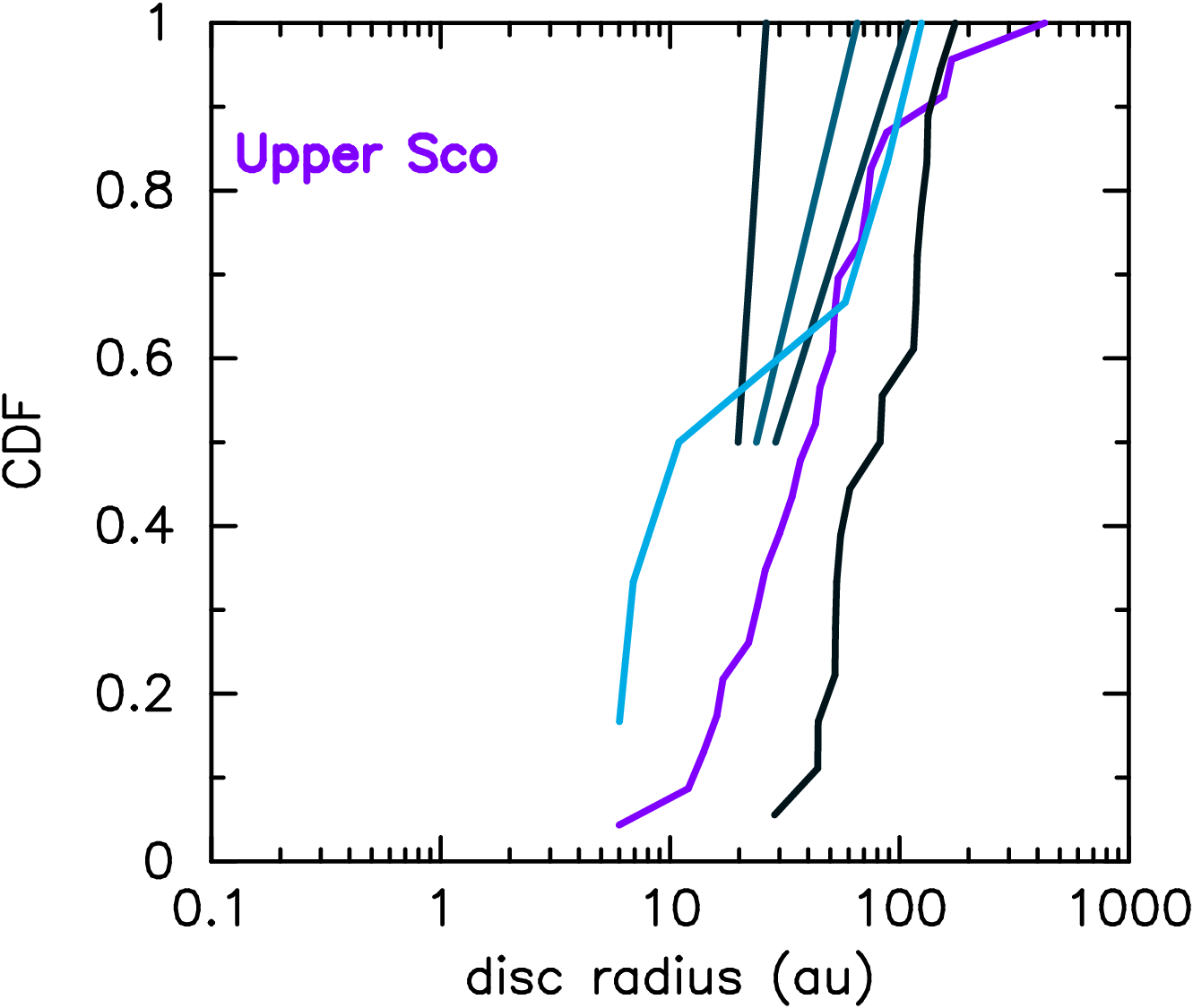}}}
\hspace*{0.3cm}\subfigure[1\,Myr]{\label{high_dens_100au_viscous-d}\rotatebox{270}{\includegraphics[scale=0.23]{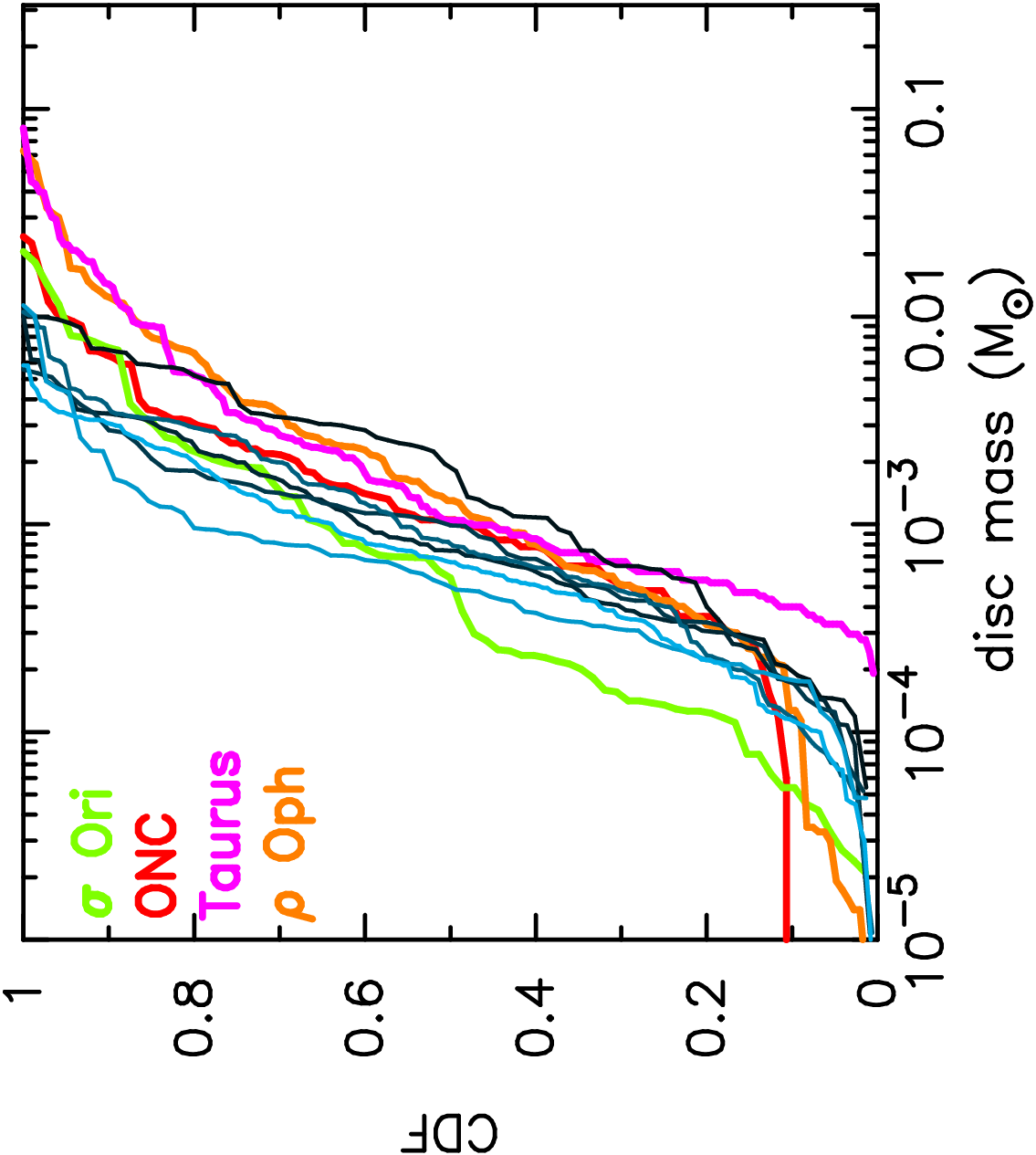}}}
\hspace*{0.3cm}\subfigure[5\,Myr]{\label{high_dens_100au_viscous-e}\rotatebox{270}{\includegraphics[scale=0.23]{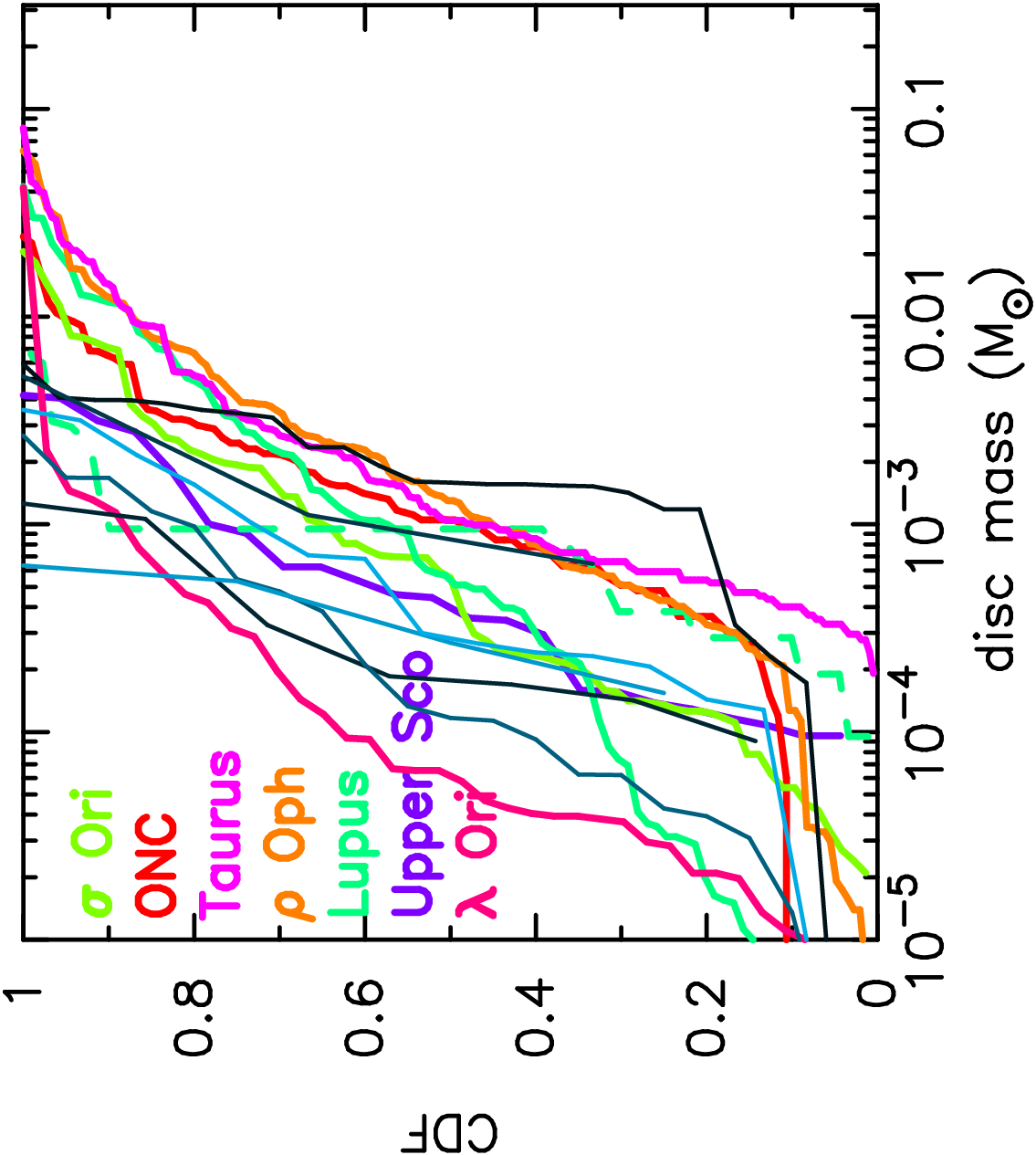}}}
\hspace*{0.3cm}\subfigure[10\,Myr]{\label{high_dens_100au_viscous-f}\rotatebox{270}{\includegraphics[scale=0.23]{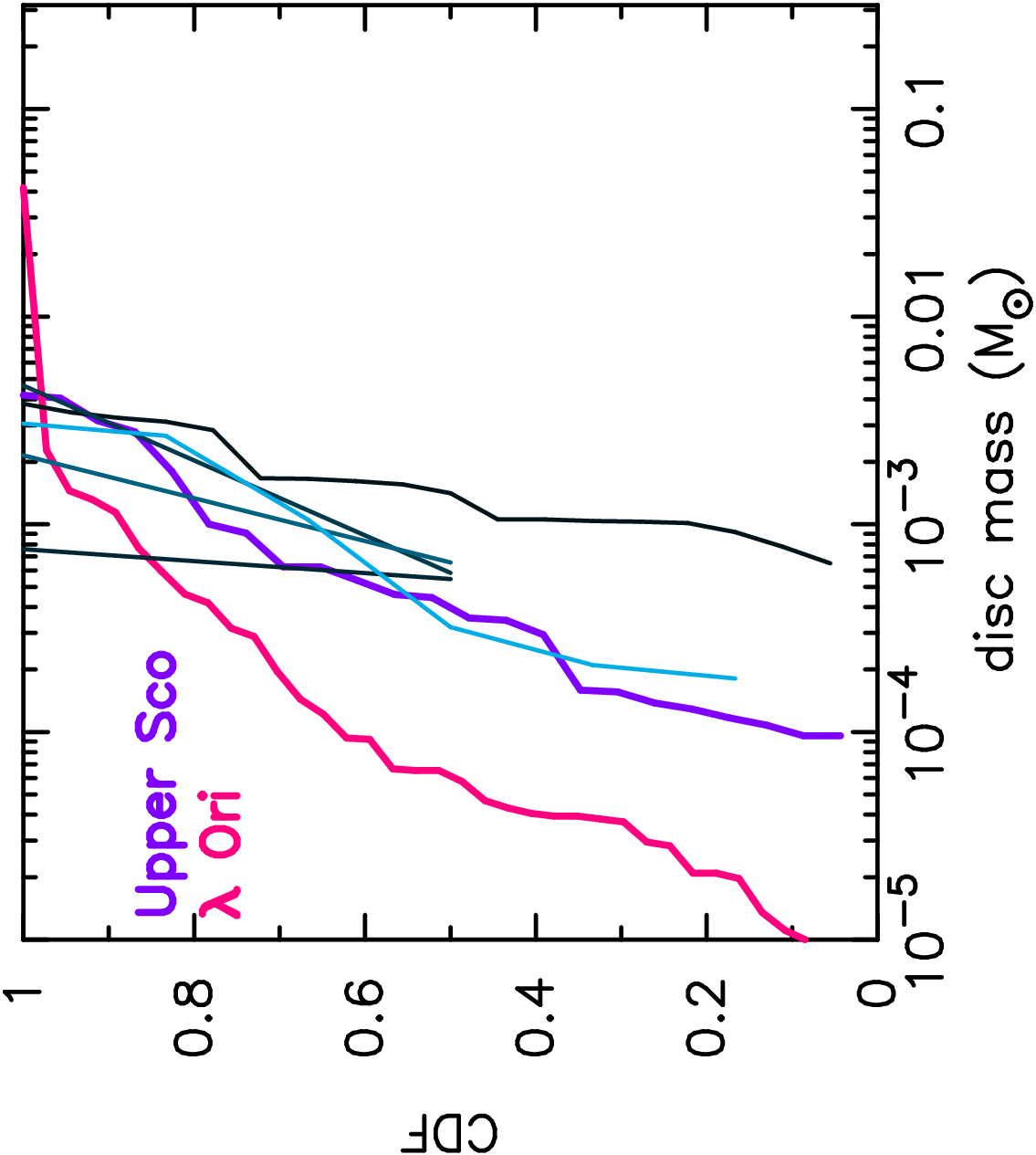}}}
\hspace*{0.3cm}\subfigure[5\,Myr]{\label{high_dens_100au_viscous-g}\rotatebox{270}{\includegraphics[scale=0.23]{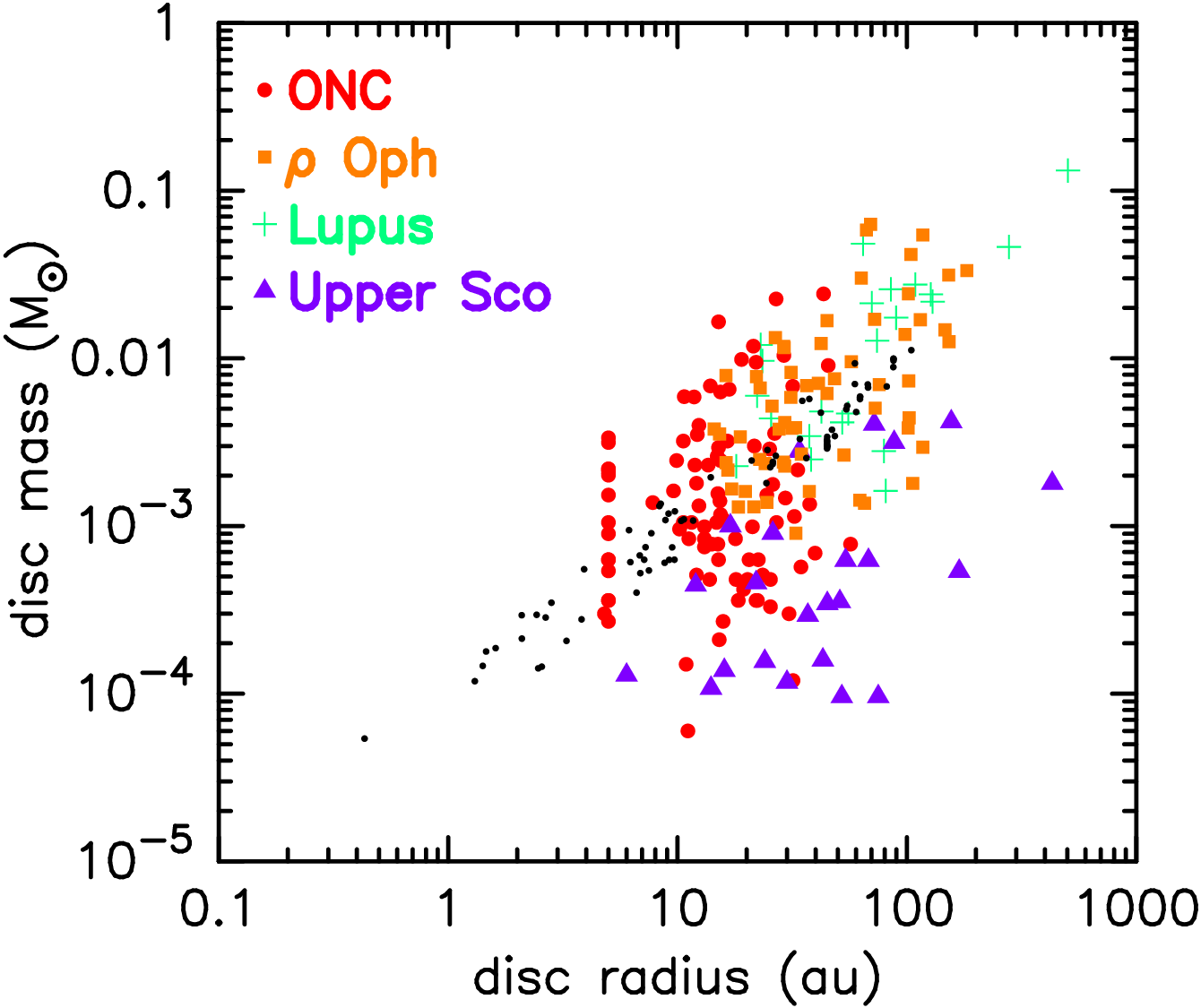}}}
\hspace*{0.3cm}\subfigure[5\,Myr]{\label{high_dens_100au_viscous-h}\rotatebox{270}{\includegraphics[scale=0.23]{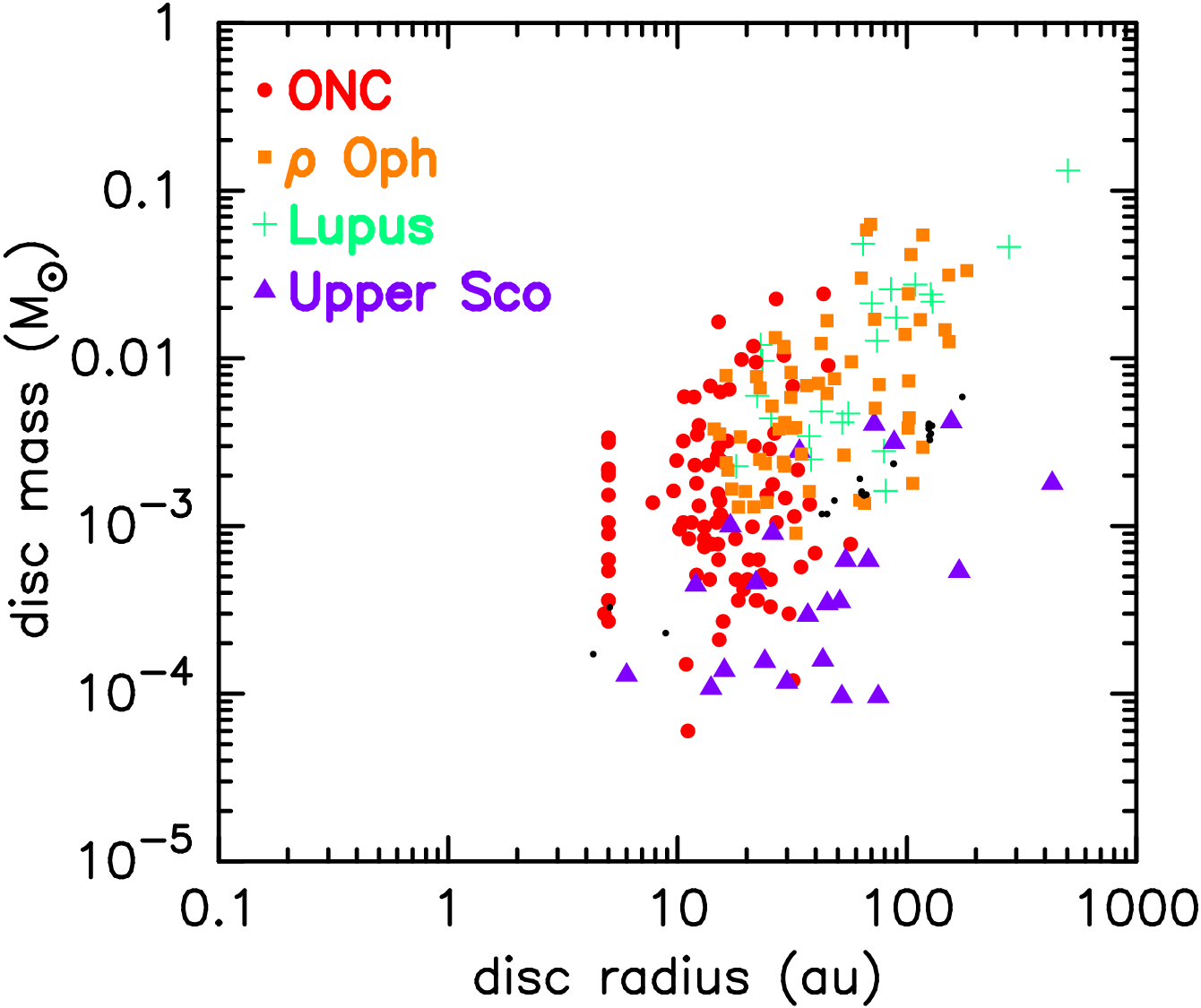}}}
\hspace*{0.3cm}\subfigure[Disc fractions]{\label{high_dens_100au_viscous-i}\rotatebox{270}{\includegraphics[scale=0.23]{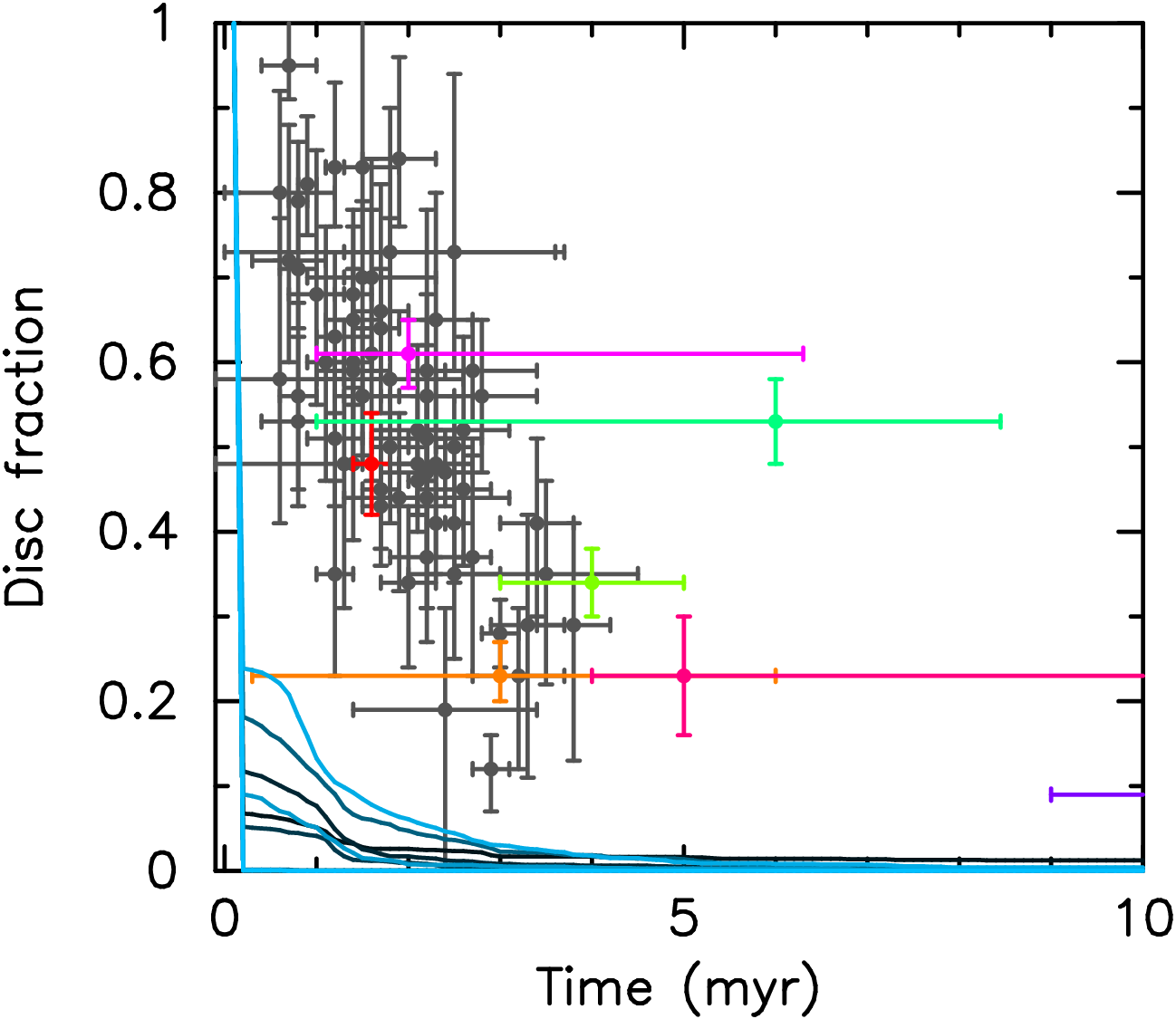}}}

\caption[bf]{Results for simulations in which the initial disc radii are drawn from a delta function with $r_d = 100$\,au, and the discs are allowed to evolve inwards due to external photoevaporation, and outwards due to viscous spreading.  The star-forming regions have high initial stellar density ($\tilde{\rho} \sim 1000$\,M$_\odot$\,pc$^{-3}$). In panels (a)--(c) we show the cumulative distributions of disc radii at different ages, and in panels (d)--(f) we show the cumulative distributions of disc masses. The blue lines of varying hues show the results from the ten individual $N$-body simulations. The mint green dashed line represents the respective gas distribution in the Lupus star-forming region, whereas solid lines represent the dust distributions. Dotted lines indicate an alternative dust radius for $\rho$~Oph (orange line) and Lupus (mint green line); see Table~\ref{obs_info} for further details. In panels (g) and (h) we show the disc mass versus disc radius for four observed star-forming regions at 1 and 5\,Myr, and data from one post-processed $N$-body simulation are shown by the black points. We plot the disc fractions (defined as when the discs have non-zero mass) as a function of time in our $N$-body simulations in panel (i), with observational data taken from \citet{Richert18} and \citet{Ribas15}.}
\label{high_dens_100au_viscous}
  \end{center}
\end{figure*}

In Fig.~\ref{high_dens_100au_viscous} we show the evolution of discs where the inital radii are all $r_{\rm disc} = 100$\,au. In four out of ten simulations, the discs are all destroyed in the first 0.5\,Myr, and in the remaining six simulations only between 5 and 20\,per cent of discs remain (Fig.~\ref{high_dens_100au_viscous-i}).

After 1\,Myr, the disc radii from the simulations are smaller than the observed data (Fig.~\ref{high_dens_100au_viscous-a}), although at 5\,Myr several simulations are consistent with both the ONC, and other regions (Taurus, $\rho$~Oph and Upper~Sco), with the caveat that there are very few discs remaining at this time. 

Interestingly, at all ages several of the disc mass distributions from the simulations are consistent with the observed distributions (Fig.~\ref{high_dens_100au_viscous-d}--\ref{high_dens_100au_viscous-f}). However, the individual disc mass versus disc radii plots (Fig.~\ref{high_dens_100au_viscous-g}~and~\ref{high_dens_100au_viscous-h}), and the disc fractions (Fig.~\ref{high_dens_100au_viscous-i}), show that after 1\,Myr, very few discs remain, due to the high $G_0$ fields in these dense regions.

Finally, we note that simulations with these densities could facilitate truncation of discs due to direct dynamical encounters, although such densities need to be maintained over several Myr \citep{Vincke18,Winter18b}. However, in our simulations such high densities are not maintained for more than a $\sim$Myr (Fig.~\ref{G0_densities}).

\subsection{Low stellar density}

The present day density of nearby star-forming regions spans a wide range \citep{Bressert10}, from several M$_\odot$\,pc$^{-3}$ to several 100M$_\odot$\,pc$^{-3}$. In particular, of the observed regions, Taurus has a very low density (less than 5\,M$_\odot$\,pc$^{-3}$). We therefore run simulations with an initial median stellar density $\tilde{\rho} \sim 10$M$_\odot$\,pc$^{-3}$. These simulations have radiation fields much lower than the high-density simulations ($100 - 1000G_0$), but still high due to the relatively high numbers of OB stars.

Again, for simulations in which the initial disc distribution is a delta function with $r_{\rm disc} = 10$\,au, the evolution of the disc radii distribution moves inwards as the simulations progress, but when the disc radii are in the same regime as the observed distributions, the disc masses are too high, and when the mass distributions are consistent with the observations, the disc radii are too small. Furthermore, none of the simulations have disc fractions consistent with the observed disc fractions in \citet{Richert18}.

\begin{figure*}
  \begin{center}
\setlength{\subfigcapskip}{10pt}
\hspace*{0.3cm}\subfigure[1\,Myr]{\label{low_dens_100au_viscous-a}\rotatebox{270}{\includegraphics[scale=0.23]{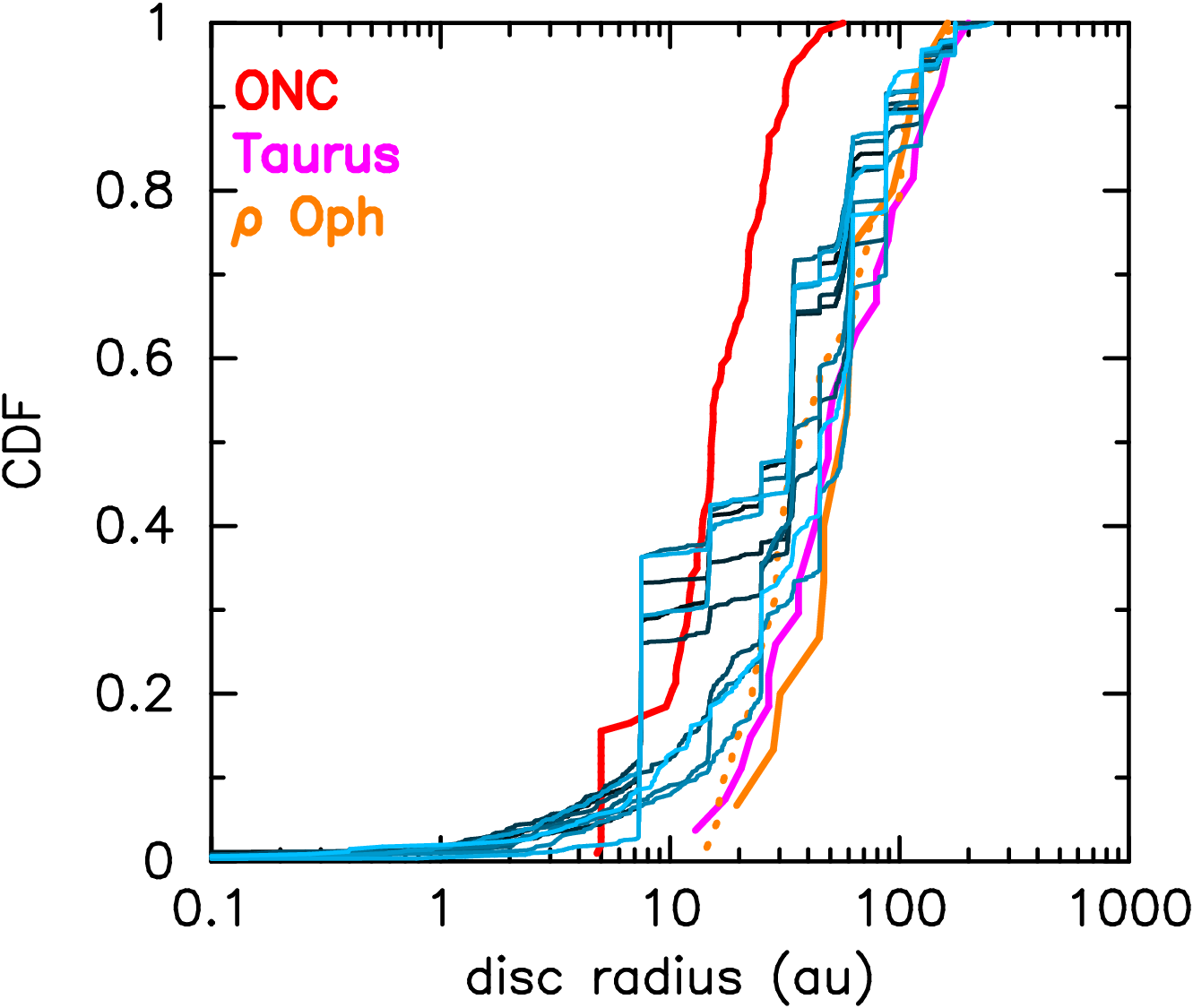}}}
\hspace*{0.3cm}\subfigure[5\,Myr]{\label{low_dens_100au_viscous-b}\rotatebox{270}{\includegraphics[scale=0.23]{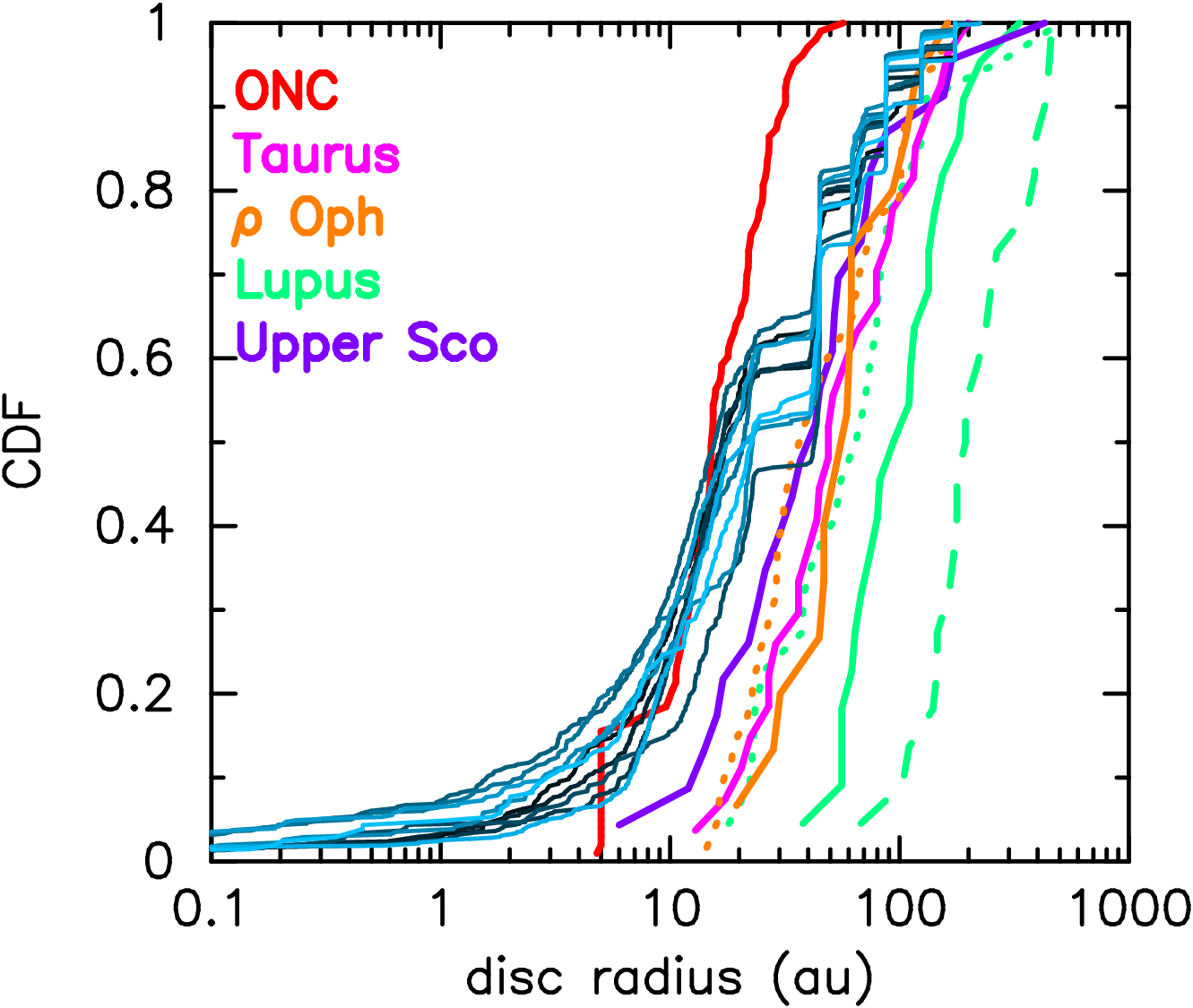}}}
\hspace*{0.3cm}\subfigure[10\,Myr]{\label{low_dens_100au_viscous-c}\rotatebox{270}{\includegraphics[scale=0.23]{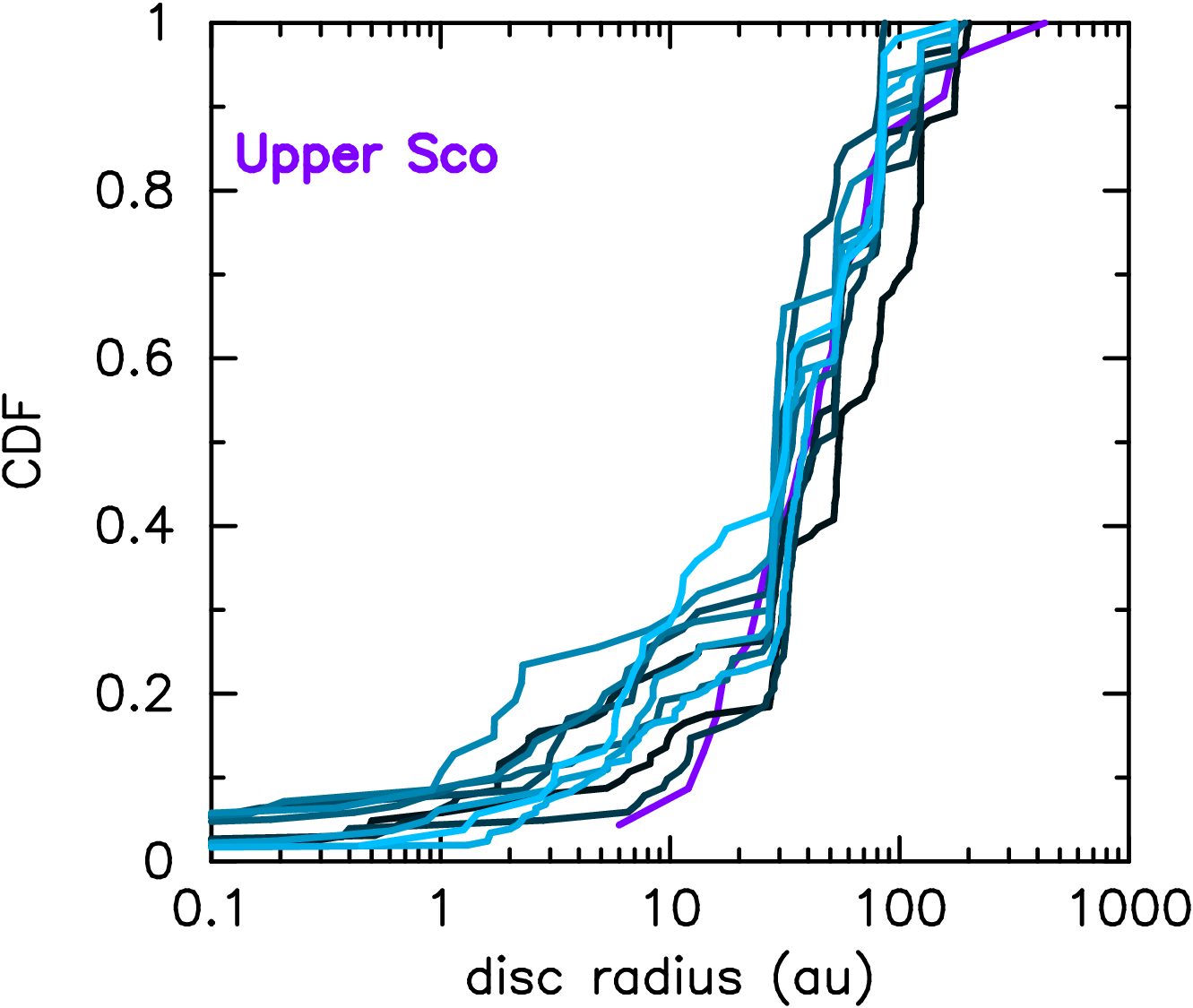}}}
\hspace*{0.3cm}\subfigure[1\,Myr]{\label{low_dens_100au_viscous-d}\rotatebox{270}{\includegraphics[scale=0.23]{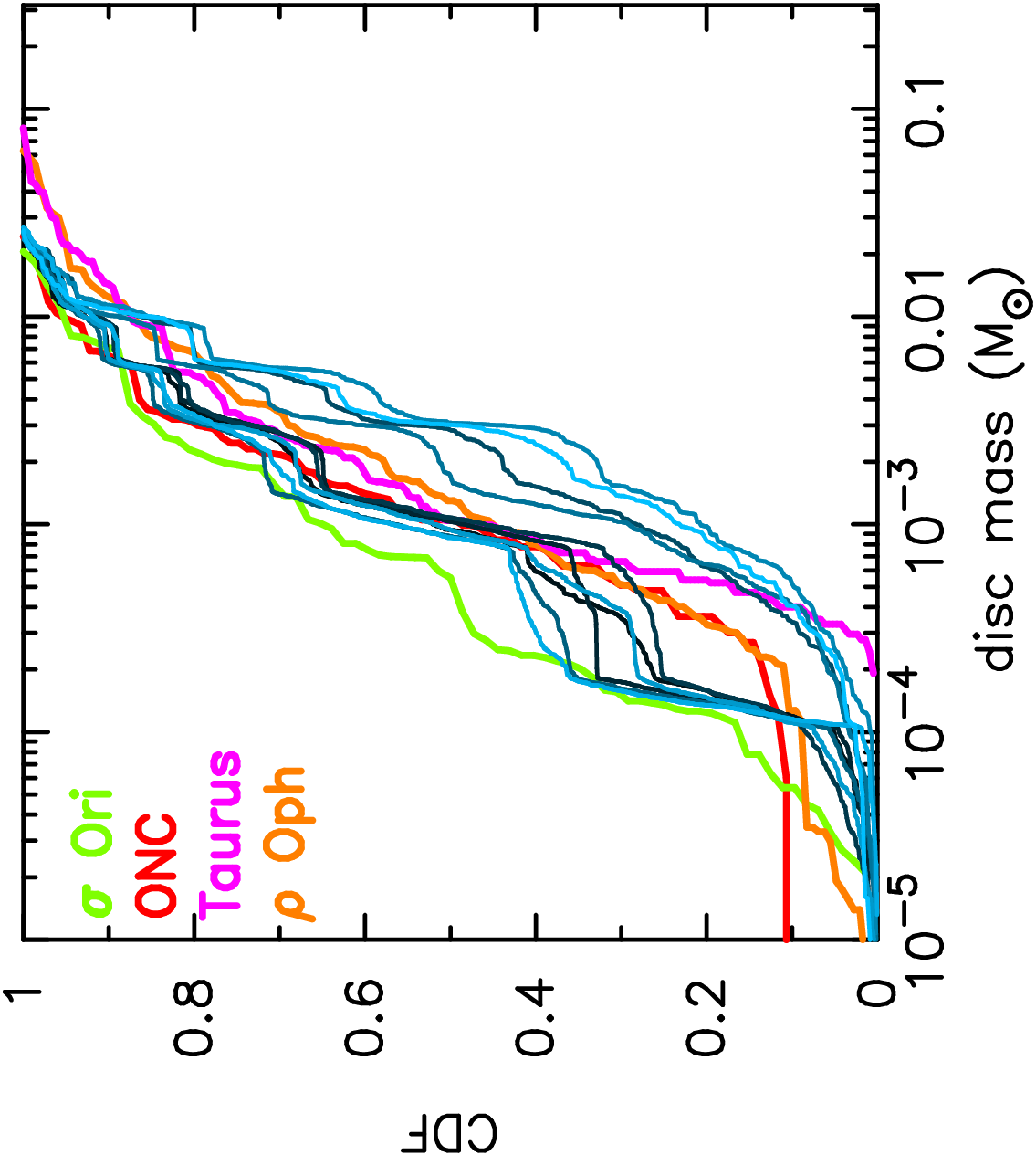}}}
\hspace*{0.3cm}\subfigure[5\,Myr]{\label{low_dens_100au_viscous-e}\rotatebox{270}{\includegraphics[scale=0.23]{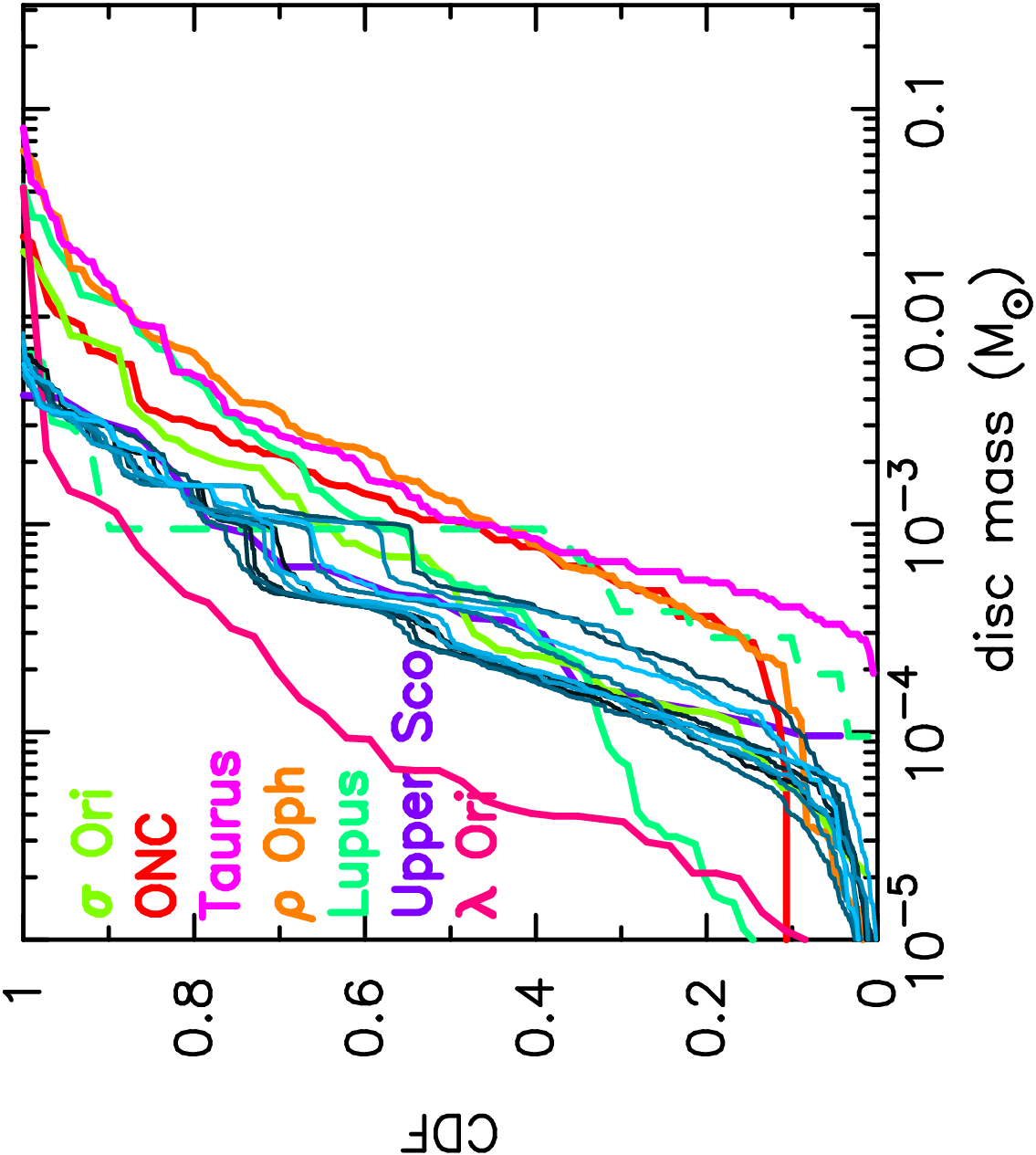}}}
\hspace*{0.3cm}\subfigure[10\,Myr]{\label{low_dens_100au_viscous-f}\rotatebox{270}{\includegraphics[scale=0.23]{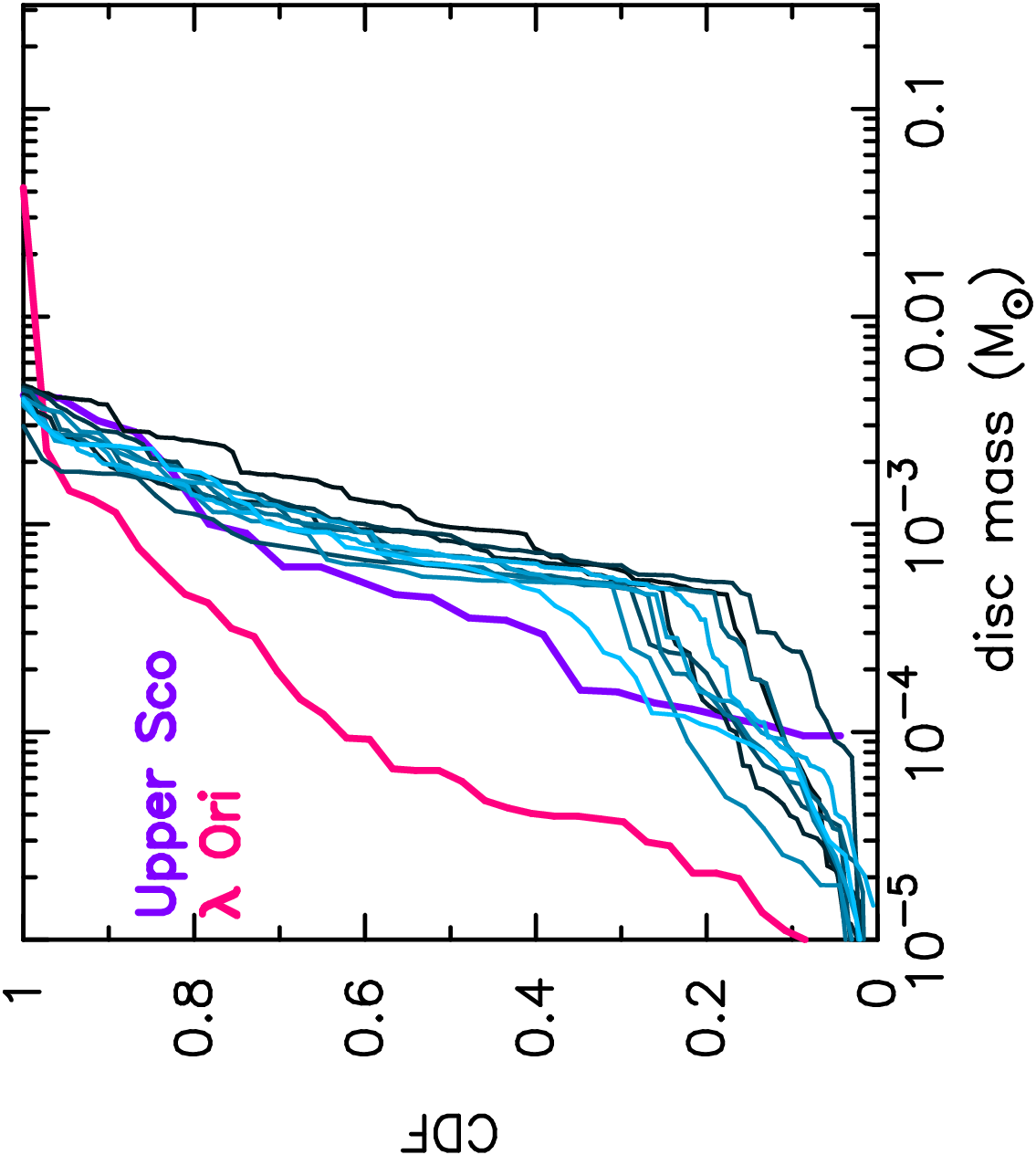}}}
\hspace*{0.3cm}\subfigure[Mass--radius, 1\,Myr]{\label{low_dens_100au_viscous-g}\rotatebox{270}{\includegraphics[scale=0.23]{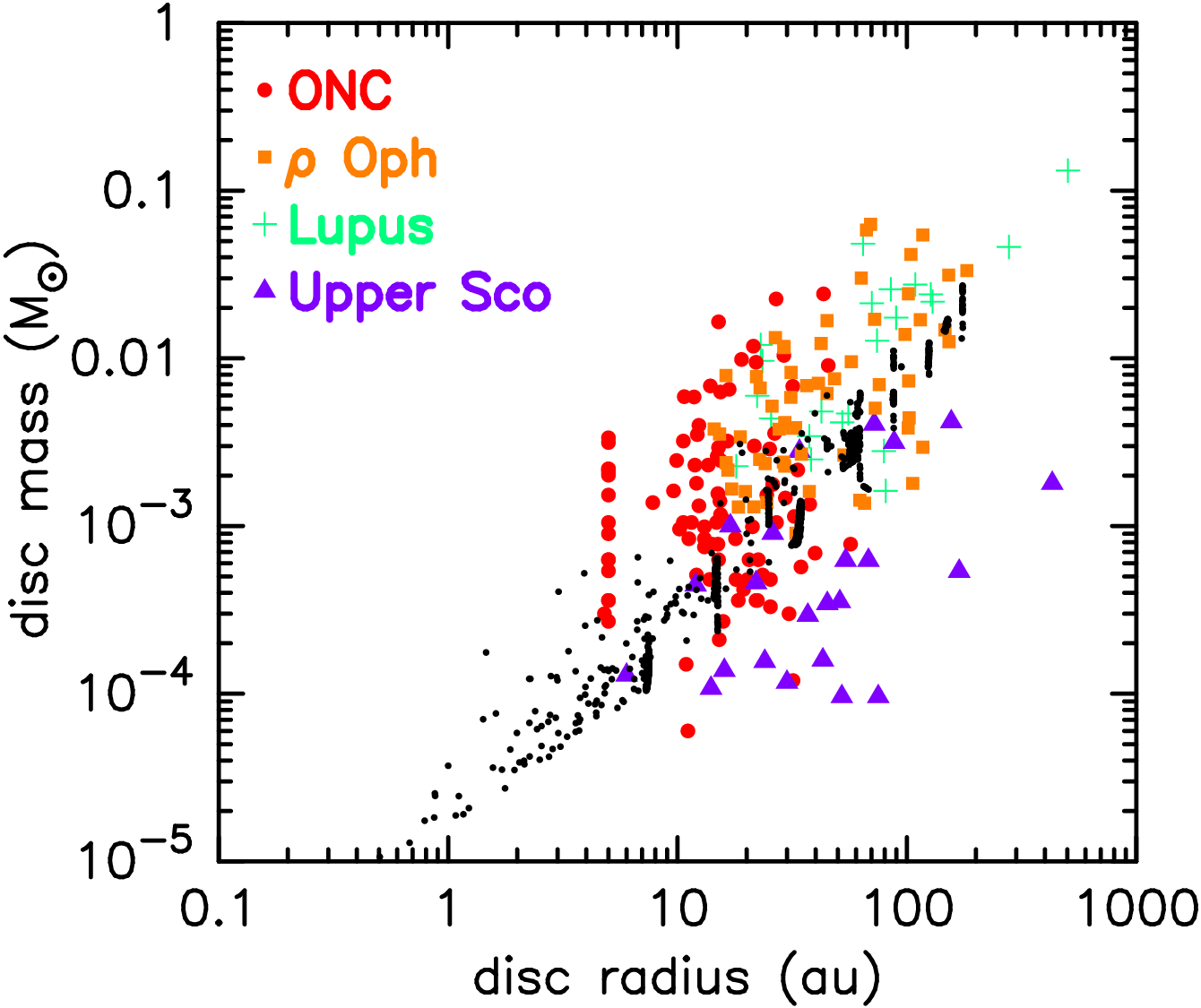}}}
\hspace*{0.3cm}\subfigure[Mass--radius, 5\,Myr]{\label{low_dens_100au_viscous-h}\rotatebox{270}{\includegraphics[scale=0.23]{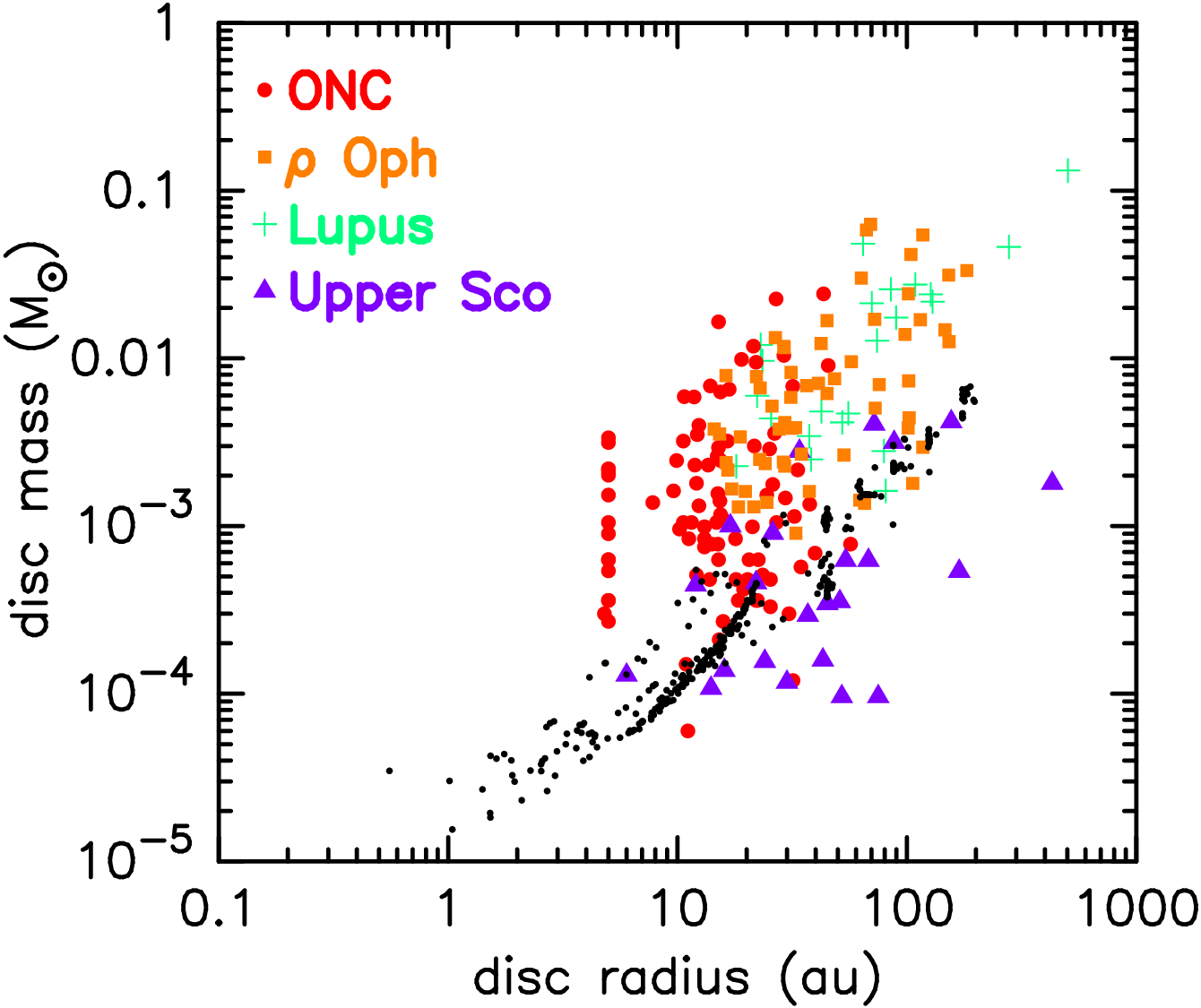}}}
\hspace*{0.3cm}\subfigure[Disc fractions]{\label{low_dens_100au_viscous-i}\rotatebox{270}{\includegraphics[scale=0.23]{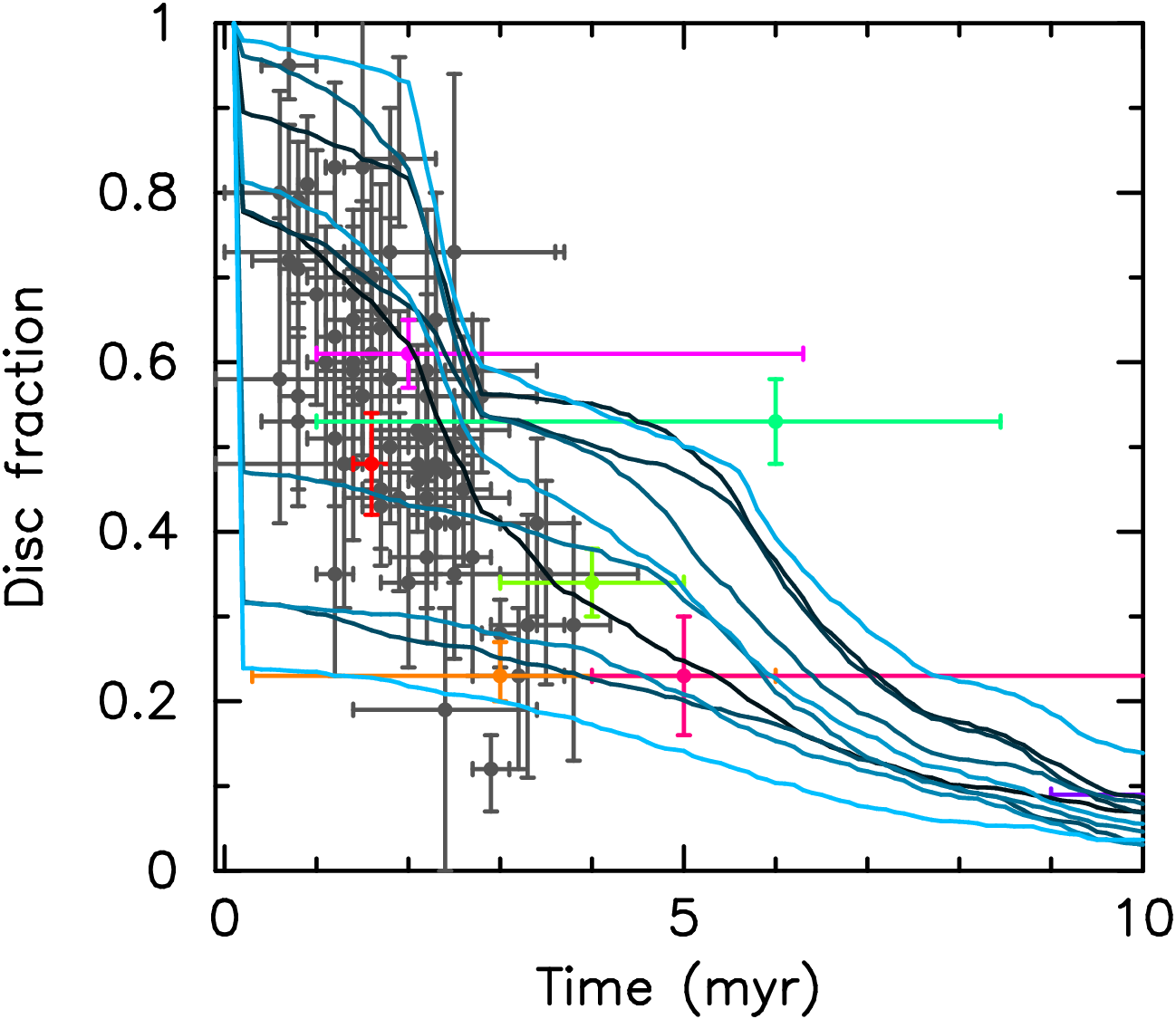}}}

\caption[bf]{Results for simulations in which the initial disc radii are drawn from a delta function with $r_{\rm disc} = 100$\,au, and the discs are allowed to evolve inwards due to external photoevaporation, and outwards due to viscous spreading.  The star-forming regions have low initial stellar density ($\tilde{\rho} \sim 10$\,M$_\odot$\,pc$^{-3}$). In panels (a)--(c) we show the cumulative distributions of disc radii at different ages, and in panels (d)--(f) we show the cumulative distributions of disc masses. The blue lines of varying hues show the results from the ten individual $N$-body simulations. The mint green dashed line represents the respective gas distribution in the Lupus star-forming region, whereas solid lines represent the dust distributions. Dotted lines indicate an alternative dust radius for $\rho$~Oph (orange line) and Lupus (mint green line); see Table~\ref{obs_info} for further details. In panels (g) and (h) we show the disc mass versus disc radius for four observed star-forming regions at 1 and 5\,Myr, and data from one post-processed $N$-body simulation are shown by the black points. We plot the disc fractions (defined as when the discs have non-zero mass) as a function of time in our $N$-body simulations in panel (i), with observational data taken from \citet{Richert18} and \citet{Ribas15}.}
\label{low_dens_100au_viscous}
  \end{center}
\end{figure*}

We therefore just focus on the simulations where the initial disc radii are drawn from a delta function with $r_{\rm disc} = 100$\,au. In Fig.~\ref{low_dens_100au_viscous} we show the evolution of the disc radii and masses in these simulations. After 1--5\,Myr, the disc radii distributions from the simulations are consistent with the observations of discs in Taurus, and $\rho$~Oph, but not the ONC (Fig.~\ref{low_dens_100au_viscous-a}--\ref{low_dens_100au_viscous-b}). At later ages (10\,Myr, Fig.~\ref{low_dens_100au_viscous-c}), the radii distributions from the simulations are consistent with the observations of Upper~Sco.

The mass distributions from the simulations are consistent with those observed in Taurus and $\rho$~Oph after 1\,Myr (Fig.~\ref{low_dens_100au_viscous-d}), but at later times the masses are too low. The continued destruction of discs also preferentially removes low-mass discs, so after 10\,Myr the cumulative mass distributions of discs remaining in the simulations have actually evolved to higher masses (Fig.~\ref{low_dens_100au_viscous-f}, though we reiterate that this is not because the discs have actually gained mass).

The simulations exhibit a wide range of disc fractions (Fig.~\ref{low_dens_100au_viscous-i}), encompassing the entire range of observed disc fractions at varying ages. The individual disc mass versus disc radius distributions (Figs.~\ref{low_dens_100au_viscous-g}~and~\ref{low_dens_100au_viscous-h}) show that the surviving population is consistent with the observations at 1\,Myr, but consistent only with Upper Sco at 5\,Myr.

\subsection{$N = 150$ stars}

Our analysis thus far has been based on star-forming regions with $N = 1500$ stars, which from a typical sampling of the IMF leads to around five massive stars that produce ionising photons. We now present the results for star-forming regions that contain $N = 150$ stars, and from random sampling of the IMF these regions contain far fewer massive stars (in some instances, no stars more massive than around 5\,M$_\odot$). Many of the star-forming regions with which we compare observational data (e.g. Taurus, $\rho$~Oph, Lupus, $\sigma$~Ori) only contain a few hundred stars.

We set the radii of the star-forming regions to be such that the initial median stellar density is the same as in our moderate density simulations, i.e.\,\,$\tilde{\rho} \sim 100$\,M$_\odot$\,pc$^{-3}$. We present the results for simulations in which the initial disc radii are drawn from a delta function with $r_{\rm disc} = 100$\,au.

\begin{figure*}
  \begin{center}
\setlength{\subfigcapskip}{10pt}
\hspace*{0.3cm}\subfigure[1\,Myr]{\label{mod_dens_lowN_100au_viscous-a}\rotatebox{270}{\includegraphics[scale=0.23]{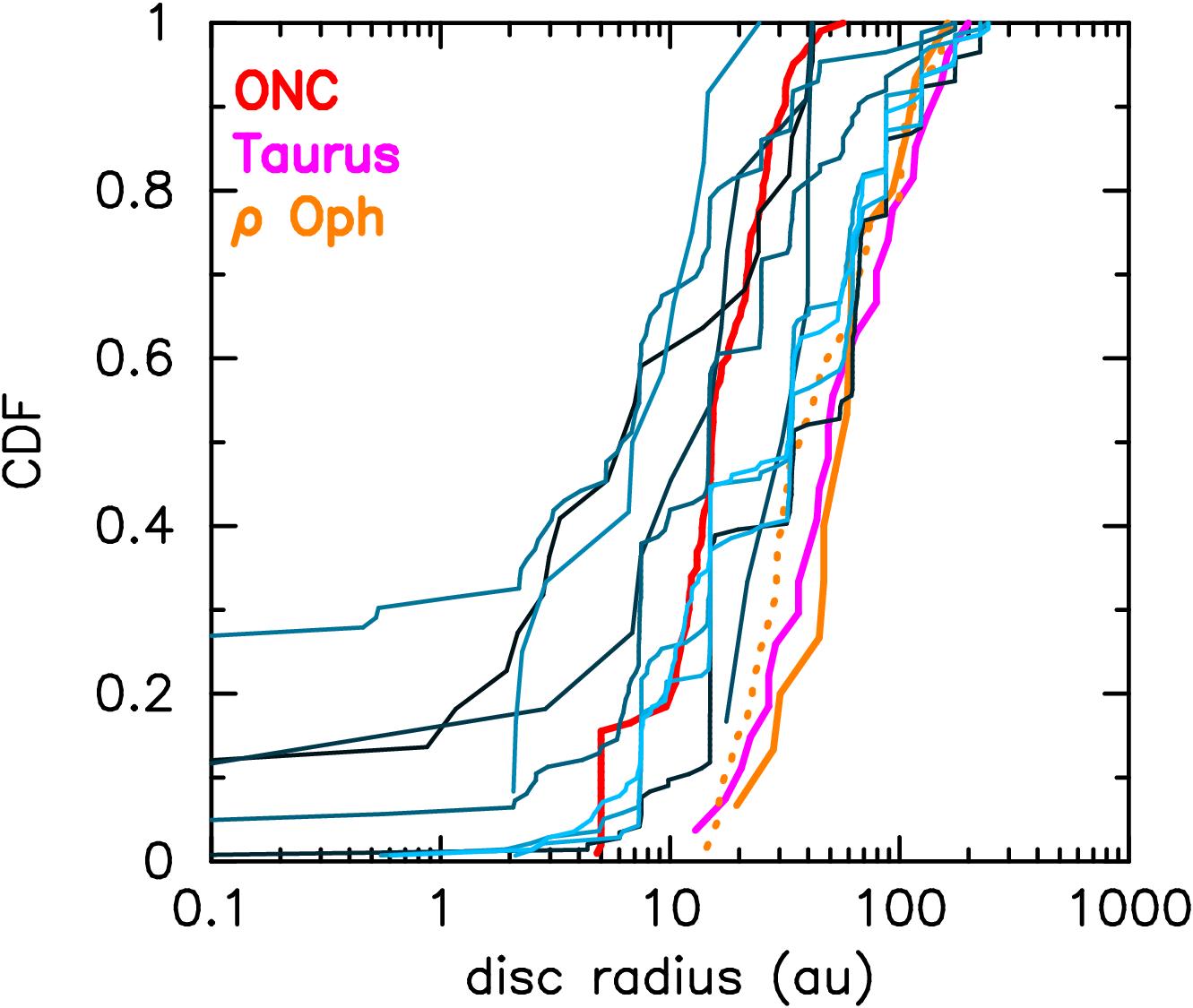}}}
\hspace*{0.3cm}\subfigure[5\,Myr]{\label{mod_dens_lowN_100au_viscous-b}\rotatebox{270}{\includegraphics[scale=0.23]{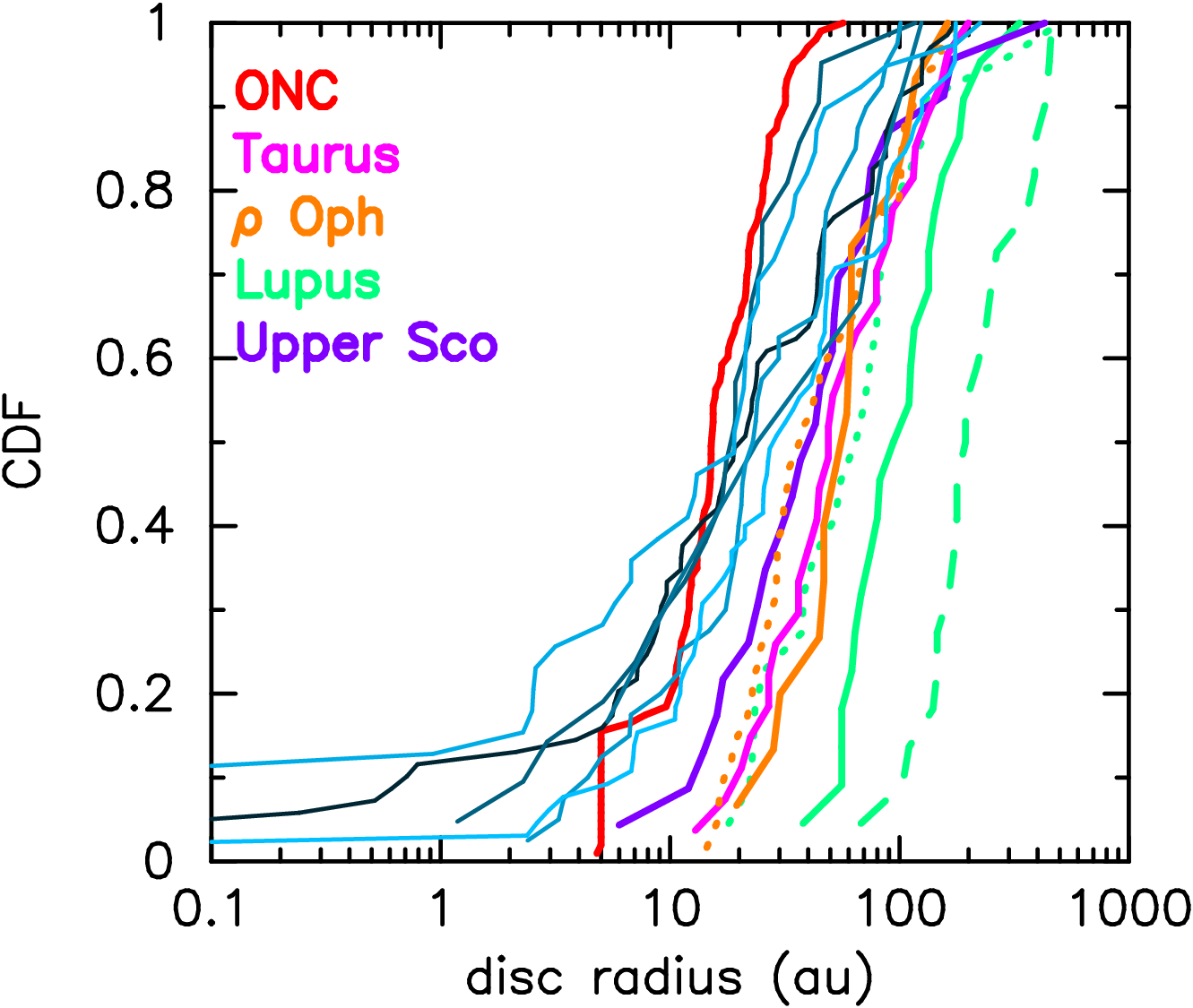}}}
\hspace*{0.3cm}\subfigure[10\,Myr]{\label{mod_dens_lowN_100au_viscous-c}\rotatebox{270}{\includegraphics[scale=0.23]{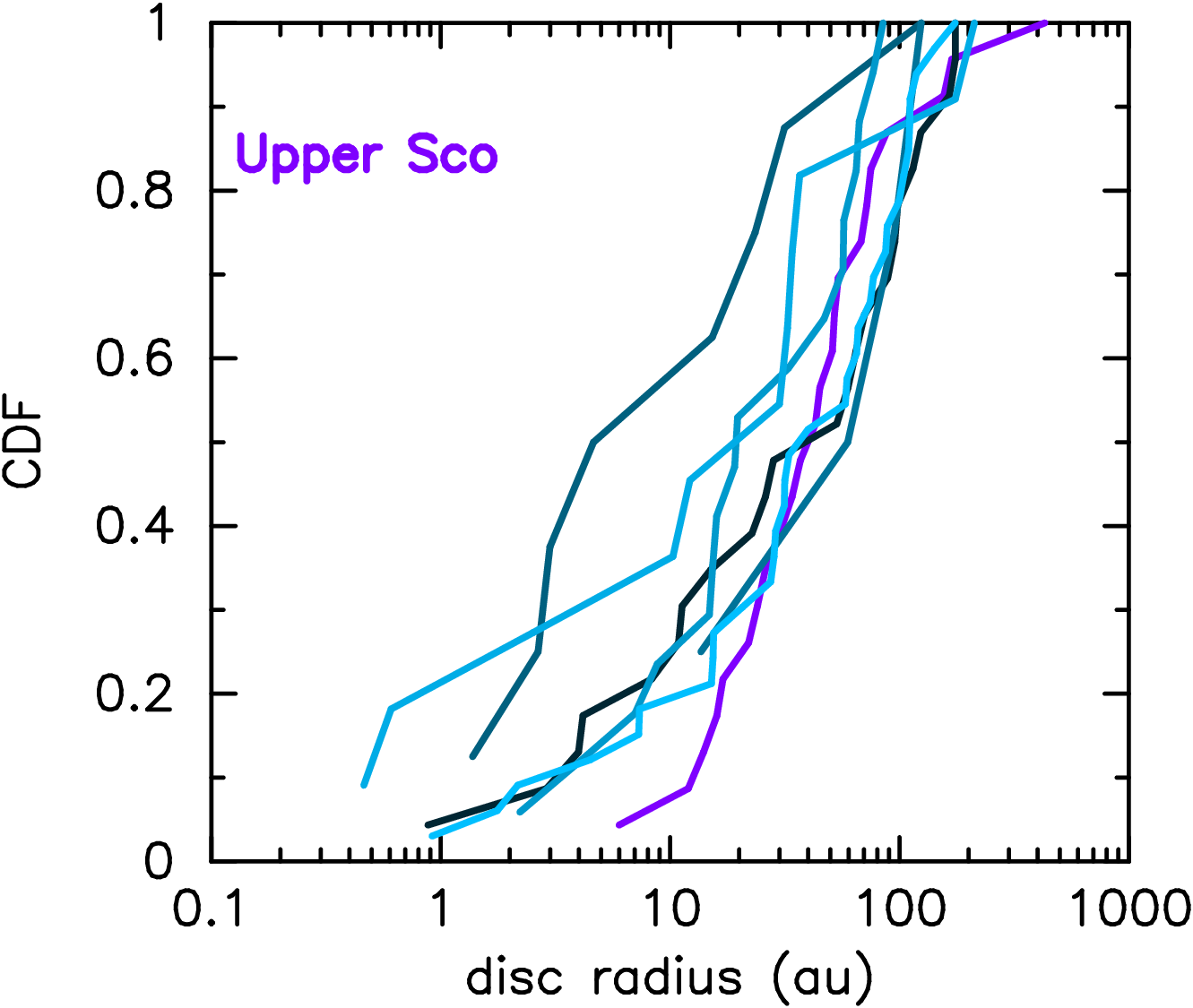}}}
\hspace*{0.3cm}\subfigure[1\,Myr]{\label{mod_dens_lowN_100au_viscous-d}\rotatebox{270}{\includegraphics[scale=0.23]{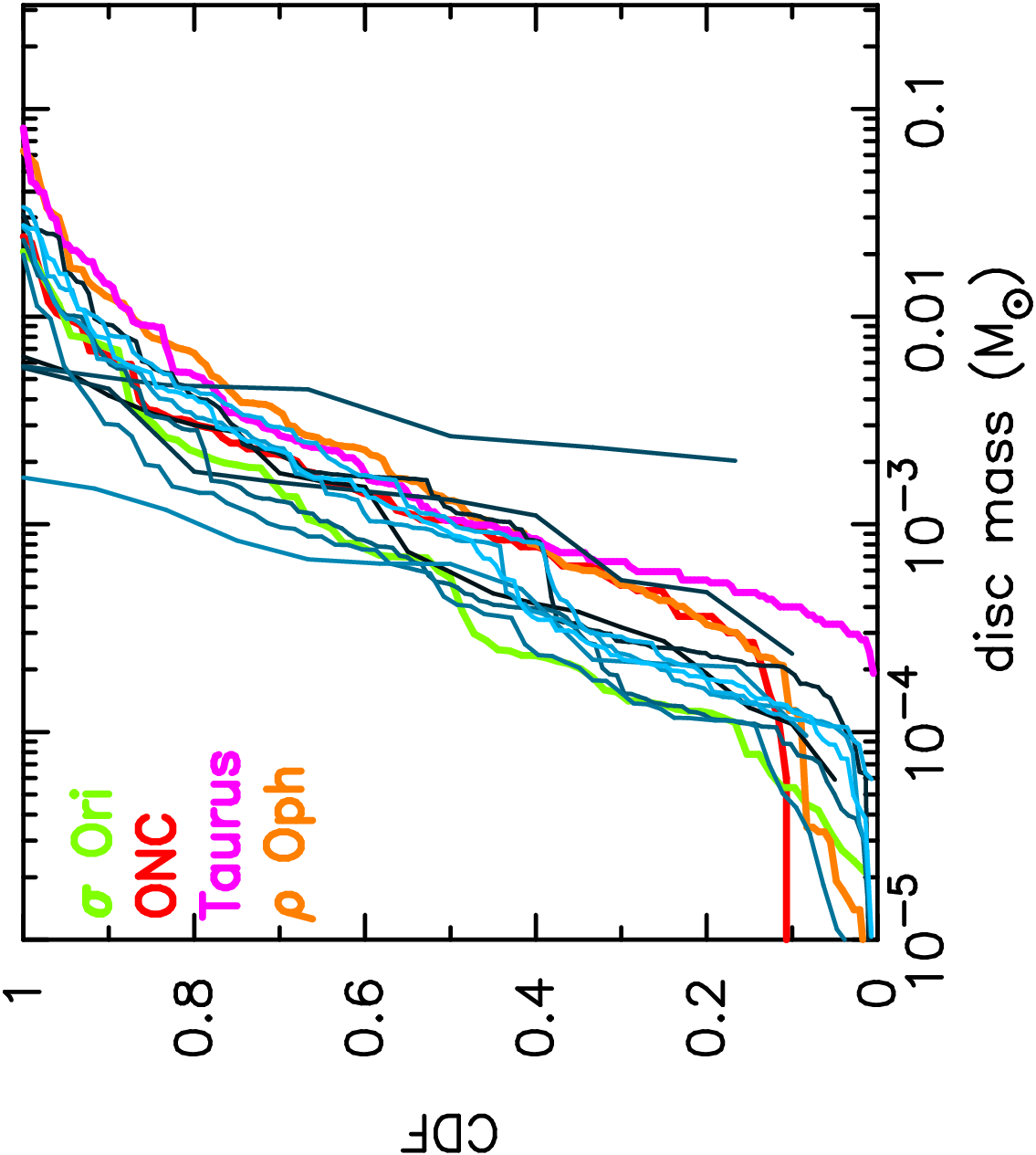}}}
\hspace*{0.3cm}\subfigure[5\,Myr]{\label{mod_dens_lowN_100au_viscous-e}\rotatebox{270}{\includegraphics[scale=0.23]{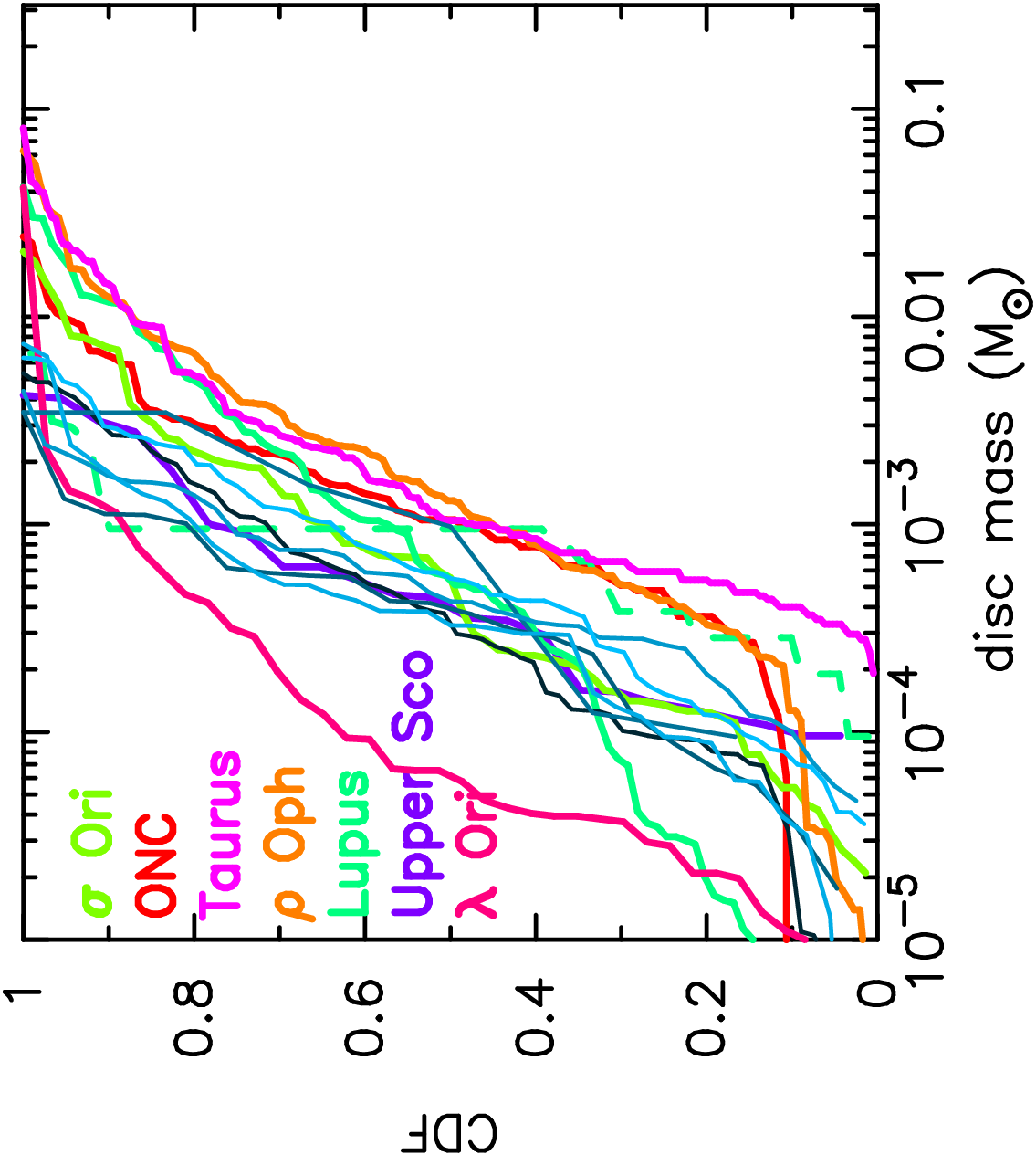}}}
\hspace*{0.3cm}\subfigure[10\,Myr]{\label{mod_dens_lowN_100au_viscous-f}\rotatebox{270}{\includegraphics[scale=0.23]{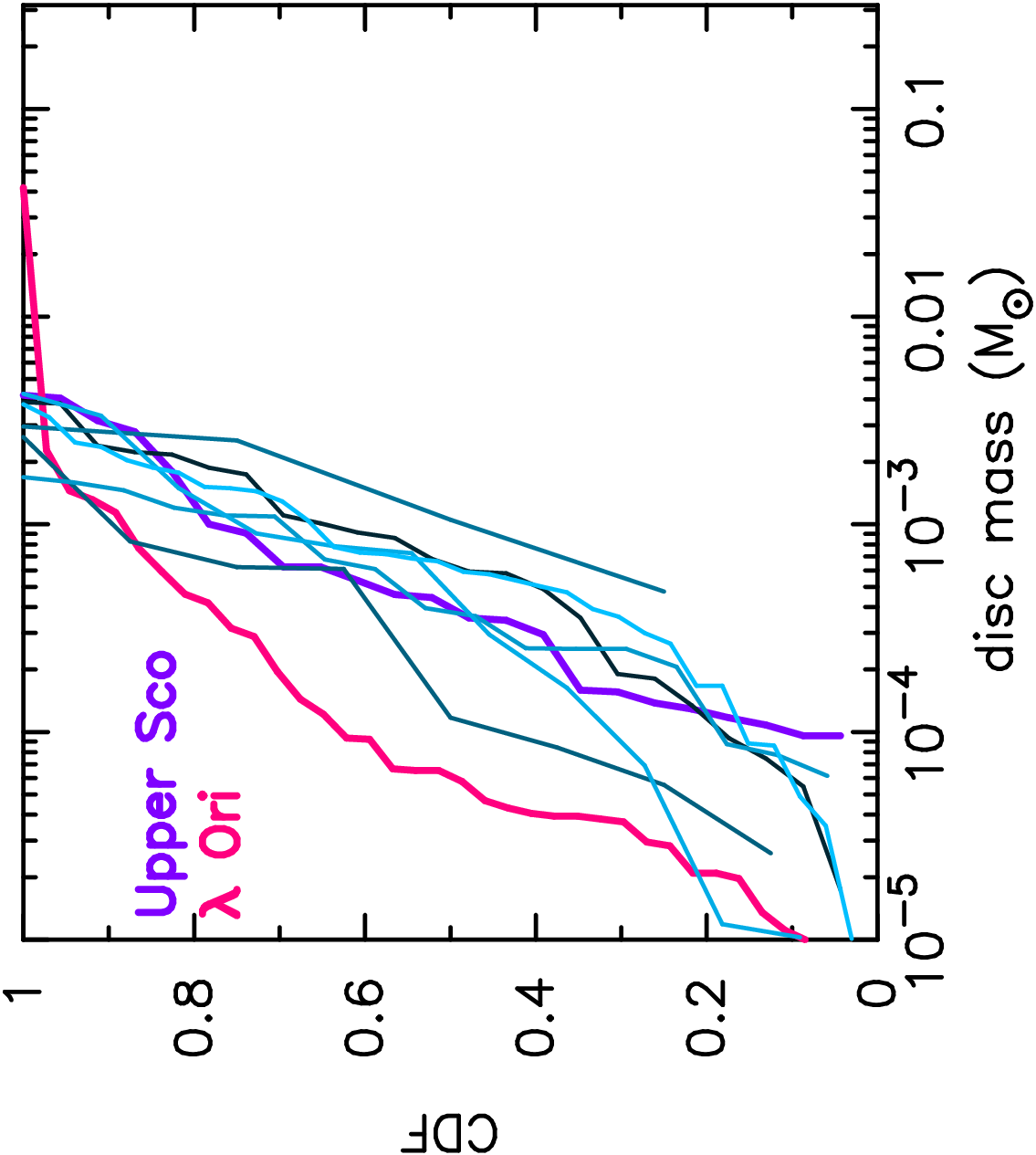}}}
\hspace*{0.3cm}\subfigure[Mass--radius, 1\,Myr]{\label{mod_dens_lowN_100au_viscous-g}\rotatebox{270}{\includegraphics[scale=0.23]{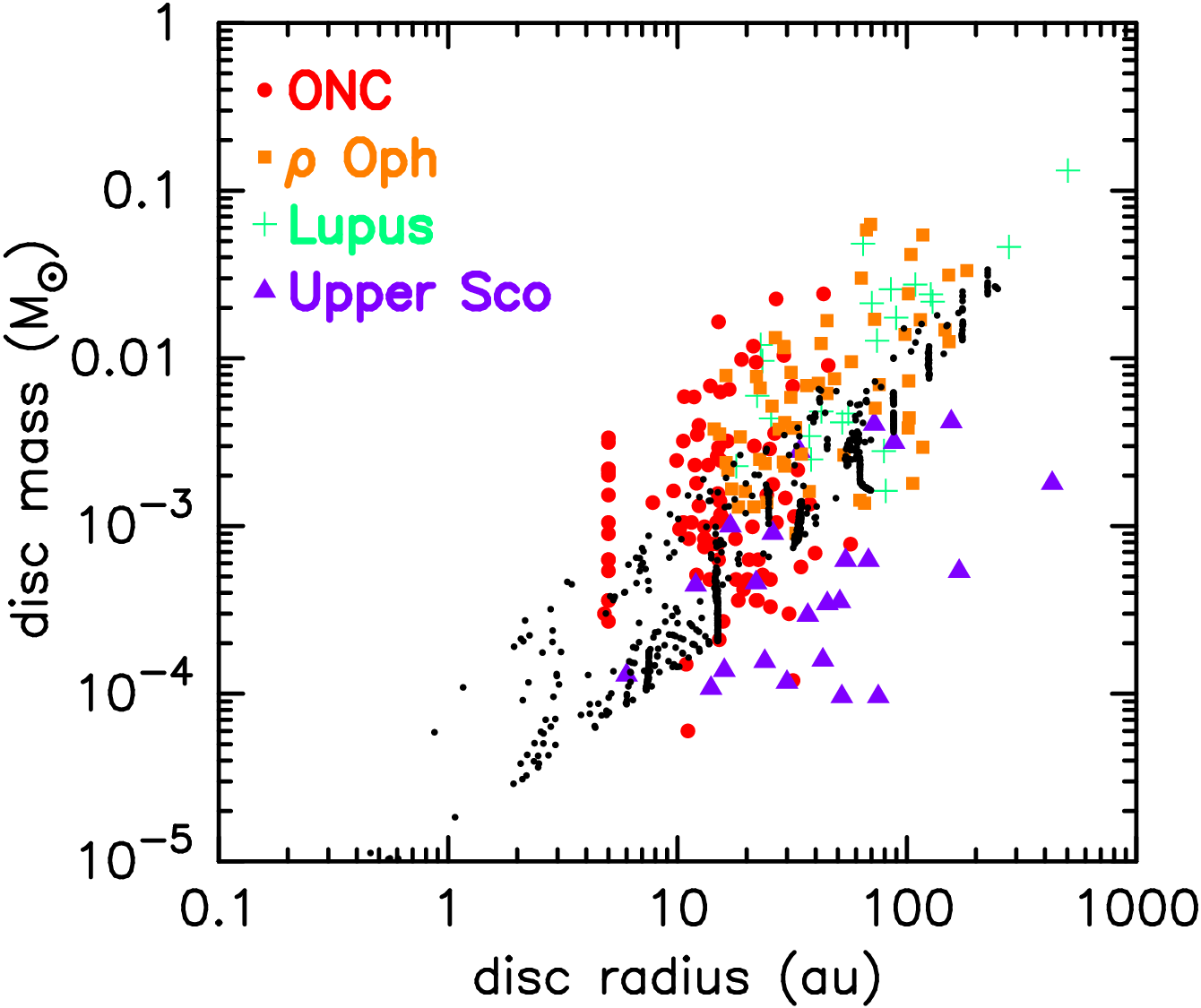}}}
\hspace*{0.3cm}\subfigure[Mass--radius, 5\,Myr]{\label{mod_dens_lowN_100au_viscous-h}\rotatebox{270}{\includegraphics[scale=0.23]{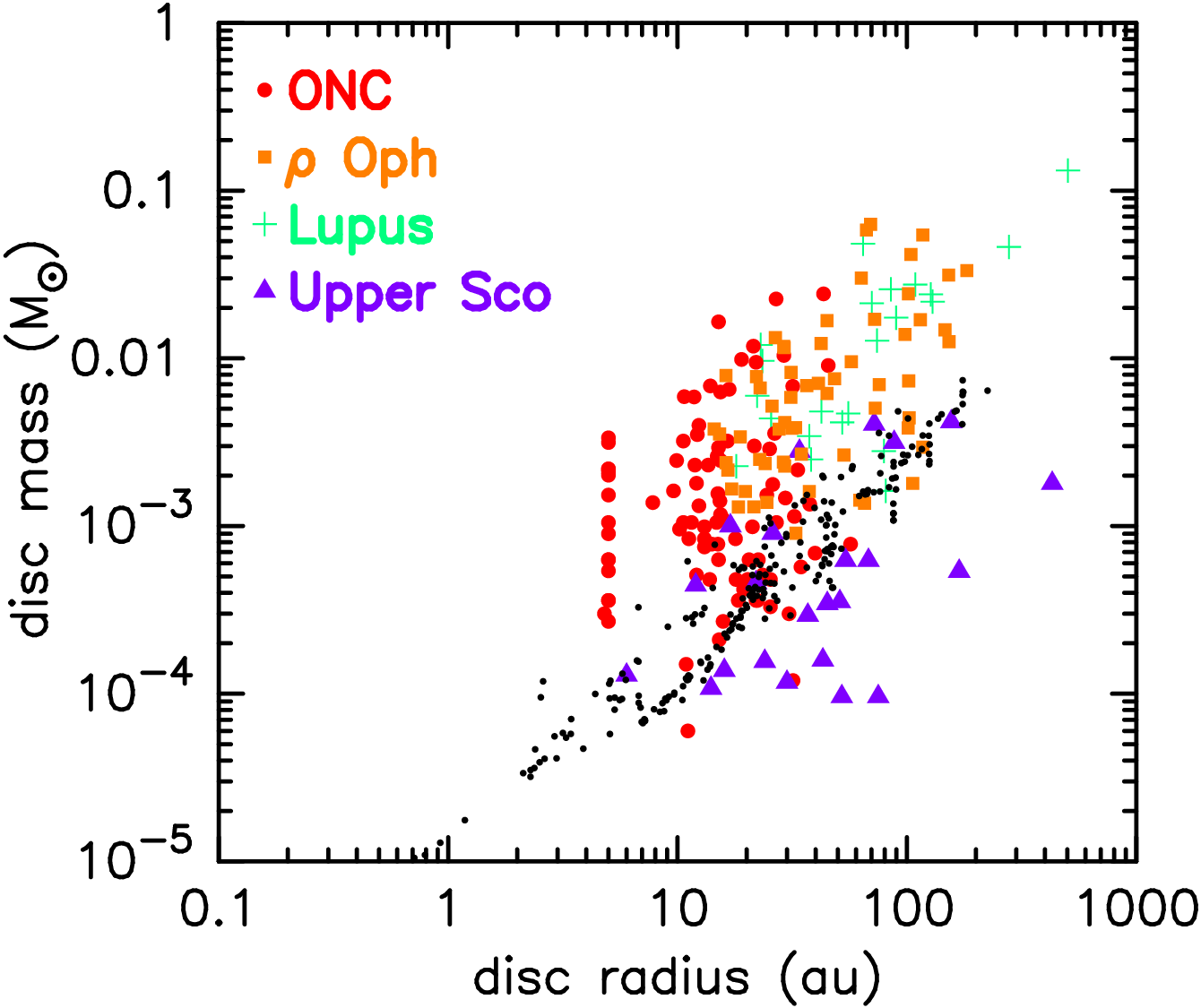}}}
\hspace*{0.3cm}\subfigure[Disc fractions]{\label{mod_dens_lowN_100au_viscous-i}\rotatebox{270}{\includegraphics[scale=0.23]{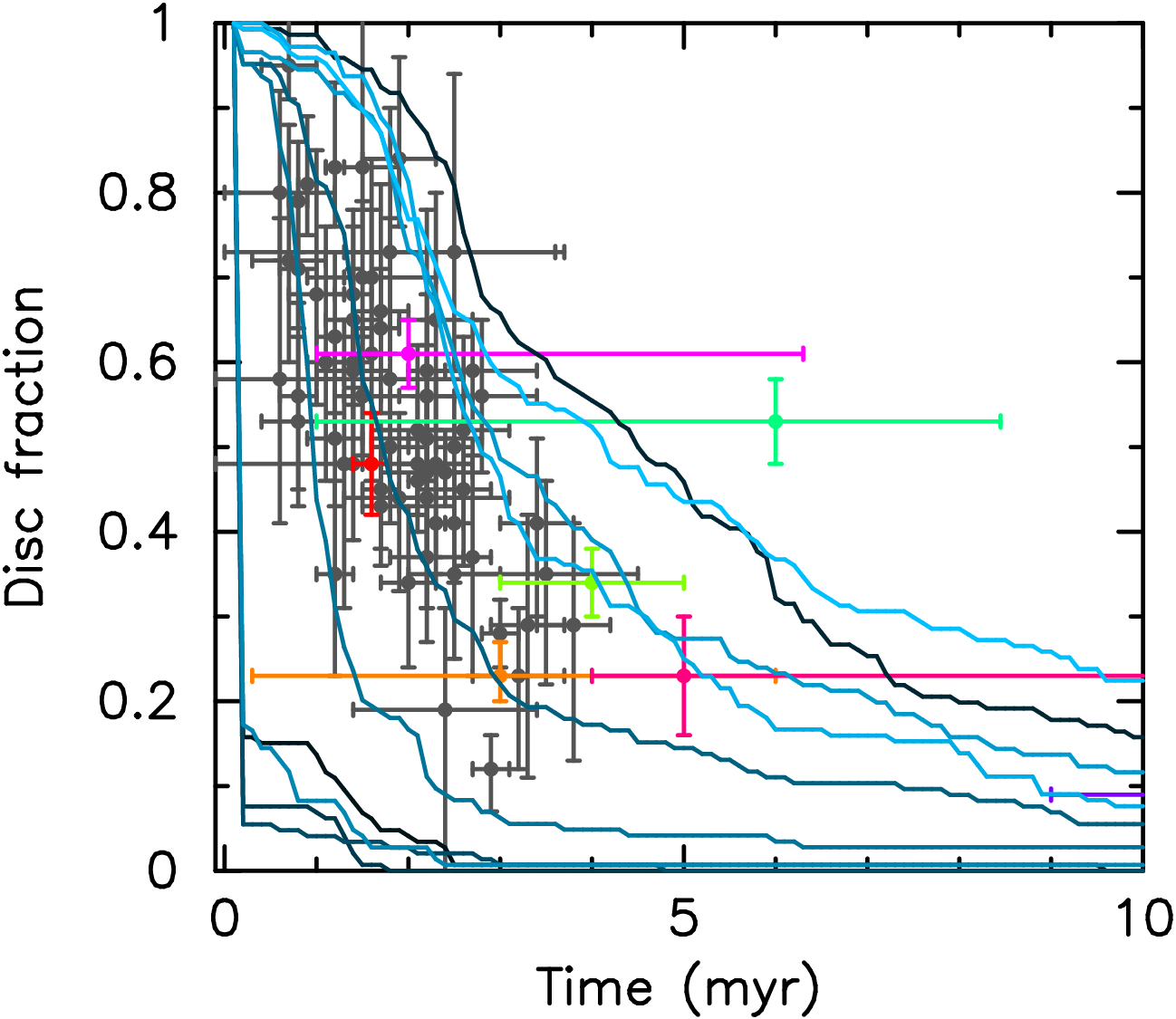}}}

\caption[bf]{Results for simulations in which the initial disc radii are drawn from a delta function with $r_{\rm disc} = 100$\,au, and the discs are allowed to evolve inwards due to external photoevaporation, and outwards due to viscous spreading.  The star-forming regions have moderate initial stellar density ($\tilde{\rho} \sim 100$\,M$_\odot$\,pc$^{-3}$), but with fewer stars ($N = 150$). In panels (a)--(c) we show the cumulative distributions of disc radii at different ages, and in panels (d)--(f) we show the cumulative distributions of disc masses. The blue lines of varying hues show the results from the ten individual $N$-body simulations. The mint green dashed line represents the respective gas distribution in the Lupus star-forming region, whereas solid lines represent the dust distributions. Dotted lines indicate an alternative dust radius for $\rho$~Oph (orange line) and Lupus (mint green line); see Table~\ref{obs_info} for further details. In panels (g) and (h) we show the disc mass versus disc radius for four observed star-forming regions at 1 and 5\,Myr, and data from all ten post-processed $N$-body simulation are shown by the black points. We plot the disc fractions (defined as when the discs have non-zero mass) as a function of time in our $N$-body simulations in panel (i), with observational data taken from \citet{Richert18} and \citet{Ribas15}.}
\label{mod_dens_lowN_100au_viscous}
  \end{center}
\end{figure*}

We find that the disc radii distributions from our simulations are partially consistent with the observed regions (in particular Taurus and $\rho$~Oph) for discs $r_{\rm disc} > \sim 50$\,au, but there are too many discs with very small radii in our simulations (Figs.~\ref{mod_dens_lowN_100au_viscous-a}~and~\ref{mod_dens_lowN_100au_viscous-b}). 

After 1\,Myr of evolution, the disc mass distributions from the simulations are similar to many of those in the observed star-forming regions (Fig.~\ref{mod_dens_lowN_100au_viscous-d}), but then subsequent mass-loss causes the distributions in the simulations to evolve to lower masses than are observed (Fig.~\ref{mod_dens_lowN_100au_viscous-e}). As with previous simulations, the surviving disc population at 10\,Myr is consistent with the observations of Upper~Sco, but not $\lambda$~Ori (Fig.~\ref{mod_dens_lowN_100au_viscous-f}).

As in the other simulations, the plots of individual disc mass versus disc radius are broadly consistent with some of the observed values (Figs.~\ref{mod_dens_lowN_100au_viscous-g}~and~\ref{mod_dens_lowN_100au_viscous-h}); in this instance we have combined all ten simulations due to the low number of stars in each individual simulation.

Finally, we note that in four out of ten simulations the fraction of discs is reduced within the first Myr, such that the simulations are not consistent with any observed disc fraction (see Fig.~\ref{mod_dens_lowN_100au_viscous-i}). However, the simulations with lower FUV and EUV radiation fields retain a substantial number of discs, and are consistent with the observed disc fractions from \citet{Richert18}.

\section{Discussion}

Before discussing our results, we deem it prudent to discuss the caveats in our models, as we are wary of overinterpreting some of the evolution we see in our simulated disc populations. We will then attempt to draw conclusions based on the evolution of the disc radii distributions, disc mass distributions, and disc fractions. 

The observed disc fractions are usually calculated by determining if a star has an infrared excess, thought to be the signature of a disc with micron-sized dust grains. Given that most of the mass lost from a disc due to photoevaporation is gas, comparing the disc fractions in our simulations with the observed fractions is unlikely to be a straightforward comparison.

A similar caveat exists when comparing the dust mass distributions. We have taken the observed disc masses, assumed a 1:100 dust to gas ratio, and then adjusted the observed distributions accordingly. The mass lost due to FUV photoevaporation is predominantly gas, and so again conclusions drawn from these comparisons should be made with caution.

During the photoevaporation of the gas, the dust grains will likely be coagulating into larger grains that are subsequently prone to inward migration \citep{Weidenschilling77b,Birnstiel10} and settling about the disc mid-plane \citep{Goldreich73,Drazkowska22}. It is therefore unclear whether the dust radius (as probed by the majority of the cited observational samples) accurately traces the gas radius.

Observations indicate that the gas and dust radii may be similar \citep[e.g.][]{Ansdell18}, although gas radii are likely to be higher by a factor of around two \citep{Sanchis21,Long22} and simulations show the dust radius moves with the gas radius during disc evolution (both inward evolution due to external photoevaporation, and outward evolution due to viscous spreading, \citep{Sellek20}).

Our initial disc mass distributions assume that the initial mass is 10\,per cent of the host star mass, and we have not tested any other distributions in our analysis. We believe this is a reasonable assumption \citep[discs in the early phases (Class I and Class II) of protostellar evolution are thought to have masses up to 0.1\,M$_\star$, e.g.][]{Weidenschilling77,Beckwith90,Eisner06,Andrews05,Sheehan17}, although we note that other work assumes lower initial disc masses \citep[e.g.][]{Parker21b}, with lower masses leading to greater destruction of the disc population.

In the majority of our simulations, our initial disc radii are drawn from a delta function, with either $r_{\rm disc} = 10$\,au or $r_{\rm disc} = 100$\,au. \citet{Coleman22} assume an initial radii distribution of the form
\begin{equation}
r_{\rm disc} = 200\,{\rm au}\left(\frac{M_\star}{\rm M_\odot}\right)^{0.3}.
\end{equation}
We note that in this distribution the smallest discs would have radii $r_{\rm disc} \geq 100$\,au, assuming our lower limit to the stellar IMF of $M_\star = 0.1$\,M$_\odot$, and hence be even more susceptible to destruction from photoevaporation in our simulations.

Despite using a delta function as an input, the variable $G_0$ field, and different disc mass for each individual star means that the disc radius distribution rapidly changes to resemble a similar shape to the observed distributions. We find that in the absence of viscous evolution, the change in the disc radius due to external photoevaporation never produces any of the observed disc radii distributions, with the surviving disc radii always being smaller than those in the observed star-forming regions.

We add a further caveat here that our prescription for the change in disc radius due to photoevaporation assumes the disc radius changes in proportion to the reduction in disc mass \citep[following][]{Haworth18b}, but  other implementations of the change in disc radius could potentially produce different results.

We find that assuming an initial disc radius distribution where the discs are $r_{\rm disc} = 10$\,au never produces the observed disc radii distributions, with the simulated disc radii all becoming too small. Conversely, these discs are less susceptible to mass-loss from photoevaporation, and their masses are always too high. As the observed ONC disc radii distribution \citep{Eisner18} has a median radius of $\sim$20\,au, we note that the ONC disc distribution must have evolved from a distribution with larger radii, unless the discs are initially much less massive than 0.1\,M$_\star$.

The simulations that are most consistent with the observed disc radii distributions are those where the disc radii are $r_{\rm disc} = 100$\,au and the initial stellar density is $\tilde{\rho} \sim 100$\,M$_\odot$\,pc$^{-3}$ (Fig.~\ref{mod_dens_100au}). The density in these simulation remains fairly constant over time \citep{Parker21a}, but we note that the observed present-day density in the ONC is slightly higher, at around 400\,M$_\odot$\,pc$^{-3}$ \citep{King12a}. Analyses of the spatial and kinematic distributions of stars in the ONC suggest a much higher initial density \citep{Allison10,Allison11,Parker14b,Schoettler20}, which would place our results in tension with those dynamical considerations.

Simulations with more dense initial conditions for the ONC ($\tilde{\rho} \sim 1000$\,M$_\odot$\,pc$^{-3}$, Fig.~\ref{high_dens_100au_viscous}) would only be consistent with  the disc radii observations if a fairly substantial number of discs with small radii ($<5$\,au) exist, currently beyond the resolution of the observations. We note that these high densities are inconsistent with the observed disc fraction in the ONC, although this carries the significant caveat that we are mainly modelling the loss of gas from the discs, and the disc fractions are calculated based on infrared excess from micron-sized dust particles. 

Low density simulations ($\tilde{\rho} \sim 10$\,M$_\odot$\,pc$^{-3}$, Fig.~\ref{low_dens_100au_viscous}) produce radii (and disc mass) distributions that are consistent with the observations of Taurus and $\rho$~Oph, as well as Upper Sco at later ages, but not the ONC.

Many of the observed star-forming regions contain only several hundred stars, and many of these regions do not contain very massive stars. However, \citet{Haworth17} show that discs in Lupus have been affected by external photoevaporation and in Fig.~\ref{mod_dens_lowN_100au_viscous} we show the results from simulations with $N = 150$ stars, rather than $N = 1500$ stars. Interestingly, we cannot reproduce the disc radii distribution observed in Lupus with our models, suggesting that our models are missing a crucial aspect of disc evolution \citep{Coleman22}. \citet{Haworth17} suggest the extended diffuse emission around IM~Lup could be due to external photoevaporation from the OB association adjacent to this region. The radiation field in Lupus is still low ($\sim 4G_0$), so any mass-loss due to photoevaporation is minimal -- \citet{Haworth17} estimate a photoevaporation rate of $\dot{M} \sim 10^{-8}$\,$M_\odot$\,yr$^{-1}$.

\section{Conclusions}

We present $N$-body simulations of star-forming regions of different stellar densities in which we calculate the EUV and FUV radiation fields from intermediate - massive stars ($>$5\,M$_\odot$). We then perform a post-processing analysis to calculate the mass lost from prototoplanetary discs depending on the strength of the radiation fields at each point in time in the simulations. We then calculate the inward evolution of the disc radii, and the outward evolution due to viscous spreading. We compare the radii and mass distributions to recent observations. Our conclusions are the following.

(i) Our simulated disc radii distributions are only consistent with the observational data if we allow viscous spreading to occur following any inward evolution of the disc radius due to external photoevaporation.

(ii) The initial disc distribution in the ONC cannot be the same as that currently observed, and we extend this conclusion to state that the initial disc radii distribution \emph{must be} dominated by discs with radii $>>$10\,au.

(iii) The simulations that best reproduce the observed disc radius distribution in the ONC have only moderate ($\tilde{\rho} \sim 100$\,M$_\odot$\,pc$^{-3}$) stellar densities, in tension with dynamical models for the ONC that predict much higher stellar densities. This tension can only be resolved if the external photoevaporation only destroys the gas component of discs and the observed discs in the ONC are mostly gas-free.

(iv) Our models fail to reproduce the relatively large radii observed in the Lupus star-forming region. This could be due to a lack of external photoevaporation in this star-forming region, although recent analysis tailored to Lupus suggests external photoevaporation has occurred to the discs in this region \citep{Haworth17}.

\section*{Acknowledgements}

We are grateful to the anonymous referee, whose comments and suggestions greatly improved the original manuscript. RJP acknowledges support from the Royal Society in the form of a Dorothy Hodgkin Fellowship.

\section*{Data Availability}

The data generated in this study will be made available upon reasonable request to the corresponding author.

\bibliographystyle{mnras}  
\bibliography{general_ref}

\begin{thebibliography}{}
\makeatletter
\relax
\def\mn@urlcharsother{\let\do\@makeother \do\$\do\&\do\#\do\^\do\_\do\%\do\~}
\def\mn@doi{\begingroup\mn@urlcharsother \@ifnextchar [ {\mn@doi@}
  {\mn@doi@[]}}
\def\mn@doi@[#1]#2{\def\@tempa{#1}\ifx\@tempa\@empty \href
  {http://dx.doi.org/#2} {doi:#2}\else \href {http://dx.doi.org/#2} {#1}\fi
  \endgroup}
\def\mn@eprint#1#2{\mn@eprint@#1:#2::\@nil}
\def\mn@eprint@arXiv#1{\href {http://arxiv.org/abs/#1} {{\tt arXiv:#1}}}
\def\mn@eprint@dblp#1{\href {http://dblp.uni-trier.de/rec/bibtex/#1.xml}
  {dblp:#1}}
\def\mn@eprint@#1:#2:#3:#4\@nil{\def\@tempa {#1}\def\@tempb {#2}\def\@tempc
  {#3}\ifx \@tempc \@empty \let \@tempc \@tempb \let \@tempb \@tempa \fi \ifx
  \@tempb \@empty \def\@tempb {arXiv}\fi \@ifundefined
  {mn@eprint@\@tempb}{\@tempb:\@tempc}{\expandafter \expandafter \csname
  mn@eprint@\@tempb\endcsname \expandafter{\@tempc}}}

\bibitem[\protect\citeauthoryear{Adams}{Adams}{2010}]{Adams10}
Adams F.~C.,  2010, ARA\&A, 48, 47

\bibitem[\protect\citeauthoryear{Adams, Hollenbach, Laughlin  \& Gorti}{Adams
  et~al.}{2004}]{Adams04}
Adams F.~C.,  Hollenbach D.,  Laughlin G.,   Gorti U.,  2004, ApJ, 611, 360

\bibitem[\protect\citeauthoryear{Allison \& Goodwin}{Allison \&
  Goodwin}{2011}]{Allison11}
Allison R.~J.,  Goodwin S.~P.,  2011, MNRAS, 415, 1967

\bibitem[\protect\citeauthoryear{Allison, Goodwin, Parker, {Portegies Zwart}
  \& de Grijs}{Allison et~al.}{2010}]{Allison10}
Allison R.~J.,  Goodwin S.~P.,  Parker R.~J.,  {Portegies Zwart} S.~F.,   de
  Grijs R.,  2010, MNRAS, 407, 1098

\bibitem[\protect\citeauthoryear{{Alves}, {Cleeves}, {Girart}, {Zhu}, {Franco},
  {Zurlo}  \& {Caselli}}{{Alves} et~al.}{2020}]{Alves20}
{Alves} F.~O.,  {Cleeves} L.~I.,  {Girart} J.~M.,  {Zhu} Z.,  {Franco} G.
  A.~P.,  {Zurlo} A.,   {Caselli} P.,  2020, \mn@doi [\apjl]
  {10.3847/2041-8213/abc550}, \href
  {https://ui.adsabs.harvard.edu/abs/2020ApJ...904L...6A} {904, L6}

\bibitem[\protect\citeauthoryear{{Andrews} \& {Williams}}{{Andrews} \&
  {Williams}}{2005}]{Andrews05}
{Andrews} S.~M.,  {Williams} J.~P.,  2005, \mn@doi [ApJL] {10.1086/427325},
  \href {http://adsabs.harvard.edu/abs/2005ApJ...619L.175A} {619, L175}

\bibitem[\protect\citeauthoryear{{Andrews}, {Wilner}, {Hughes}, {Qi}  \&
  {Dullemond}}{{Andrews} et~al.}{2010}]{Andrews10}
{Andrews} S.~M.,  {Wilner} D.~J.,  {Hughes} A.~M.,  {Qi} C.,   {Dullemond}
  C.~P.,  2010, \mn@doi [ApJ] {10.1088/0004-637X/723/2/1241}, \href
  {https://ui.adsabs.harvard.edu/abs/2010ApJ...723.1241A} {723, 1241}

\bibitem[\protect\citeauthoryear{{Andrews}, {Rosenfeld}, {Kraus}  \&
  {Wilner}}{{Andrews} et~al.}{2013}]{Andrews13}
{Andrews} S.~M.,  {Rosenfeld} K.~A.,  {Kraus} A.~L.,   {Wilner} D.~J.,  2013,
  \mn@doi [\apj] {10.1088/0004-637X/771/2/129}, \href
  {https://ui.adsabs.harvard.edu/abs/2013ApJ...771..129A} {771, 129}

\bibitem[\protect\citeauthoryear{{Ansdell} et~al.,}{{Ansdell}
  et~al.}{2016}]{Ansdell16}
{Ansdell} M.,  et~al., 2016, \mn@doi [\apj] {10.3847/0004-637X/828/1/46}, \href
  {https://ui.adsabs.harvard.edu/abs/2016ApJ...828...46A} {828, 46}

\bibitem[\protect\citeauthoryear{{Ansdell}, {Williams}, {Manara}, {Miotello},
  {Facchini}, {van der Marel}, {Testi}  \& {van Dishoeck}}{{Ansdell}
  et~al.}{2017}]{Ansdell17}
{Ansdell} M.,  {Williams} J.~P.,  {Manara} C.~F.,  {Miotello} A.,  {Facchini}
  S.,  {van der Marel} N.,  {Testi} L.,   {van Dishoeck} E.~F.,  2017, \mn@doi
  [\aj] {10.3847/1538-3881/aa69c0}, \href
  {http://adsabs.harvard.edu/abs/2017AJ....153..240A} {153, 240}

\bibitem[\protect\citeauthoryear{{Ansdell} et~al.,}{{Ansdell}
  et~al.}{2018}]{Ansdell18}
{Ansdell} M.,  et~al., 2018, \mn@doi [\apj] {10.3847/1538-4357/aab890}, \href
  {https://ui.adsabs.harvard.edu/abs/2018ApJ...859...21A} {859, 21}

\bibitem[\protect\citeauthoryear{{Ansdell} et~al.,}{{Ansdell}
  et~al.}{2020}]{Ansdell20}
{Ansdell} M.,  et~al., 2020, \mn@doi [\aj] {10.3847/1538-3881/abb9af}, \href
  {https://ui.adsabs.harvard.edu/abs/2020AJ....160..248A} {160, 248}

\bibitem[\protect\citeauthoryear{{Armitage}}{{Armitage}}{2000}]{Armitage00}
{Armitage} P.~J.,  2000, A\&A, 362, 968

\bibitem[\protect\citeauthoryear{{Barenfeld}, {Carpenter}, {Sargent}, {Isella}
  \& {Ricci}}{{Barenfeld} et~al.}{2017}]{Barenfeld17}
{Barenfeld} S.~A.,  {Carpenter} J.~M.,  {Sargent} A.~I.,  {Isella} A.,
  {Ricci} L.,  2017, \mn@doi [\apj] {10.3847/1538-4357/aa989d}, \href
  {https://ui.adsabs.harvard.edu/abs/2017ApJ...851...85B} {851, 85}

\bibitem[\protect\citeauthoryear{Bastian, Covey  \& Meyer}{Bastian
  et~al.}{2010}]{Bastian10}
Bastian N.,  Covey K.~R.,   Meyer M.~R.,  2010, ARA\&A, 48, 339

\bibitem[\protect\citeauthoryear{{Bayo} et~al.,}{{Bayo} et~al.}{2011}]{Bayo11}
{Bayo} A.,  et~al., 2011, \mn@doi [A\&A] {10.1051/0004-6361/201116617}, 536,
  A63

\bibitem[\protect\citeauthoryear{{Beccari} et~al.,}{{Beccari}
  et~al.}{2017}]{Beccari17}
{Beccari} G.,  et~al., 2017, \mn@doi [\aap] {10.1051/0004-6361/201730432},
  \href {https://ui.adsabs.harvard.edu/abs/2017A&A...604A..22B} {604, A22}

\bibitem[\protect\citeauthoryear{{Beckwith}, {Sargent}, {Chini}  \&
  {Guesten}}{{Beckwith} et~al.}{1990}]{Beckwith90}
{Beckwith} S. V.~W.,  {Sargent} A.~I.,  {Chini} R.~S.,   {Guesten} R.,  1990,
  \mn@doi [\aj] {10.1086/115385}, \href
  {https://ui.adsabs.harvard.edu/abs/1990AJ.....99..924B} {99, 924}

\bibitem[\protect\citeauthoryear{{Bell}, {Naylor}, {Mayne}, {Jeffries}  \&
  {Littlefair}}{{Bell} et~al.}{2013}]{Bell13}
{Bell} C.~P.~M.,  {Naylor} T.,  {Mayne} N.~J.,  {Jeffries} R.~D.,
  {Littlefair} S.~P.,  2013, \mn@doi [MNRAS] {10.1093/mnras/stt1075}, \href
  {http://adsabs.harvard.edu/abs/2013MNRAS.434..806B} {434, 806}

\bibitem[\protect\citeauthoryear{{Birnstiel}, {Dullemond}  \&
  {Brauer}}{{Birnstiel} et~al.}{2010}]{Birnstiel10}
{Birnstiel} T.,  {Dullemond} C.~P.,   {Brauer} F.,  2010, \mn@doi [\aap]
  {10.1051/0004-6361/200913731}, \href
  {https://ui.adsabs.harvard.edu/abs/2010A&A...513A..79B} {513, A79}

\bibitem[\protect\citeauthoryear{Bonnell, Smith, Davies  \& Horne}{Bonnell
  et~al.}{2001}]{Bonnell01b}
Bonnell I.~A.,  Smith K.~W.,  Davies M.~B.,   Horne K.,  2001, MNRAS, 322, 859

\bibitem[\protect\citeauthoryear{Bressert et~al.,}{Bressert
  et~al.}{2010}]{Bressert10}
Bressert E.,  et~al., 2010, MNRAS, 409, L54

\bibitem[\protect\citeauthoryear{{Buckner} et~al.,}{{Buckner}
  et~al.}{2019}]{Buckner19}
{Buckner} A. S.~M.,  et~al., 2019, \mn@doi [\aap]
  {10.1051/0004-6361/201832936}, \href
  {https://ui.adsabs.harvard.edu/abs/2019A&A...622A.184B} {622, A184}

\bibitem[\protect\citeauthoryear{{Caballero}, {de Burgos}, {Alonso-Floriano},
  {Cabrera-Lavers}, {Garc{\'\i}a-{\'A}lvarez}  \& {Montes}}{{Caballero}
  et~al.}{2019}]{Caballero19}
{Caballero} J.~A.,  {de Burgos} A.,  {Alonso-Floriano} F.~J.,  {Cabrera-Lavers}
  A.,  {Garc{\'\i}a-{\'A}lvarez} D.,   {Montes} D.,  2019, \mn@doi [\aap]
  {10.1051/0004-6361/201935987}, \href
  {https://ui.adsabs.harvard.edu/abs/2019A&A...629A.114C} {629, A114}

\bibitem[\protect\citeauthoryear{{Cai}, {Kouwenhoven}, {Portegies Zwart}  \&
  {Spurzem}}{{Cai} et~al.}{2017}]{Cai17a}
{Cai} M.~X.,  {Kouwenhoven} M.~B.~N.,  {Portegies Zwart} S.~F.,   {Spurzem} R.,
   2017, \mn@doi [\mnras] {10.1093/mnras/stx1464}, \href
  {http://adsabs.harvard.edu/abs/2017MNRAS.470.4337C} {470, 4337}

\bibitem[\protect\citeauthoryear{Cartwright \& Whitworth}{Cartwright \&
  Whitworth}{2004}]{Cartwright04}
Cartwright A.,  Whitworth A.~P.,  2004, MNRAS, 348, 589

\bibitem[\protect\citeauthoryear{{Cieza} et~al.,}{{Cieza}
  et~al.}{2019}]{Cieza19}
{Cieza} L.~A.,  et~al., 2019, \mn@doi [\mnras] {10.1093/mnras/sty2653}, \href
  {https://ui.adsabs.harvard.edu/abs/2019MNRAS.482..698C} {482, 698}

\bibitem[\protect\citeauthoryear{{Cleeves}, {{\"O}berg}, {Wilner}, {Huang},
  {Loomis}, {Andrews}  \& {Czekala}}{{Cleeves} et~al.}{2016}]{Cleeves16}
{Cleeves} L.~I.,  {{\"O}berg} K.~I.,  {Wilner} D.~J.,  {Huang} J.,  {Loomis}
  R.~A.,  {Andrews} S.~M.,   {Czekala} I.,  2016, \mn@doi [\apj]
  {10.3847/0004-637X/832/2/110}, \href
  {https://ui.adsabs.harvard.edu/abs/2016ApJ...832..110C} {832, 110}

\bibitem[\protect\citeauthoryear{{Coleman} \& {Haworth}}{{Coleman} \&
  {Haworth}}{2022}]{Coleman22}
{Coleman} G. A.~L.,  {Haworth} T.~J.,  2022, arXiv e-prints, \href
  {https://ui.adsabs.harvard.edu/abs/2022arXiv220402303C} {p. arXiv:2204.02303}

\bibitem[\protect\citeauthoryear{{Concha-Ram{\'\i}rez}, {Wilhelm}, {Portegies
  Zwart}  \& {Haworth}}{{Concha-Ram{\'\i}rez} et~al.}{2019a}]{ConchaRamirez19}
{Concha-Ram{\'\i}rez} F.,  {Wilhelm} M. J.~C.,  {Portegies Zwart} S.,
  {Haworth} T.~J.,  2019a, MNRAS, \href
  {https://ui.adsabs.harvard.edu/abs/2019arXiv190703760C} {p. arXiv:1907.03760}

\bibitem[\protect\citeauthoryear{{Concha-Ram{\'\i}rez}, {Vaher}  \& {Portegies
  Zwart}}{{Concha-Ram{\'\i}rez} et~al.}{2019b}]{ConchaRamirez19a}
{Concha-Ram{\'\i}rez} F.,  {Vaher} E.,   {Portegies Zwart} S.,  2019b, \mn@doi
  [\mnras] {10.1093/mnras/sty2721}, \href
  {https://ui.adsabs.harvard.edu/abs/2019MNRAS.482..732C} {482, 732}

\bibitem[\protect\citeauthoryear{{Cox}}{{Cox}}{2000}]{Cox00}
{Cox} A.~N.,  2000, {Allen's astrophysical quantities}

\bibitem[\protect\citeauthoryear{{Da Rio}, {Robberto}, {Soderblom}, {Panagia},
  {Hillenbrand}, {Palla}  \& {Stassun}}{{Da Rio} et~al.}{2010}]{DaRio10}
{Da Rio} N.,  {Robberto} M.,  {Soderblom} D.~R.,  {Panagia} N.,  {Hillenbrand}
  L.~A.,  {Palla} F.,   {Stassun} K.~G.,  2010, \mn@doi [ApJ]
  {10.1088/0004-637X/722/2/1092}, \href
  {https://ui.adsabs.harvard.edu/abs/2010ApJ...722.1092D} {722, 1092}

\bibitem[\protect\citeauthoryear{{Daffern-Powell} \& {Parker}}{{Daffern-Powell}
  \& {Parker}}{2020}]{DaffernPowell20}
{Daffern-Powell} E.~C.,  {Parker} R.~J.,  2020, \mn@doi [\mnras]
  {10.1093/mnras/staa575}, \href
  {https://ui.adsabs.harvard.edu/abs/2020MNRAS.493.4925D} {493, 4925}

\bibitem[\protect\citeauthoryear{{Drazkowska} et~al.,}{{Drazkowska}
  et~al.}{2022}]{Drazkowska22}
{Drazkowska} J.,  et~al., 2022, arXiv e-prints, \href
  {https://ui.adsabs.harvard.edu/abs/2022arXiv220309759D} {p. arXiv:2203.09759}

\bibitem[\protect\citeauthoryear{{Eisner} \& {Carpenter}}{{Eisner} \&
  {Carpenter}}{2006}]{Eisner06}
{Eisner} J.~A.,  {Carpenter} J.~M.,  2006, \mn@doi [\apj] {10.1086/500637},
  \href {https://ui.adsabs.harvard.edu/abs/2006ApJ...641.1162E} {641, 1162}

\bibitem[\protect\citeauthoryear{{Eisner} et~al.,}{{Eisner}
  et~al.}{2018}]{Eisner18}
{Eisner} J.~A.,  et~al., 2018, \mn@doi [\apj] {10.3847/1538-4357/aac3e2}, \href
  {http://adsabs.harvard.edu/abs/2018ApJ...860...77E} {860, 77}

\bibitem[\protect\citeauthoryear{{Fatuzzo} \& {Adams}}{{Fatuzzo} \&
  {Adams}}{2008}]{Fatuzzo08}
{Fatuzzo} M.,  {Adams} F.~C.,  2008, \mn@doi [ApJ] {10.1086/527469}, \href
  {http://adsabs.harvard.edu/abs/2008ApJ...675.1361F} {675, 1361}

\bibitem[\protect\citeauthoryear{{Flaherty} et~al.,}{{Flaherty}
  et~al.}{2020}]{Flaherty20}
{Flaherty} K.,  et~al., 2020, \mn@doi [\apj] {10.3847/1538-4357/ab8cc5}, \href
  {https://ui.adsabs.harvard.edu/abs/2020ApJ...895..109F} {895, 109}

\bibitem[\protect\citeauthoryear{{Galli} et~al.,}{{Galli}
  et~al.}{2020}]{Galli20}
{Galli} P.~A.~B.,  et~al., 2020, \mn@doi [\aap] {10.1051/0004-6361/202038717},
  \href {https://ui.adsabs.harvard.edu/abs/2020A&A...643A.148G} {643, A148}

\bibitem[\protect\citeauthoryear{{Goldreich} \& {Ward}}{{Goldreich} \&
  {Ward}}{1973}]{Goldreich73}
{Goldreich} P.,  {Ward} W.~R.,  1973, \mn@doi [\apj] {10.1086/152291}, \href
  {https://ui.adsabs.harvard.edu/abs/1973ApJ...183.1051G} {183, 1051}

\bibitem[\protect\citeauthoryear{Gomez, Hartmann, Kenyon  \& Hewitt}{Gomez
  et~al.}{1993}]{Gomez93}
Gomez M.,  Hartmann L.,  Kenyon S.~J.,   Hewitt R.,  1993, AJ, 105, 1927

\bibitem[\protect\citeauthoryear{Goodwin \& Whitworth}{Goodwin \&
  Whitworth}{2004}]{Goodwin04a}
Goodwin S.~P.,  Whitworth A.~P.,  2004, A\&A, 413, 929

\bibitem[\protect\citeauthoryear{{Grasser} et~al.,}{{Grasser}
  et~al.}{2021}]{Grasser21}
{Grasser} N.,  et~al., 2021, \mn@doi [\aap] {10.1051/0004-6361/202140438},
  \href {https://ui.adsabs.harvard.edu/abs/2021A&A...652A...2G} {652, A2}

\bibitem[\protect\citeauthoryear{G{\"u}del et~al.,}{G{\"u}del
  et~al.}{2007}]{Guedel07}
G{\"u}del M.,  et~al., 2007, A\&A, 468, 353

\bibitem[\protect\citeauthoryear{Gutermuth, Megeath, Myers, Allen  \&
  Fazio}{Gutermuth et~al.}{2009}]{Gutermuth09}
Gutermuth R.~A.,  Megeath S.~T.,  Myers P.~C.,  Allen L.~E.,   Fazio J. L. P.
  G.~G.,  2009, ApJS, 184, 18

\bibitem[\protect\citeauthoryear{{Habing}}{{Habing}}{1968}]{Habing68}
{Habing} H.~J.,  1968, BAIN, \href
  {http://adsabs.harvard.edu/abs/1968BAN....19..421H} {19, 421}

\bibitem[\protect\citeauthoryear{{Haisch}, {Lada}  \& {Lada}}{{Haisch}
  et~al.}{2001}]{Haisch01}
{Haisch} Jr. K.~E.,  {Lada} E.~A.,   {Lada} C.~J.,  2001, \mn@doi [ApJL]
  {10.1086/320685}, \href {http://adsabs.harvard.edu/abs/2001ApJ...553L.153H}
  {553, L153}

\bibitem[\protect\citeauthoryear{{Hartmann}, {Calvet}, {Gullbring}  \&
  {D'Alessio}}{{Hartmann} et~al.}{1998}]{Hartmann98}
{Hartmann} L.,  {Calvet} N.,  {Gullbring} E.,   {D'Alessio} P.,  1998, \mn@doi
  [ApJ] {10.1086/305277}, \href
  {https://ui.adsabs.harvard.edu/abs/1998ApJ...495..385H} {495, 385}

\bibitem[\protect\citeauthoryear{{Haworth} \& {Clarke}}{{Haworth} \&
  {Clarke}}{2019}]{Haworth19}
{Haworth} T.~J.,  {Clarke} C.~J.,  2019, \mn@doi [\mnras]
  {10.1093/mnras/stz706}, \href
  {https://ui.adsabs.harvard.edu/abs/2019MNRAS.485.3895H} {485, 3895}

\bibitem[\protect\citeauthoryear{{Haworth}, {Facchini}, {Clarke}  \&
  {Cleeves}}{{Haworth} et~al.}{2017}]{Haworth17}
{Haworth} T.~J.,  {Facchini} S.,  {Clarke} C.~J.,   {Cleeves} L.~I.,  2017,
  \mn@doi [\mnras] {10.1093/mnrasl/slx037}, \href
  {https://ui.adsabs.harvard.edu/abs/2017MNRAS.468L.108H} {468, L108}

\bibitem[\protect\citeauthoryear{{Haworth}, {Clarke}, {Rahman}, {Winter}  \&
  {Facchini}}{{Haworth} et~al.}{2018}]{Haworth18b}
{Haworth} T.~J.,  {Clarke} C.~J.,  {Rahman} W.,  {Winter} A.~J.,   {Facchini}
  S.,  2018, \mn@doi [\mnras] {10.1093/mnras/sty2323}, \href
  {http://adsabs.harvard.edu/abs/2018MNRAS.481..452H} {481, 452}

\bibitem[\protect\citeauthoryear{Hillenbrand \& Hartmann}{Hillenbrand \&
  Hartmann}{1998}]{Hillenbrand98}
Hillenbrand L.~A.,  Hartmann L.~W.,  1998, ApJ, 492, 540

\bibitem[\protect\citeauthoryear{{Hollenbach}, {Yorke}  \&
  {Johnstone}}{{Hollenbach} et~al.}{2000}]{Hollenbach00}
{Hollenbach} D.~J.,  {Yorke} H.~W.,   {Johnstone} D.,  2000, Protostars and
  Planets IV, \href {http://adsabs.harvard.edu/abs/2000prpl.conf..401H} {pp
  401--428}

\bibitem[\protect\citeauthoryear{{Isella}, {Carpenter}  \& {Sargent}}{{Isella}
  et~al.}{2009}]{Isella09}
{Isella} A.,  {Carpenter} J.~M.,   {Sargent} A.~I.,  2009, \mn@doi [\apj]
  {10.1088/0004-637X/701/1/260}, \href
  {https://ui.adsabs.harvard.edu/abs/2009ApJ...701..260I} {701, 260}

\bibitem[\protect\citeauthoryear{{Jaehnig}, {Da Rio}  \& {Tan}}{{Jaehnig}
  et~al.}{2015}]{Jaehnig15}
{Jaehnig} K.~O.,  {Da Rio} N.,   {Tan} J.~C.,  2015, \mn@doi [\apj]
  {10.1088/0004-637X/798/2/126}, \href
  {http://adsabs.harvard.edu/abs/2015ApJ...798..126J} {798, 126}

\bibitem[\protect\citeauthoryear{{Jeffries}, {Littlefair}, {Naylor}  \&
  {Mayne}}{{Jeffries} et~al.}{2011}]{Jeffries11}
{Jeffries} R.~D.,  {Littlefair} S.~P.,  {Naylor} T.,   {Mayne} N.~J.,  2011,
  \mn@doi [MNRAS] {10.1111/j.1365-2966.2011.19613.x}, \href
  {http://adsabs.harvard.edu/abs/2011MNRAS.418.1948J} {418, 1948}

\bibitem[\protect\citeauthoryear{{Johnstone}, {Hollenbach}  \&
  {Bally}}{{Johnstone} et~al.}{1998}]{Johnstone98}
{Johnstone} D.,  {Hollenbach} D.,   {Bally} J.,  1998, \mn@doi [\apj]
  {10.1086/305658}, \href
  {https://ui.adsabs.harvard.edu/abs/1998ApJ...499..758J} {499, 758}

\bibitem[\protect\citeauthoryear{King, Parker, Patience  \& Goodwin}{King
  et~al.}{2012}]{King12a}
King R.~R.,  Parker R.~J.,  Patience J.,   Goodwin S.~P.,  2012, MNRAS, 421,
  2025

\bibitem[\protect\citeauthoryear{{Korchagin}, {Girard}, {Borkova}, {Dinescu}
  \& {van Altena}}{{Korchagin} et~al.}{2003}]{Korchagin03}
{Korchagin} V.~I.,  {Girard} T.~M.,  {Borkova} T.~V.,  {Dinescu} D.~I.,   {van
  Altena} W.~F.,  2003, \mn@doi [AJ] {10.1086/379138}, \href
  {http://adsabs.harvard.edu/abs/2003AJ....126.2896K} {126, 2896}

\bibitem[\protect\citeauthoryear{{Krolikowski}, {Kraus}  \&
  {Rizzuto}}{{Krolikowski} et~al.}{2021}]{Krolikowski21}
{Krolikowski} D.~M.,  {Kraus} A.~L.,   {Rizzuto} A.~C.,  2021, \mn@doi [\aj]
  {10.3847/1538-3881/ac0632}, \href
  {https://ui.adsabs.harvard.edu/abs/2021AJ....162..110K} {162, 110}

\bibitem[\protect\citeauthoryear{Kruijssen}{Kruijssen}{2012}]{Kruijssen12b}
Kruijssen J. M.~D.,  2012, MNRAS, 426, 3008

\bibitem[\protect\citeauthoryear{Lada \& Lada}{Lada \& Lada}{2003}]{Lada03}
Lada C.~J.,  Lada E.~A.,  2003, ARA\&A, 41, 57

\bibitem[\protect\citeauthoryear{Larson}{Larson}{1981}]{Larson81}
Larson R.~B.,  1981, MNRAS, 194, 809

\bibitem[\protect\citeauthoryear{Larson}{Larson}{1995}]{Larson95}
Larson R.~B.,  1995, MNRAS, 272, 213

\bibitem[\protect\citeauthoryear{{Long} et~al.,}{{Long} et~al.}{2022}]{Long22}
{Long} F.,  et~al., 2022, \mn@doi [\apj] {10.3847/1538-4357/ac634e}, \href
  {https://ui.adsabs.harvard.edu/abs/2022ApJ...931....6L} {931, 6}

\bibitem[\protect\citeauthoryear{{Luhman}}{{Luhman}}{2020}]{Luhman20}
{Luhman} K.~L.,  2020, \mn@doi [\aj] {10.3847/1538-3881/abb12f}, \href
  {https://ui.adsabs.harvard.edu/abs/2020AJ....160..186L} {160, 186}

\bibitem[\protect\citeauthoryear{{Lynden-Bell} \& {Pringle}}{{Lynden-Bell} \&
  {Pringle}}{1974}]{LyndenBell74}
{Lynden-Bell} D.,  {Pringle} J.~E.,  1974, \mn@doi [MNRAS]
  {10.1093/mnras/168.3.603}, \href
  {https://ui.adsabs.harvard.edu/abs/1974MNRAS.168..603L} {168, 603}

\bibitem[\protect\citeauthoryear{{Marks} \& {Kroupa}}{{Marks} \&
  {Kroupa}}{2012}]{Marks12}
{Marks} M.,  {Kroupa} P.,  2012, \mn@doi [A\&A] {10.1051/0004-6361/201118231},
  543, A8

\bibitem[\protect\citeauthoryear{Maschberger}{Maschberger}{2013}]{Maschberger13}
Maschberger T.,  2013, MNRAS, 429, 1725

\bibitem[\protect\citeauthoryear{{Mayne} \& {Naylor}}{{Mayne} \&
  {Naylor}}{2008}]{Mayne08}
{Mayne} N.~J.,  {Naylor} T.,  2008, \mn@doi [\mnras]
  {10.1111/j.1365-2966.2008.13025.x}, \href
  {https://ui.adsabs.harvard.edu/abs/2008MNRAS.386..261M} {386, 261}

\bibitem[\protect\citeauthoryear{{Miotello}, {Kamp}, {Birnstiel}, {Cleeves}  \&
  {Kataoka}}{{Miotello} et~al.}{2022}]{Miotello22}
{Miotello} A.,  {Kamp} I.,  {Birnstiel} T.,  {Cleeves} L.~I.,   {Kataoka} A.,
  2022, arXiv e-prints, \href
  {https://ui.adsabs.harvard.edu/abs/2022arXiv220309818M} {p. arXiv:2203.09818}

\bibitem[\protect\citeauthoryear{{Nicholson}, {Parker}, {Church}, {Davies},
  {Fearon}  \& {Walton}}{{Nicholson} et~al.}{2019}]{Nicholson19a}
{Nicholson} R.~B.,  {Parker} R.~J.,  {Church} R.~P.,  {Davies} M.~B.,  {Fearon}
  N.~M.,   {Walton} S. R.~J.,  2019, \mn@doi [\mnras] {10.1093/mnras/stz606},
  \href {https://ui.adsabs.harvard.edu/abs/2019MNRAS.485.4893N} {485, 4893}

\bibitem[\protect\citeauthoryear{{Oliveira}, {Jeffries}, {Kenyon}, {Thompson}
  \& {Naylor}}{{Oliveira} et~al.}{2002}]{Oliveira02}
{Oliveira} J.~M.,  {Jeffries} R.~D.,  {Kenyon} M.~J.,  {Thompson} S.~A.,
  {Naylor} T.,  2002, \mn@doi [\aap] {10.1051/0004-6361:20011778}, \href
  {https://ui.adsabs.harvard.edu/abs/2002A&A...382L..22O} {382, L22}

\bibitem[\protect\citeauthoryear{{Otter}, {Ginsburg}, {Ballering}, {Bally},
  {Eisner}, {Goddi}, {Plambeck}  \& {Wright}}{{Otter} et~al.}{2021}]{Otter21}
{Otter} J.,  {Ginsburg} A.,  {Ballering} N.~P.,  {Bally} J.,  {Eisner} J.~A.,
  {Goddi} C.,  {Plambeck} R.,   {Wright} M.,  2021, \mn@doi [\apj]
  {10.3847/1538-4357/ac29c2}, \href
  {https://ui.adsabs.harvard.edu/abs/2021ApJ...923..221O} {923, 221}

\bibitem[\protect\citeauthoryear{Parker}{Parker}{2014}]{Parker14e}
Parker R.~J.,  2014, MNRAS, 445, 4037

\bibitem[\protect\citeauthoryear{{Parker}}{{Parker}}{2020}]{Parker20}
{Parker} R.~J.,  2020, \mn@doi [Royal Society Open Science]
  {10.1098/rsos.201271}, \href
  {https://ui.adsabs.harvard.edu/abs/2020RSOS....701271P} {7, 201271}

\bibitem[\protect\citeauthoryear{Parker \& Quanz}{Parker \&
  Quanz}{2012}]{Parker12a}
Parker R.~J.,  Quanz S.~P.,  2012, MNRAS, 419, 2448

\bibitem[\protect\citeauthoryear{Parker, Maschberger  \& {Alves de
  Oliveira}}{Parker et~al.}{2012}]{Parker12c}
Parker R.~J.,  Maschberger T.,   {Alves de Oliveira} C.,  2012, MNRAS, 426,
  3079

\bibitem[\protect\citeauthoryear{Parker, Wright, Goodwin  \& Meyer}{Parker
  et~al.}{2014}]{Parker14b}
Parker R.~J.,  Wright N.~J.,  Goodwin S.~P.,   Meyer M.~R.,  2014, MNRAS, 438,
  620

\bibitem[\protect\citeauthoryear{{Parker}, {Nicholson}  \& {Alcock}}{{Parker}
  et~al.}{2021a}]{Parker21a}
{Parker} R.~J.,  {Nicholson} R.~B.,   {Alcock} H.~L.,  2021a, \mn@doi [\mnras]
  {10.1093/mnras/stab054}, \href
  {https://ui.adsabs.harvard.edu/abs/2021MNRAS.tmp..106P} {502, 2665}

\bibitem[\protect\citeauthoryear{{Parker}, {Alcock}, {Nicholson}, {Pani{\'c}}
  \& {Goodwin}}{{Parker} et~al.}{2021b}]{Parker21b}
{Parker} R.~J.,  {Alcock} H.~L.,  {Nicholson} R.~B.,  {Pani{\'c}} O.,
  {Goodwin} S.~P.,  2021b, \mn@doi [\apj] {10.3847/1538-4357/abf4cc}, \href
  {https://ui.adsabs.harvard.edu/abs/2021ApJ...913...95P} {913, 95}

\bibitem[\protect\citeauthoryear{{Pecaut}, {Mamajek}  \& {Bubar}}{{Pecaut}
  et~al.}{2012}]{Pecaut12}
{Pecaut} M.~J.,  {Mamajek} E.~E.,   {Bubar} E.~J.,  2012, \mn@doi [ApJ]
  {10.1088/0004-637X/746/2/154}, \href
  {http://adsabs.harvard.edu/abs/2012ApJ...746..154P} {746, 154}

\bibitem[\protect\citeauthoryear{{Picogna}, {Ercolano}, {Owen}  \&
  {Weber}}{{Picogna} et~al.}{2019}]{Picogna19}
{Picogna} G.,  {Ercolano} B.,  {Owen} J.~E.,   {Weber} M.~L.,  2019, \mn@doi
  [\mnras] {10.1093/mnras/stz1166}, \href
  {https://ui.adsabs.harvard.edu/abs/2019MNRAS.487..691P} {487, 691}

\bibitem[\protect\citeauthoryear{{Pinte}, {Dent}, {M{\'e}nard}, {Hales},
  {Hill}, {Cortes}  \& {de Gregorio-Monsalvo}}{{Pinte} et~al.}{2016}]{Pinte16}
{Pinte} C.,  {Dent} W.~R.~F.,  {M{\'e}nard} F.,  {Hales} A.,  {Hill} T.,
  {Cortes} P.,   {de Gregorio-Monsalvo} I.,  2016, \mn@doi [\apj]
  {10.3847/0004-637X/816/1/25}, \href
  {https://ui.adsabs.harvard.edu/abs/2016ApJ...816...25P} {816, 25}

\bibitem[\protect\citeauthoryear{{Portegies Zwart}, Makino, McMillan  \&
  Hut}{{Portegies Zwart} et~al.}{1999}]{Zwart99}
{Portegies Zwart} S.~F.,  Makino J.,  McMillan S. L.~W.,   Hut P.,  1999, A\&A,
  348, 117

\bibitem[\protect\citeauthoryear{{Portegies Zwart}, McMillan, Hut  \&
  Makino}{{Portegies Zwart} et~al.}{2001}]{Zwart01}
{Portegies Zwart} S.~F.,  McMillan S. L.~W.,  Hut P.,   Makino J.,  2001,
  MNRAS, 321, 199

\bibitem[\protect\citeauthoryear{Preibisch, Brown, Bridges, Guenther  \&
  Zinnecker}{Preibisch et~al.}{2002}]{Preibisch02}
Preibisch T.,  Brown A. G.~A.,  Bridges T.,  Guenther E.,   Zinnecker H.,
  2002, AJ, 124, 404

\bibitem[\protect\citeauthoryear{{Reggiani}, {Robberto}, {Da Rio}, {Meyer},
  {Soderblom}  \& {Ricci}}{{Reggiani} et~al.}{2011}]{Reggiani11b}
{Reggiani} M.,  {Robberto} M.,  {Da Rio} N.,  {Meyer} M.~R.,  {Soderblom}
  D.~R.,   {Ricci} L.,  2011, \mn@doi [A\&A] {10.1051/0004-6361/201116946},
  534, A83

\bibitem[\protect\citeauthoryear{{Ribas}, {Bouy}  \& {Mer{\'\i}n}}{{Ribas}
  et~al.}{2015}]{Ribas15}
{Ribas} {\'A}.,  {Bouy} H.,   {Mer{\'\i}n} B.,  2015, \mn@doi [\aap]
  {10.1051/0004-6361/201424846}, \href
  {https://ui.adsabs.harvard.edu/abs/2015A&A...576A..52R} {576, A52}

\bibitem[\protect\citeauthoryear{{Richert}, {Getman}, {Feigelson}, {Kuhn},
  {Broos}, {Povich}, {Bate}  \& {Garmire}}{{Richert} et~al.}{2018}]{Richert18}
{Richert} A.~J.~W.,  {Getman} K.~V.,  {Feigelson} E.~D.,  {Kuhn} M.~A.,
  {Broos} P.~S.,  {Povich} M.~S.,  {Bate} M.~R.,   {Garmire} G.~P.,  2018,
  \mn@doi [\mnras] {10.1093/mnras/sty949}, \href
  {http://adsabs.harvard.edu/abs/2018MNRAS.477.5191R} {477, 5191}

\bibitem[\protect\citeauthoryear{{Rigliaco} et~al.,}{{Rigliaco}
  et~al.}{2016}]{Rigliaco16}
{Rigliaco} E.,  et~al., 2016, \mn@doi [\aap] {10.1051/0004-6361/201527253},
  \href {https://ui.adsabs.harvard.edu/abs/2016A&A...588A.123R} {588, A123}

\bibitem[\protect\citeauthoryear{Salpeter}{Salpeter}{1955}]{Salpeter55}
Salpeter E.~E.,  1955, ApJ, 121, 161

\bibitem[\protect\citeauthoryear{S{\'a}nchez \& Alfaro}{S{\'a}nchez \&
  Alfaro}{2009}]{Sanchez09}
S{\'a}nchez N.,  Alfaro E.~J.,  2009, ApJ, 696, 2086

\bibitem[\protect\citeauthoryear{{Sanchis} et~al.,}{{Sanchis}
  et~al.}{2021}]{Sanchis21}
{Sanchis} E.,  et~al., 2021, \mn@doi [\aap] {10.1051/0004-6361/202039733},
  \href {https://ui.adsabs.harvard.edu/abs/2021A&A...649A..19S} {649, A19}

\bibitem[\protect\citeauthoryear{Scally \& Clarke}{Scally \&
  Clarke}{2001}]{Scally01}
Scally A.,  Clarke C.,  2001, MNRAS, 325, 449

\bibitem[\protect\citeauthoryear{{Schoettler}, {de Bruijne}, {Vaher}  \&
  {Parker}}{{Schoettler} et~al.}{2020}]{Schoettler20}
{Schoettler} C.,  {de Bruijne} J.,  {Vaher} E.,   {Parker} R.~J.,  2020,
  \mn@doi [\mnras] {10.1093/mnras/staa1228}, \href
  {https://ui.adsabs.harvard.edu/abs/2020MNRAS.tmp.1474S} {495, 3104}

\bibitem[\protect\citeauthoryear{{Sellek}, {Booth}  \& {Clarke}}{{Sellek}
  et~al.}{2020}]{Sellek20}
{Sellek} A.~D.,  {Booth} R.~A.,   {Clarke} C.~J.,  2020, \mn@doi [\mnras]
  {10.1093/mnras/stz3528}, \href
  {https://ui.adsabs.harvard.edu/abs/2020MNRAS.492.1279S} {492, 1279}

\bibitem[\protect\citeauthoryear{{Shakura} \& {Sunyaev}}{{Shakura} \&
  {Sunyaev}}{1973}]{Shakura73}
{Shakura} N.~I.,  {Sunyaev} R.~A.,  1973, A\&A, \href
  {https://ui.adsabs.harvard.edu/abs/1973A&A....24..337S} {500, 33}

\bibitem[\protect\citeauthoryear{{Sheehan} \& {Eisner}}{{Sheehan} \&
  {Eisner}}{2017}]{Sheehan17}
{Sheehan} P.~D.,  {Eisner} J.~A.,  2017, \mn@doi [\apj]
  {10.3847/1538-4357/aa9990}, \href
  {https://ui.adsabs.harvard.edu/abs/2017ApJ...851...45S} {851, 45}

\bibitem[\protect\citeauthoryear{{Sternberg}, {Hoffmann}  \&
  {Pauldrach}}{{Sternberg} et~al.}{2003}]{Sternberg03}
{Sternberg} A.,  {Hoffmann} T.~L.,   {Pauldrach} A.~W.~A.,  2003, \mn@doi
  [\apj] {10.1086/379506}, \href
  {http://adsabs.harvard.edu/abs/2003ApJ...599.1333S} {599, 1333}

\bibitem[\protect\citeauthoryear{{St{\"o}rzer} \& {Hollenbach}}{{St{\"o}rzer}
  \& {Hollenbach}}{1999}]{Storzer99}
{St{\"o}rzer} H.,  {Hollenbach} D.,  1999, \mn@doi [\apj] {10.1086/307055},
  \href {http://adsabs.harvard.edu/abs/1999ApJ...515..669S} {515, 669}

\bibitem[\protect\citeauthoryear{{Tazzari} et~al.,}{{Tazzari}
  et~al.}{2017}]{Tazzari17}
{Tazzari} M.,  et~al., 2017, \mn@doi [\aap] {10.1051/0004-6361/201730890},
  \href {https://ui.adsabs.harvard.edu/abs/2017A&A...606A..88T} {606, A88}

\bibitem[\protect\citeauthoryear{{Tripathi}, {Andrews}, {Birnstiel}  \&
  {Wilner}}{{Tripathi} et~al.}{2017}]{Tripathi17}
{Tripathi} A.,  {Andrews} S.~M.,  {Birnstiel} T.,   {Wilner} D.~J.,  2017,
  \mn@doi [\apj] {10.3847/1538-4357/aa7c62}, \href
  {https://ui.adsabs.harvard.edu/abs/2017ApJ...845...44T} {845, 44}

\bibitem[\protect\citeauthoryear{{Vacca}, {Garmany}  \& {Shull}}{{Vacca}
  et~al.}{1996}]{Vacca96}
{Vacca} W.~D.,  {Garmany} C.~D.,   {Shull} J.~M.,  1996, \mn@doi [\apj]
  {10.1086/177020}, \href {http://adsabs.harvard.edu/abs/1996ApJ...460..914V}
  {460, 914}

\bibitem[\protect\citeauthoryear{{Vincke} \& {Pfalzner}}{{Vincke} \&
  {Pfalzner}}{2018}]{Vincke18}
{Vincke} K.,  {Pfalzner} S.,  2018, \mn@doi [\apj] {10.3847/1538-4357/aae7d1},
  \href {https://ui.adsabs.harvard.edu/abs/2018ApJ...868....1V} {868, 1}

\bibitem[\protect\citeauthoryear{{Weidenschilling}}{{Weidenschilling}}{1977a}]{Weidenschilling77}
{Weidenschilling} S.~J.,  1977a, \mn@doi [Ap\&SS] {10.1007/BF00642464}, \href
  {https://ui.adsabs.harvard.edu/abs/1977Ap&SS..51..153W} {51, 153}

\bibitem[\protect\citeauthoryear{{Weidenschilling}}{{Weidenschilling}}{1977b}]{Weidenschilling77b}
{Weidenschilling} S.~J.,  1977b, \mn@doi [\mnras] {10.1093/mnras/180.2.57},
  \href {https://ui.adsabs.harvard.edu/abs/1977MNRAS.180...57W} {180, 57}

\bibitem[\protect\citeauthoryear{{Winter}, {Clarke}, {Rosotti}  \&
  {Booth}}{{Winter} et~al.}{2018a}]{Winter18a}
{Winter} A.~J.,  {Clarke} C.~J.,  {Rosotti} G.,   {Booth} R.~A.,  2018a,
  \mn@doi [\mnras] {10.1093/mnras/sty012}, \href
  {https://ui.adsabs.harvard.edu/abs/2018MNRAS.475.2314W} {475, 2314}

\bibitem[\protect\citeauthoryear{{Winter}, {Clarke}, {Rosotti}, {Ih},
  {Facchini}  \& {Haworth}}{{Winter} et~al.}{2018b}]{Winter18b}
{Winter} A.~J.,  {Clarke} C.~J.,  {Rosotti} G.,  {Ih} J.,  {Facchini} S.,
  {Haworth} T.~J.,  2018b, \mn@doi [\mnras] {10.1093/mnras/sty984}, \href
  {http://adsabs.harvard.edu/abs/2018MNRAS.478.2700W} {478, 2700}

\bibitem[\protect\citeauthoryear{{Wright}}{{Wright}}{2020}]{Wright20}
{Wright} N.~J.,  2020, \mn@doi [\nar] {10.1016/j.newar.2020.101549}, \href
  {https://ui.adsabs.harvard.edu/abs/2020NewAR..9001549W} {90, 101549}

\bibitem[\protect\citeauthoryear{{Wright}, {Bouy}, {Drew}, {Sarro}, {Bertin},
  {Cuillandre}  \& {Barrado}}{{Wright} et~al.}{2016}]{Wright16}
{Wright} N.~J.,  {Bouy} H.,  {Drew} J.~E.,  {Sarro} L.~M.,  {Bertin} E.,
  {Cuillandre} J.-C.,   {Barrado} D.,  2016, \mn@doi [MNRAS]
  {10.1093/mnras/stw1148}, \href
  {http://adsabs.harvard.edu/abs/2016MNRAS.460.2593W} {460, 2593}

\bibitem[\protect\citeauthoryear{{Wright}, {Goodwin}, {Jeffries}, {Kounkel}  \&
  {Zari}}{{Wright} et~al.}{2022}]{Wright22}
{Wright} N.~J.,  {Goodwin} S.,  {Jeffries} R.~D.,  {Kounkel} M.,   {Zari} E.,
  2022, arXiv e-prints, \href
  {https://ui.adsabs.harvard.edu/abs/2022arXiv220310007W} {p. arXiv:2203.10007}

\bibitem[\protect\citeauthoryear{{Zapatero Osorio}, {B{\'e}jar}, {Pavlenko},
  {Rebolo}, {Allende Prieto}, {Mart{\'\i}n}  \& {Garc{\'\i}a
  L{\'o}pez}}{{Zapatero Osorio} et~al.}{2002}]{Osorio02a}
{Zapatero Osorio} M.~R.,  {B{\'e}jar} V.~J.~S.,  {Pavlenko} Y.,  {Rebolo} R.,
  {Allende Prieto} C.,  {Mart{\'\i}n} E.~L.,   {Garc{\'\i}a L{\'o}pez} R.~J.,
  2002, \mn@doi [\aap] {10.1051/0004-6361:20020046}, \href
  {https://ui.adsabs.harvard.edu/abs/2002A&A...384..937Z} {384, 937}

\makeatother
\end{thebibliography}

\appendix
\section{The ONC disc radius distribution as simulation input}
\label{appendix:eisner}

\begin{figure*}
  \begin{center}
\setlength{\subfigcapskip}{10pt}
\hspace*{0.3cm}\subfigure[1\,Myr]{\label{mod_dens_eisner-a}\rotatebox{270}{\includegraphics[scale=0.23]{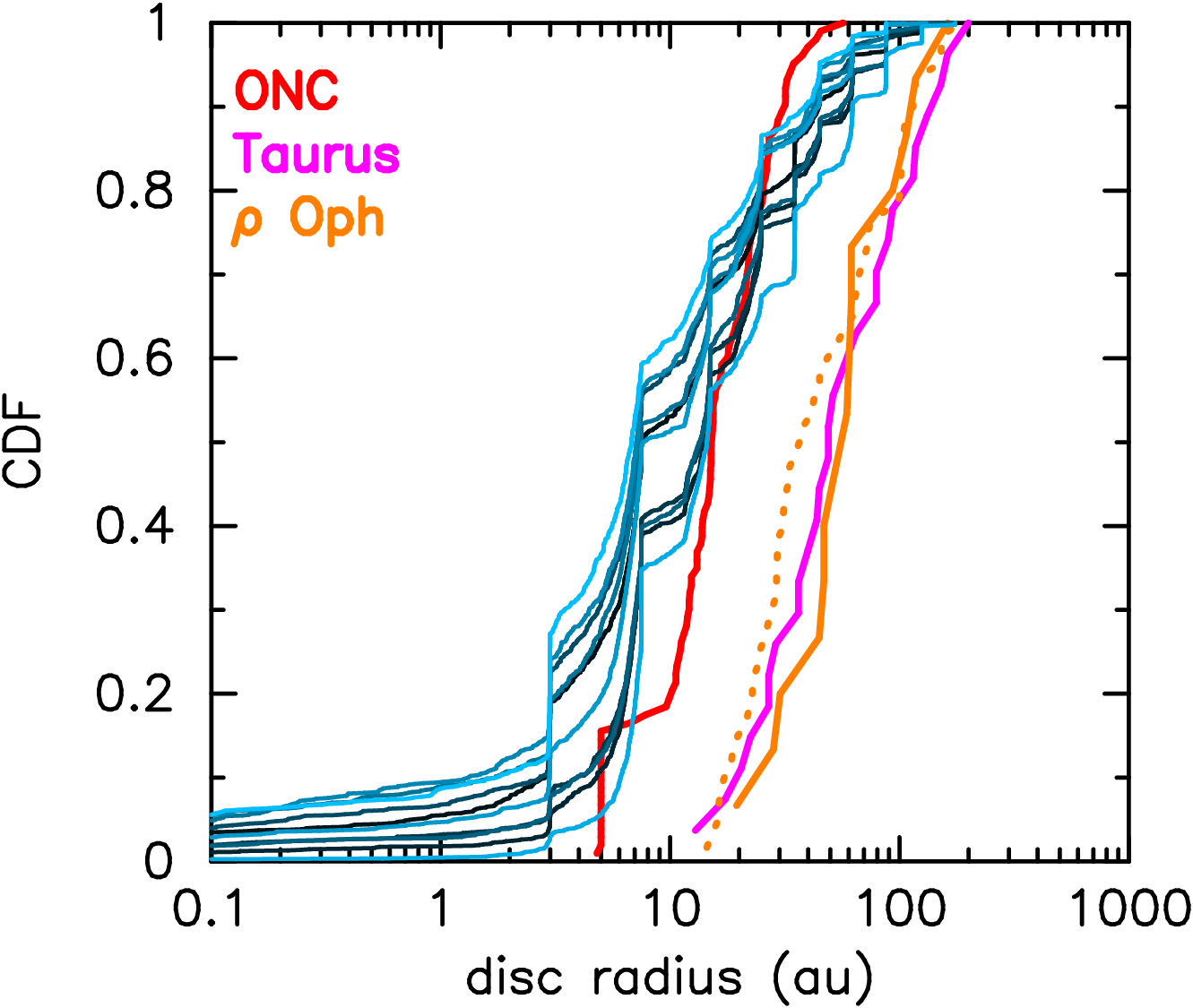}}}
\hspace*{0.3cm}\subfigure[5\,Myr]{\label{mod_dens_eisner-b}\rotatebox{270}{\includegraphics[scale=0.23]{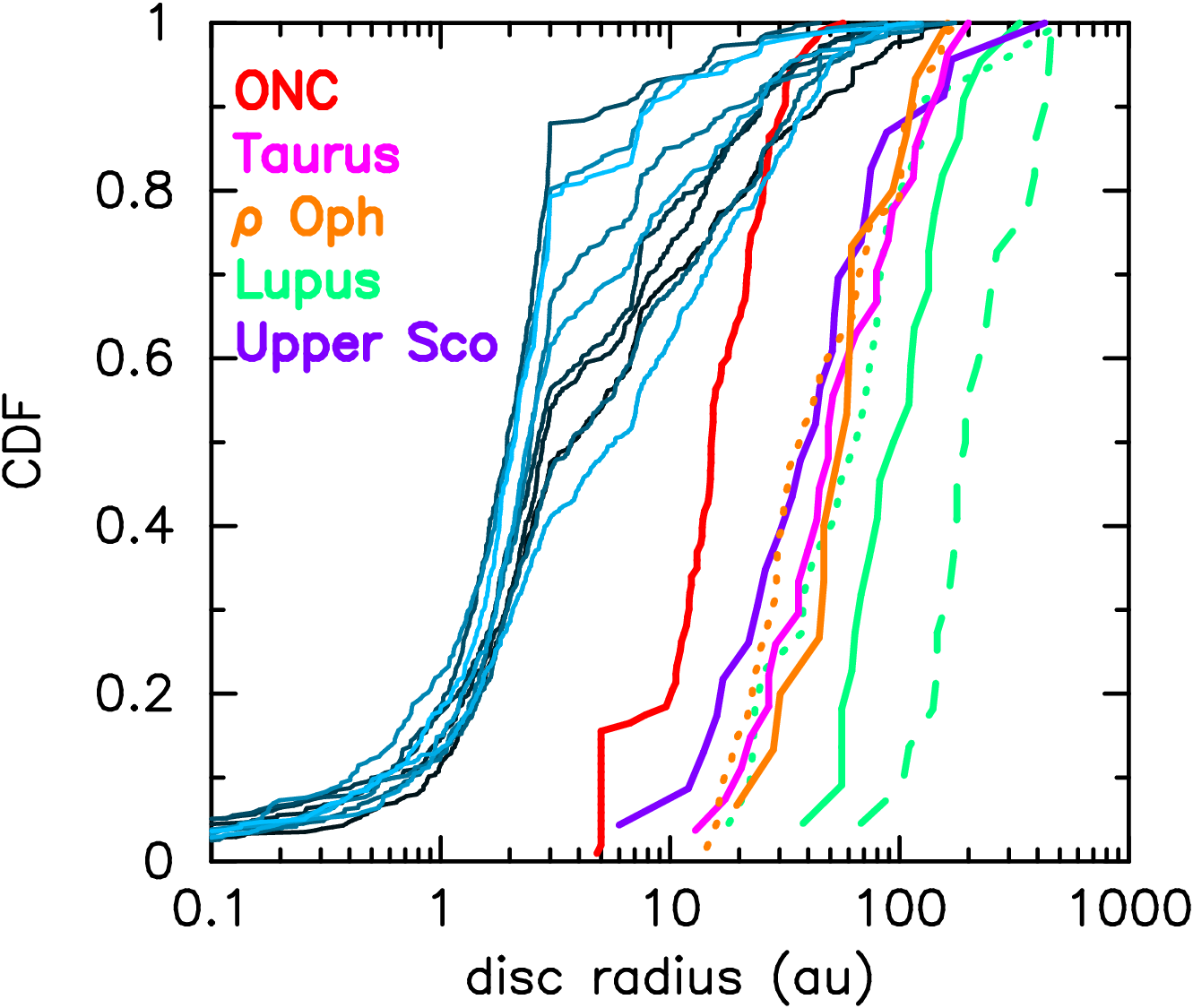}}}
\hspace*{0.3cm}\subfigure[10\,Myr]{\label{mod_dens_eisner-c}\rotatebox{270}{\includegraphics[scale=0.23]{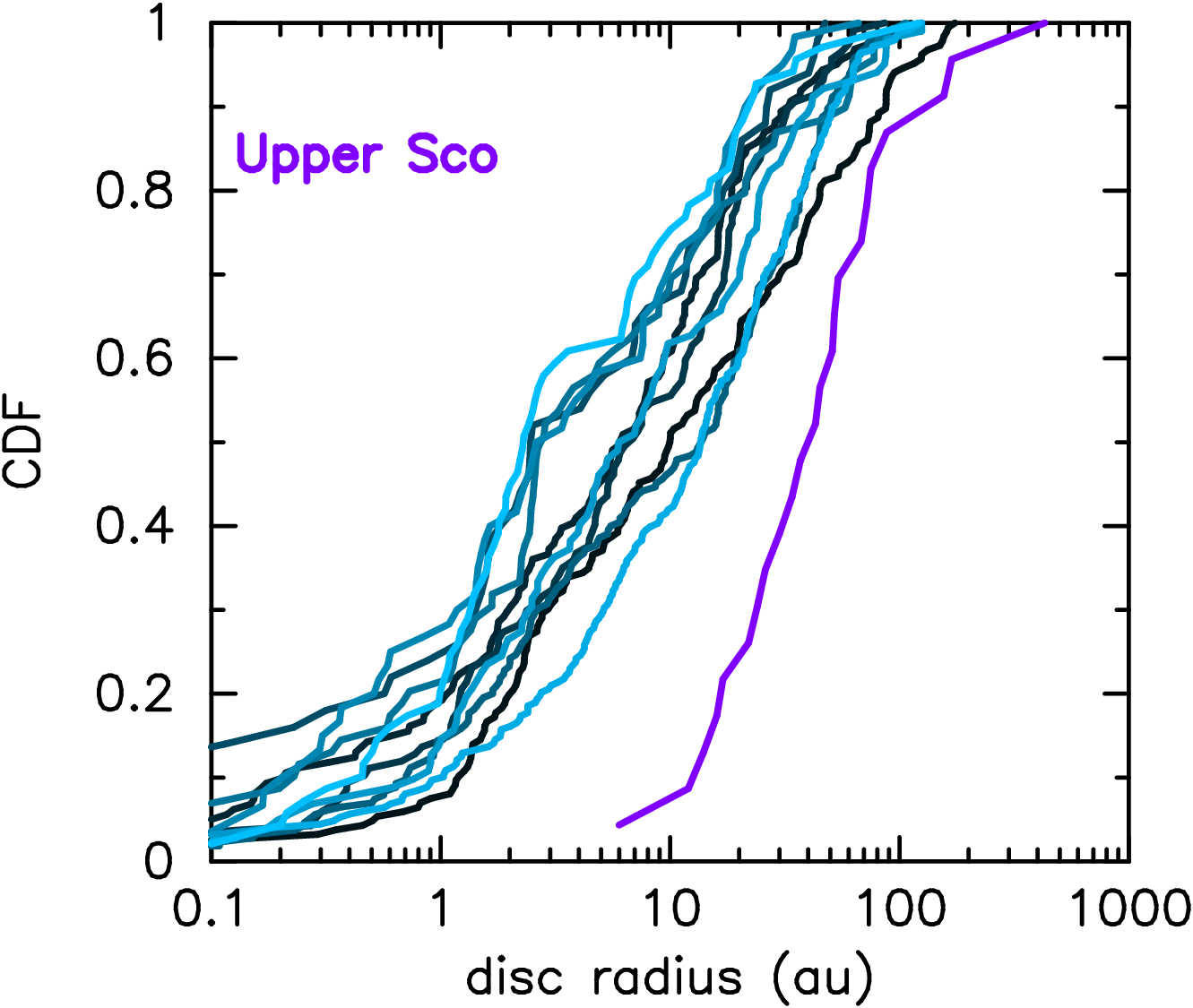}}}
\hspace*{0.3cm}\subfigure[1\,Myr]{\label{mod_dens_eisner-d}\rotatebox{270}{\includegraphics[scale=0.23]{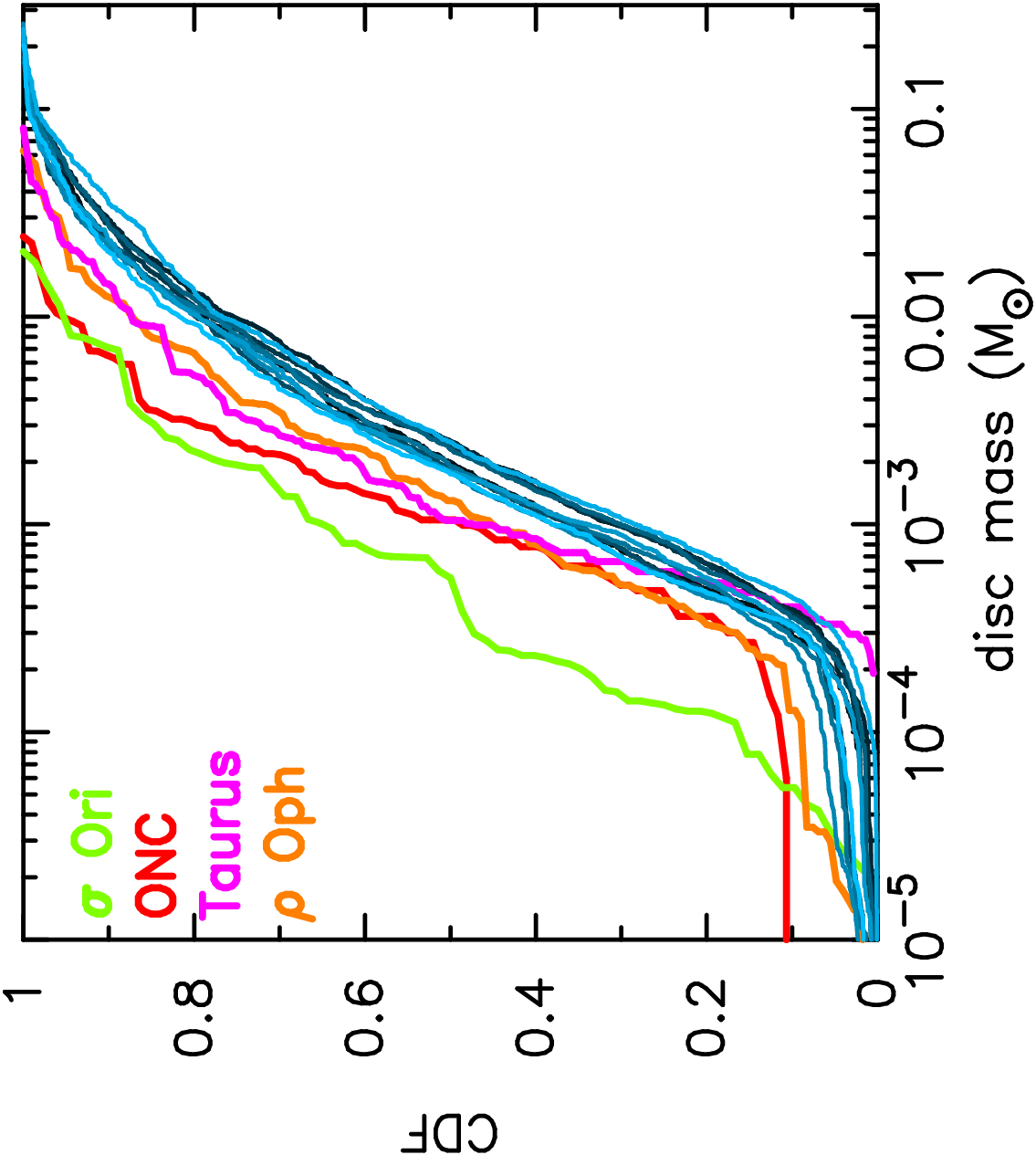}}}
\hspace*{0.3cm}\subfigure[5\,Myr]{\label{mod_dens_eisner-e}\rotatebox{270}{\includegraphics[scale=0.23]{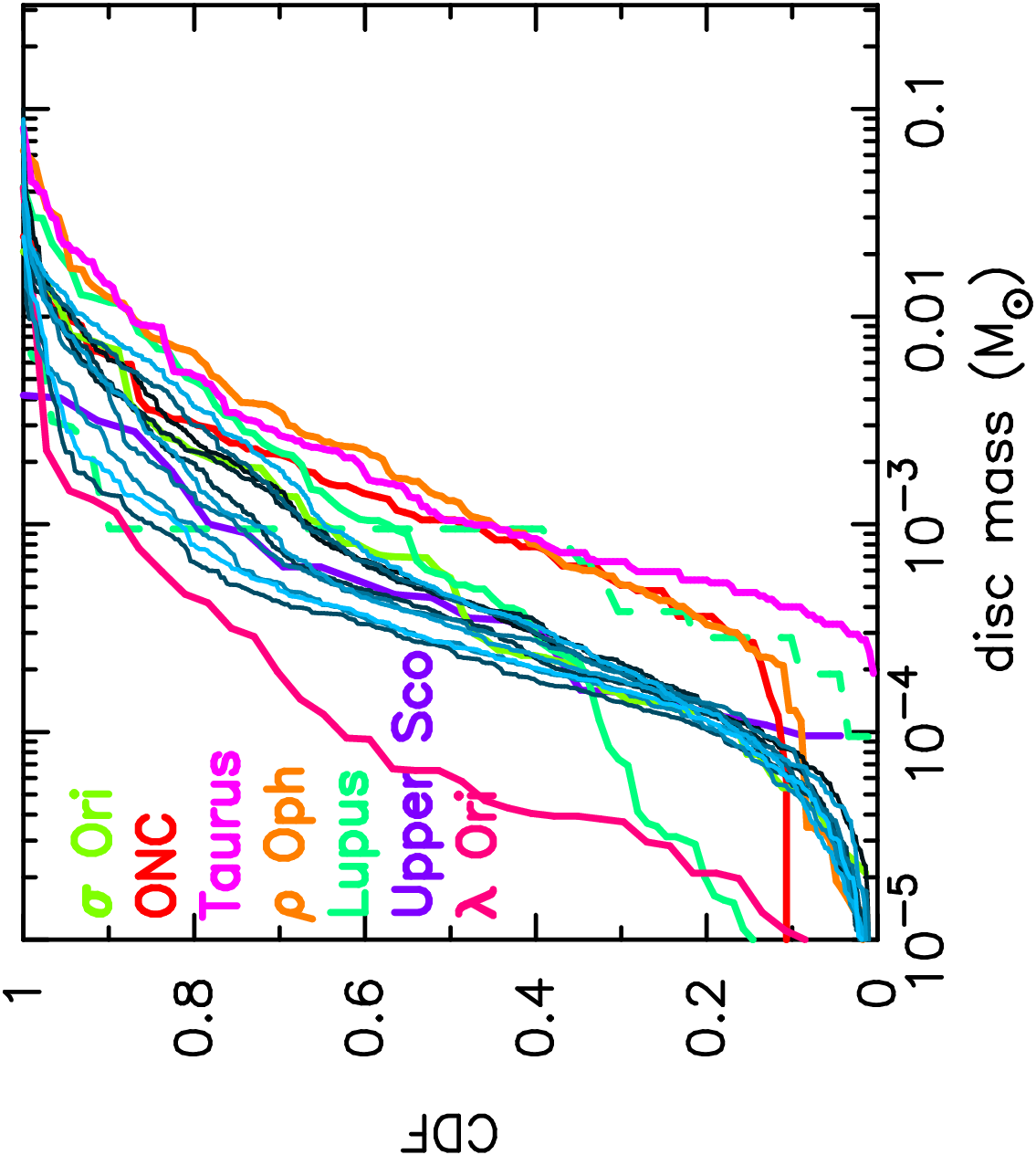}}}
\hspace*{0.3cm}\subfigure[10\,Myr]{\label{mod_dens_eisner-f}\rotatebox{270}{\includegraphics[scale=0.23]{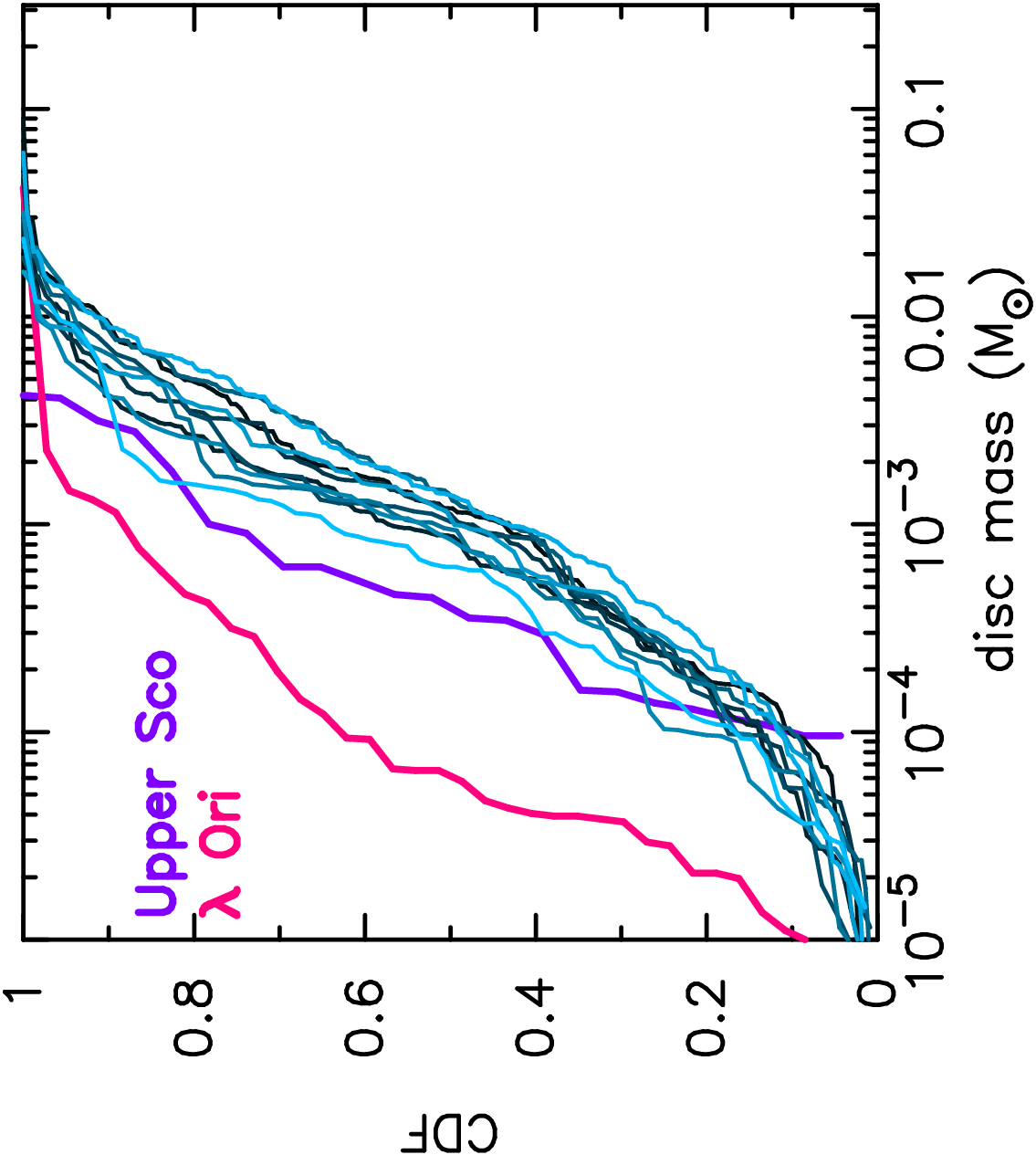}}}
\hspace*{0.3cm}\subfigure[Mass--radius, 1\,Myr]{\label{mod_dens_eisner-g}\rotatebox{270}{\includegraphics[scale=0.23]{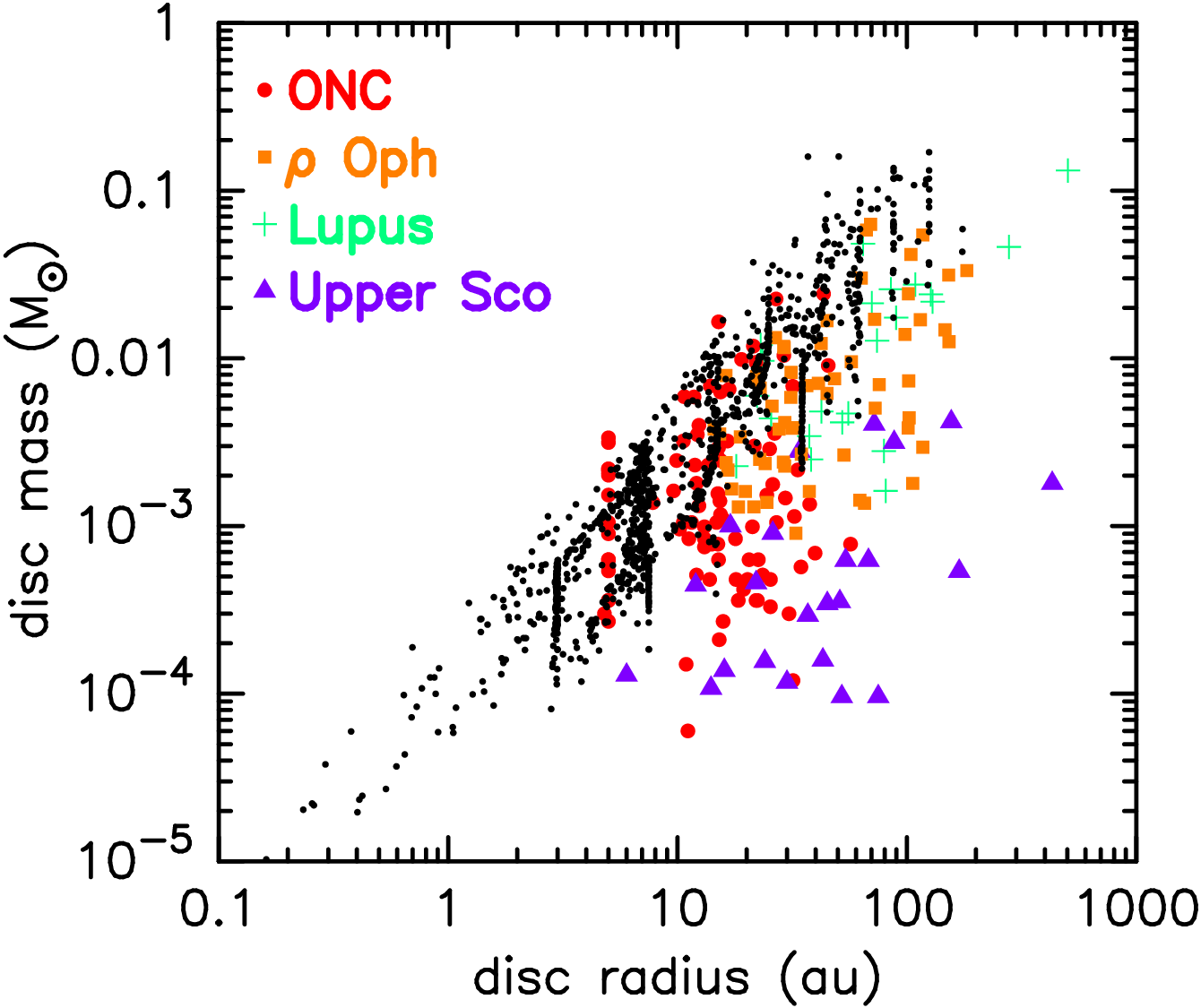}}}
\hspace*{0.3cm}\subfigure[Mass--radius, 5\,Myr]{\label{mod_dens_eisner-h}\rotatebox{270}{\includegraphics[scale=0.23]{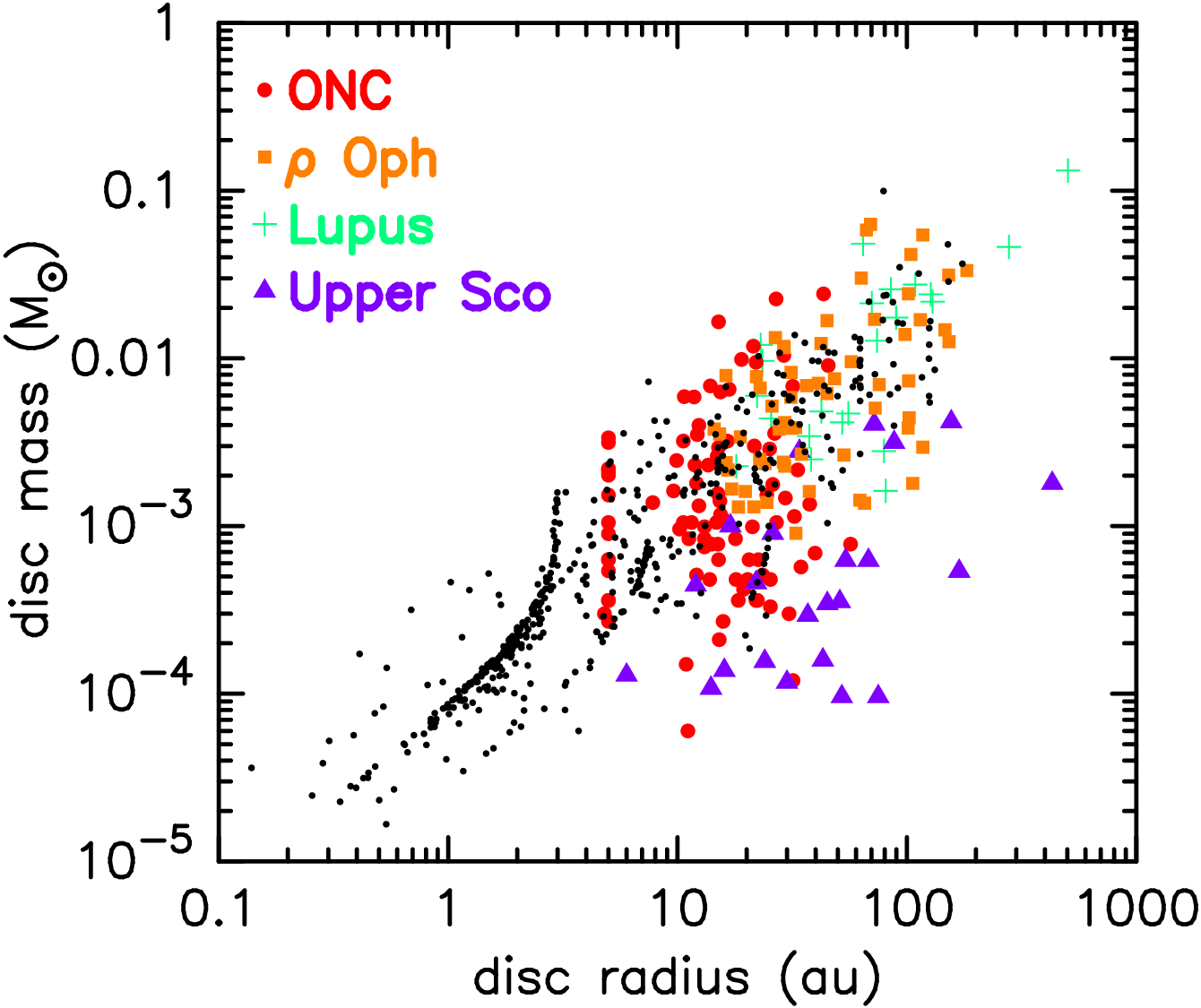}}}
\hspace*{0.3cm}\subfigure[Disc fractions]{\label{mod_dens_eisner-i}\rotatebox{270}{\includegraphics[scale=0.23]{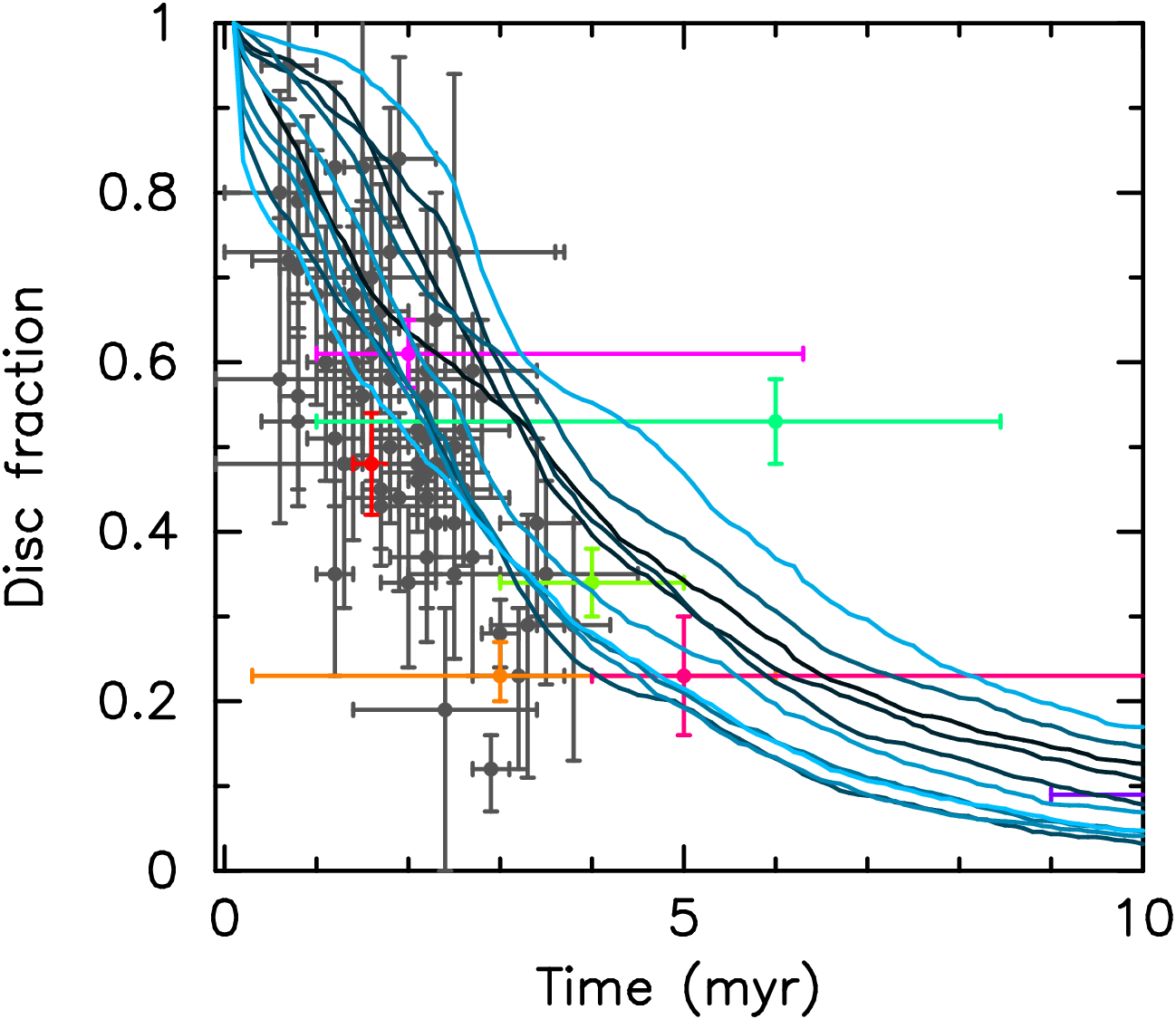}}}
\caption[bf]{Results for simulations in which the initial disc radii are drawn from a Gaussian designed to mimic the observed \citet{Eisner18} radius distribution, and the discs are allowed to evolve inwards due to external photoevaporation, and outwards due to viscous spreading.  The star-forming regions have moderate initial stellar density ($\tilde{\rho} \sim 100$\,M$_\odot$\,pc$^{-3}$). In panels (a)--(c) we show the cumulative distributions of disc radii at different ages, and in panels (d)--(f) we show the cumulative distributions of disc masses. The blue lines of varying hues show the results from the ten individual $N$-body simulations. The mint green dashed line represents the respective gas distribution in the Lupus star-forming region, whereas solid lines represent the dust distributions. Dotted lines indicate an alternative dust radius for $\rho$~Oph (orange line) and Lupus (mint green line); see Table~\ref{obs_info} for further details. In panels (g) and (h) we show the disc mass versus disc radius for four observed star-forming regions at 1 and 5\,Myr, and data from one post-processed $N$-body simulation are shown by the black points. We plot the disc fractions (defined as when the discs have non-zero mass) as a function of time in our $N$-body simulations in panel (i), with observational data taken from \citet{Richert18} and \citet{Ribas15}.}
\label{mod_dens_eisner}
  \end{center}
\end{figure*}

In Fig.~\ref{mod_dens_eisner} we show the results for regions with moderate stellar density ($\tilde{\rho} \sim 100$\,M$_\odot$\,pc$^{-3}$) where the initial disc radii are drawn from a similar distribution to that observed for the ONC \citep{Eisner18}. As the median radius in this distribution is around 16\,au, the distributions evolve in a very similar way to the simulations in which we set all of the disc radii to be $r_{\rm disc} = 10$\,au.

\section{Simulations with $\alpha = 10^{-4}$}
\label{appendix_viscosity}

We run a set of our default simulations ($N_\star = 1500$, $\tilde{\rho} \sim 100$\,M$_\odot$\,pc$^{-3}$, $r_{\rm disc} = 100$\,au with viscous evolution). However, for the turbulent mixing parameter we adopt $\alpha = 10^{-4}$, instead of $\alpha = 10^{-2}$.

\begin{figure*}
  \begin{center}
\setlength{\subfigcapskip}{10pt}
\hspace*{0.3cm}\subfigure[1\,Myr]{\label{mod_dens_alpha_e-4-a}\rotatebox{270}{\includegraphics[scale=0.23]{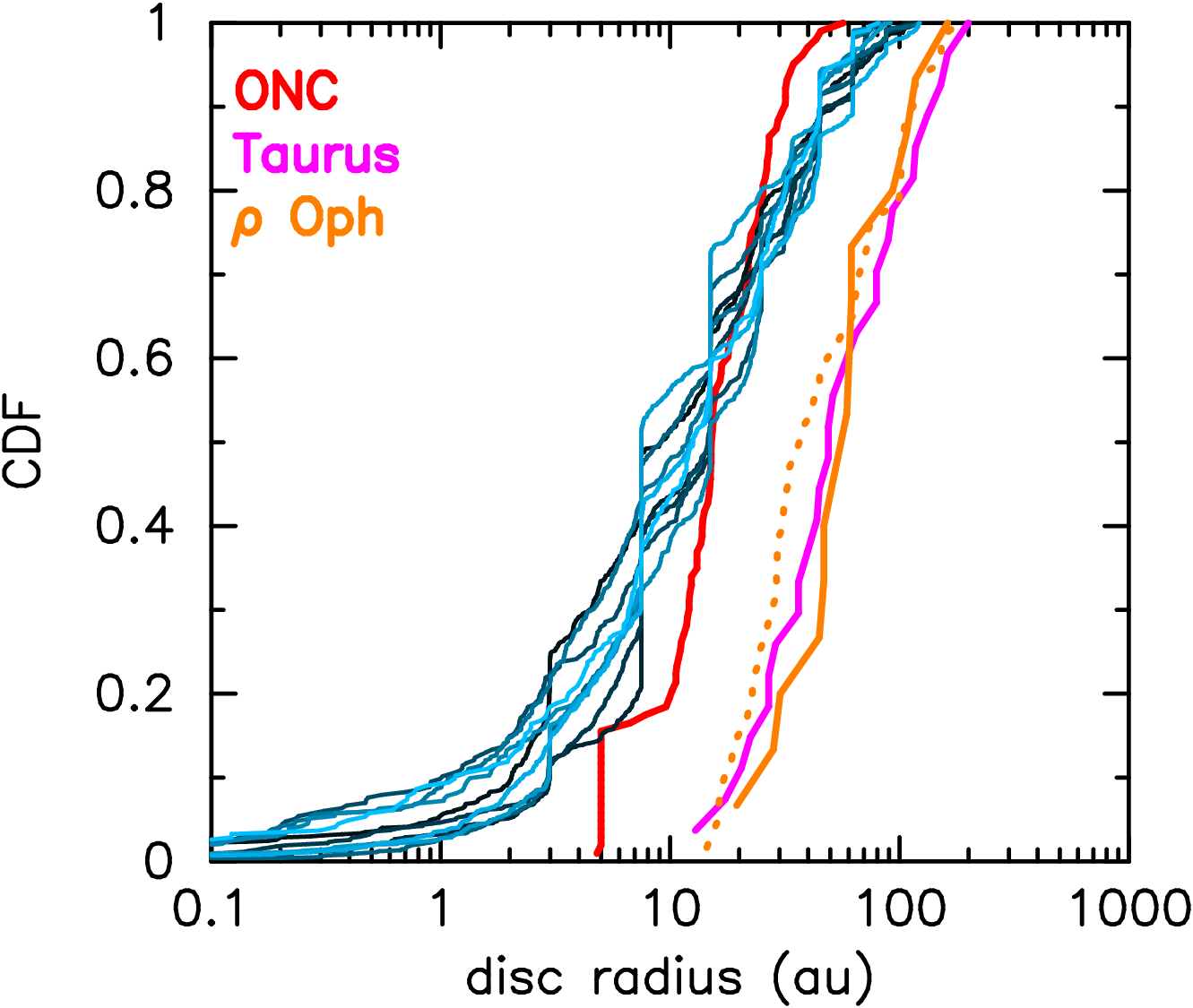}}}
\hspace*{0.3cm}\subfigure[5\,Myr]{\label{mod_dens_alpha_e-4-b}\rotatebox{270}{\includegraphics[scale=0.23]{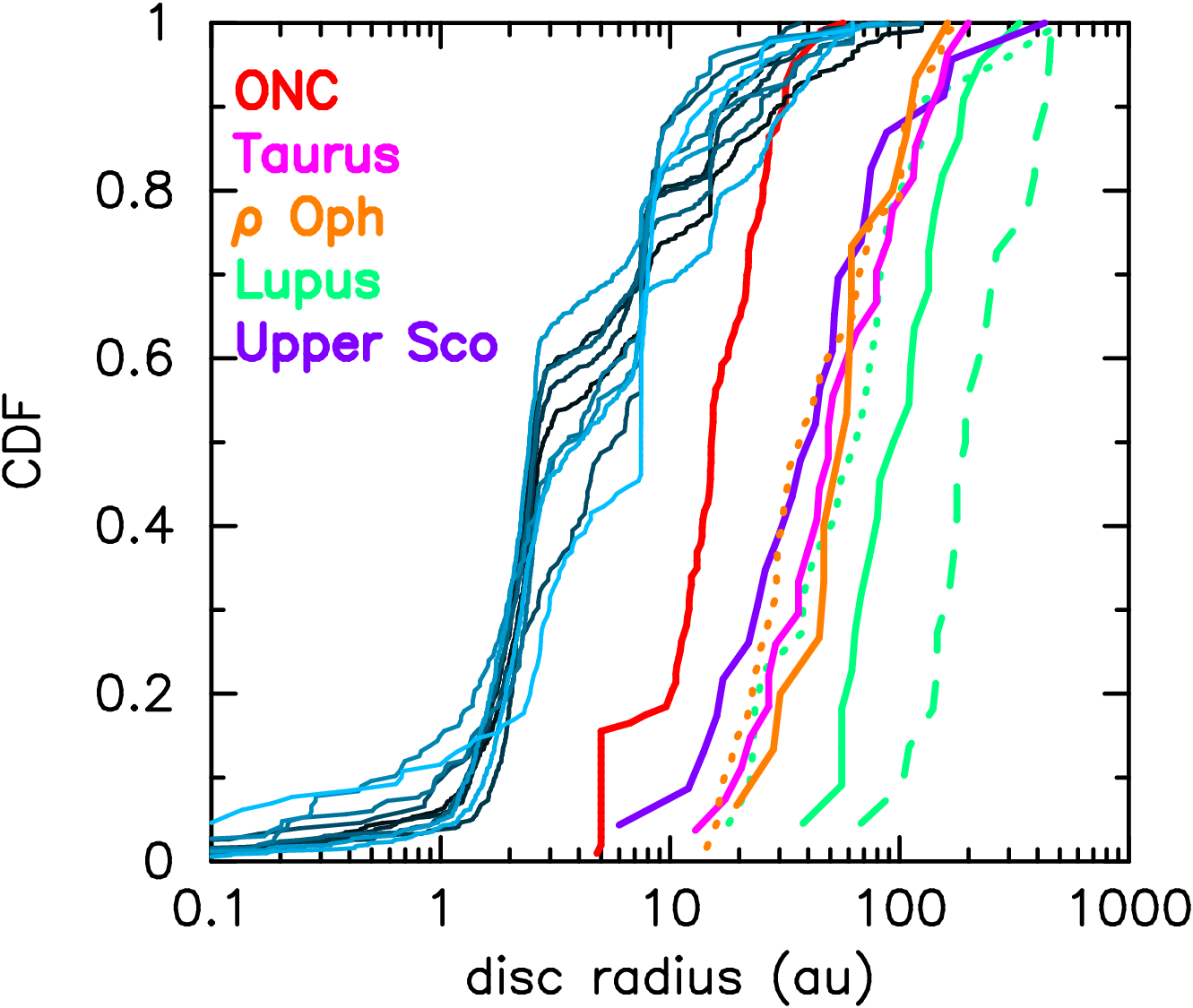}}}
\hspace*{0.3cm}\subfigure[10\,Myr]{\label{mod_dens_alpha_e-4-c}\rotatebox{270}{\includegraphics[scale=0.23]{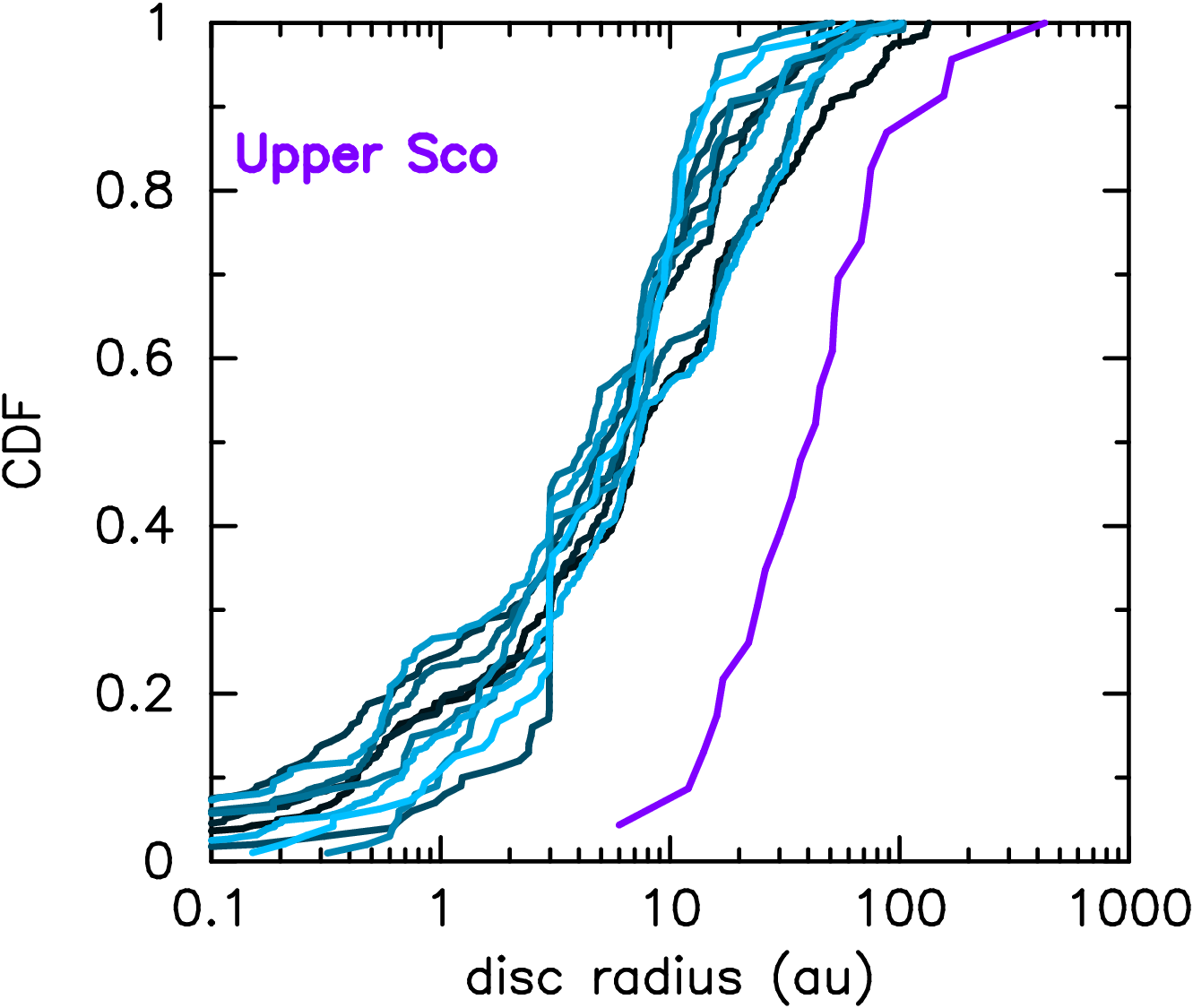}}}
\hspace*{0.3cm}\subfigure[1\,Myr]{\label{mod_dens_alpha_e-4-d}\rotatebox{270}{\includegraphics[scale=0.23]{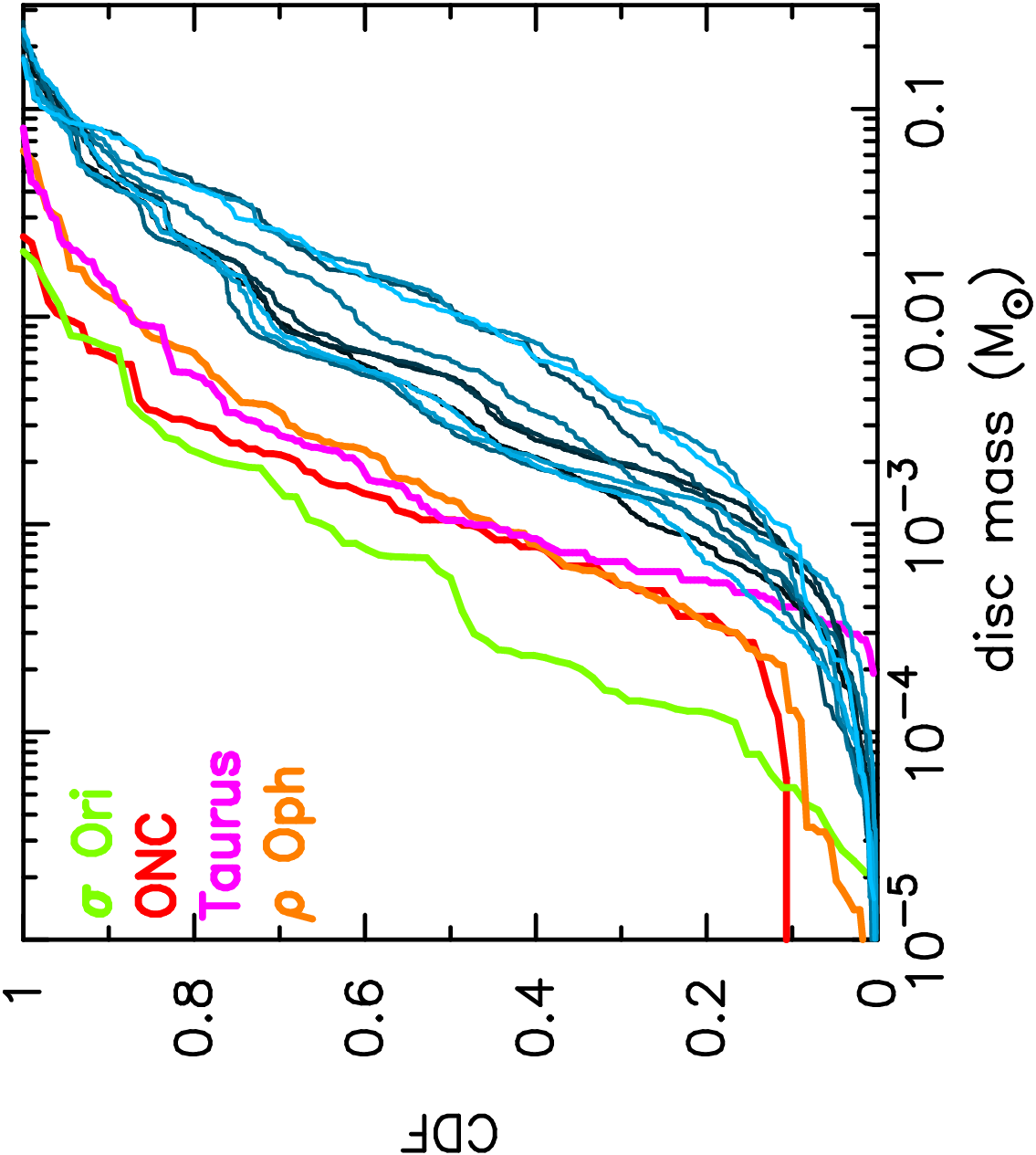}}}
\hspace*{0.3cm}\subfigure[5\,Myr]{\label{mod_dens_alpha_e-4-e}\rotatebox{270}{\includegraphics[scale=0.23]{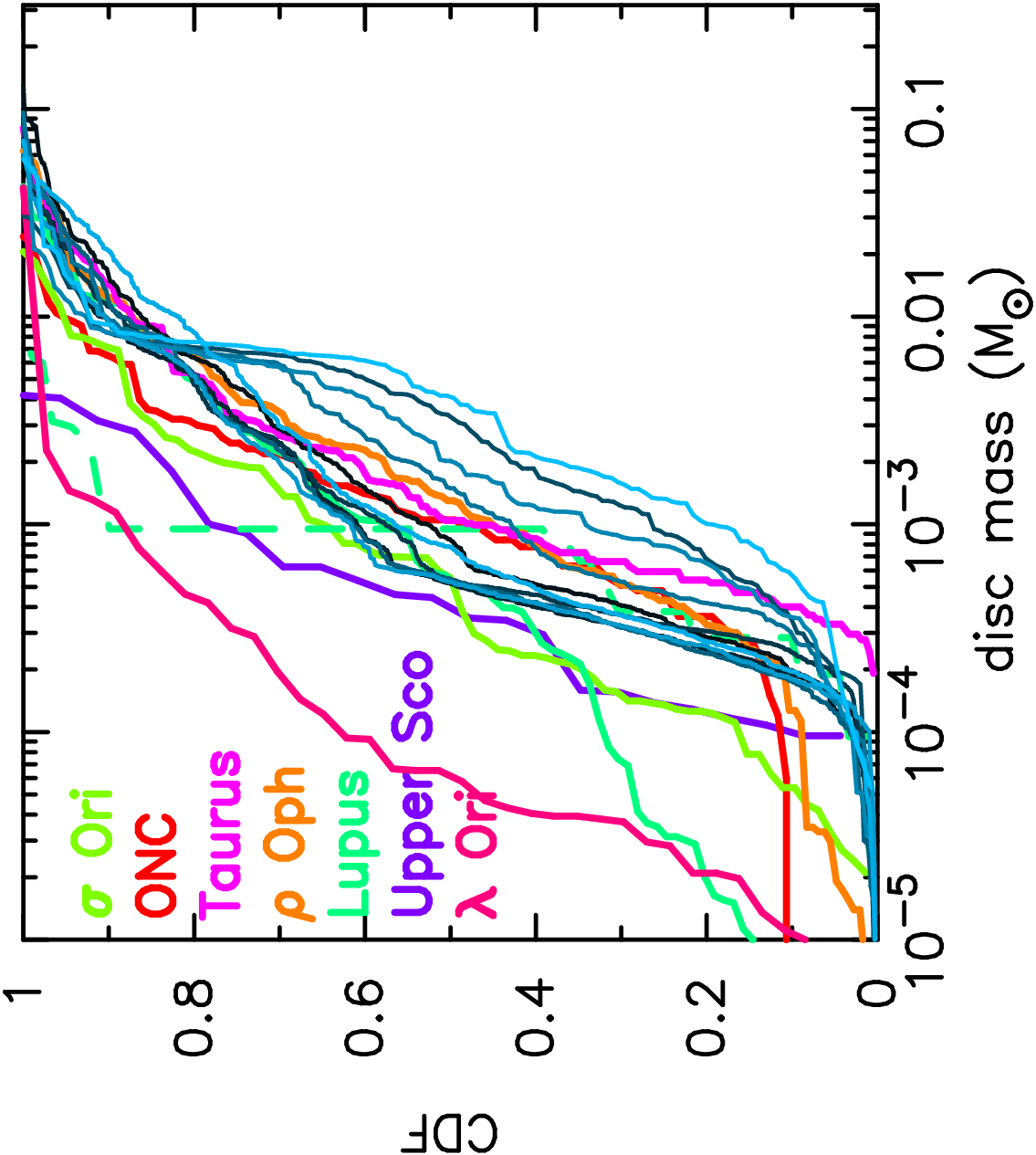}}}
\hspace*{0.3cm}\subfigure[10\,Myr]{\label{mod_dens_alpha_e-4-f}\rotatebox{270}{\includegraphics[scale=0.23]{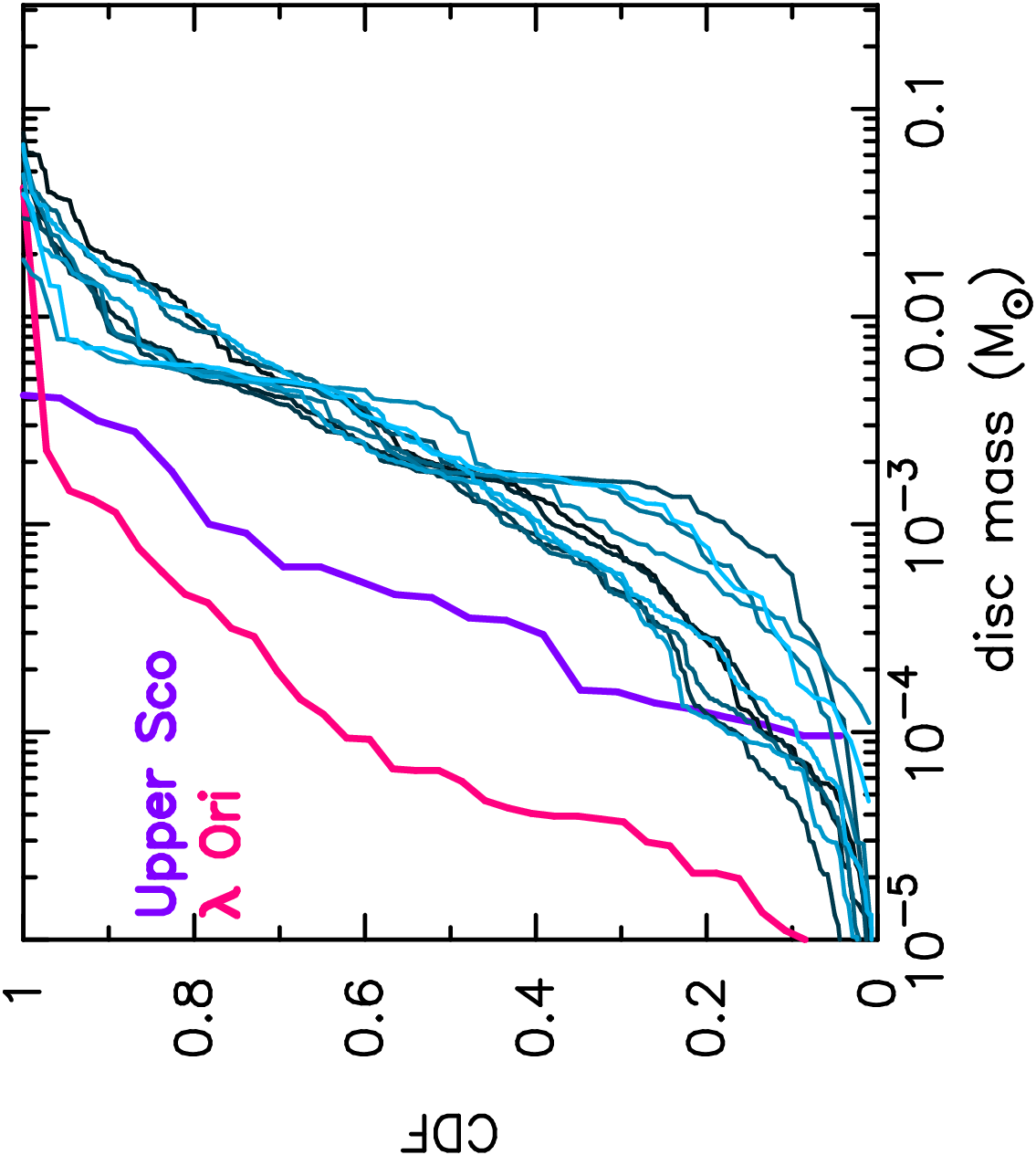}}}
\hspace*{0.3cm}\subfigure[Mass--radius, 1\,Myr]{\label{mod_dens_alpha_e-4-g}\rotatebox{270}{\includegraphics[scale=0.23]{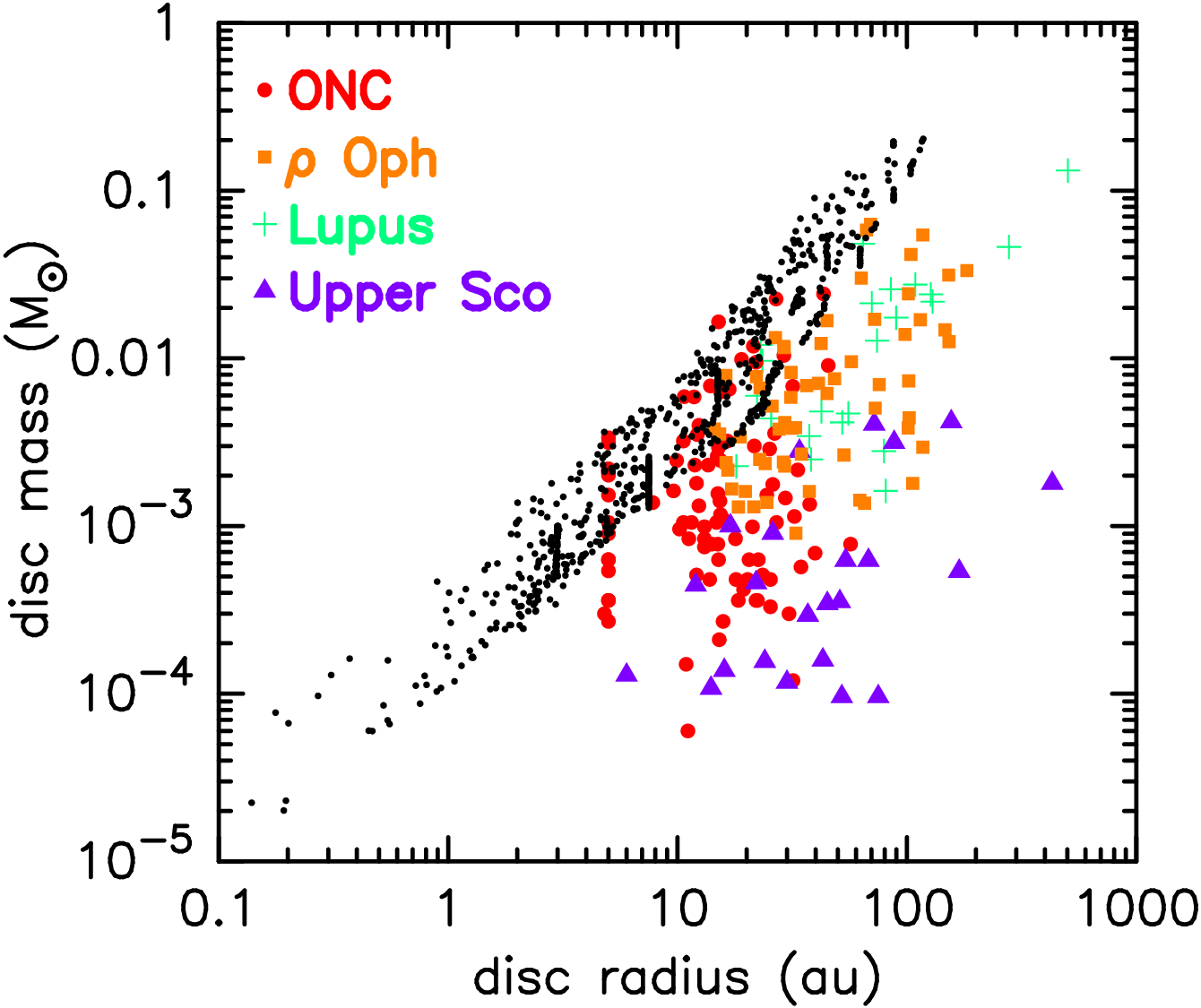}}}
\hspace*{0.3cm}\subfigure[Mass--radius, 5\,Myr]{\label{mod_dens_alpha_e-4-h}\rotatebox{270}{\includegraphics[scale=0.23]{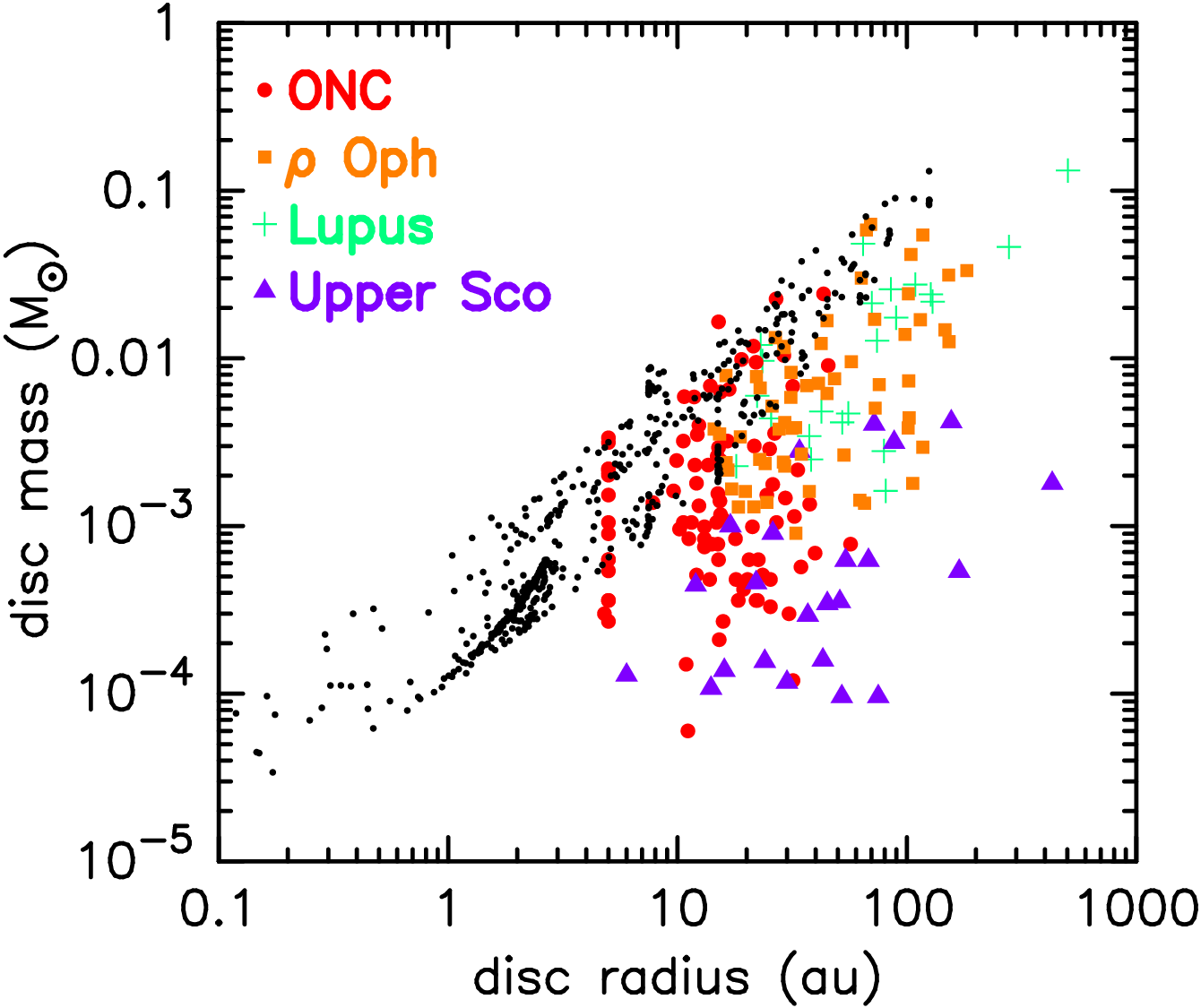}}}
\hspace*{0.3cm}\subfigure[Disc fractions]{\label{mod_dens_alpha_e-4-i}\rotatebox{270}{\includegraphics[scale=0.23]{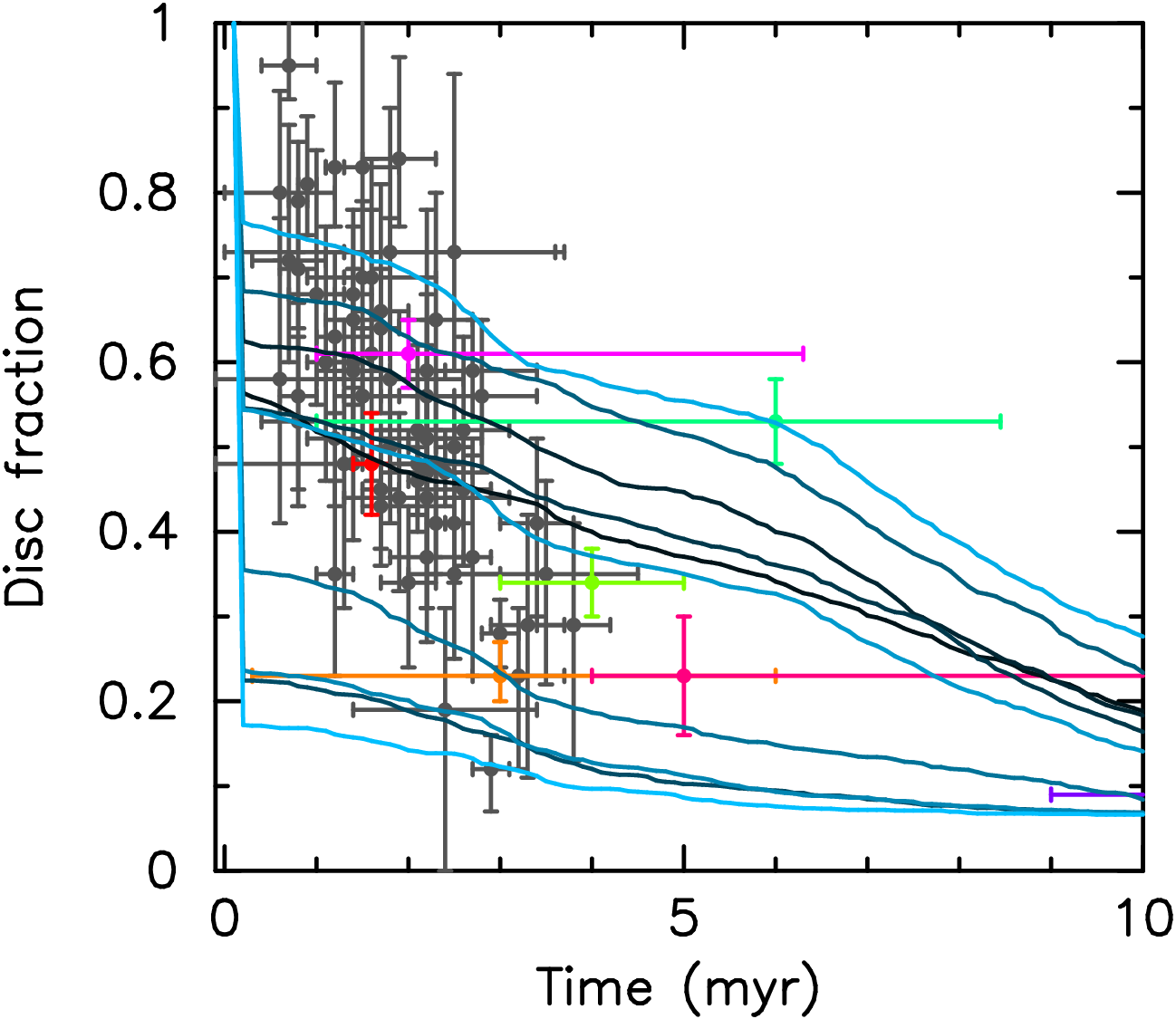}}}
\caption[bf]{Results for simulations in which the initial disc radii are all 100\,au and the discs are allowed to evolve inwards due to external photoevaporation, and outwards due to viscous spreading, but where the viscosity parameter is now $\alpha = 1\times10^{-4}$, rather than $\alpha = 1\times10^{-2}$.  The star-forming regions have moderate initial stellar density ($\tilde{\rho} \sim 100$\,M$_\odot$\,pc$^{-3}$). In panels (a)--(c) we show the cumulative distributions of disc radii at different ages, and in panels (d)--(f) we show the cumulative distributions of disc masses. The blue lines of varying hues show the results from the ten individual $N$-body simulations. The mint green dashed line represents the respective gas distribution in the Lupus star-forming region, whereas solid lines represent the dust distributions. Dotted lines indicate an alternative dust radius for $\rho$~Oph (orange line) and Lupus (mint green line); see Table~\ref{obs_info} for further details. In panels (g) and (h) we show the disc mass versus disc radius for four observed star-forming regions at 1 and 5\,Myr, and data from one post-processed $N$-body simulation are shown by the black points. We plot the disc fractions (defined as when the discs have non-zero mass) as a function of time in our $N$-body simulations in panel (i), with observational data taken from \citet{Richert18} and \citet{Ribas15}.}
\label{mod_dens_alpha_e-4}
  \end{center}
\end{figure*}

We show the evolution of the disc radii, disc masses and disc fractions in Fig.~\ref{mod_dens_alpha_e-4}. The effect of reducing $\alpha$ is that there is very little viscous spreading, and inward evolution of the disc due to photoevaporation dominates instead. As a result, the disc radii and disc mass distributions, and disc fractions, are very similar to our simulations where solely inward evolution of the disc due to photoevaporation occurs (compare Fig.~\ref{mod_dens_alpha_e-4} with Fig.~\ref{mod_dens_100au_no_viscous} -- with very little difference between the two sets of plots).

\label{lastpage}

\end{document}